%% file: ms.tex
\documentclass[%
 reprint,
 amsmath,amssymb,
 aps,
 showkeys,
]{revtex4-2}

\usepackage{graphicx}
\usepackage{dcolumn}
\usepackage{bm}

\usepackage[utf8]{inputenc}
\usepackage[T1]{fontenc}
\usepackage{fancyhdr}
\usepackage{lastpage}
\usepackage{amsmath}
\usepackage{amssymb}
\usepackage{graphicx}
\usepackage{subcaption}
\usepackage{soul}
\usepackage{comment}
\usepackage{float}
\usepackage{color}
\usepackage{listings}
\usepackage{amsfonts}
\usepackage{multirow}
\usepackage{hyperref}
\usepackage{ragged2e}
\usepackage{url}
\usepackage{xurl}
\usepackage{breakurl}
\DeclareCaptionJustification{justified}{\justifying}
\begin{document}

\preprint{APS/123-QED}

\title{Universality in Kinetic Models of Circadian Rhythms in \textit{Arabidopsis thaliana}}

\author{Yian Xu}
 \altaffiliation[]{Trinity University, Physics \& Astronomy, San Antonio, Texas, 78212, United States}
 
\author{Masoud Asadi-Zeydabadi}
\altaffiliation{University of Colorado Denver, Physics, Denver, Colorado, 80203, United States}

\author{Randall Tagg}
\altaffiliation{University of Colorado Denver, Physics, Denver, Colorado, 80203, United States}

\author{Orrin Shindell}
\email{oshindel@trinity.edu}
\altaffiliation{Trinity University, Physics \& Astronomy, San Antonio, Texas, 78212, United States}

\date{\today}

\begin{abstract}
Biological evolution has endowed the plant \textit{Arabidopsis thaliana} with genetically regulated circadian rhythms. 
A number of authors have published kinetic models for these oscillating chemical reactions based on a network of interacting genes.  
To investigate the hypothesis that the \textit{Arabidopsis} circadian dynamical system is poised near a Hopf bifurcation like some other biological oscillators, we varied the kinetic parameters in the models and searched for bifurcations. 
Finding that each model does exhibit a supercritical Hopf bifurcation, we performed a weakly nonlinear analysis near the bifurcation points to derive the Stuart-Landau amplitude equation. 
To illustrate a common dynamical structure, we scaled the numerical solutions to the models with the asymptotic solutions to the Stuart-Landau equation to collapse the circadian oscillations onto two universal curves -- one for amplitude, and one for frequency. 
However, some models are close to bifurcation while others are far, some models are post-bifurcation while others are pre-bifurcation, and kinetic parameters that lead to a bifurcation in some models do not lead to a bifurcation in others. 
Future kinetic modeling can make use of our analysis to ensure models are consistent with each other and with the dynamics of the \textit{Arabidopsis} circadian rhythm.
\end{abstract}


\maketitle


\section{Introduction}
Adapting to the 24-hour light-dark cycle caused by the rotation of the Earth, plants have evolved endogenous circadian rhythms that control many of their biological functions \citep{millar2016intracellular}. 
Circadian rhythms are genetically regulated chemical reactions inside cells that cause chemical concentrations to rise and fall with daily periodicity. 
Over the past fifteen years, eleven papers have proposed chemical kinetic models to govern the circadian oscillations in the laboratory plant \textit{Arabidopsis thaliana} \citep{locke2005modelling,locke2005extension,locke2006experimental,zeilinger2006novel,pokhilko2010data,pokhilko2012clock,pokhilko2013modelling,fogelmark2014rethinking,ohara2015extended,foo2016kernel,de2016compact}.
These sets of differential equations specify gene-interactions that were deduced through genetic experiments and include chemical reaction rate constants that were estimated by fits to experimental time series data \citep{bujdoso2013mathematical,chew2014mathematical,johansson2019move}. 

As knowledge about the genetic regulation of circadian rhythms has increased, the models have become larger and more complicated: the original model of a two-gene feedback loop has seven differential equations and 28 parameters \citep{locke2005modelling}, while the largest model of 12 genes has 35 differential equations with 122 parameters \citep{fogelmark2014rethinking}. 
Recent efforts have aimed to reduce the mathematical complexity of these models while retaining their dynamical features \citep{foo2016kernel,de2016compact,tokuda2019reducing,foo2020simplified}. 

Another body of literature studying circadian rhythms in \textit{Arabidopsis} has focused on the spatiotemporal patterns formed when the genetic expression of individual cells are coupled together in the tissues of a live plant \citep{fukuda2007synchronization,wenden2012spontaneous,fukuda2013controlling,endo2014tissue,takahashi2015hierarchical,endo2016tissue,gould2018coordination}. 
Some studies employed phenomenological models like the Stuart-Landau amplitude equation \citep{fukuda2007synchronization} and the related Kuramoto coupled phase oscillator model  \citep{kuramoto2003chemical,takahashi2015hierarchical,gould2018coordination}. 
These coarse-grained descriptions are useful because they reduce the complicated dynamics of interacting gene networks with many rate constants to simple dynamical forms that contain only one or a few parameters. 
The parameters can then be fit to experimental results, a process recently proposed as a tool in agricultural engineering projects \citep{anpo2018plant}.

In the present work, we employ a weakly nonlinear analysis method, the \textit{Reductive Perturbation Method} (RPM) of \citet{kuramoto2003chemical}, to cast the \textit{Arabidopsis} models into a two-dimensional form that is universally valid in systems poised near a Hopf bifurction.
The success of this approach is based on the fact that the published models are situated near supercritical Hopf bifurcation points in parameter space, a fact that may have biological significance: a nonlinear oscillator tuned near a Hopf bifurcation exhibits a resonance response when it is forced near its natural frequency \citep{mora2011biological,munoz2018colloquium}.
This mechanism confers sensitivity to some other biological oscillators, like the hair cells of the cochlea in the ears of humans  \citep{eguiluz2000essential,hudspeth2010critique} and frogs \citep{ospeck2001evidence}.

Near a Hopf bifurcation, the nonlinear equations governing the chemical oscillations may be linearized about a fixed point to give an approximate two-dimensional oscillating solution with a complex amplitude. 
The complex amplitude is governed by the Stuart-Landau equation whose parameter values are derived from a higher order expansion of the nonlinear system, and are therefore functions of the rate constants without free parameters \citep{hassard1981theory,guckenheimer1983local,kuramoto2003chemical}. 

We perform the RPM calculation on the published circadian rhythms models for two reasons. 
The first is to show how to calculate the parameters appearing in the Stuart-Landau equation directly from the kinetic rate constants in the chemical kinetic models. 
This calculation suggests experimental routes to drive bifurcations in circadian rhythms by manipulating kinetic parameters, a useful method in pattern formation experiments and plant engineering projects. 
The second reason is to quantify the essential dynamical features that models should aim to capture in addition to the oscillation period: the eigenvalues of the linearized system indicate whether the model is post-bifurcation or pre-bifurcation; the dependence of the eigenvalues on the kinetic rate constants show which rate constants lead to bifurcation; and the error between the perturbation solution and numerical solution to the full kinetic equations provides a measure of the  proximity of the model to Hopf bifurcation.
These dynamical quantities should be consistent between models and with the circadian rhythms of the organism. 

Our paper is organized into two sections. 
In the Methods and Results section we summarize the \textit{Reductive Perturbation Method} and show how to apply it to one of the circadian rhythms models. 
Then we illustrate the universal structure underlying the kinetic models. 
Finally, in the Discussion we point out inconsistencies between the kinetic models and suggest experimental and modeling approaches that could give additional information about the chemical dynamics of the \textit{Arabidopsis} circadian rhythm.

\section{Methods and Results}
\subsection{Hopf Bifurcations in the Circadian Rhythms Models}
The \textit{Arabidopsis} circadian rhythms models we considered are sets of coupled, nonlinear, first-order, ordinary differential equations. 
All of these models explicitly incorporate time dependence as a 24-hour periodic function. 
To investigate the dynamics of the endogenous chemical oscillations, we chose to make the differential equations autonomous by assuming perpetual darkness \citep{tokuda2019reducing} or perpetual illumination \citep{fukuda2013controlling}.
The autonomous equations may be expressed in the general form,
\begin{equation}
\label{eq.GeneralODEs}
    \frac{d\textbf{x}}{dt} = \textbf{f}(\textbf{x};\mu)
\end{equation}
where $\textbf{x}$ is an $n$-dimensional vector of chemical concentrations associated with the circadian reactions, $\textbf{f}$ is a nonlinear $n$-dimensional vector-valued function specifying the chemical reactions in a model, and $\mu$ is a function of one of the rate constants in the reaction equations.
Using the kinetic rate constants as bifurcation parameters differs from previous work that studied time delay models for circadian rhythms and used the time delay constant as a Hopf bifurcation parameter \citep{xiao2008genetic}.

Each of the circadian rhythms models we studied exhibits a supercritical Hopf bifurcation. Near $\mu = 0$, with $\mu$ suitably defined, the system of equations Eq. \eqref{eq.GeneralODEs} possesses a fixed point, i.e., a constant solution $\textbf{X}_0$ that satisfies 
\begin{equation}
    \textbf{f}(\textbf{X}_0;\mu) = \textbf{0}
\end{equation}
At the bifurcation point $\mu=0$, the fixed point switches linear stability: in the pre-bifurcation region $\mu<0$, $\textbf{X}_0$ is stable, and in the post-bifurcation region $\mu>0$, $\textbf{X}_0$ is unstable. 
When the system is post-bifurcation, in addition to an unstable fixed point, it possess a stable limit cycle, i.e., a linearly stable periodic solution $\textbf{X}(t)$ that satisfies
\begin{equation}
    \frac{d{\textbf{X}}}{dt} = \textbf{f}(\textbf{X};\mu), \quad \textbf{X}(t) = \textbf{X}(t + T)
\end{equation} 
for some period $T$.
The limit-cycle dynamics near a Hopf bifurcation belong to a dynamical universality class: for $\mu \gtrapprox 0$, the amplitude of the limit cycle oscillations scales in proportion to $\sqrt{\mu}$ and the frequency in proportion to $\mu$.
These properties enable an approximation to the amplitude and frequency of the limit cycle that is valid near the bifurcation point. 

For a concrete example from the literature, we consider here the original model (L2005a) presented in \citet{locke2005modelling}, which assumes a single negative feedback interaction between two core circadian genes:
\textit{LATE ELONGATED HYPO-COTYL} (\textit{LHY}), which is partially redundant with \textit{CIRCADIAN CLOCK ASSOCIATED 1}, and \textit{TIMING OF CAB EXPRESSION 1} (\textit{TOC1}) \citep{alabadi2001reciprocal}. 
A bifurcation diagram for L2005a is displayed in Fig. 1(f) in \citet{tokuda2019reducing}, which used the set of reaction rates given in the caption of Fig. 4 of \citet{locke2005modelling}. 
We use this same parameter set in the analysis presented in this subsection and the next. 
As the transcription rate of \textit{LHY} is varied in perpetual darkness, a supercritical Hopf bifurcation occurs.

\begin{figure*}
    \begin{subfigure}{0.5\textwidth}
        \includegraphics[width=\textwidth]{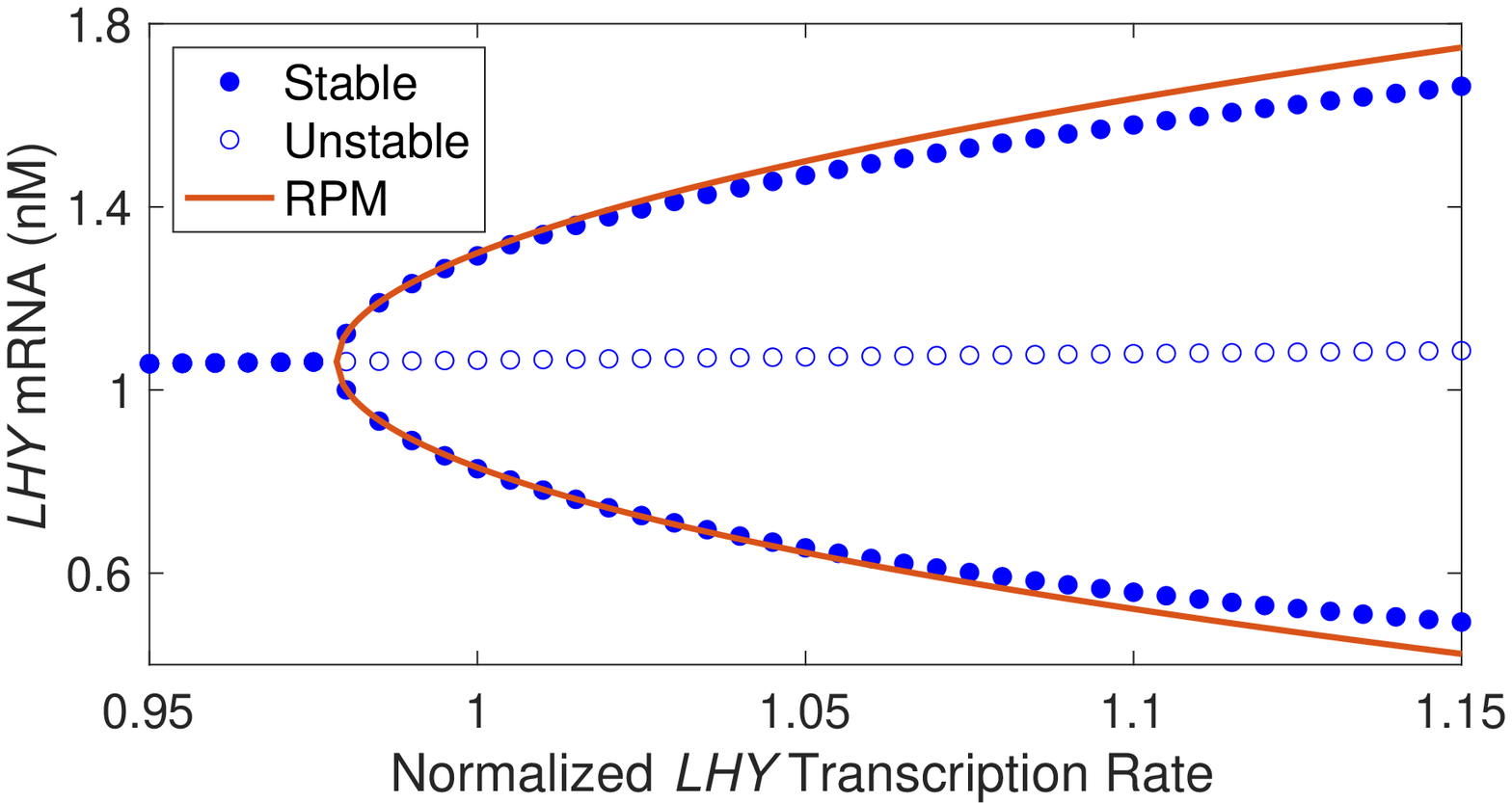}
        \caption{\centering}
        \label{BifDiagram:amp}
    \end{subfigure}\hfill%
    \begin{subfigure}{0.5\linewidth}
        \includegraphics[width=\textwidth]{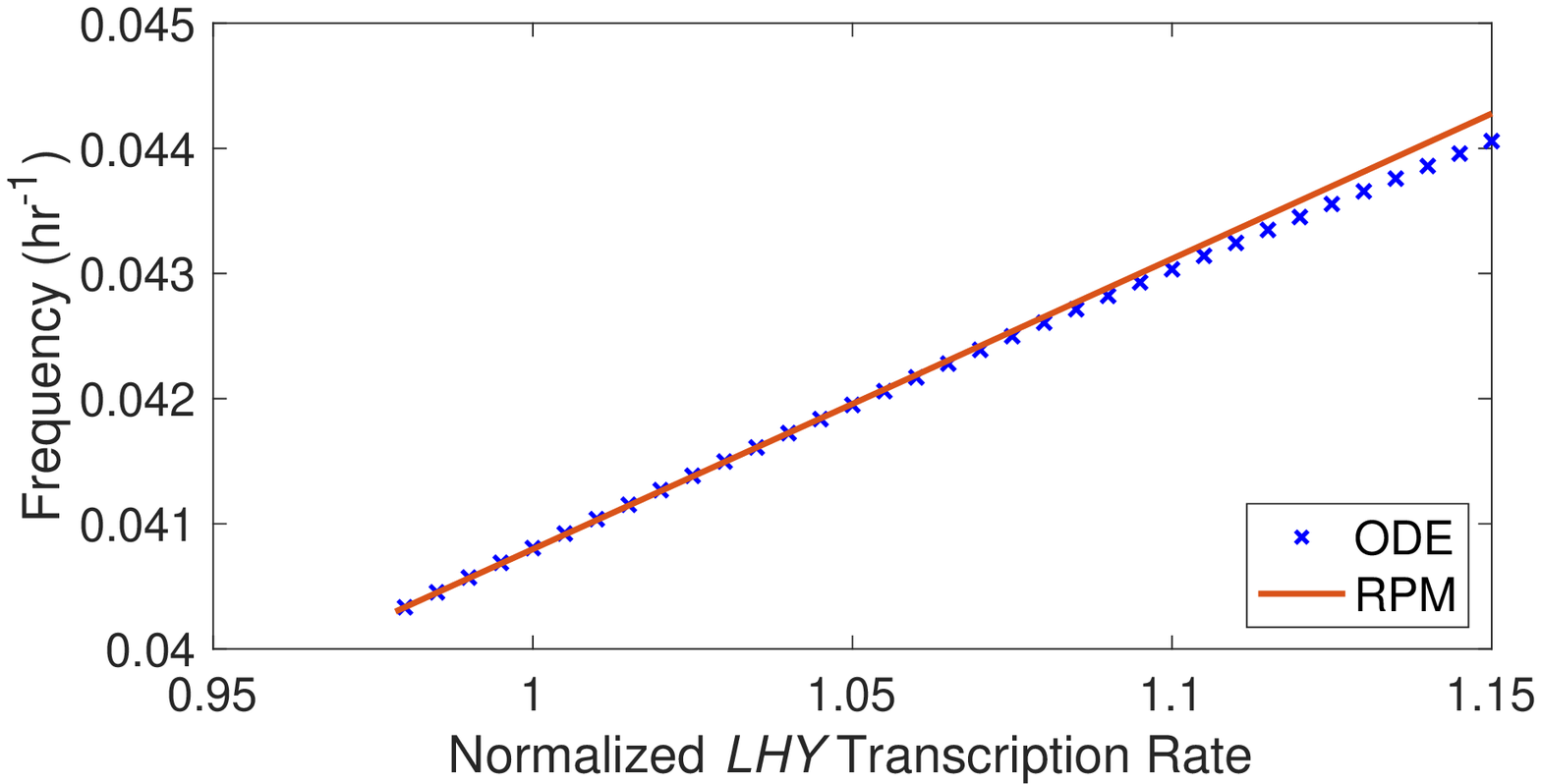}
        \caption{\centering}
        \label{BifDiagram:freq}
    \end{subfigure}%
    \caption{Bifurcation in (a) concentration and (b) frequency of circadian oscillations in a model (L2005a) poised near a supercritical Hopf bifurcation. 
    The transcription rate of \textit{LHY} is normalized such that the estimated biological value is unity; the bifurcation occurs at 0.9785. 
    (a) The central branch is the fixed point, which is stable to the left of 0.9785 (closed circles) and unstable to the right (open circles). 
    The upper and lower branches are the maximum and minimum values of \textit{LHY} mRNA limit cycle oscillations, calculated numerically (closed blue circles) and perturbatively (orange line). 
    (b) The limit cycle frequency is calculated numerically (blue hash marks) and perturbatively (orange line).}
    \label{BifurcationDiagram}
\end{figure*}

In Fig. \ref{BifDiagram:amp}, we compare the bifurcation diagram for the levels of \textit{LHY} mRNA calculated analytically with RPM (discussed in more detail in the next subsection) to the numerical solution obtained using the differential equation solver MATLAB ODE15s \citep{shampine1997matlab}. 
The \textit{LHY} transcription rate is normalized so that a value of unity corresponds to the fitted parameter value $ 7.5038\,\textrm{nM/h}$ \citep{tokuda2019reducing}. 
In Fig. \ref{BifDiagram:freq}, we compare the frequency of oscillations calculated analytically with the results obtained numerically. 
At the biological value of the \textit{LHY} transcription rate, the RPM calculation matches the amplitude and frequency determined by the numerical solution to the system of differential equations to within 0.86\% for the amplitude and 0.45\% for the frequency. 
In the Supplementary Materials, we show bifurcation diagrams for concentration and frequency of the other post-bifurcation models; as a measure of how close each model is poised to bifurcation we calculate the percent differences in amplitude and frequency between the numerical calculation and the RPM calculation at the fitted parameter values reported for each model.

\subsection{Reductive Perturbation Method and the Stuart-Landau Amplitude Equation}
In dynamical systems that undergo a Hopf bifurcation, an approximate two-dimensional periodic solution, valid near the bifurcation point, may be derived \citep{hassard1981theory,kuramoto2003chemical}. This stems from the fact that one pair of complex conjugate eigenvalues of the Jacobian matrix have positive real parts, while all the other eigenvalues have negative real parts.
The two eigenmodes associated with the former dominate the long-time behavior of the system, as the modes associated with the latter decay to zero.

\begin{figure*}
    \begin{subfigure}{0.62\textwidth}
        \includegraphics[width=\textwidth]{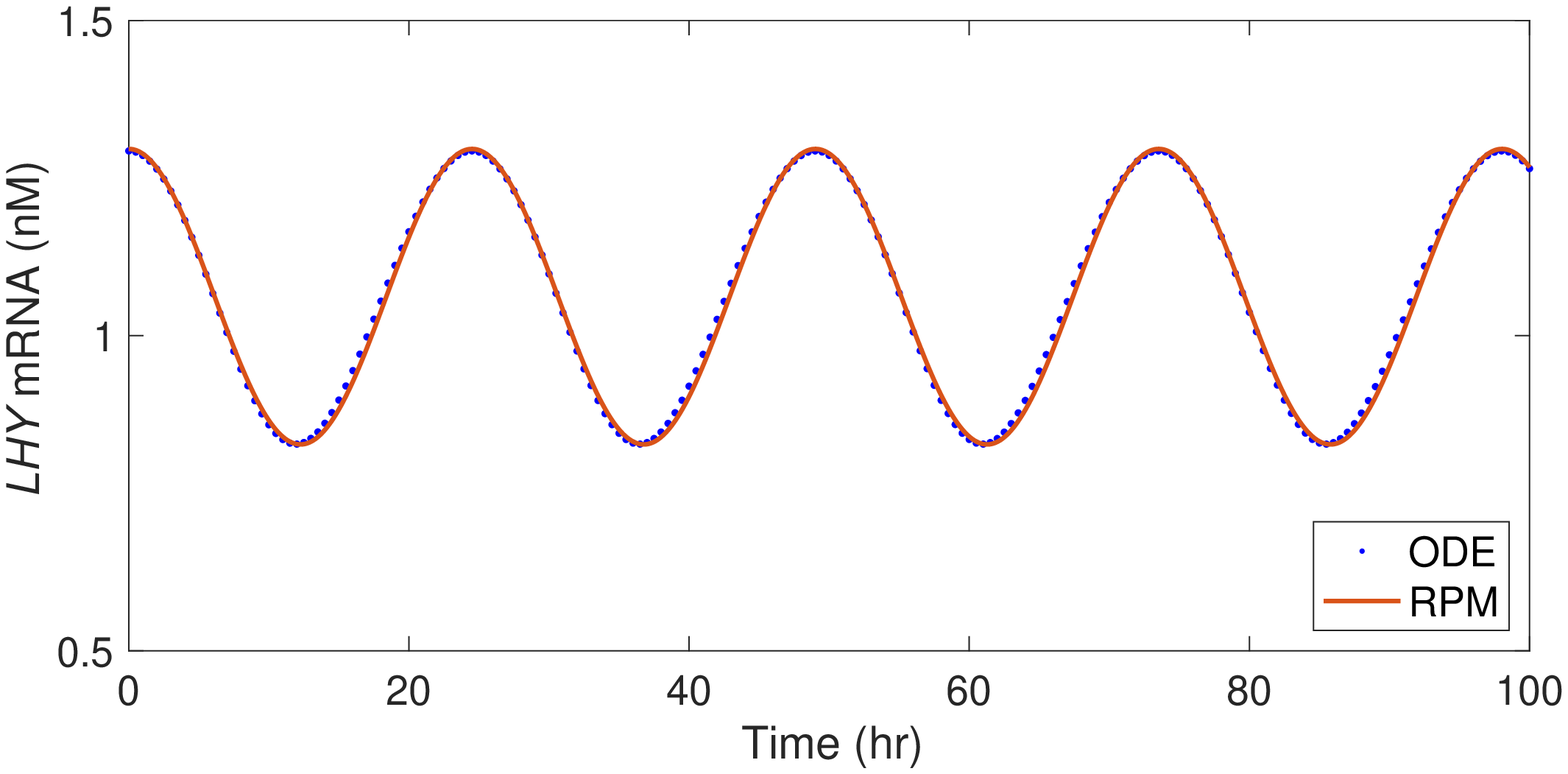}
        \caption{\centering}
        \label{TimeSeries}
    \end{subfigure}%
    \begin{subfigure}{0.38\textwidth}
        \includegraphics[width=\textwidth]{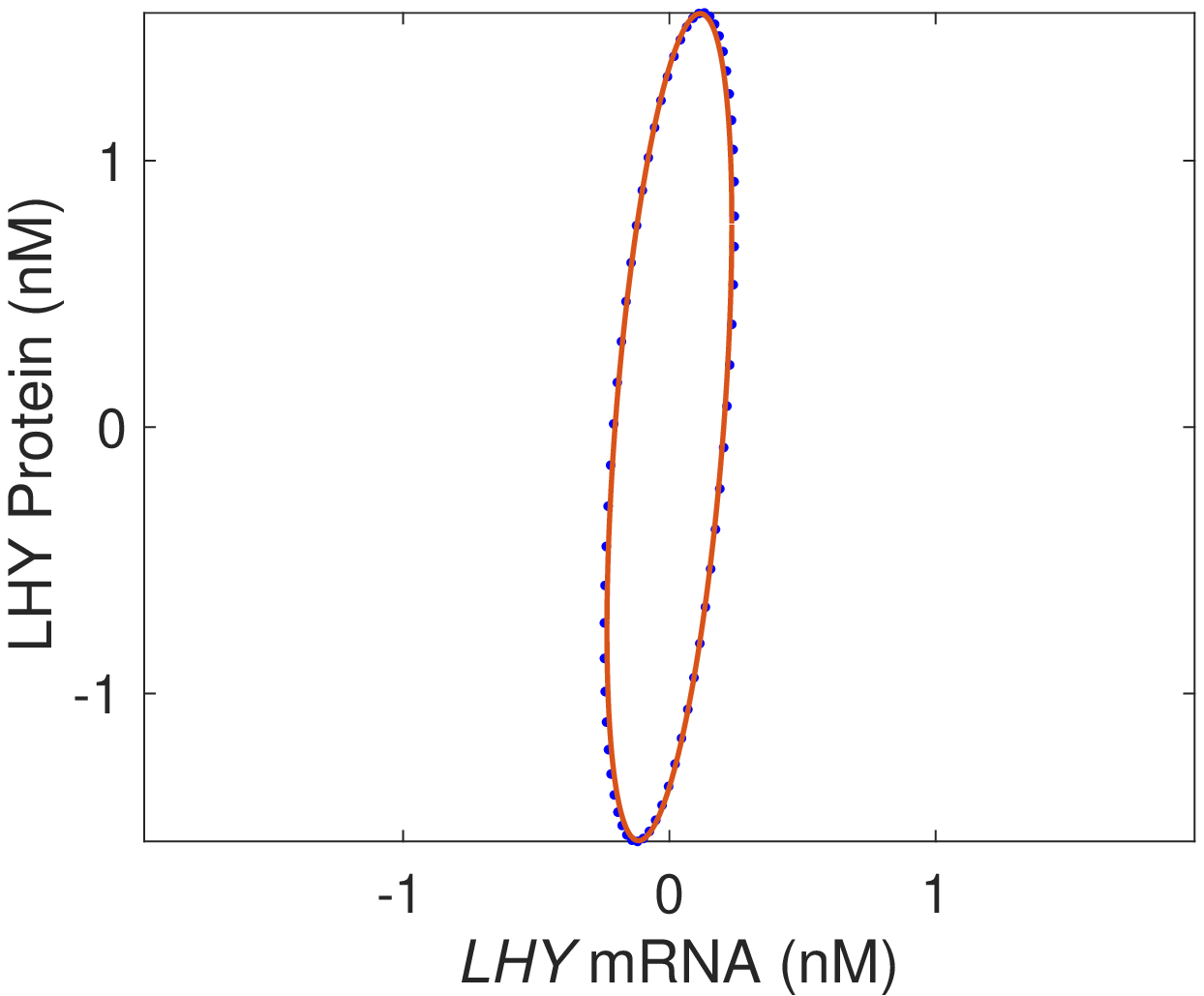}
        \caption{\centering}
        \label{fig:PhaseDiagrams}
    \end{subfigure}
    \caption{Time series (a) and phase space plot (b) are calculated numerically (ODE) and perturbatively (RPM) for a model (L2005a) poised near a supercritical Hopf bifurcation.}
    \label{Locke2005a:example}
\end{figure*}

The system of equations Eq. \eqref{eq.GeneralODEs} may be linearized about the fixed point $\textbf{X}_0$, to give
\begin{equation}
    \label{eq.LinearizedSystem}
    \begin{split}
        \frac{d\textbf{u}}{dt} & = \textbf{L}\textbf{u}, \quad \textbf{u} =  \textbf{x} - \textbf{X}_0, \quad L_{ij} = \left.\frac{\partial f_i}{\partial x_j}\right|_{(\textbf{X}_0,\mu)}
    \end{split}
\end{equation}
where $\textbf{u}$ is the deviation from the fixed point and $\textbf{L}$ is the Jacobian matrix. 
At the bifurcation point $\mu = 0$, the Jacobian matrix possesses $n$ eigenvalues with $n$ associated eigenvectors (the systems we studied had no repeated eigenvalues), with two of the eigenvalues purely imaginary. 
Thus, at bifurcation, 
\begin{equation}
    \label{eq.CriticalEigenvalueEQ}
    \textbf{L}_0\textbf{U} = i\omega_0 \textbf{U}
\end{equation}
where $\textbf{L}_0$ is the Jacobian matrix evaluated at $\mu=0$ and $\textbf{U}$ is the eigenvector corresponding to the eigenvalue $i \omega_0$, with the complex conjugate eigenvector $\overline{\textbf{U}}$ corresponding to the eigenvalue $-i\omega_0$. 
Near the bifurcation point in the post-bifurcation region, the linearized system has two eigenvalues with positive real parts that dominate the dynamics of system, as the remaining $n-2$ eigenvalues have strictly negative real parts. 
There is the approximate eigenvalue equation
\begin{equation}
    \left(\textbf{L}_0 + \mu \textbf{L}_1\right)\textbf{U} = \left(i\omega_0 + \mu\lambda_{1}\right) \textbf{U},\quad \lambda_{1}= \sigma_1 + i\omega_1,\quad \sigma_1 >0
\end{equation}
where $\textbf{L}_1$ is the first-order term of the Taylor expansion of $\textbf{L}$ about $\mu = 0$, and $\lambda_{1} $ is the first-order term in the Taylor series about $\mu=0$ of the eigenvalue of $\textbf{L}$ whose zeroth-order term is $i\omega_0$, with $\sigma_1$ the real part and $\omega_1$ the imaginary part. 
For $\mu \gtrapprox 0$, the solution $\textbf{x}(t)$ takes the approximate form
\begin{equation}
    \label{eq.RPMresult}
        \textbf{x} = \textbf{X}_0 + \sqrt{\mu}\left[W(t)\textbf{U}e^{i\omega_0t} + \overline{W}(t)\overline{\textbf{U}}e^{-i\omega_0t}\right]
\end{equation}
where $W$ is a complex amplitude, with $\overline{W}$ the complex conjugate, that evolves according to the Stuart-Landau equation
\begin{equation}
    \label{eq.Stuart-Landau}
            \frac{dW}{dt} = \mu\left(\lambda_{1} W - g\left|W\right|^2W\right),\quad g = g' + ig'',\quad g'>0
\end{equation}
The complex number $g$ is a function of the higher order expansion coefficients of Eq. \eqref{eq.GeneralODEs}. We refer to \citet{kuramoto2003chemical} for the details leading to Eqs. \eqref{eq.RPMresult} and \eqref{eq.Stuart-Landau}. Eq. \eqref{eq.Stuart-Landau} can be split into two equations by setting $W(t) = R(t)e^{i\Theta(t)}$, with $R(t)$ and $\Theta(t)$ both real, which lead to the asymptotic quantities
\begin{equation}
    \label{eq.NatualScales}
    \begin{split}
        R_s & \equiv \lim_{t\rightarrow\infty}R = \sqrt{\frac{\sigma_1}{g'}},\\
        \omega_s & \equiv \lim_{{t\rightarrow\infty}}\frac{d\Theta}{dt} = \left(\omega_1 - g''R_s^2\right)
    \end{split}
\end{equation}
The limit cycle in the post-bifurcation region is thus approximately given by
\begin{equation}
\label{eq.AsympLimitCycle}
\textbf{X} = \textbf{X}_0 + \sqrt{\mu}R_s\left[\textbf{U}e^{i\left(\omega_0 + \mu\omega_s\right)t} + \overline{\textbf{U}} e^{-i\left(\omega_0 + \mu\omega_s\right)t}\right]
\end{equation}
which describes an elliptical orbit in a two-dimensional subspace of $\mathbb{R}^n$. 
In Fig. \ref{TimeSeries}, we compare the time series given by the limit cycle prediction of Eq. \eqref{eq.AsympLimitCycle} to the numerically computed solution for the mRNA concentration of the central gene \textit{LHY}.  

The phase differences between different chemical species in the circadian system, which are important to biological function \citep{locke2005modelling}, can be estimated directly from the eigenvector $\textbf{U}$.
In Fig. \ref{fig:PhaseDiagrams}, we plot phase space diagrams for LHY protein and \textit{LHY} mRNA oscillations. 
The phase difference between the pair of chemical species deviates from those obtained from numerical solutions; as a fraction of $2\pi$ the absolute value of the difference in phase difference is 0.003.
In the Supplementary Materials, we show time series and phase space plots for the other post-bifurcation models.

\subsection{Scaling and Data Collapse}
The parameters appearing in the Stuart-Landau equation, which result from the values of the kinetic rate parameters in the circadian reactions, set natural scales for the chemical oscillations in the circadian rhythms models. 
Exactly at the bifurcation point, the system exhibits zero amplitude oscillations about the fixed point $\textbf{X}_0$ with frequency $\omega_0$. 
As the system deviates from the bifurcation point with increasing $\mu$, the limit cycle amplitudes of the oscillating chemicals increase in proportion $\sqrt{\mu}R_s$ while their frequency changes from $\omega_0$ by $\mu \omega_s$. 
There is an arbitrariness in the calculation of these parameters, however, as the eigenvectors in Eq. \eqref{eq.CriticalEigenvalueEQ} are unique only up to a multiplicative constant. 
To set an exact scale, we chose to normalize the eigenvectors by the component with the largest modulus, which sets $\sqrt{\mu}2R_s$ as an upper bound for the chemical oscillation amplitudes. 
With this definition, Eq. \eqref{eq.AsympLimitCycle} implies that the amplitude $A$ and frequency $\omega$ for the largest amplitude chemical species in the limit cycle regime can be collapsed onto parameter-free curves near $\mu = 0$:
\begin{equation}
\label{eq.Scaling}
    \begin{split}
       \frac{A}{2R_s} & = \sqrt{\mu},\\
        \frac{\omega - \omega_0}{\omega_s} & = \mu
    \end{split}
\end{equation}

\begin{figure*}
    \includegraphics[width=\textwidth]{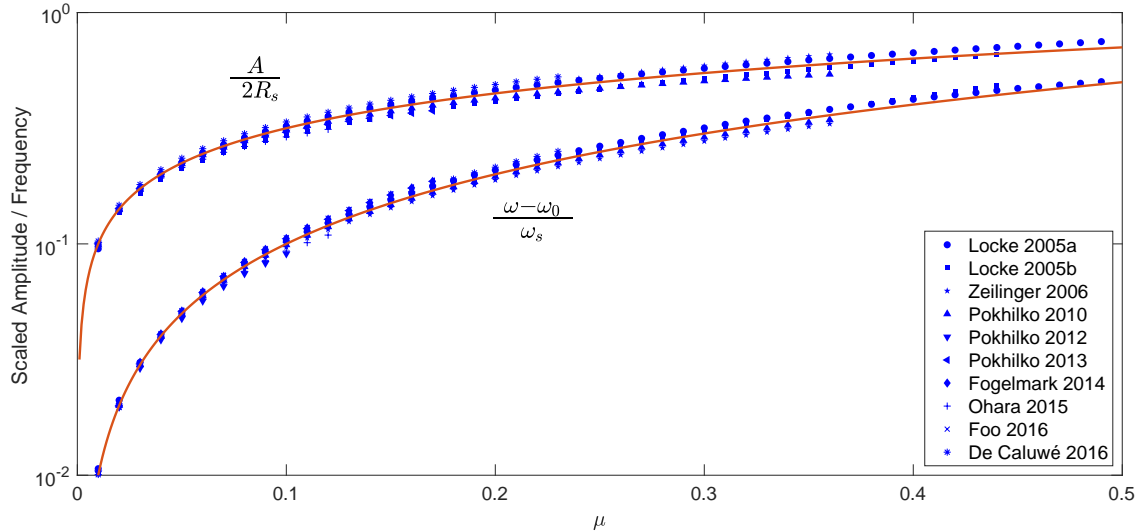}
    \caption{Amplitude (upper) and frequency (lower) of limit cycle oscillations for ten models of \textit{Arabidopsis} circadian rhythms are collapsed onto universal functions of the bifurcation parameter $\mu$. 
    The limit cycle amplitude and frequency calculated numerically with ODE solvers are scaled with the asymptotic solutions to the Stuart-Landau equation. 
    Data for each model is shown up to the value of $\mu$ that they diverge from one of the universal curves by 10\%.}
    \label{Results:LL_plots}
\end{figure*}

In Fig. \ref{Results:LL_plots}, we show the amplitude $A$ and frequency $\omega$ of circadian oscillations in perpetual illumination for ten of the eleven models determined using MATLAB ODE solvers \citep{shampine1997matlab} scaled by ${R}_s$ and ${\omega}_s$ into the forms of Eqs. \eqref{eq.Scaling}. 
We neglected the model given in \citet{locke2006experimental} because the reported parameter set gave two pairs of complex eigenvalues with positive real parts. 
In the Supplementary Materials, we collapse data from the same ten models in perpetual darkness, and show the collapsed data for each model separately.

For each system, we had to decide which rate parameter to use to define $\mu$, as each system contains more than one parameter whose variation leads to a Hopf bifurcation. 
To choose which was appropriate, we referenced research that suggests the circadian rhythms in \textit{Arabidopsis} are sensitive to the degradation rates of mRNA \citep{yakir2007circadian}. 
Thus we confined ourselves to using chemical degradation rates as bifurcation parameters. 

Of the models we analyzed, the majority employ Michaelis-Menten kinetics \citep{murray2013mathematical} to govern the degradation rates of mRNA and protein. 
A few models use constant degradation rates. 
We varied the maximum degradation rates in the Michaelis-Menten kinetics or the constant degradation rates one at a time while keeping all other parameters constant. 
Whenever the variation of one of these parameters $m$ led to a Hopf bifurcation, we defined the parameter value at bifurcation as $m_c$.  
Motivated by the original work of Stuart \citep{stuart1958non}, we further defined a dimensionless bifurcation parameter,
\begin{equation}
    \mu_m = \frac{m_c - m}{m_c}
\end{equation}
For each model, we investigated the amplitude and frequency scaling (Eqs. \eqref{eq.Scaling}) for each $\mu_m$. 
As each $\mu_m$ increased from zero, the scaled quantities deviated from the universal curves.
For the purpose of illustration, we present results in Fig. \eqref{Results:LL_plots} using the bifurcation parameter $\mu$ defined to be the $\mu_m$ that was largest when the one of the two scaled quantities deviated from the predicted value by ten percent. 
We note that we did not find a single degradation rate that led to a bifurcation in all the models.

\section{Discussion}
In this work, we demonstrated that the published dynamical systems models for circadian rhythms in \textit{Arabidopsis thaliana} possess supercritical Hopf bifurcations. 
We further employed the \textit{Reductive Perturbation Method} (RPM) of \citet{kuramoto2003chemical} to derive  an approximate two-dimensional form for the chemical oscillations of the models in the post-bifurcation region of parameter space with a complex amplitude governed by the Stuart-Landau equation. 
By scaling the amplitude and frequency of the numerical limit cycle solutions with the asymptotic solutions to the Stuart-Landau equation, we showed that all the models possess a common phase space near the Hopf bifurcation. 

There are nevertheless significant differences between the models that warrant discussion. 
Each of the models we investigated contain many rate parameters that were fit to experimental data. 
The results of these fits vary. 
Some of the systems we studied are situated on the pre-bifurcation side of a Hopf bifurcation, and others on the post-bifurcation side.
Moreover, rate constants that lead to a Hopf bifurcation when varied in some models do not lead to a Hopf bifurcation when varied in other models. 
We can suggest a possible experimental path to address whether \textit{Arabidopsis} circadian oscillations are post-bifurcation or pre-bifurcation; namely, by driving a plant with a periodic light signal at the free-running frequency over a range of weak intensities. 
The oscillation amplitude of a pre-bifurcation system should increase linearly with the light intensity while the oscillation amplitude of a post-bifurcation system should not change. 
In the Supplementary Materials, we show numerical amplitude response curves for both sinusoidal and square-wave forcing.

A second way the models differ is in how close each is to the Hopf bifurcation with their given parameter values, which we measured by comparing the numerical solution to the RPM approximation.
The limit cycle amplitude in the model from \citet{locke2005modelling} presented in the Methods and Results section matched the RPM calculation to within 0.8\% while the limit cycle amplitude from \citet{pokhilko2010data} shown in Supplementary Materials differed from the RPM calculation by 45.02\%. 
Despite all the differences in the models, including that the genetic network architecture and numerical values of rate constants differed significantly, we showed the dynamical structure of all the models near the bifurcation point is the same. 
This should be useful in future efforts to fit experimental data close to a supercritical Hopf bifurcation. 
An experiment that measures a bifurcation curve by varying an external parameter could yield the eigenvalues, eigenvectors, and parameters appearing in Eq. (\ref{eq.AsympLimitCycle}). Those quantities could then be used to verify the accuracy of the parameter choices in a particular kinetic model.

Our method of analysis may find application in agricultural engineering projects that use coarse-grained models as valid approximations to weakly nonlinear dynamics to describe an organism-level response to an external stimulus. 
By using RPM, we are able to arrive at a coarse-grained form, i.e., the Stuart-Landau equation, as a function of the underlying details of the circadian gene network. 
The driven Stuart-Landau equation predicts a resonance response of the oscillations that is most pronounced at the Hopf bifurcation point \citep{eguiluz2000essential}. 
This resonance behavior near bifurcation suggests a possible engineering program where the chemical kinetics of the organism is tuned, perhaps by varying temperature, to be near the bifurcation point and then driven with a small amplitude (i.e., a low power) sinusoidal light source. 
 
Finally, in recent experiments the circadian oscillations in the cyanobacterium \textit{Synechoccus} \textit{elongatus} were shown to exhibit a supercritical Hopf bifurcation with temperature as the bifurcation parameter \citep{murayama2017low}. 
It has long been noted that commonalities exist in the circadian rhythms of disparate organisms \citep{pittendrigh1960circadian}. 
We tentatively suggest that close proximity to a supercritical Hopf bifurcation may be an additional property favored in the evolutionary development of circadian rhythms. 
If this suggestion is true, then as we have demonstrated by using the \textit{Reductive Perturbation Method}, circadian rhythms can be cast into a generic mathematical form given by the Stuart-Landau equation.

\begin{acknowledgements}
The authors would like to thank Harry L. Swinney for a critical reading of the manuscript and helpful conversations. 
This research was funded by Trinity University with a Murchison Fellowship to Y.X. and start-up funds to O.S.
\end{acknowledgements}

\hypersetup{breaklinks=true}
\section*{Code availability}
The MATLAB code to replicate the calculations in this work is available on GitHub at \url{https://github.com/oshindel/Reductive-Perturbation-Method-A.-thaliana}.

\input{ms.bbl}

\end{document}


\preprint{APS/123-QED}

\title{Universality in Kinetic Models of Circadian Rhythms in \textit{Arabidopsis thaliana}}

\author{Yian Xu}
 \altaffiliation[]{Trinity University, Physics \& Astronomy, San Antonio, Texas, 78212, United States}
 
\author{Masoud Asadi-Zeydabadi}
\altaffiliation{University of Colorado Denver, Physics, Denver, Colorado, 80203, United States}

\author{Randall Tagg}
\altaffiliation{University of Colorado Denver, Physics, Denver, Colorado, 80203, United States}

\author{Orrin Shindell}
\email{oshindel@trinity.edu}
\altaffiliation{Trinity University, Physics \& Astronomy, San Antonio, Texas, 78212, United States}

\date{\today}

\maketitle

\section*{Supplementary Materials}

\subsection*{Supplementary Tables}
\vspace{0.3cm}
In Table \ref{tab:pre-post-models}, we show which models are pre-bifurcation or post-bifurcation with the sets of optimal parameter values reported in the original papers. \vspace{0.3cm}

\noindent In Tables \ref{tab.SLParametersLL} and \ref{tab.SLParametersDD} we show: 1) the kinetic parameter used as the Hopf bifurcation parameter (BP) to produce the data in Fig. 3 of the main text and Figs. \ref{7-panel:Locke2005a}-\ref{DD-cycle-models}, 2) the chemical species with the largest modulus used to scale the eigenvectors of the Jacobian matrix at the bifurcation point used to calculate the natural scales, 3) the optimal value of the BP given in the original paper, 4) the critical value of the BP where the Hopf bifurcation occurs, 5) the frequency of zero amplitude oscillations at the bifurcation point, 6) the value of the complex number $g$ in the Stuart-Landau equation, 7) the value of the first order Taylor expansion term, $\lambda_1$, of the eigenvalue of the Jacobian matrix near the bifurcation point. \vspace{0.3cm}

\noindent In Tables \ref{tab:Eigenvectors1_LL}-\ref{tab:Eigenvectors3_DD}, we show the several elements of the eigenvectors of the Jacobian matrix at the bifurcation point, normalized by the largest modulus entry, that correspond to LHY and TOC1 protein and \textit{LHY} and \textit{TOC1} mRNA concentrations.\vspace{0.3cm}

\noindent In Table \ref{tab:my_label}, we show which MATLAB ODE solver was used to find the numerical solution for each model and we include comments about our analysis particular to each model.\vspace{0.3cm}

\noindent \textbf{S1} List of Post-bifurcation and Pre-bifurcation models.\vspace{0.3cm}

\noindent \textbf{S17} Stuart-Landau parameter values of all models under perpetual illumination.\vspace{0.3cm}

\noindent \textbf{S18} Stuart-Landau parameter values of all models under perpetual darkness.\vspace{0.3cm}

\noindent \textbf{S19} Eigenvector entries of the mRNA and proteins of \textit{LHY/CCA1} and \textit{TOC1} genes of Locke et al. 2005a, Locke et al. 2005b, Zeilinger et al. 2006, Pokhilko et al. 2010 models under perpetual illumination.\vspace{0.3cm}

\noindent \textbf{S20} Eigenvector entries of the mRNA and proteins of \textit{LHY/CCA1} and \textit{TOC1} genes of Pokhilko et al. 2012, Pokhilko et al. 2013, Fogelmark et al. 2014, Ohara et al. 2015 models under perpetual illumination.\vspace{0.3cm}

\noindent \textbf{S21} Eigenvector entries of the mRNA and proteins of \textit{LHY/CCA1} and \textit{TOC1} genes of Foo et al. 2016 and De Caluw{\'e} et al. 2016 models under perpetual illumination.\vspace{0.3cm}

\noindent \textbf{S22} Eigenvector entries of the mRNA and proteins of \textit{LHY/CCA1} and \textit{TOC1} genes of Locke et al. 2005a, Locke et al. 2005b, Zeilinger et al. 2006, Pokhilko et al. 2010 models under perpetual darkness.\vspace{0.3cm}

\noindent \textbf{S23} Eigenvector entries of the mRNA and proteins of \textit{LHY/CCA1} and \textit{TOC1} genes of Pokhilko et al. 2012, Pokhilko et al. 2013, Fogelmark et al. 2014, Ohara et al. 2015 models under perpetual darkness.\vspace{0.3cm}

\noindent \textbf{S24} Eigenvector entries of the mRNA and proteins of \textit{LHY/CCA1} and \textit{TOC1} genes of Foo et al. 2016 and De Caluw{\'e} et al. 2016 models under perpetual darkness.\vspace{0.3cm}

\noindent \textbf{S25} Model idiosyncrasies.\vspace{0.3cm}

\subsection*{Supplementary Figures}
\begin{figure}[H]
    \centering
    \begin{subfigure}{0.5\textwidth}
      \includegraphics[width=\textwidth]{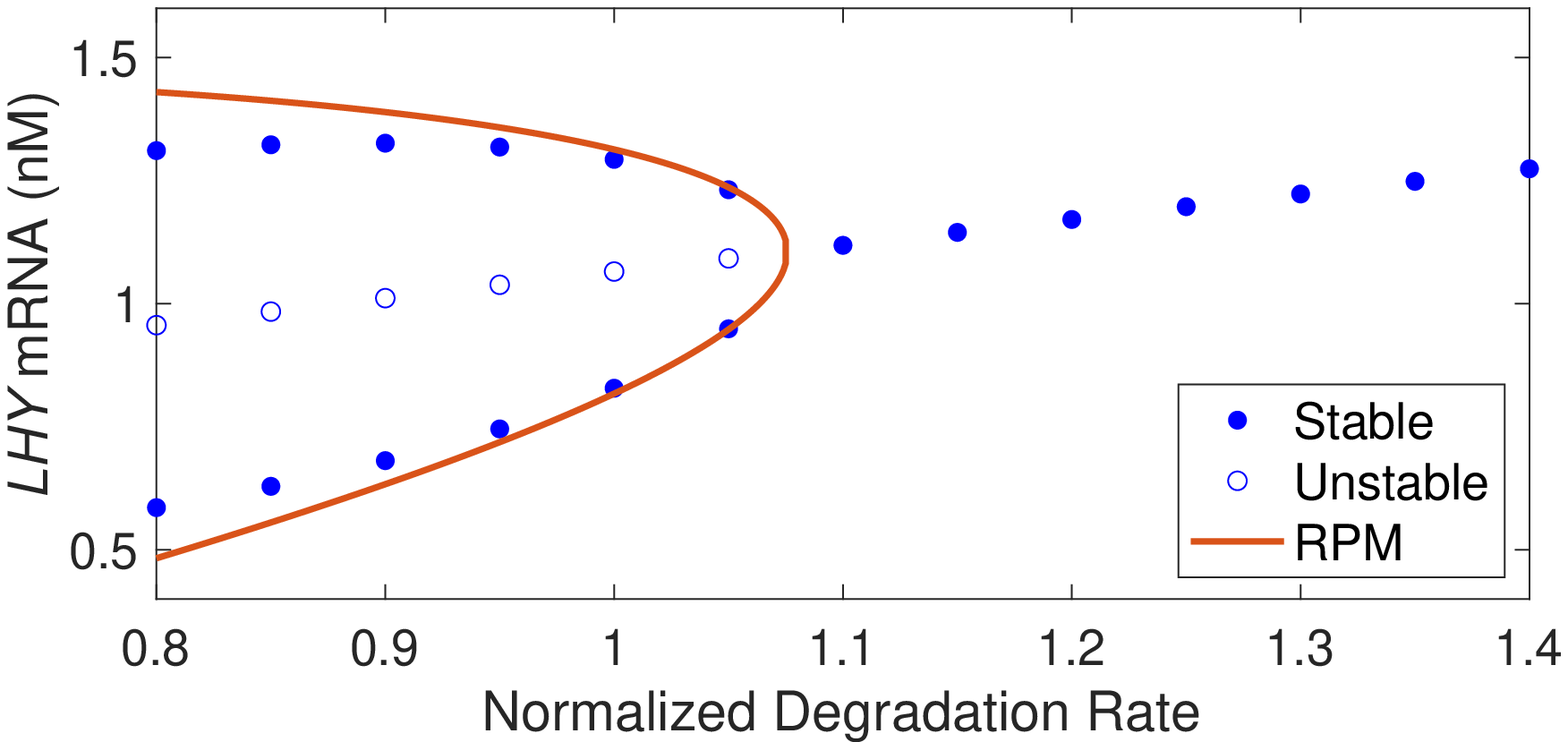}
      \caption{}
      \label{Locke2005a:bif-amp}
    \end{subfigure}%
    \begin{subfigure}{0.5\textwidth}
      \includegraphics[width=\textwidth]{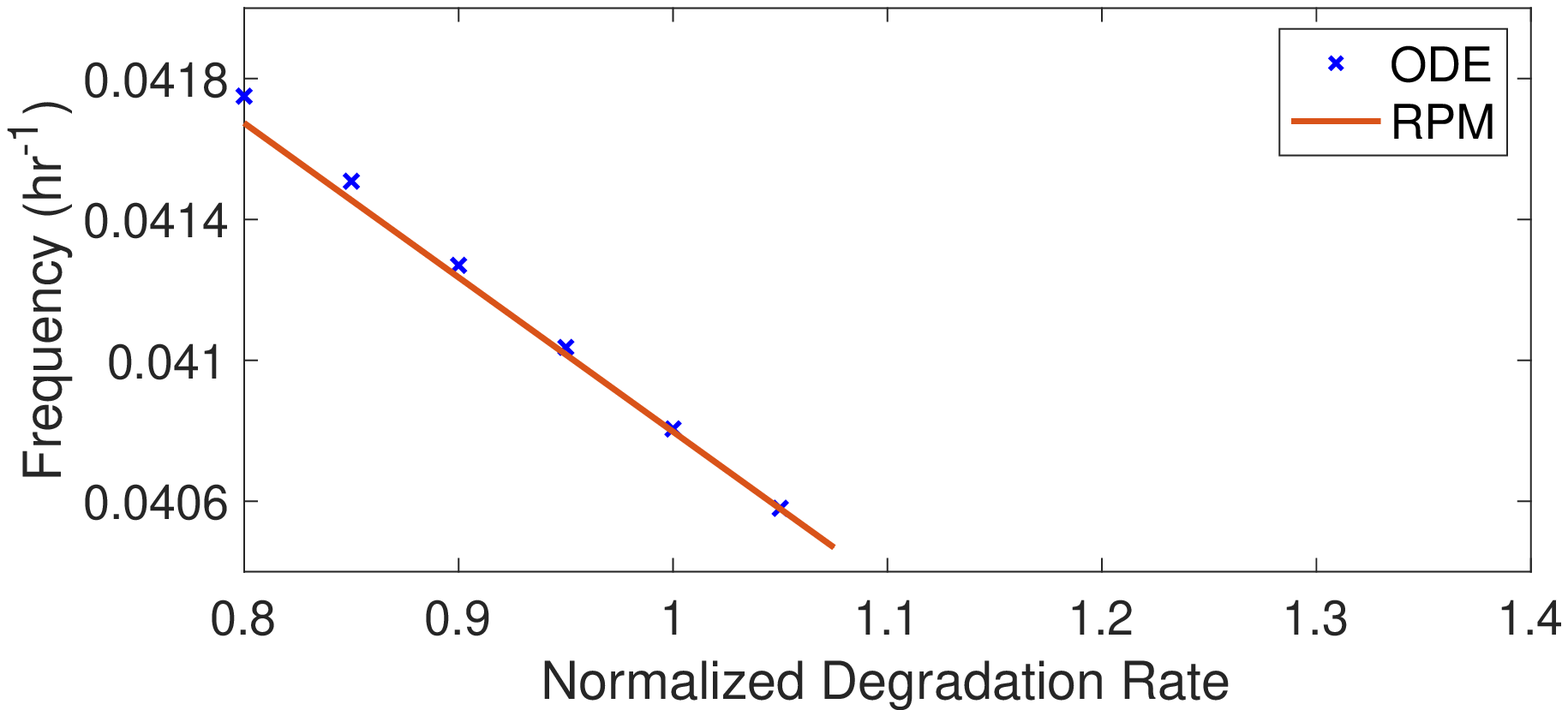}
      \caption{}
      \label{Locke2005a:bif-freq}
    \end{subfigure}
    \begin{subfigure}{0.5\textwidth}
      \includegraphics[width=\textwidth]{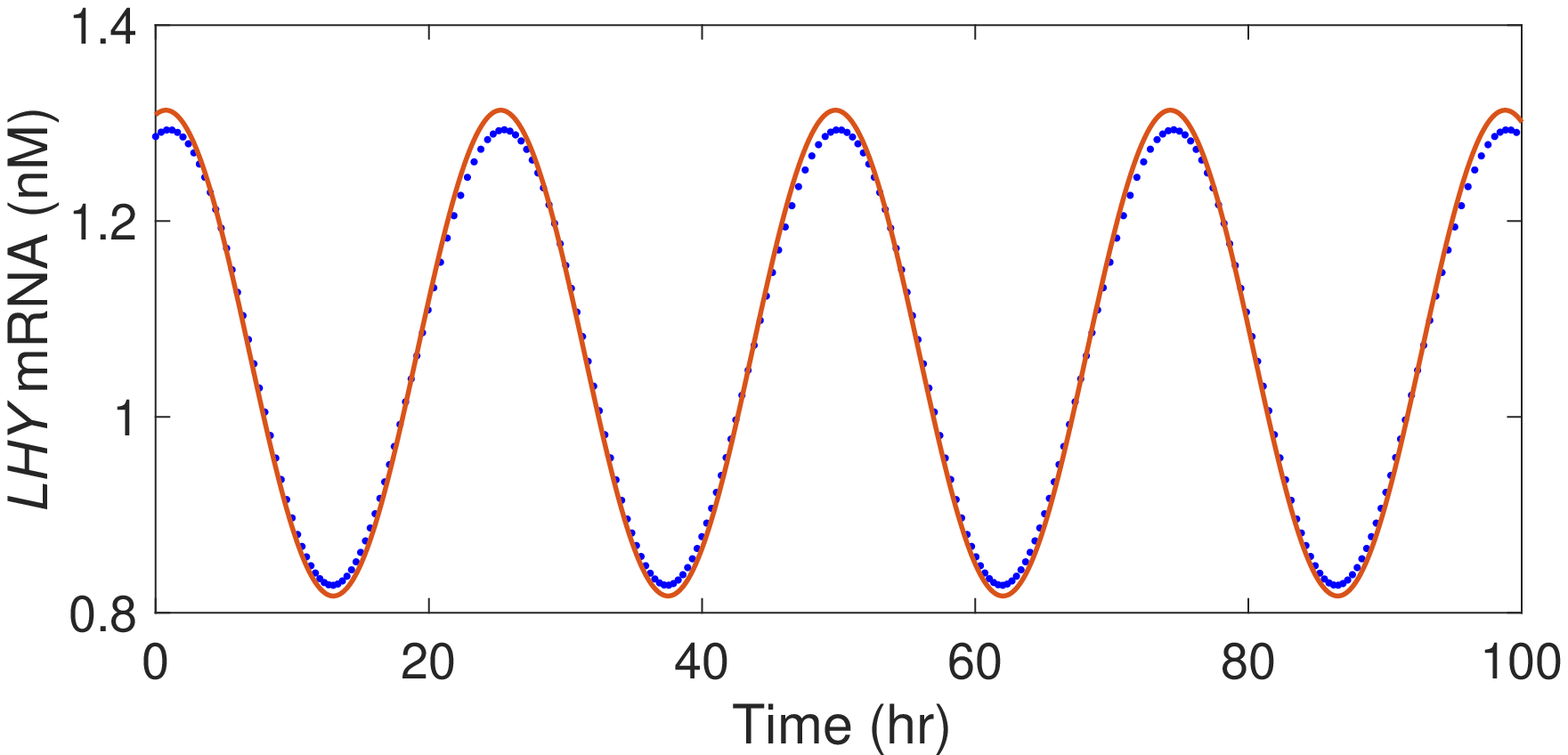}
      \caption{}
      \label{Locke2005a:series-LHY}
    \end{subfigure}%
    \begin{subfigure}{0.5\textwidth}
      \includegraphics[width=\textwidth]{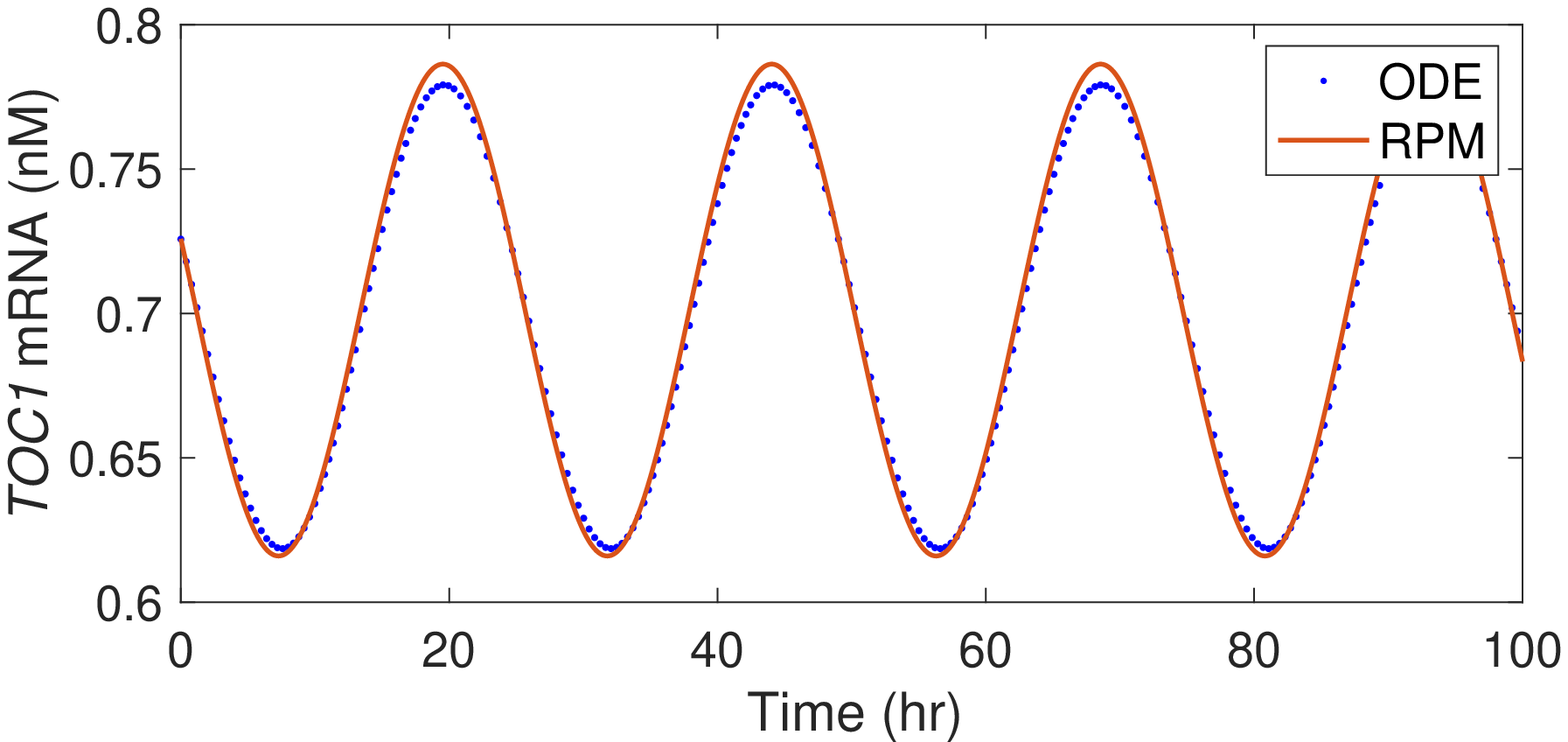}
      \caption{}
      \label{Locke2005a:series-TOC1}
    \end{subfigure}
    \begin{subfigure}{0.333\textwidth}
      \includegraphics[width=\textwidth]{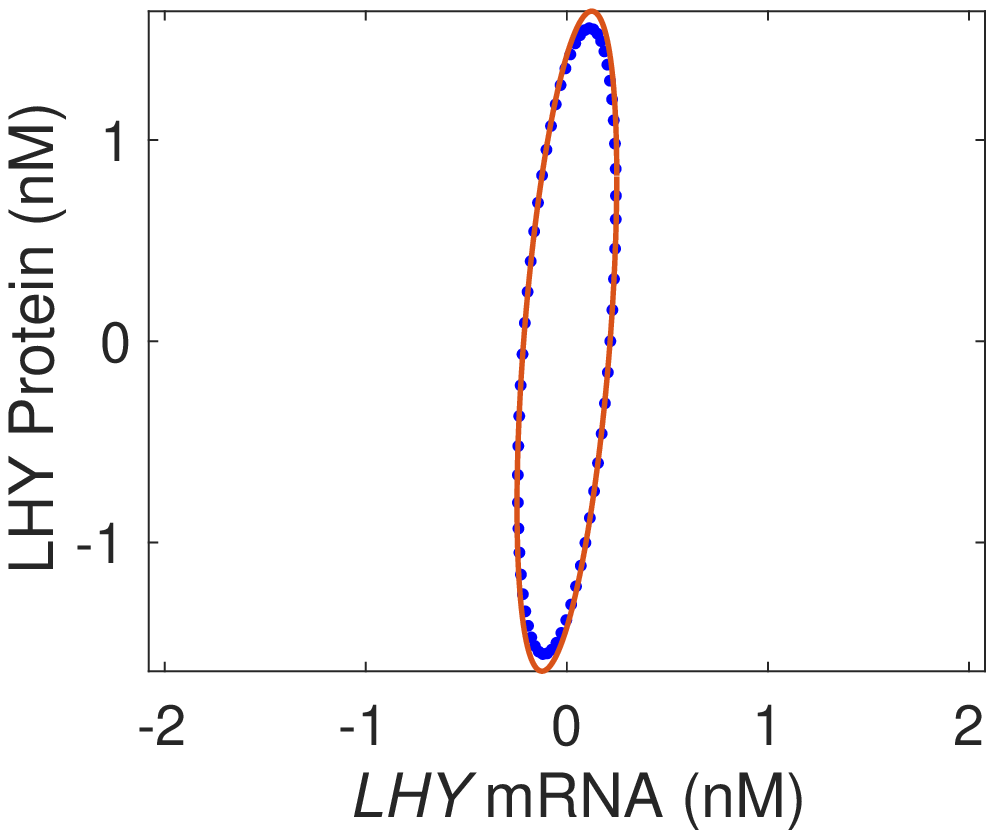}
      \caption{}
      \label{Locke2005a:phase-LHY}
    \end{subfigure}%
    \begin{subfigure}{0.333\textwidth}
      \includegraphics[width=\textwidth]{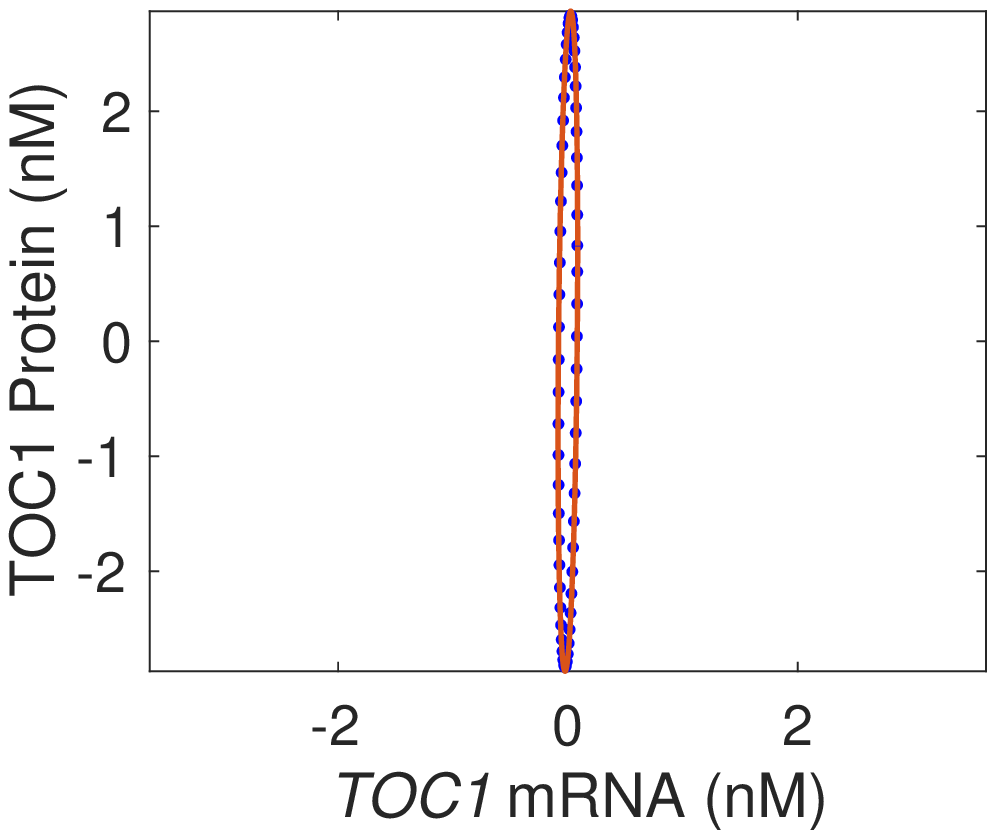}
      \caption{}
      \label{Locke2005a:phase-TOC1}
    \end{subfigure}%
    \begin{subfigure}{0.333\textwidth}
      \includegraphics[width=\textwidth]{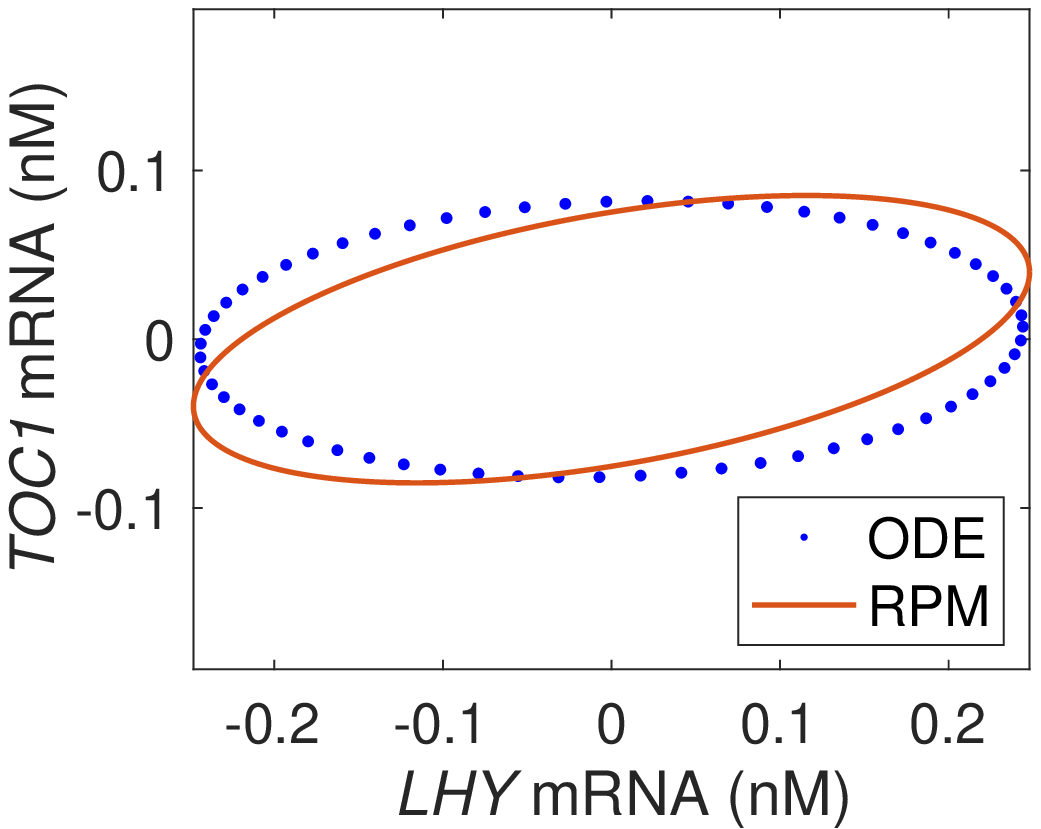}
      \caption{}
      \label{Locke2005a:phase-mRNA}
    \end{subfigure}
    \caption{A supercritical Hopf bifurcation occurs in L2005a model under perpetual illumination.  Bifurcation diagrams for (a) concentration of \textit{LHY} mRNA and (b) frequency of oscillation, time series generated from both ODE and RPM for concentrations of (c) \textit{LHY} mRNA and (d) \textit{TOC1} mRNA, and (e) - (g) phase diagrams of pairs of LHY and TOC1 protein in the cytoplasm and \textit{LHY} and \textit{TOC1} mRNA oscillations are shown.  The degradation rate in (a) and (b) are normalized so that the biological value given in the original paper is unity.  The amplitude of limit cycle oscillation calculated with RPM matches the numerical solution of the system of ODEs with 6.69 percent difference at biological values; and frequency with 0.006 percent difference. 
    As fractions of $2\pi$, the absolute values of differences in phase difference are 0.006 for the pair (\textit{LHY} mRNA, LHY protein), 0.001 for the pair (\textit{TOC1} mRNA, TOC1 protein), and 0.063 for the pair (\textit{LHY} mRNA, \textit{TOC1} mRNA).}
    \label{7-panel:Locke2005a}
\end{figure}

\begin{figure}[H]
    \centering
    \begin{subfigure}{0.5\textwidth}
      \includegraphics[width=\textwidth]{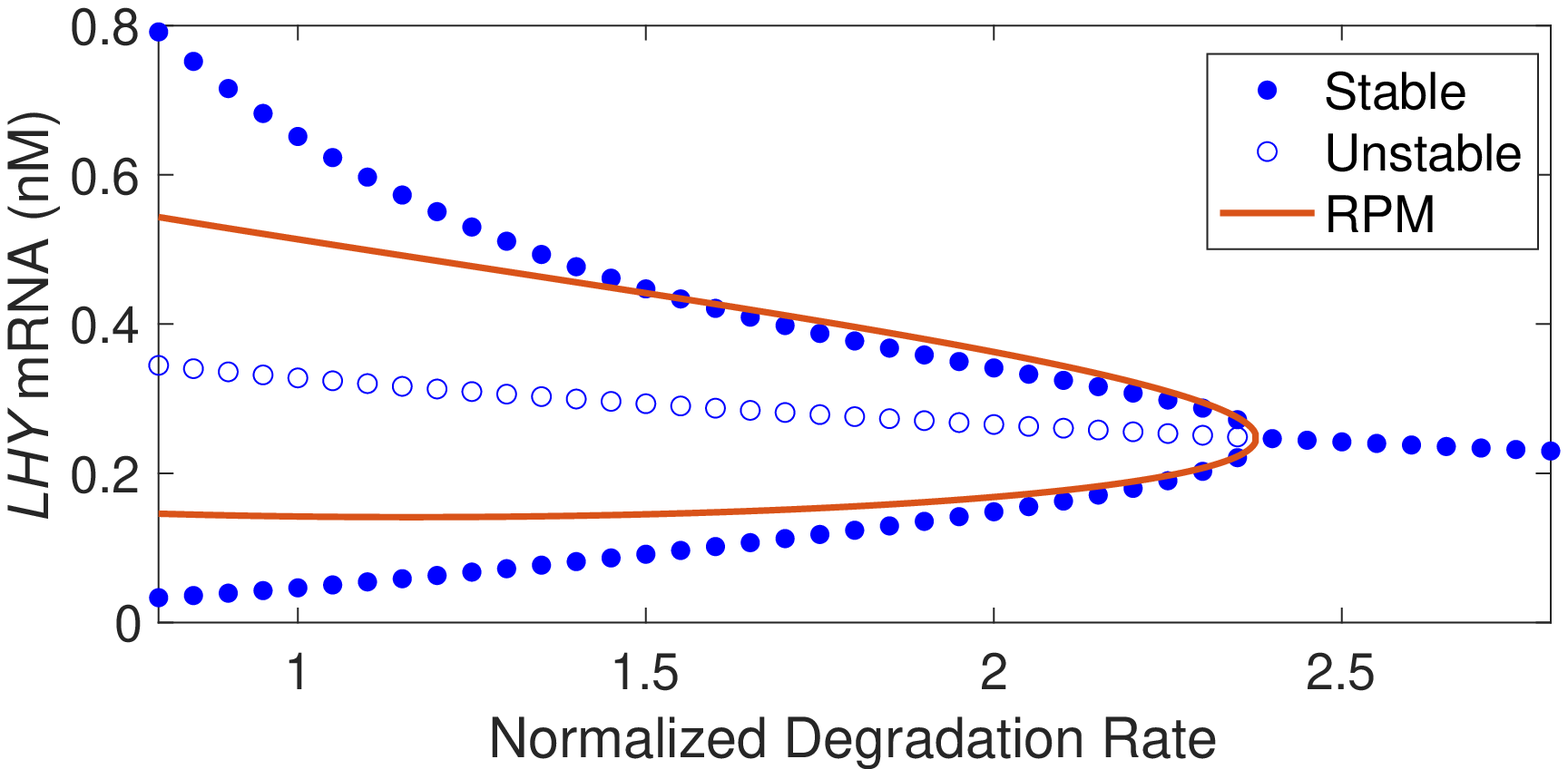}
      \caption{}
      \label{Locke2005b:bif-amp}
    \end{subfigure}%
    \begin{subfigure}{0.5\textwidth}
      \includegraphics[width=\textwidth]{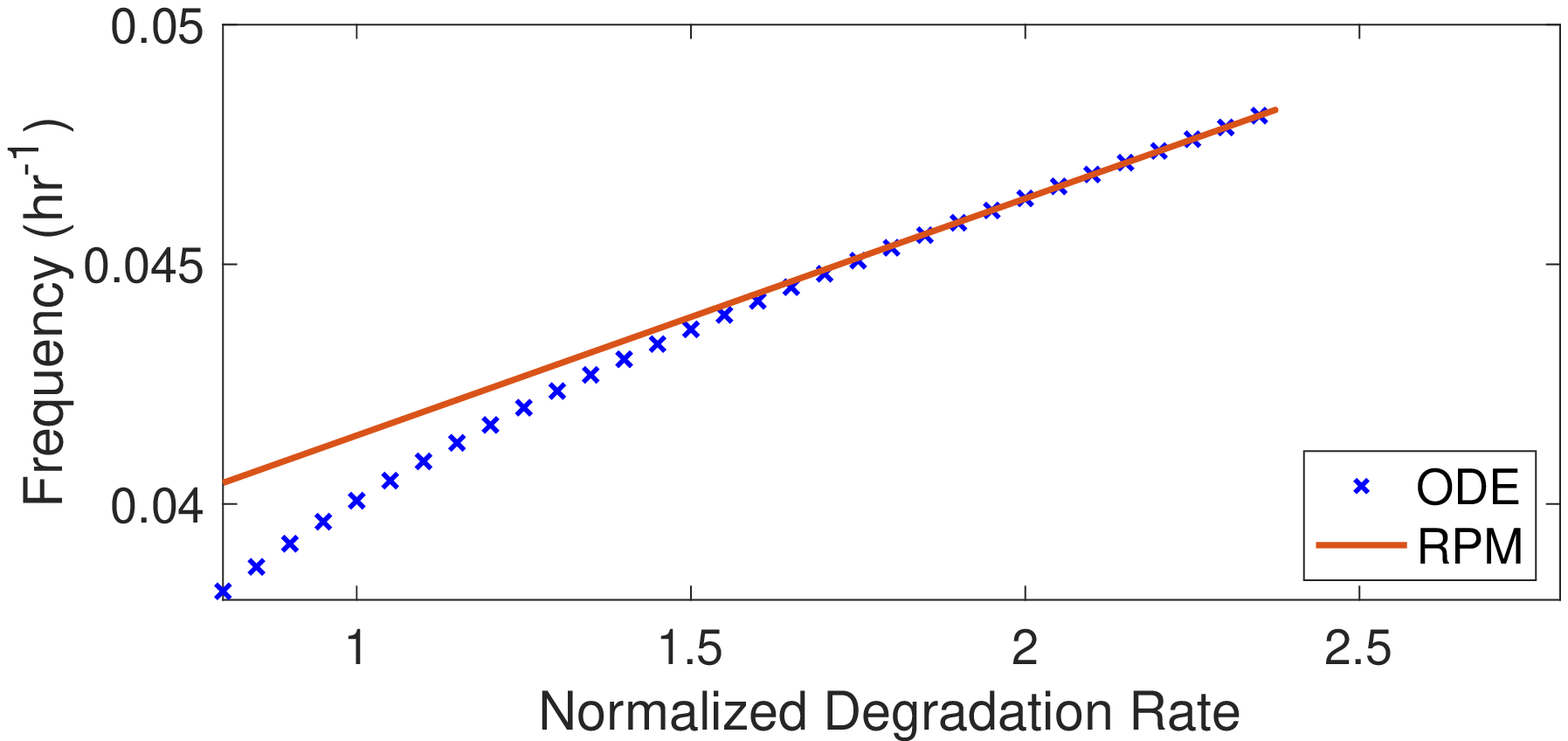}
      \caption{}
      \label{Locke2005b:bif-freq}
    \end{subfigure}
    \begin{subfigure}{0.5\textwidth}
      \includegraphics[width=\textwidth]{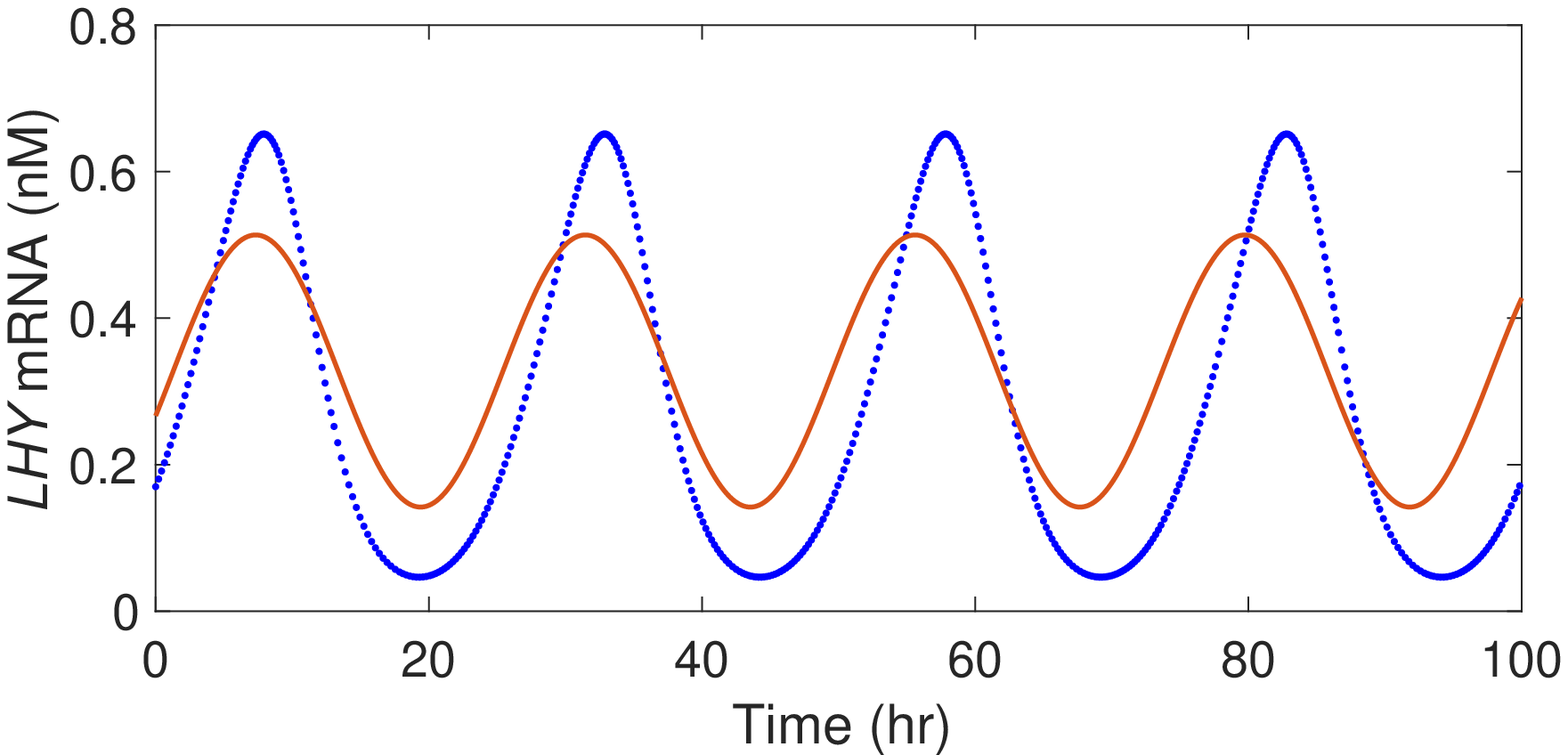}
      \caption{}
      \label{Locke2005b:series-LHY}
    \end{subfigure}%
    \begin{subfigure}{0.5\textwidth}
      \includegraphics[width=\textwidth]{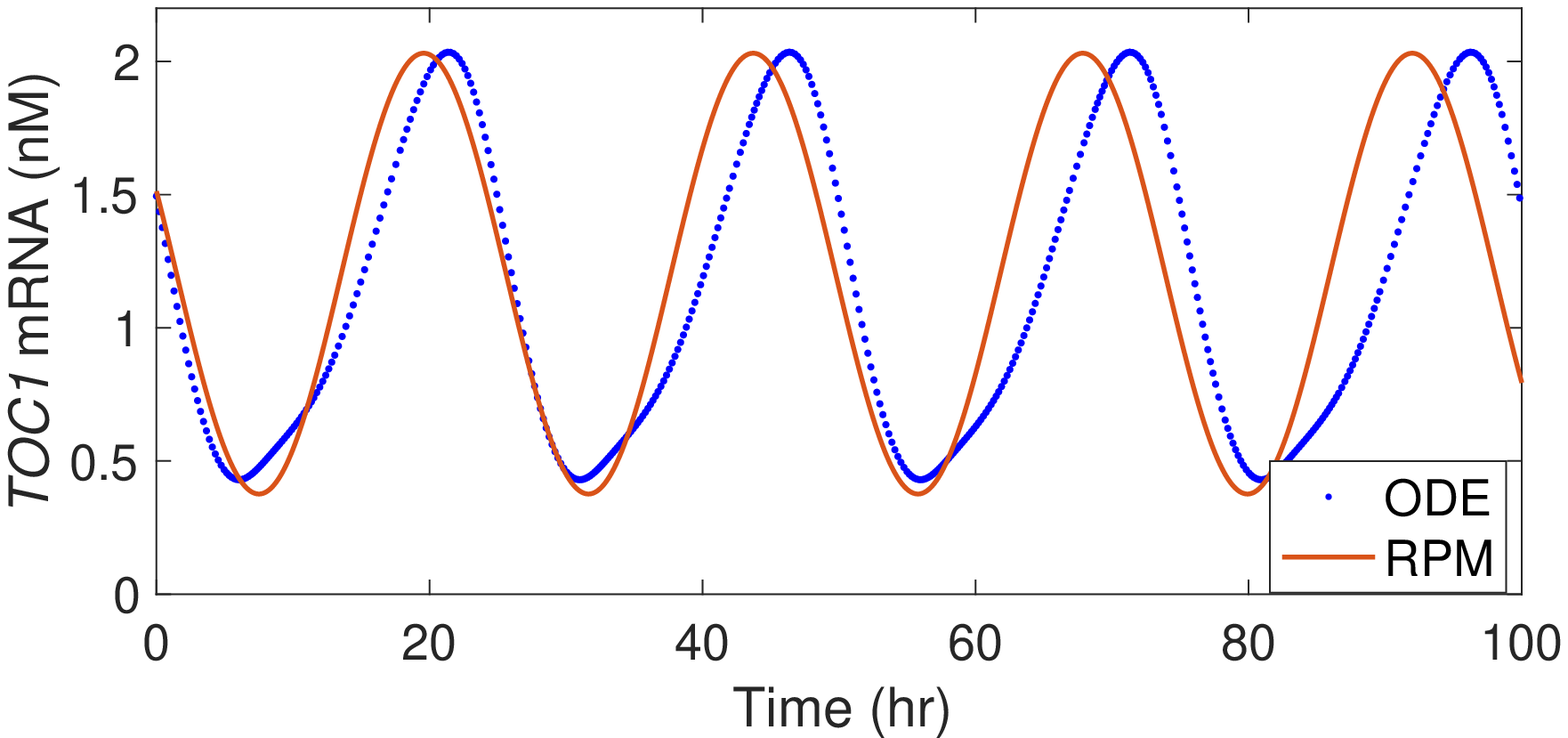}
      \caption{}
      \label{Locke2005b:series-TOC1}
    \end{subfigure}
    \begin{subfigure}{0.333\textwidth}
      \includegraphics[width=\textwidth]{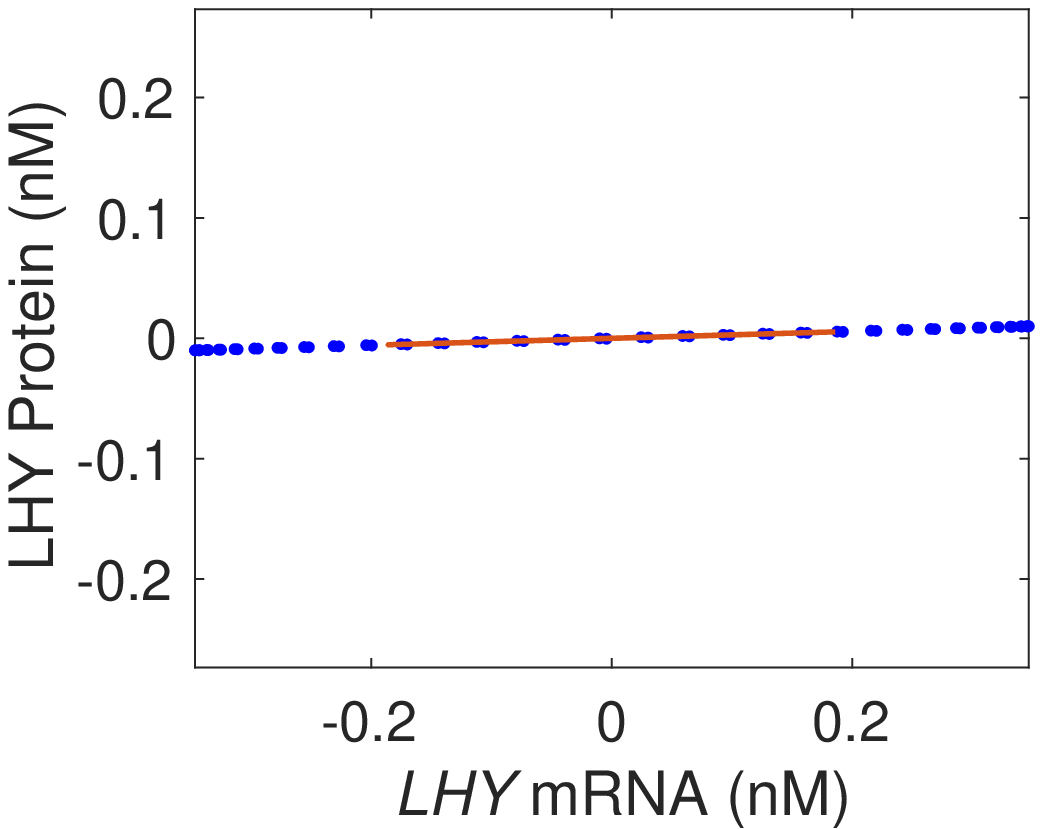}
      \caption{}
      \label{Locke2005b:phase-LHY}
    \end{subfigure}%
    \begin{subfigure}{0.333\textwidth}
      \includegraphics[width=\textwidth]{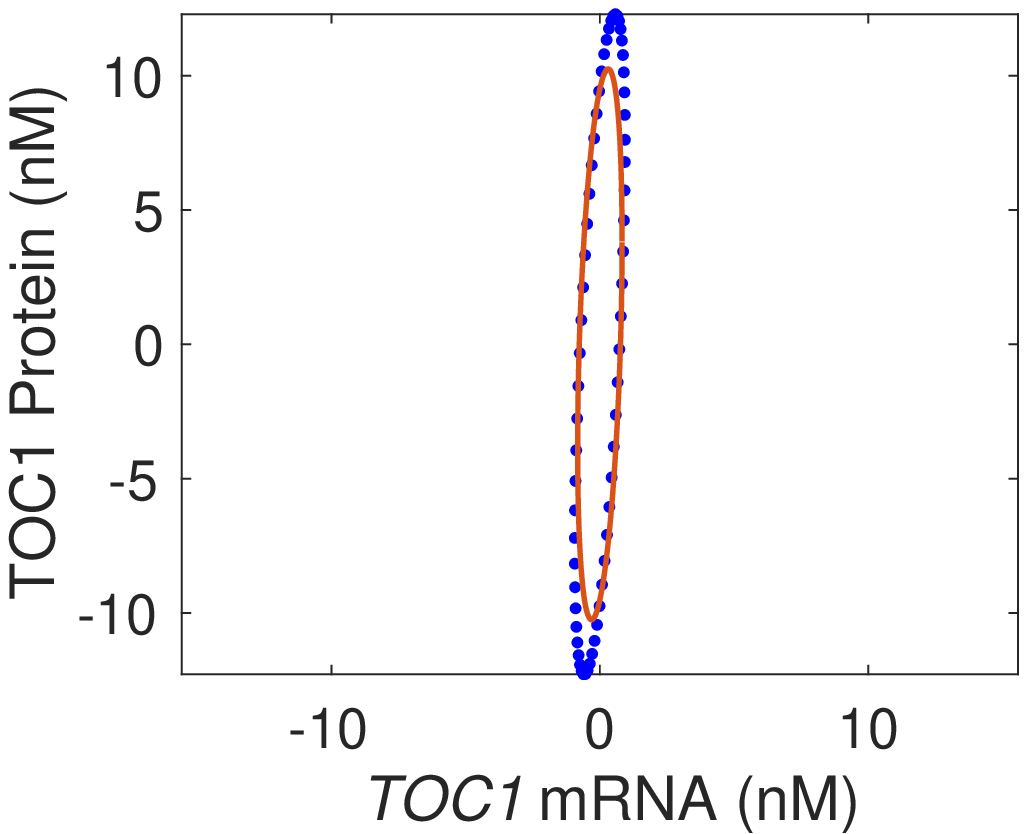}
      \caption{}
      \label{Locke2005b:phase-TOC1}
    \end{subfigure}%
    \begin{subfigure}{0.333\textwidth}
      \includegraphics[width=\textwidth]{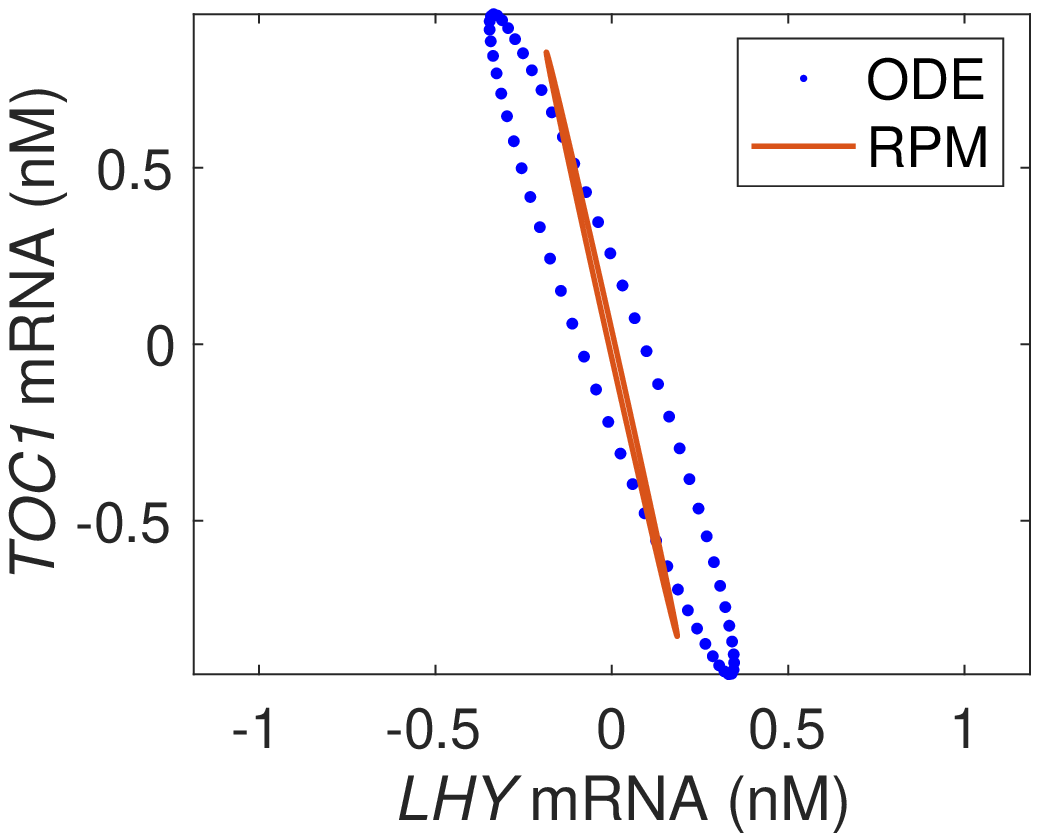}
      \caption{}
      \label{Locke2005b:phase-mRNA}
    \end{subfigure}
    \caption{
    A supercritical Hopf bifurcation occurs in L2005b model under perpetual illumination.  Bifurcation diagrams for (a) concentration of \textit{LHY} mRNA and (b) frequency of oscillation, time series generated from both ODE and RPM for concentrations of (c) \textit{LHY} mRNA and (d) \textit{TOC1} mRNA, and (e) - (g) phase diagrams of pairs of LHY and TOC1 protein in the cytoplasm and \textit{LHY} and \textit{TOC1} mRNA oscillations are shown.  The degradation rate in (a) and (b) are normalized so that the biological value given in the original paper is unity.  The amplitude of limit cycle oscillation calculated with RPM matches the numerical solution of the system of ODEs with 38.60 percent difference; and frequency with 3.24 percent difference. 
    As fractions of $2\pi$, the absolute values of differences in phase difference are 0.002 for the pair (\textit{LHY} mRNA, LHY protein), 0.045 for the pair (\textit{TOC1} mRNA, TOC1 protein), and 0.035 for the pair (\textit{LHY} mRNA, \textit{TOC1} mRNA).}
    \label{7-panel:Locke2005b}
\end{figure}

\begin{figure}[H]
    \centering
    \begin{subfigure}{0.5\textwidth}
      \includegraphics[width=\textwidth]{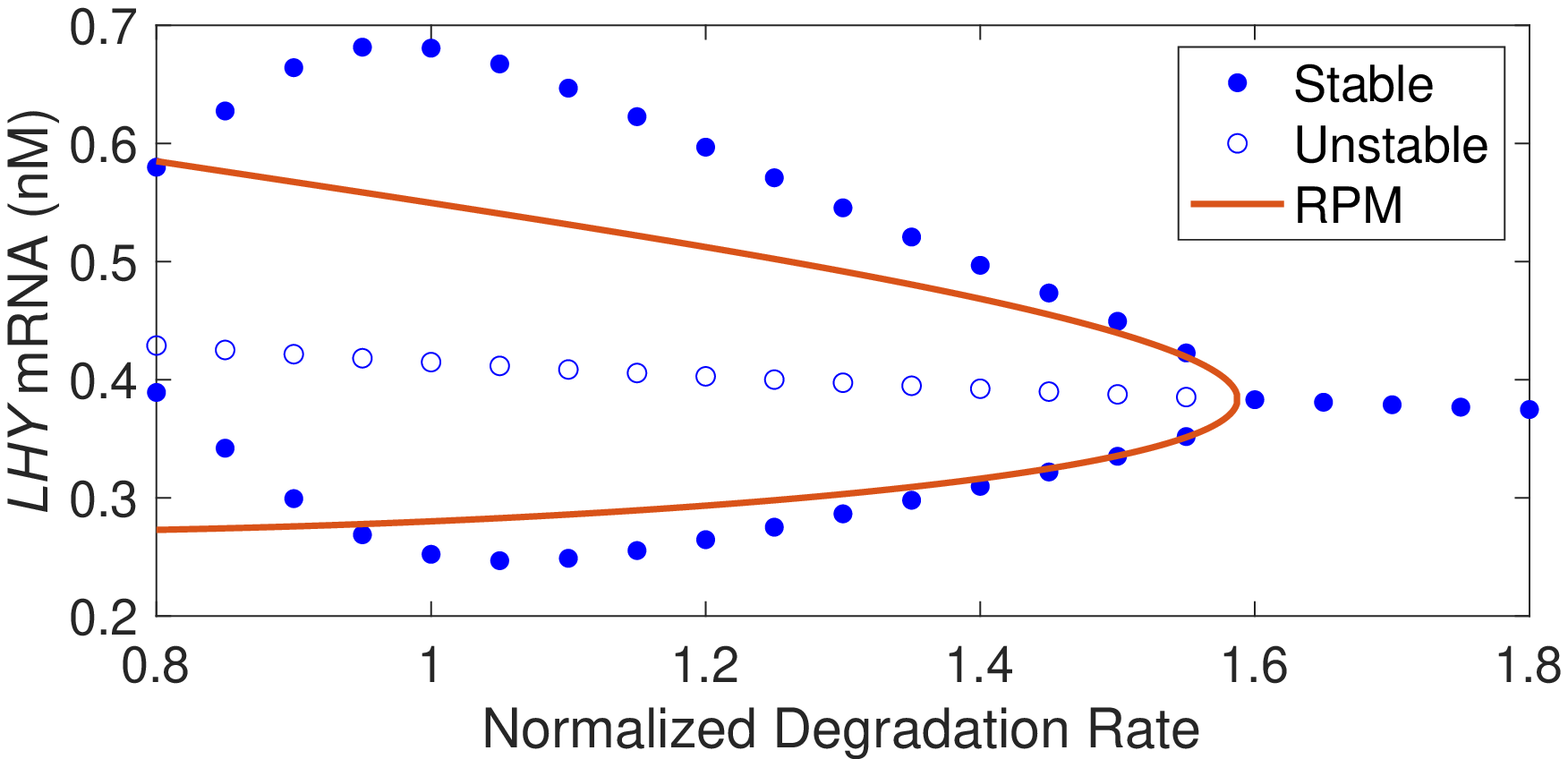}
      \caption{}
      \label{Zeilinger2006:bif-amp}
    \end{subfigure}%
    \begin{subfigure}{0.5\textwidth}
      \includegraphics[width=\textwidth]{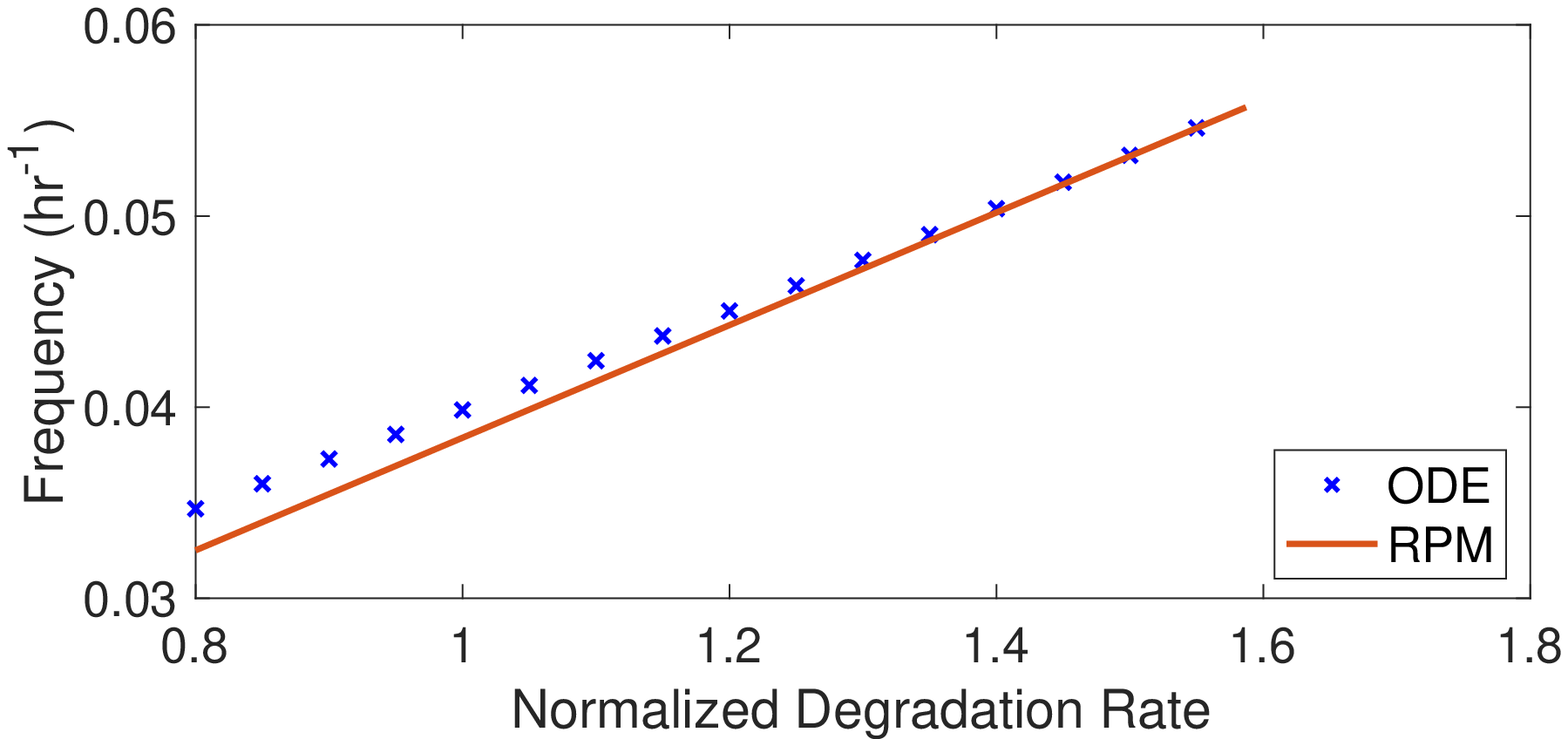}
      \caption{}
      \label{Zeilinger2006:bif-freq}
    \end{subfigure}
    \begin{subfigure}{0.5\textwidth}
      \includegraphics[width=\textwidth]{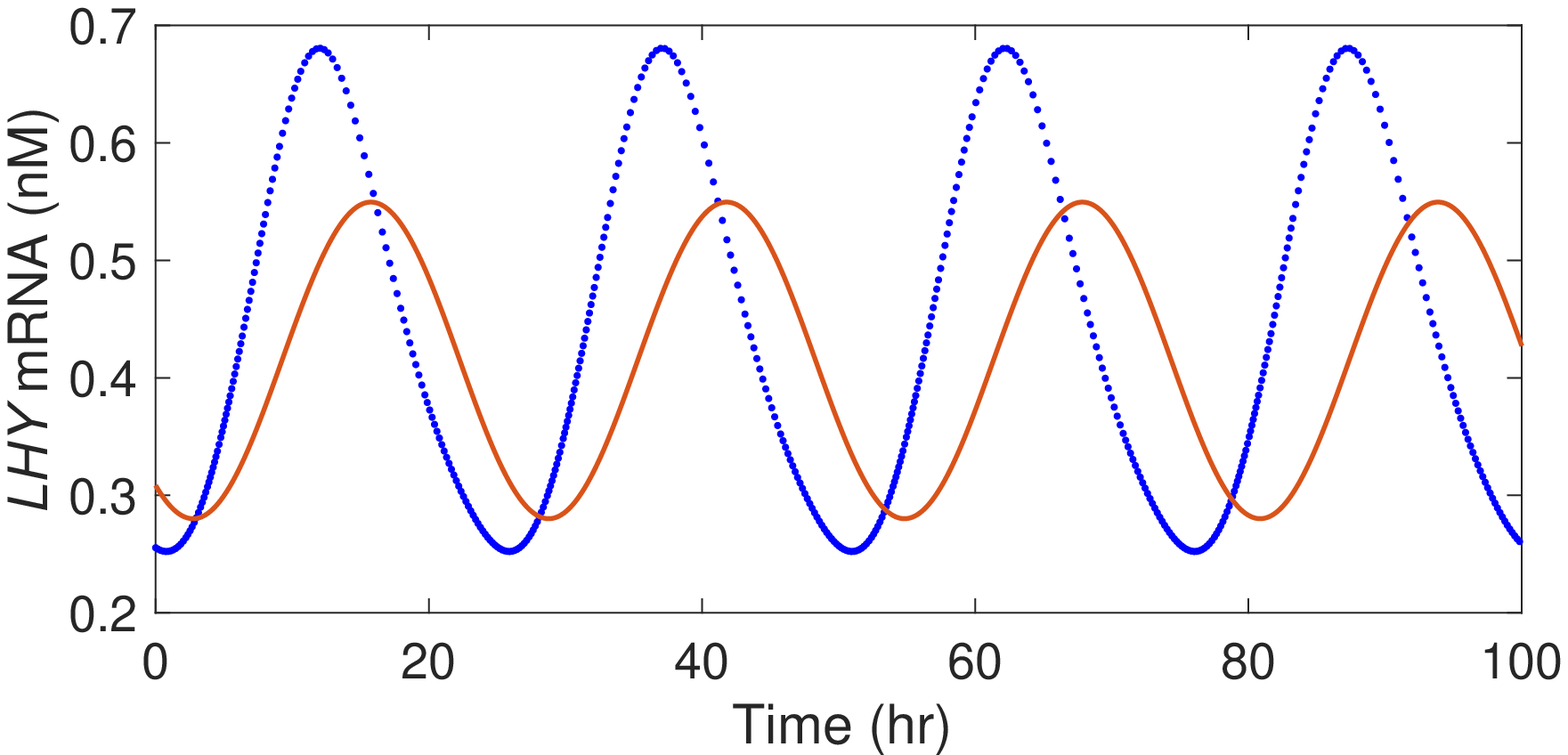}
      \caption{}
      \label{Zeilinger2006:series-LHY}
    \end{subfigure}%
    \begin{subfigure}{0.5\textwidth}
      \includegraphics[width=\textwidth]{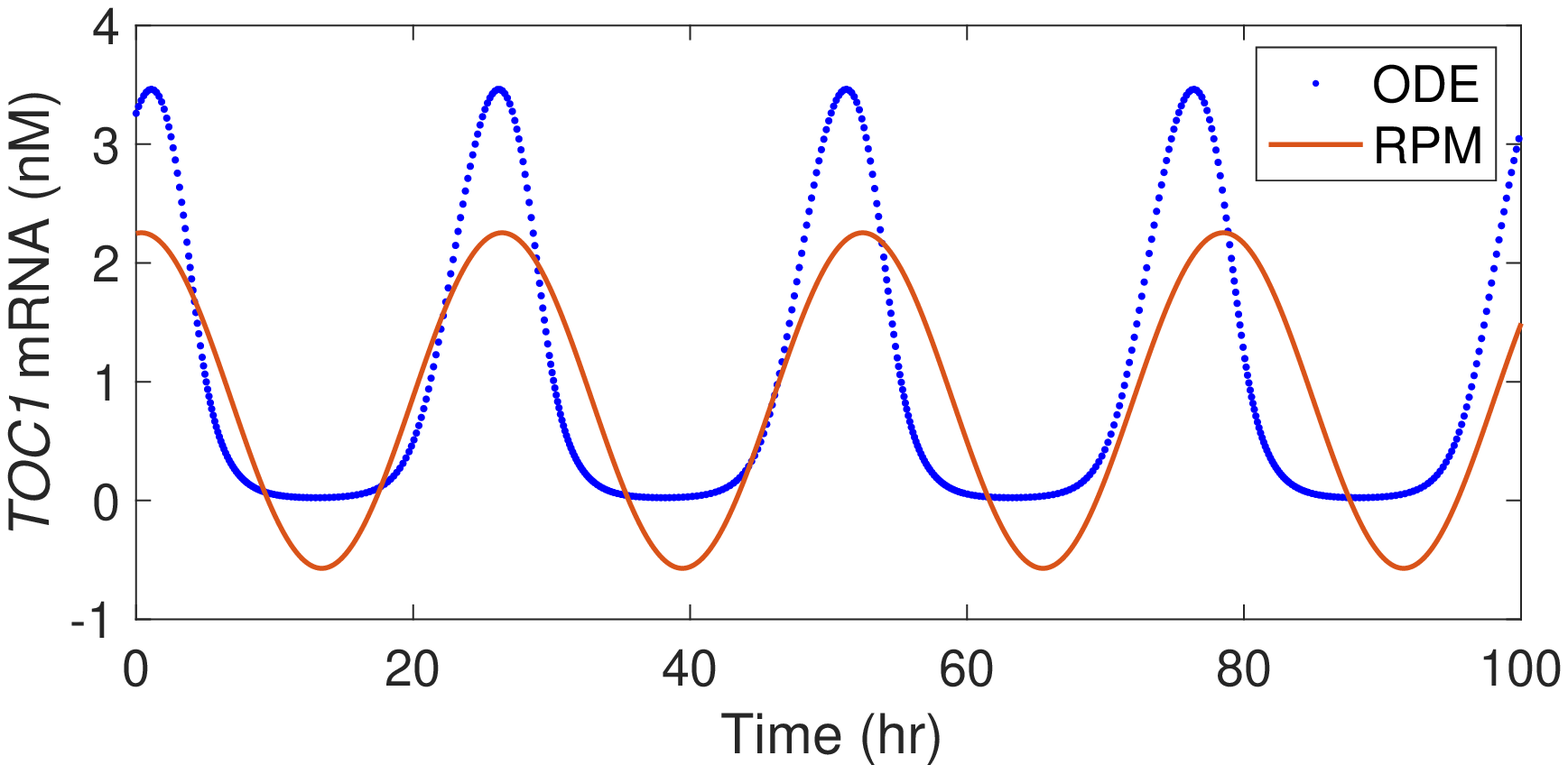}
      \caption{}
      \label{Zeilinger2006:series-TOC1}
    \end{subfigure}
    \begin{subfigure}{0.333\textwidth}
      \includegraphics[width=\textwidth]{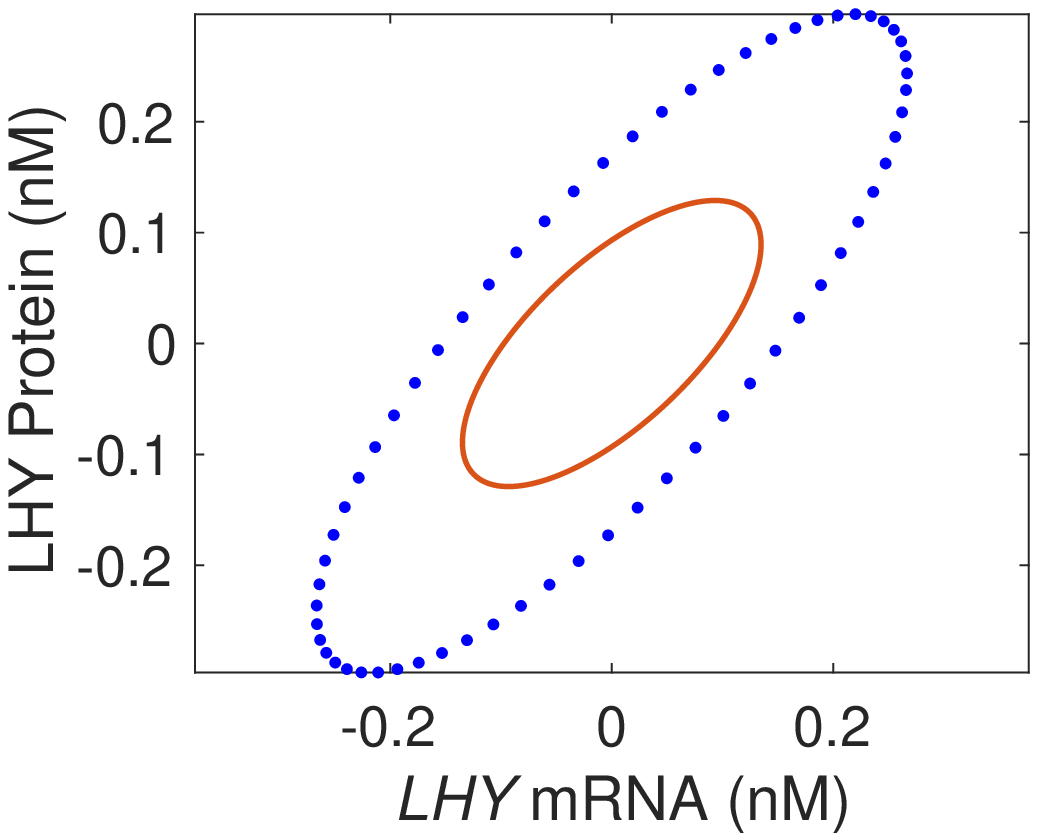}
      \caption{}
      \label{Zeilinger2006:phase-LHY}
    \end{subfigure}%
    \begin{subfigure}{0.333\textwidth}
      \includegraphics[width=\textwidth]{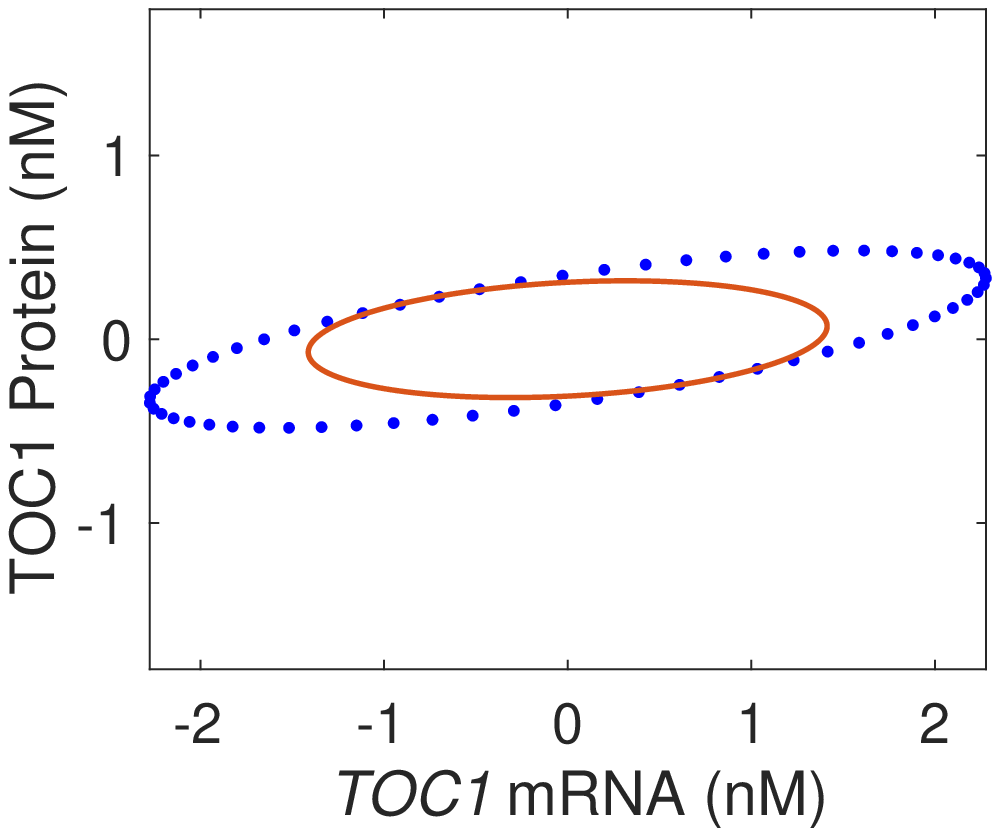}
      \caption{}
      \label{Zeilinger2006:phase-TOC1}
    \end{subfigure}%
    \begin{subfigure}{0.333\textwidth}
      \includegraphics[width=\textwidth]{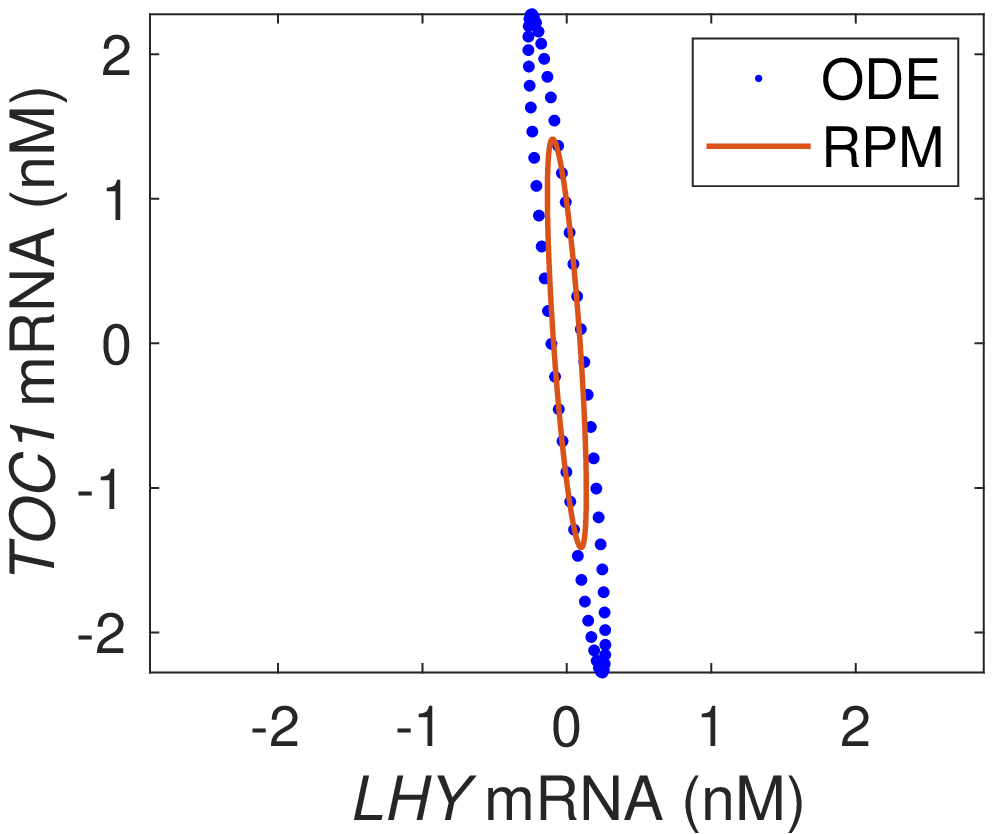}
      \caption{}
      \label{Zeilinger2006:phase-mRNA}
    \end{subfigure}
    \caption{
    A supercritical Hopf bifurcation occurs in Z2006 model under perpetual illumination.  Bifurcation diagrams for (a) concentration of \textit{LHY} mRNA and (b) frequency of oscillation, time series generated from both ODE and RPM for concentrations of (c) \textit{LHY} mRNA and (d) \textit{TOC1} mRNA, and (e) - (g) phase diagrams of pairs of LHY and TOC1 protein in the cytoplasm and \textit{LHY} and \textit{TOC1} mRNA oscillations are shown.  The degradation rate in (a) and (b) are normalized so that the biological value given in the original paper is unity.  The amplitude of limit cycle oscillation calculated with RPM matches the numerical solution of the system of ODEs with 37.09 percent difference; and frequency with 3.79 percent difference. 
    As fractions of $2\pi$, the absolute values of differences in phase difference are 0.032 for the pair (\textit{LHY} mRNA, LHY protein), 0.085 for the pair (\textit{TOC1} mRNA, TOC1 protein), and 0.054 for the pair (\textit{LHY} mRNA, \textit{TOC1} mRNA).}
    \label{7-panel:Zeilinger2006}
\end{figure}

\begin{figure}[H]
    \centering
    \begin{subfigure}{0.5\textwidth}
      \includegraphics[width=\textwidth]{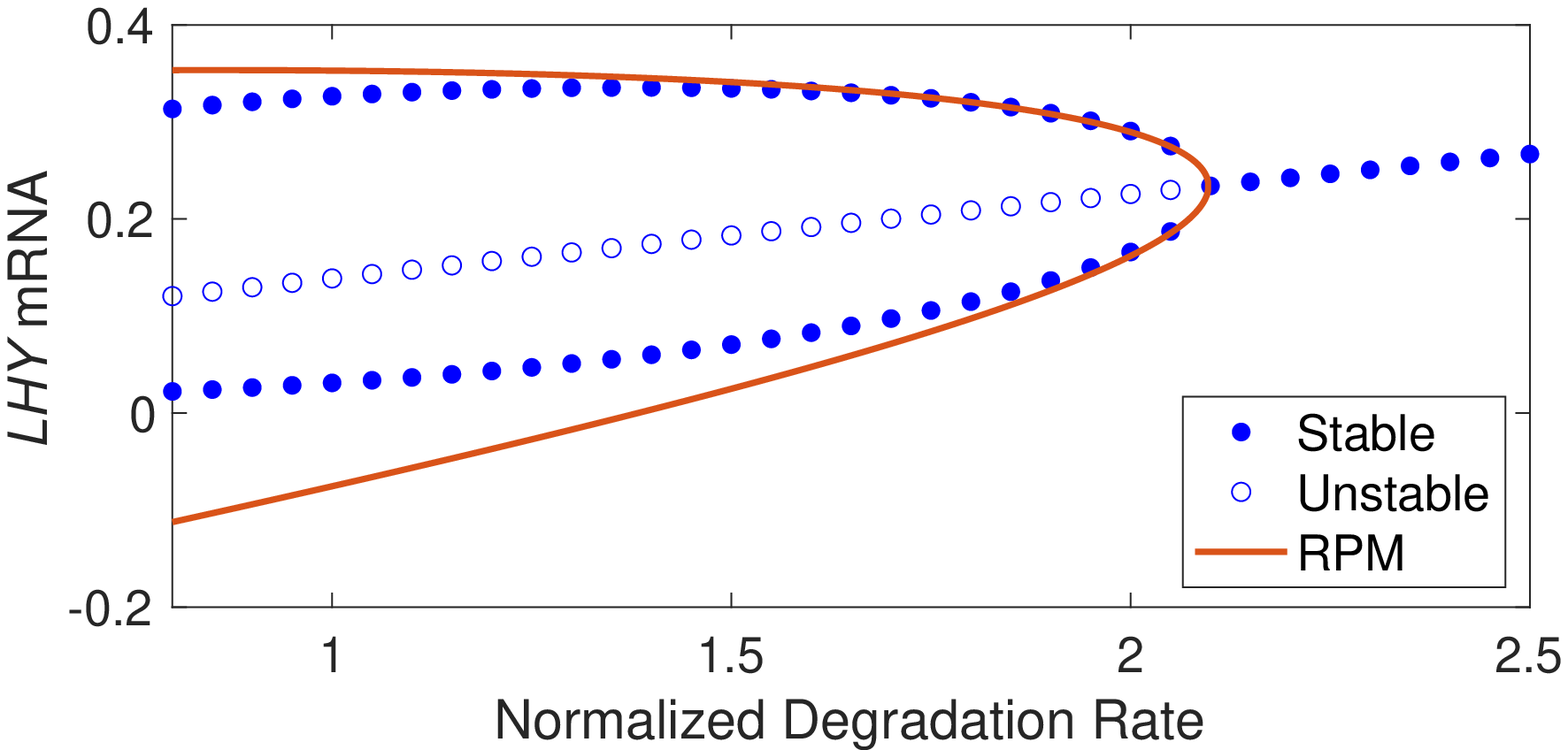}
      \caption{}
      \label{Pokhilko2010:bif-amp}
    \end{subfigure}%
    \begin{subfigure}{0.5\textwidth}
      \includegraphics[width=\textwidth]{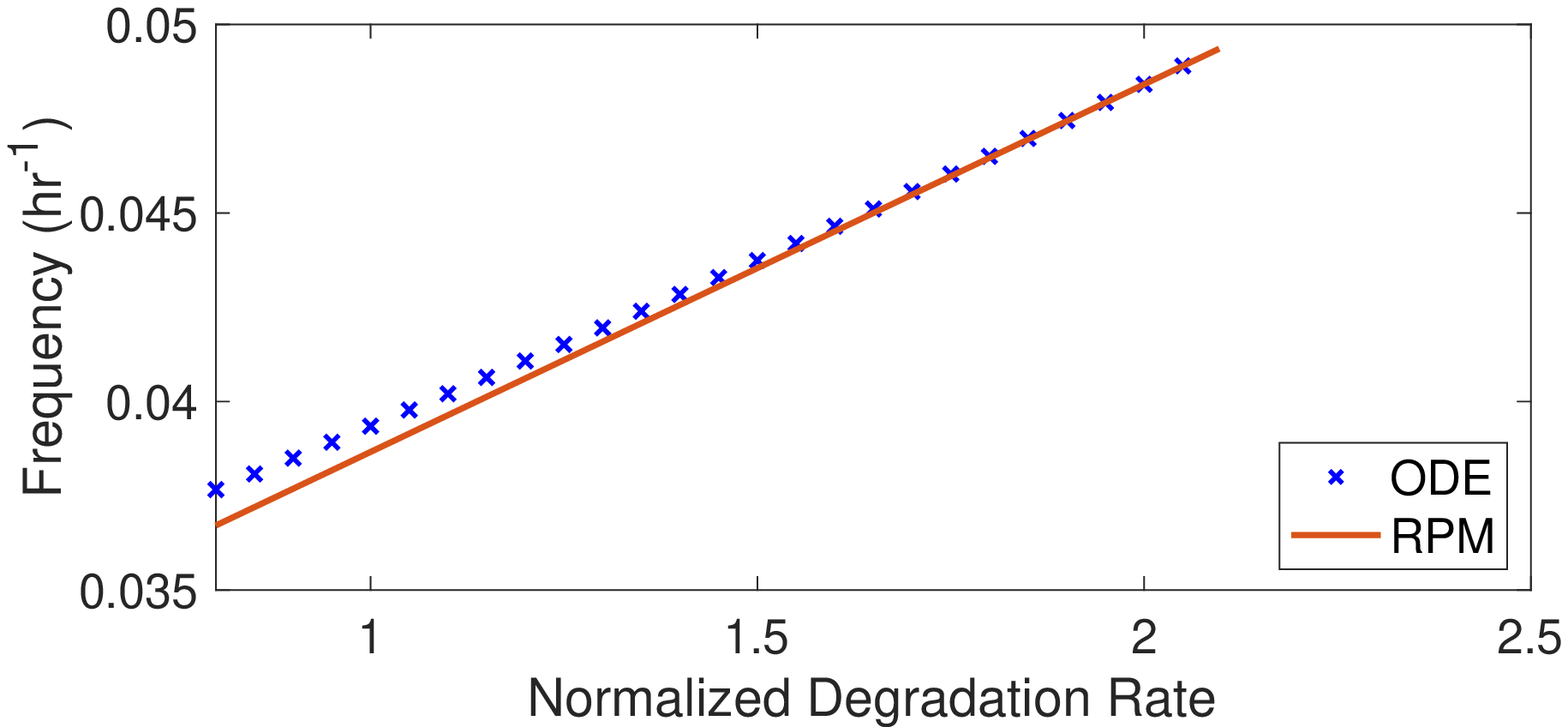}
      \caption{}
      \label{Pokhilko2010:bif-freq}
    \end{subfigure}
    \begin{subfigure}{0.5\textwidth}
      \includegraphics[width=\textwidth]{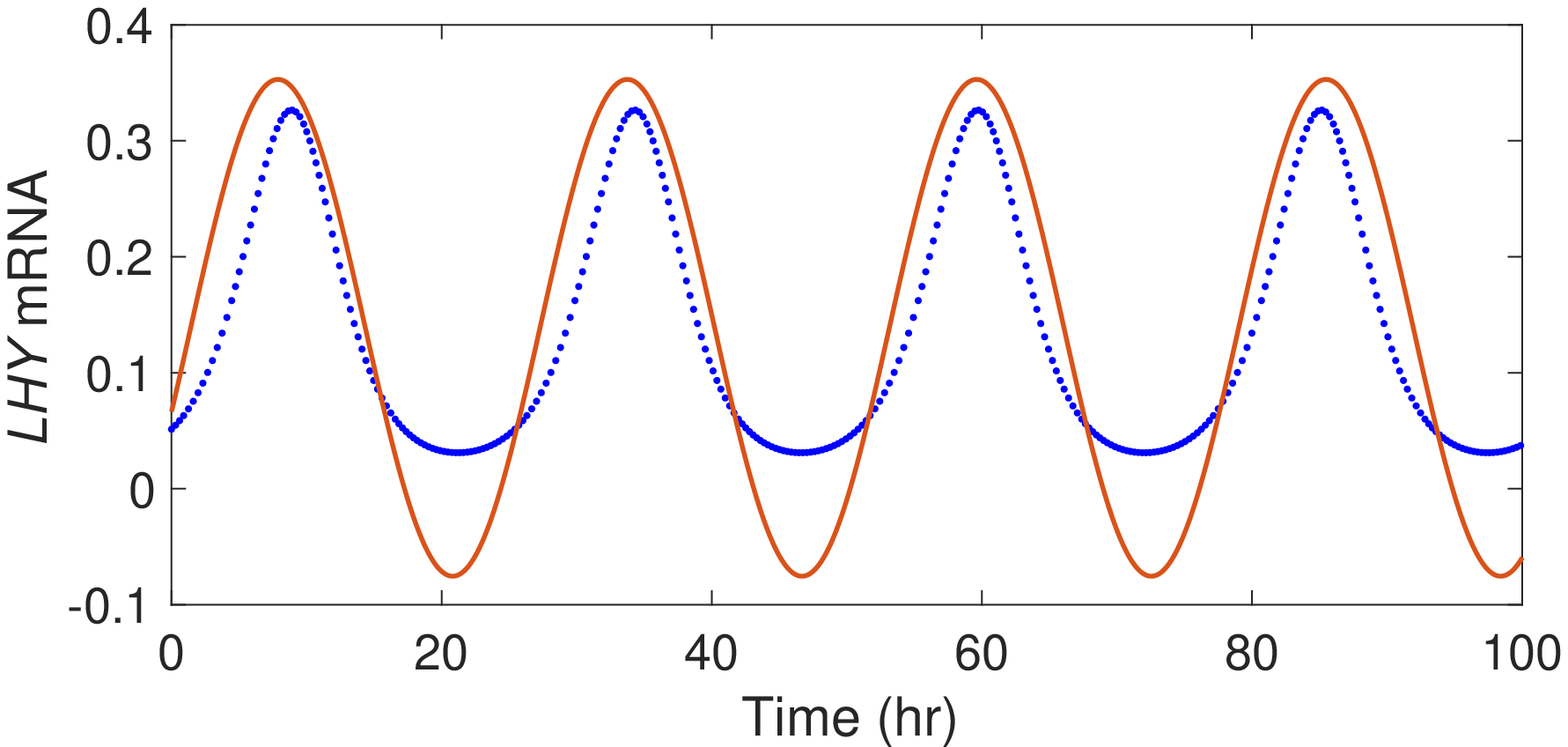}
      \caption{}
      \label{Pokhilko2010:series-LHY}
    \end{subfigure}%
    \begin{subfigure}{0.5\textwidth}
      \includegraphics[width=\textwidth]{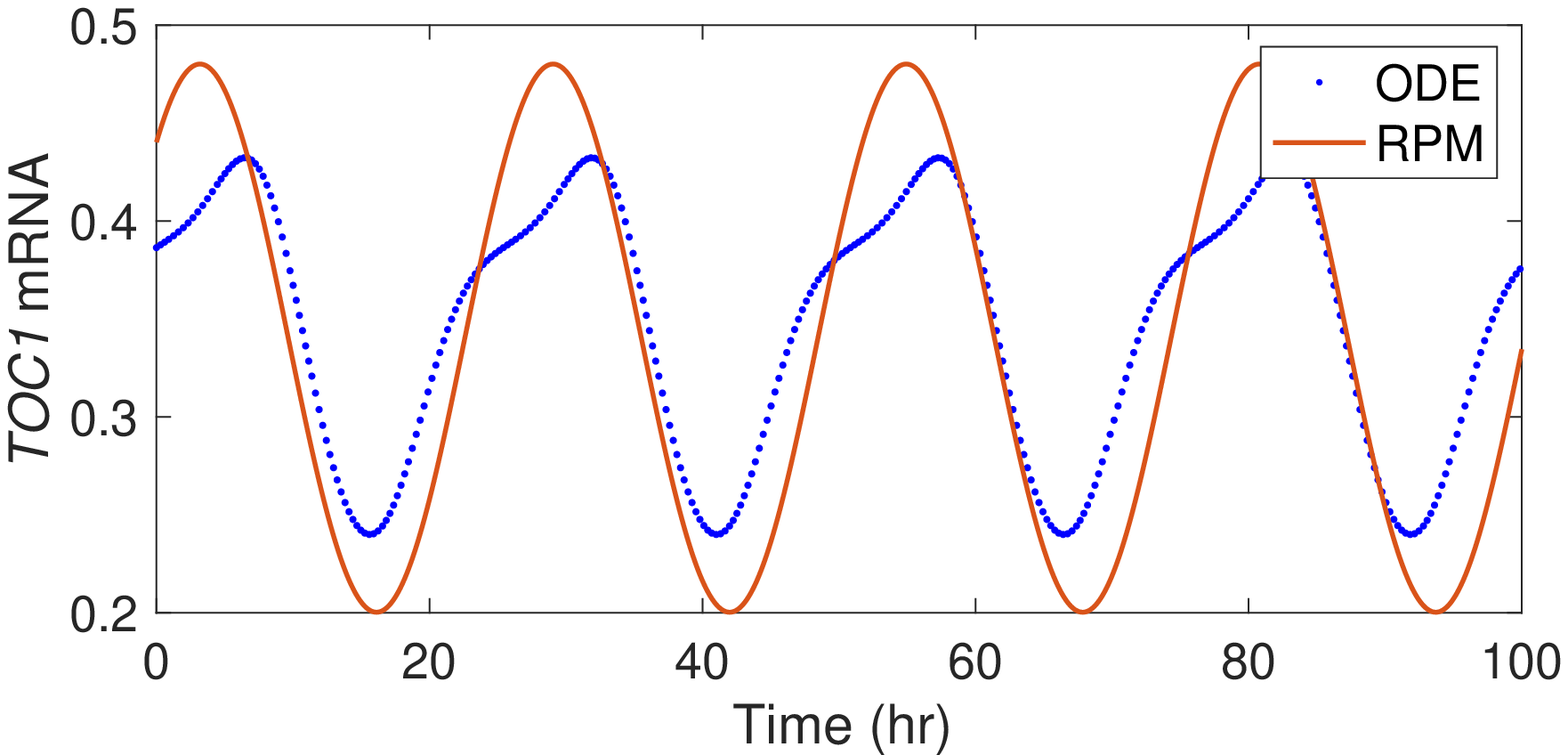}
      \caption{}
      \label{Pokhilko2010:series-TOC1}
    \end{subfigure}
    \begin{subfigure}{0.333\textwidth}
      \includegraphics[width=\textwidth]{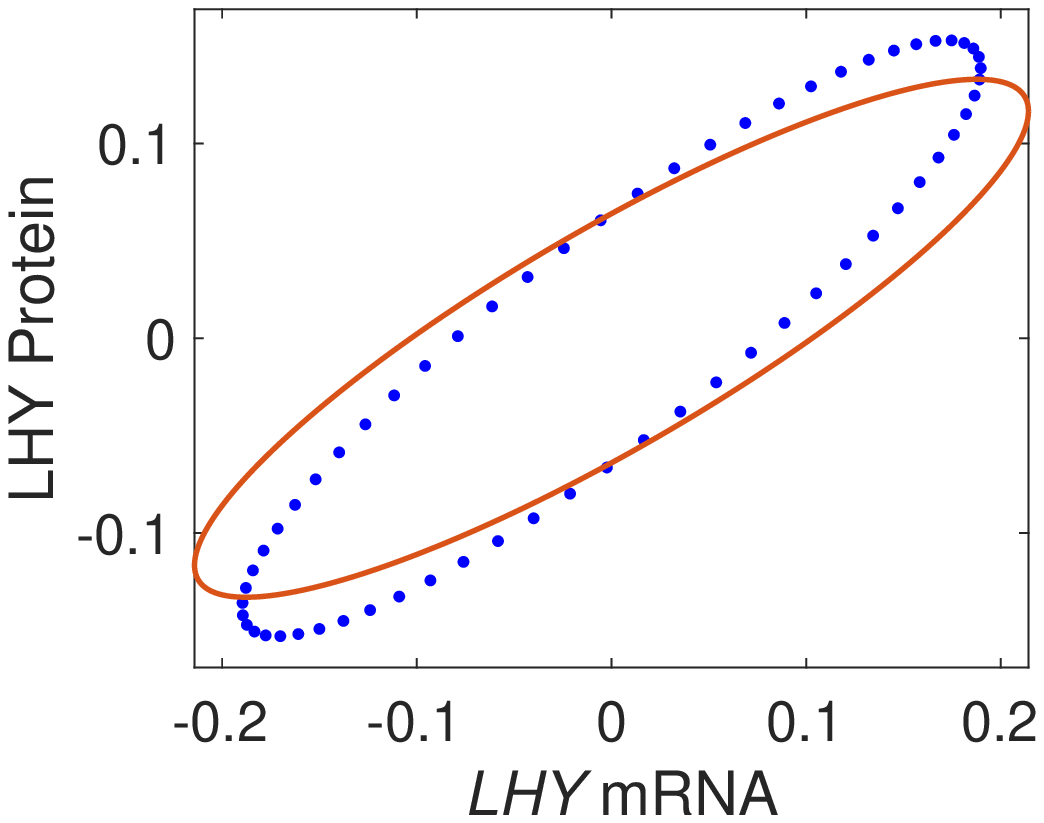}
      \caption{}
      \label{Pokhilko2010:phase-LHY}
    \end{subfigure}%
    \begin{subfigure}{0.333\textwidth}
      \includegraphics[width=\textwidth]{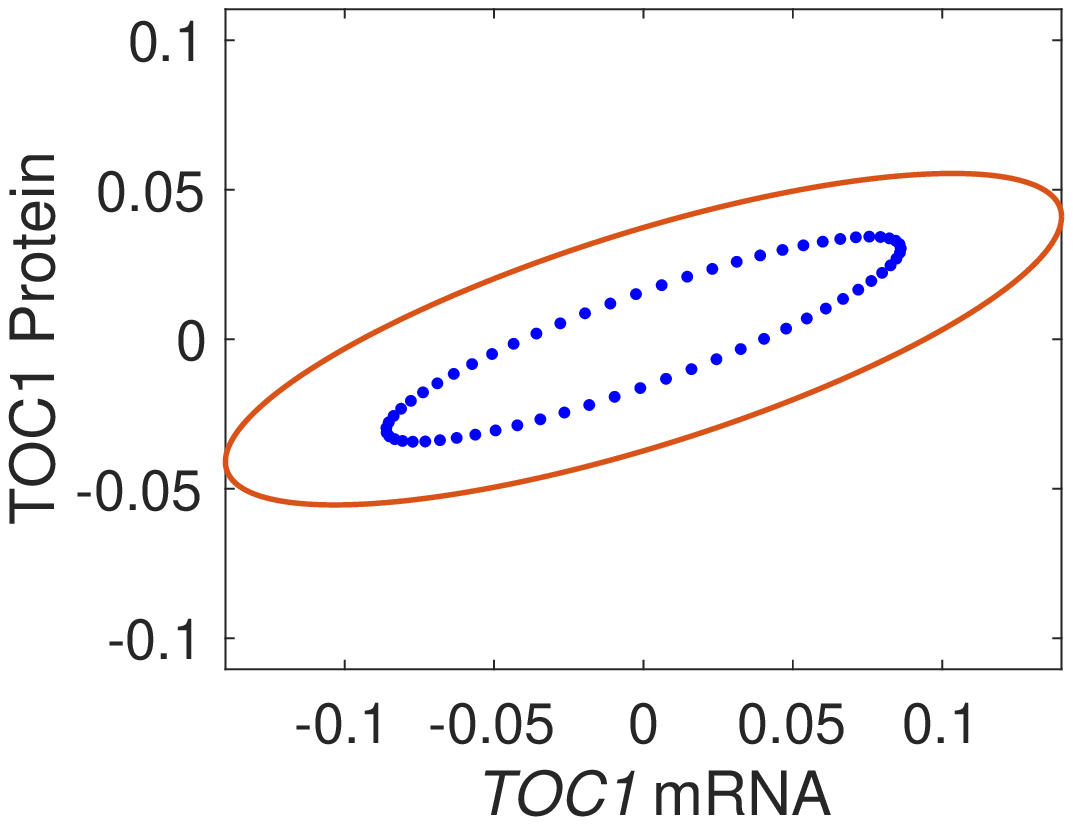}
      \caption{}
      \label{Pokhilko2010:phase-TOC1}
    \end{subfigure}%
    \begin{subfigure}{0.333\textwidth}
      \includegraphics[width=\textwidth]{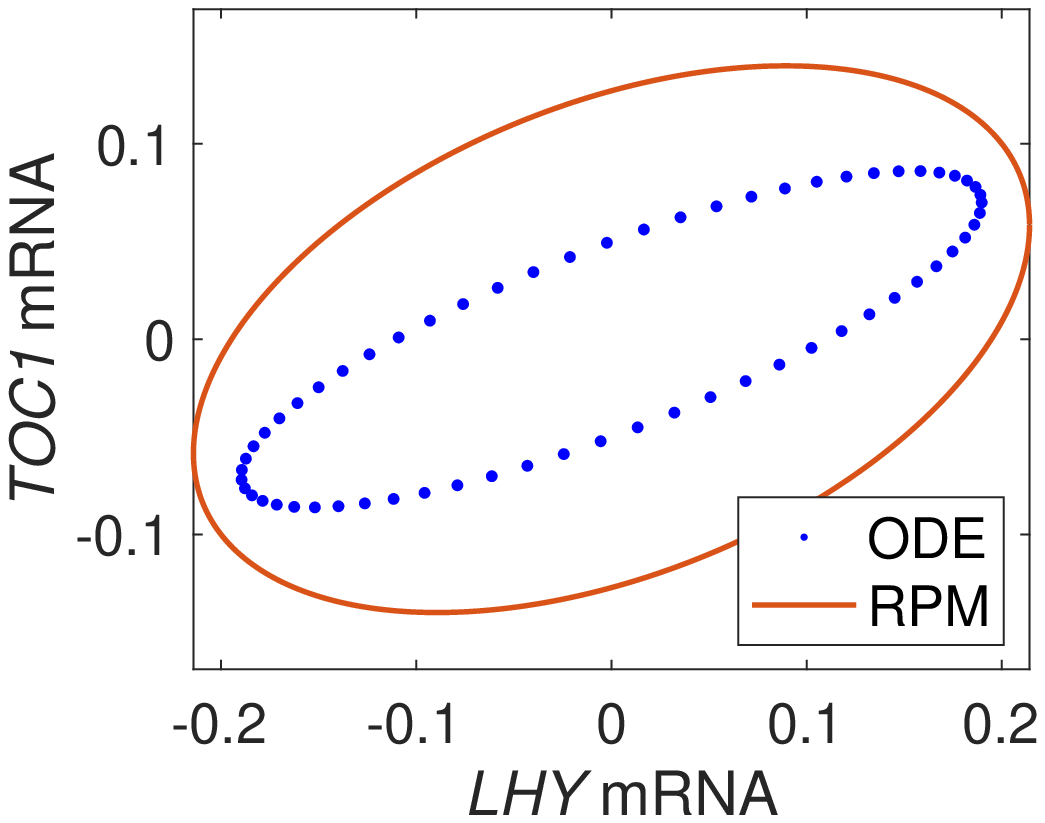}
      \caption{}
      \label{Pokhilko2010:phase-mRNA}
    \end{subfigure}
    \caption{
    A supercritical Hopf bifurcation occurs in P2010 model under perpetual illumination.  Bifurcation diagrams for (a) concentration of \textit{LHY} mRNA and (b) frequency of oscillation, time series generated from both ODE and RPM for concentrations of (c) \textit{LHY} mRNA and (d) \textit{TOC1} mRNA, and (e) - (g) phase diagrams of pairs of LHY and TOC1 protein in the cytoplasm and \textit{LHY} and \textit{TOC1} mRNA oscillations are shown.  The degradation rate in (a) and (b) are normalized so that the biological value given in the original paper is unity.  The amplitude of limit cycle oscillation calculated with RPM matches the numerical solution of the system of ODEs with 45.02 percent difference; and frequency with 1.53 percent difference. 
    As fractions of $2\pi$, the absolute values of differences in phase difference are 0.011 for the pair (\textit{LHY} mRNA, LHY protein), 0.041 for the pair (\textit{TOC1} mRNA, TOC1 protein), and 0.082 for the pair (\textit{LHY} mRNA, \textit{TOC1} mRNA).}
    \label{7-panel:Pokhilko2010}
\end{figure}

\begin{figure}[H]
    \centering
    \begin{subfigure}{0.5\textwidth}
      \includegraphics[width=\textwidth]{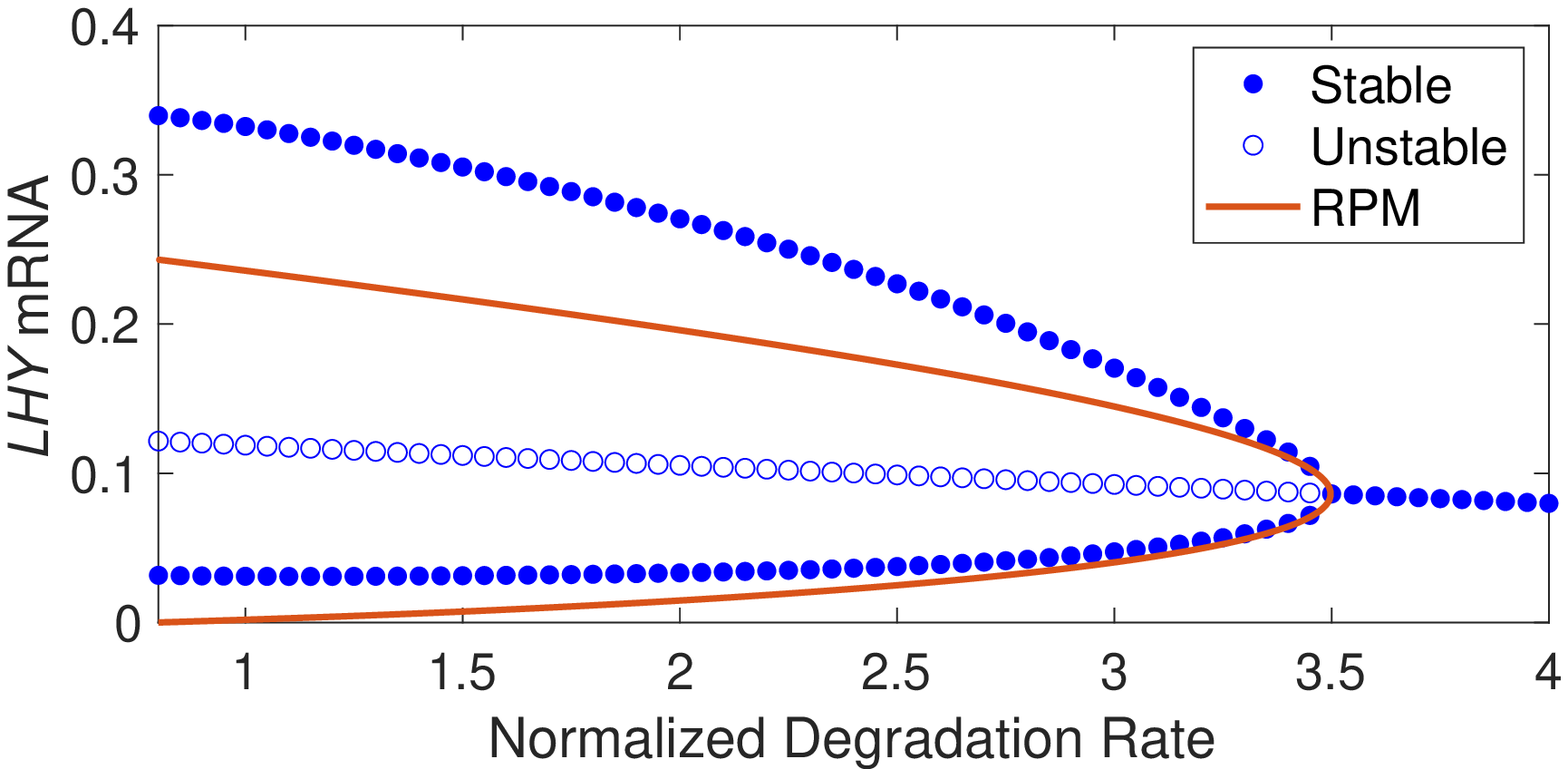}
      \caption{}
      \label{Pokhilko2012:bif-amp}
    \end{subfigure}%
    \begin{subfigure}{0.5\textwidth}
      \includegraphics[width=\textwidth]{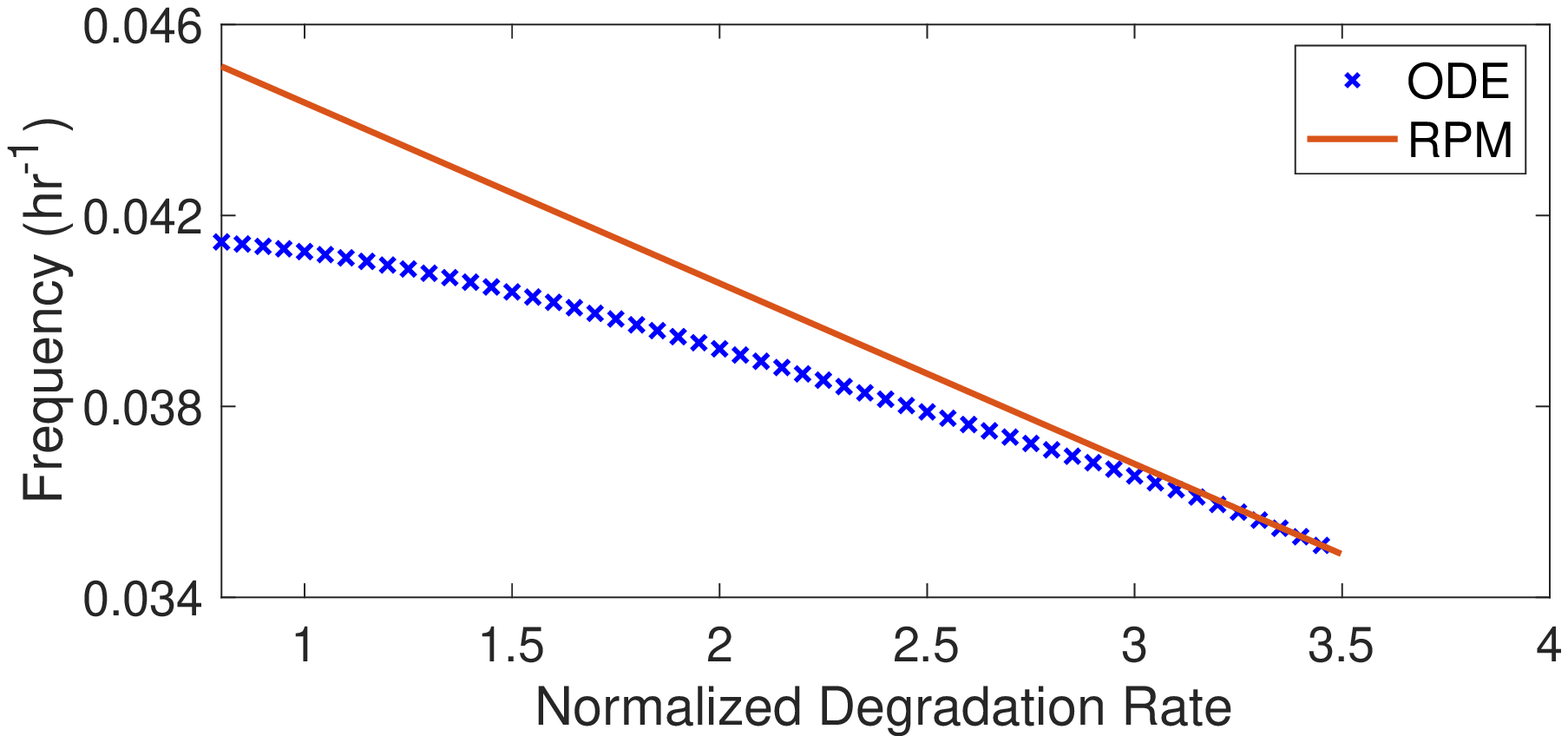}
      \caption{}
      \label{Pokhilko2012:bif-freq}
    \end{subfigure}
    \begin{subfigure}{0.5\textwidth}
      \includegraphics[width=\textwidth]{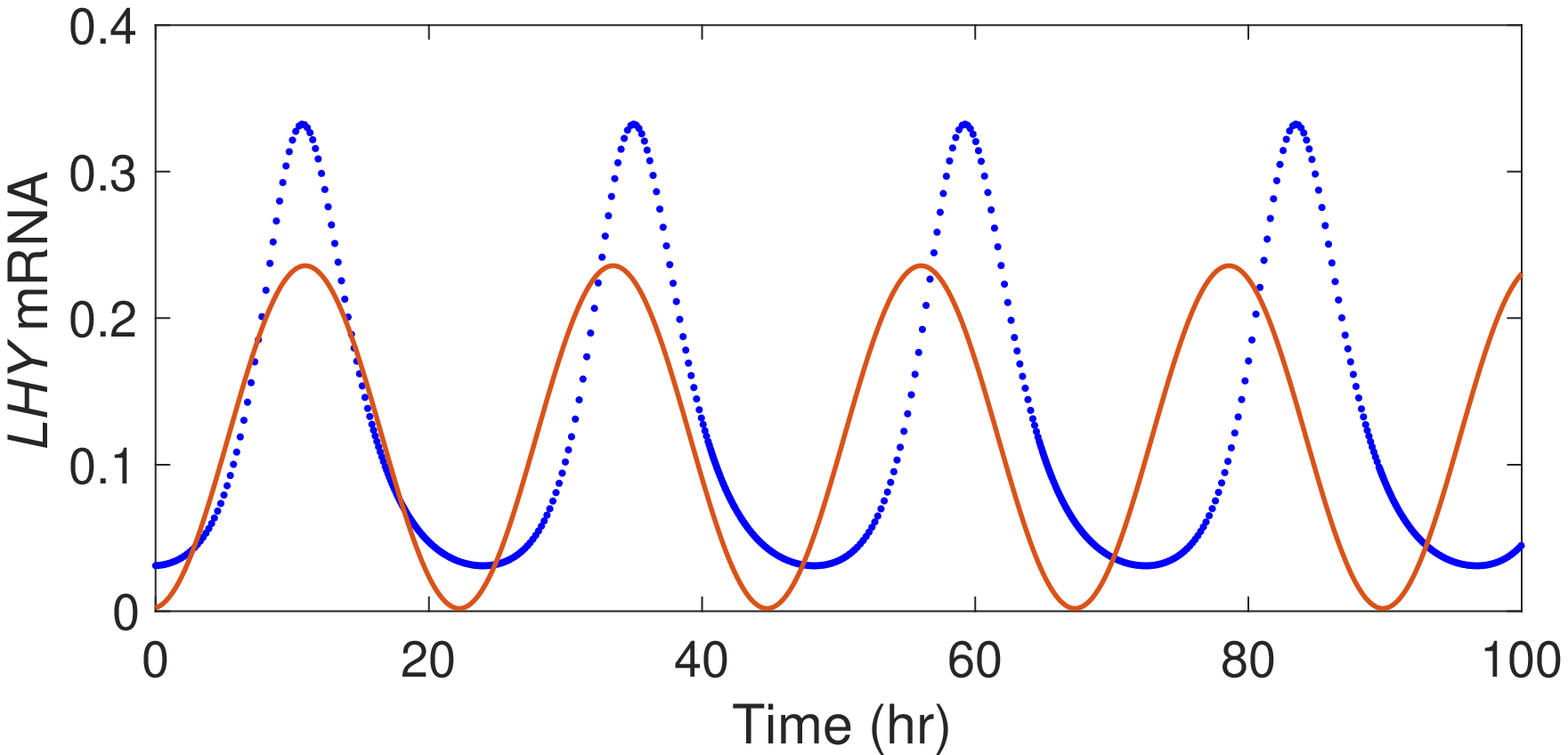}
      \caption{}
      \label{Pokhilko2012:series-LHY}
    \end{subfigure}%
    \begin{subfigure}{0.5\textwidth}
      \includegraphics[width=\textwidth]{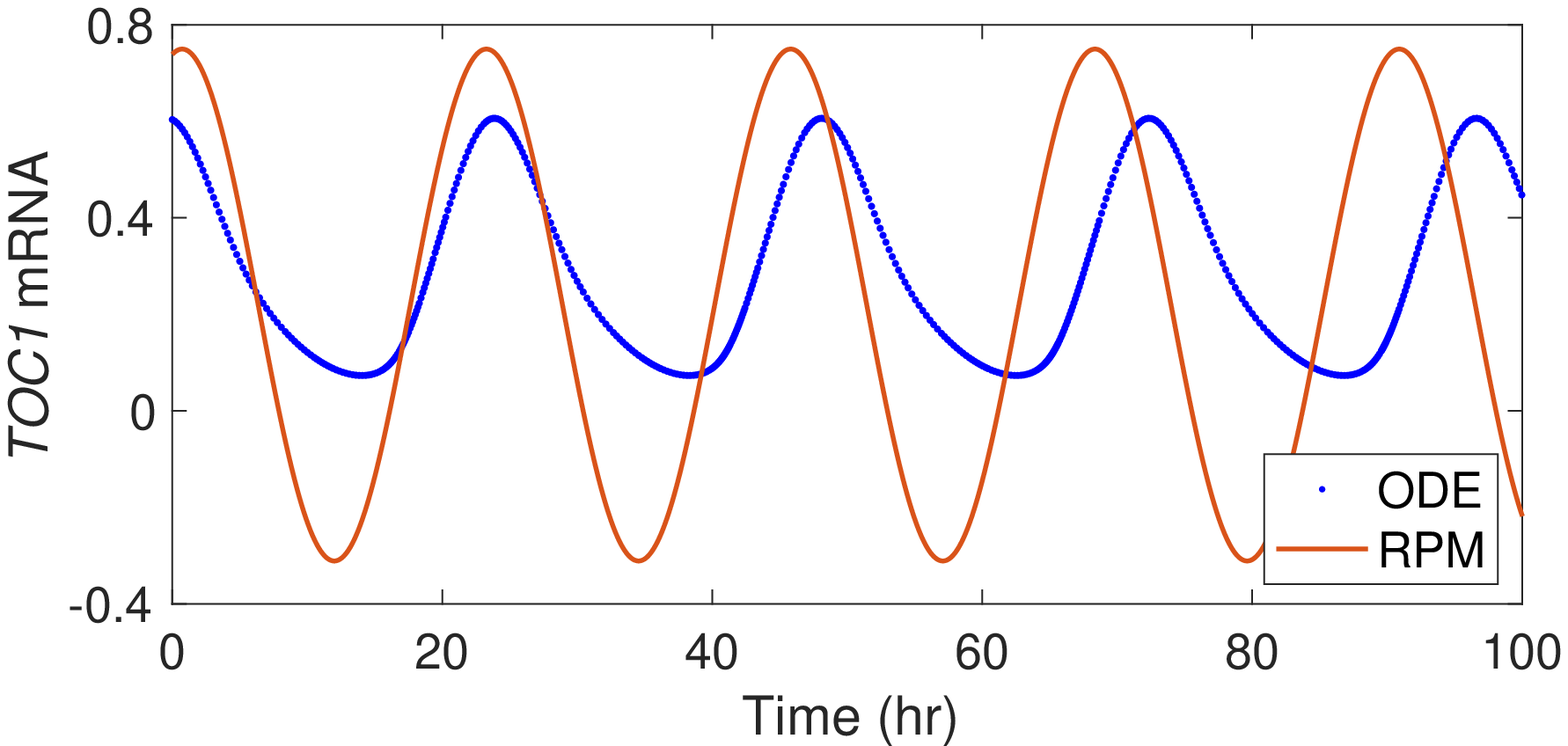}
      \caption{}
      \label{Pokhilko2012:series-TOC1}
    \end{subfigure}
    \begin{subfigure}{0.333\textwidth}
      \includegraphics[width=\textwidth]{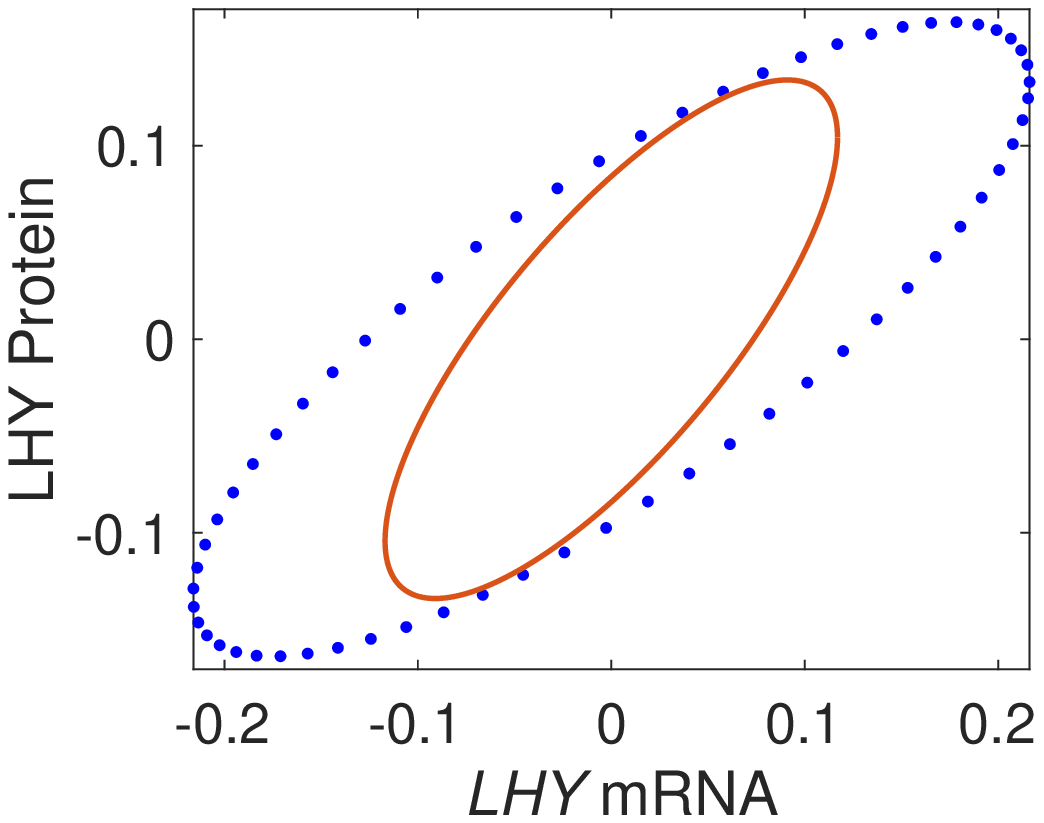}
      \caption{}
      \label{Pokhilko2012:phase-LHY}
    \end{subfigure}%
    \begin{subfigure}{0.333\textwidth}
      \includegraphics[width=\textwidth]{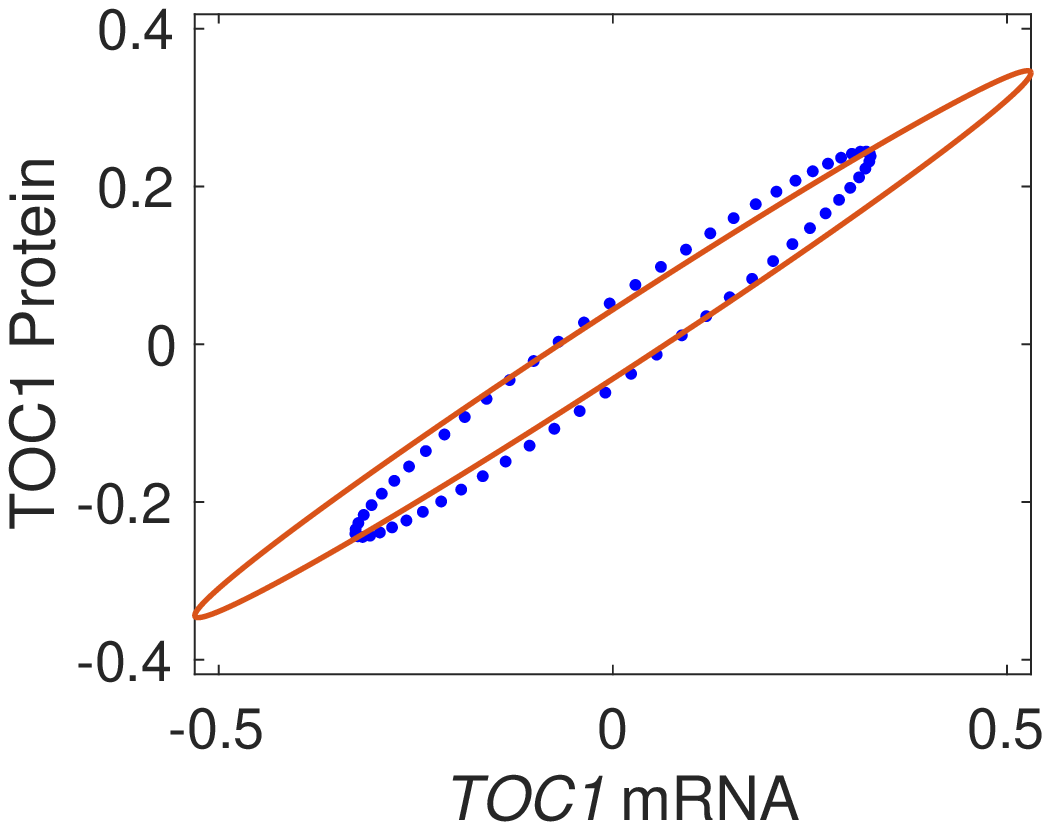}
      \caption{}
      \label{Pokhilko2012:phase-TOC1}
    \end{subfigure}%
    \begin{subfigure}{0.333\textwidth}
      \includegraphics[width=\textwidth]{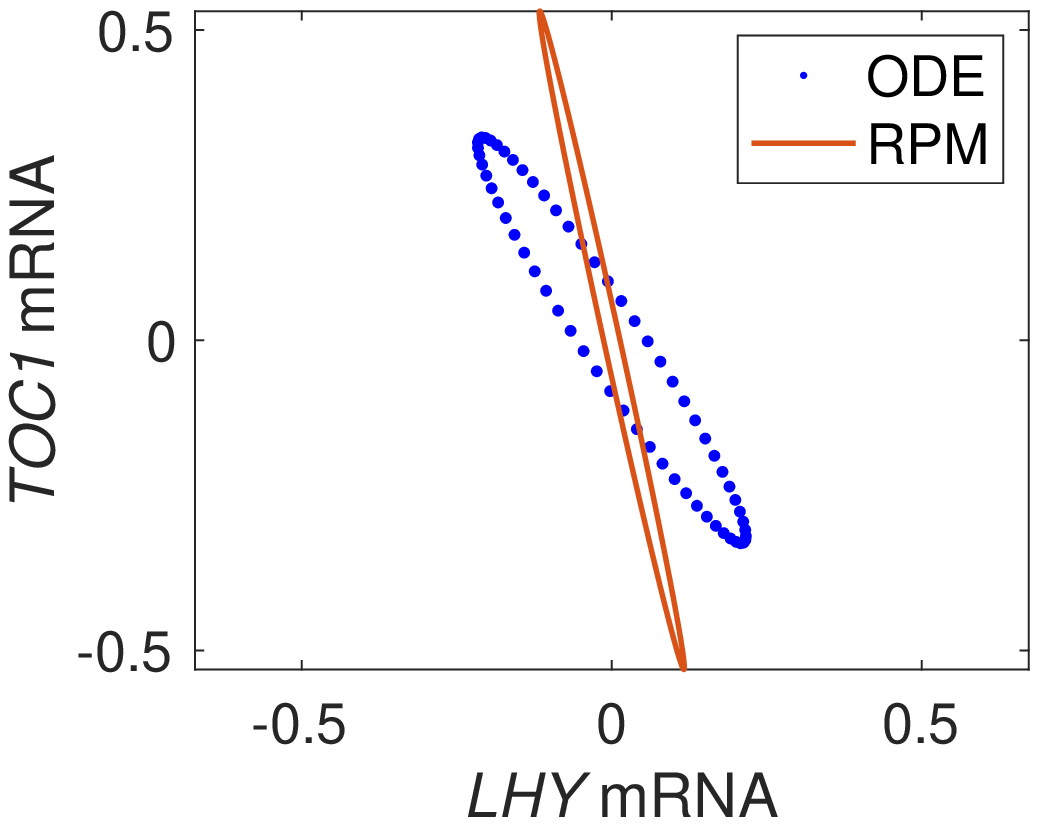}
      \caption{}
      \label{Pokhilko2012:phase-mRNA}
    \end{subfigure}
    \caption{
    A supercritical Hopf bifurcation occurs in P2012 model under perpetual illumination.  Bifurcation diagrams for (a) concentration of \textit{LHY} mRNA and (b) frequency of oscillation, time series generated from both ODE and RPM for concentrations of (c) \textit{LHY} mRNA and (d) \textit{TOC1} mRNA, and (e) - (g) phase diagrams of pairs of LHY and TOC1 protein in the cytoplasm and \textit{LHY} and \textit{TOC1} mRNA oscillations are shown.  The degradation rate in (a) and (b) are normalized so that the biological value given in the original paper is unity.  The amplitude of limit cycle oscillation calculated with RPM matches the numerical solution of the system of ODEs with 21.80 percent difference; and frequency with 7.77 percent difference. 
    As fractions of $2\pi$, the absolute values of differences in phase difference are 0.009 for the pair (\textit{LHY} mRNA, LHY protein), 0.016 for the pair (\textit{TOC1} mRNA, TOC1 protein), and 0.024 for the pair (\textit{LHY} mRNA, \textit{TOC1} mRNA).}
    \label{7-panel:Pokhilko2012}
\end{figure}

\begin{figure}[H]
    \centering
    \begin{subfigure}{0.5\textwidth}
      \includegraphics[width=\textwidth]{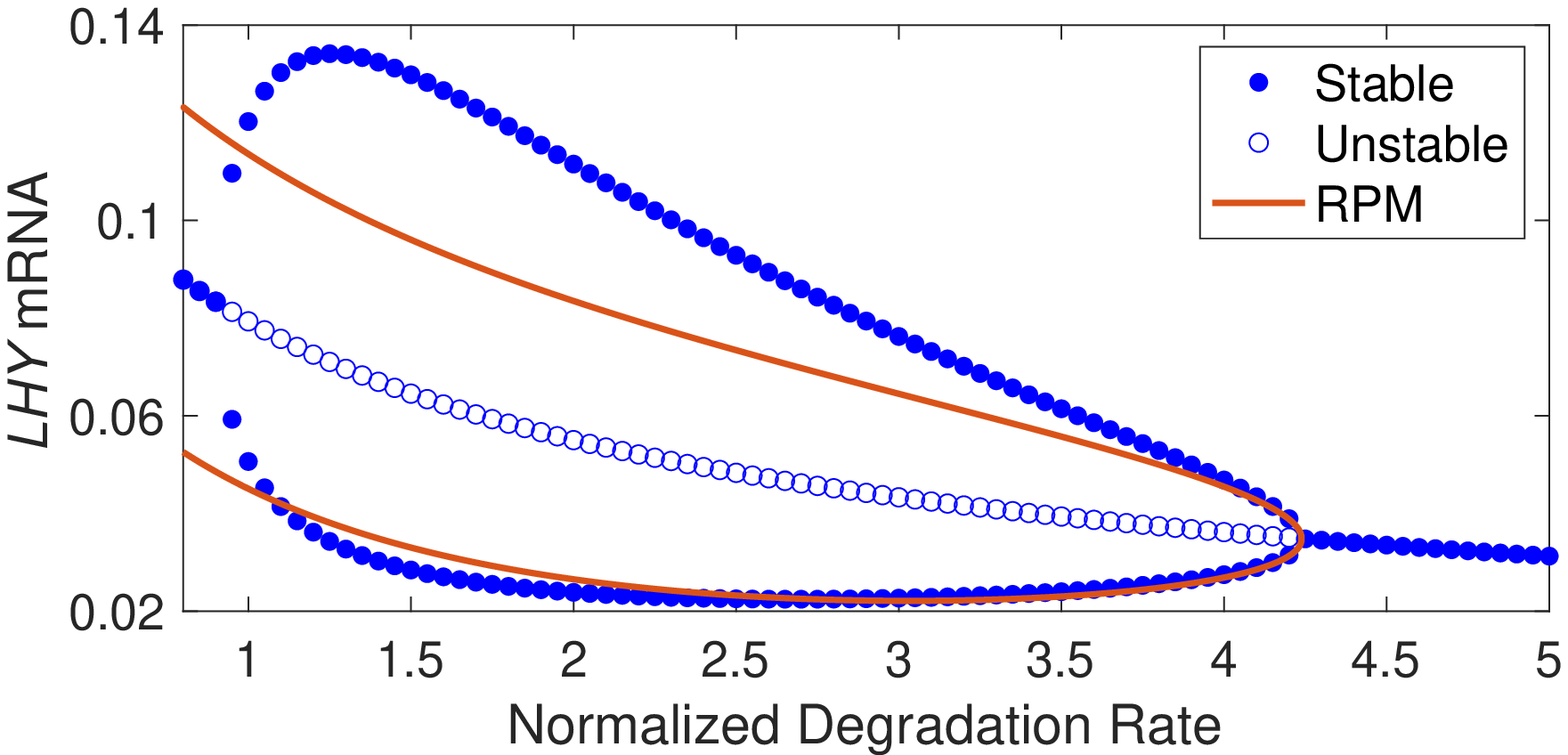}
      \caption{}
      \label{Fogelmark2014:bif-amp}
    \end{subfigure}%
    \begin{subfigure}{0.5\textwidth}
      \includegraphics[width=\textwidth]{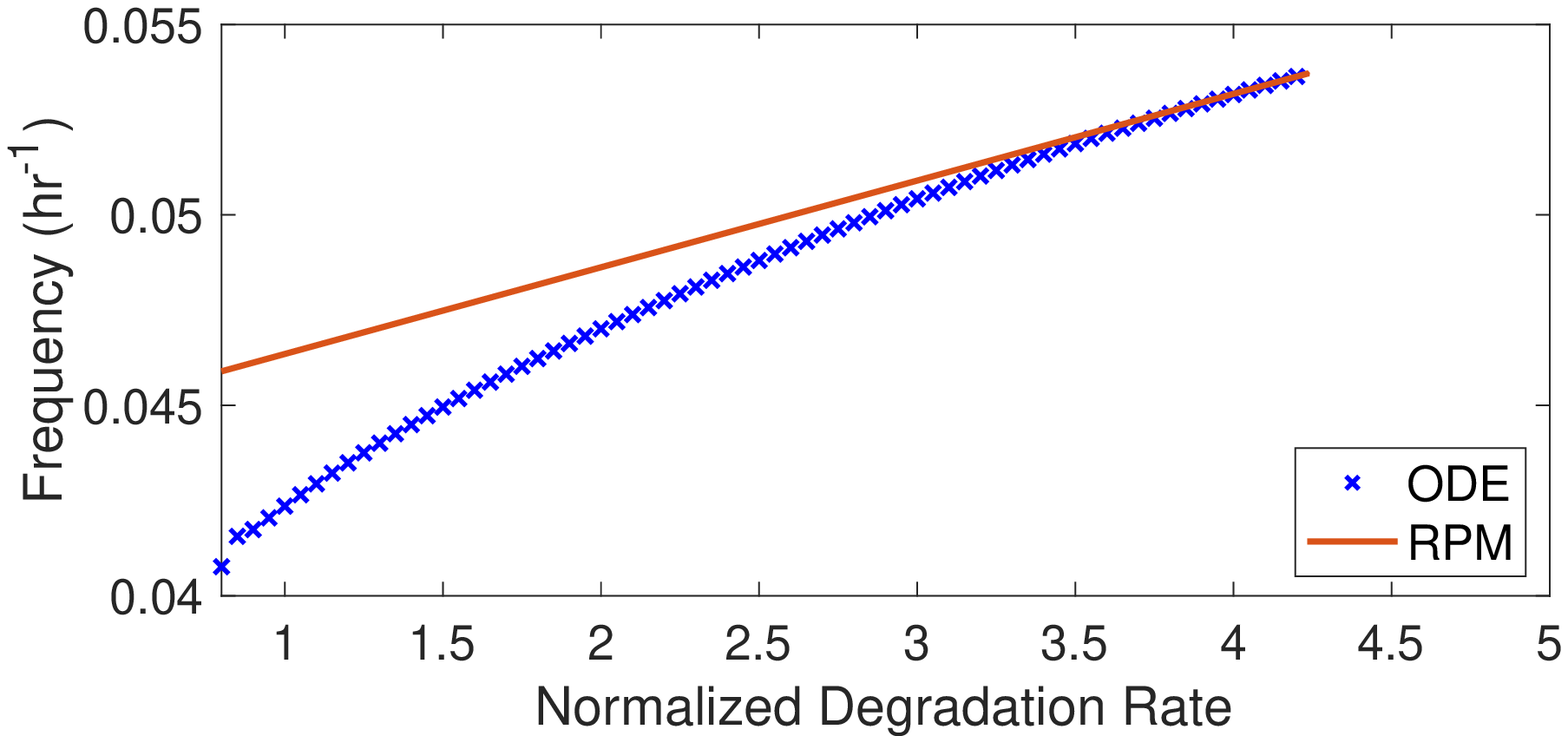}
      \caption{}
      \label{Fogelmark2014:bif-freq}
    \end{subfigure}
    \begin{subfigure}{0.5\textwidth}
      \includegraphics[width=\textwidth]{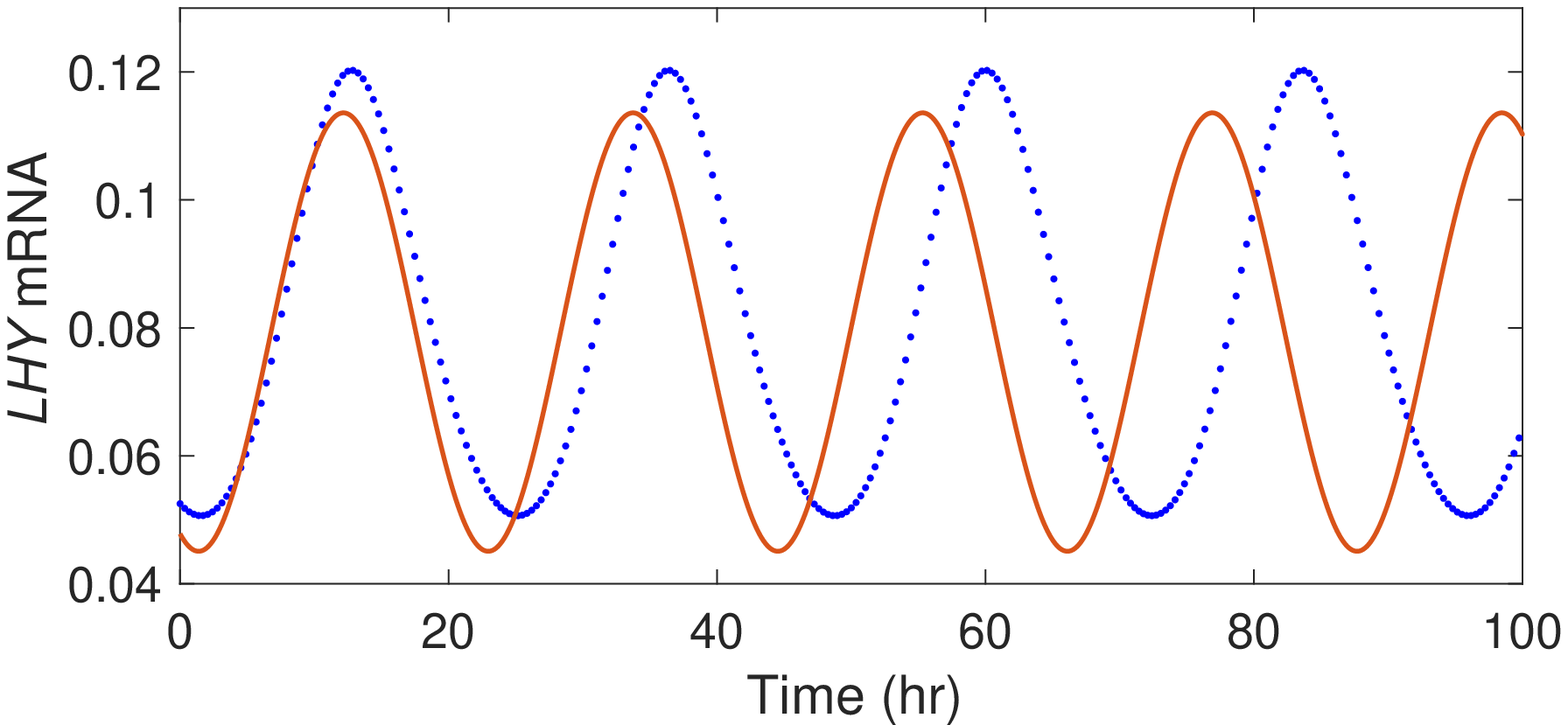}
      \caption{}
      \label{Fogelmark2014:series-LHY}
    \end{subfigure}%
    \begin{subfigure}{0.5\textwidth}
      \includegraphics[width=\textwidth]{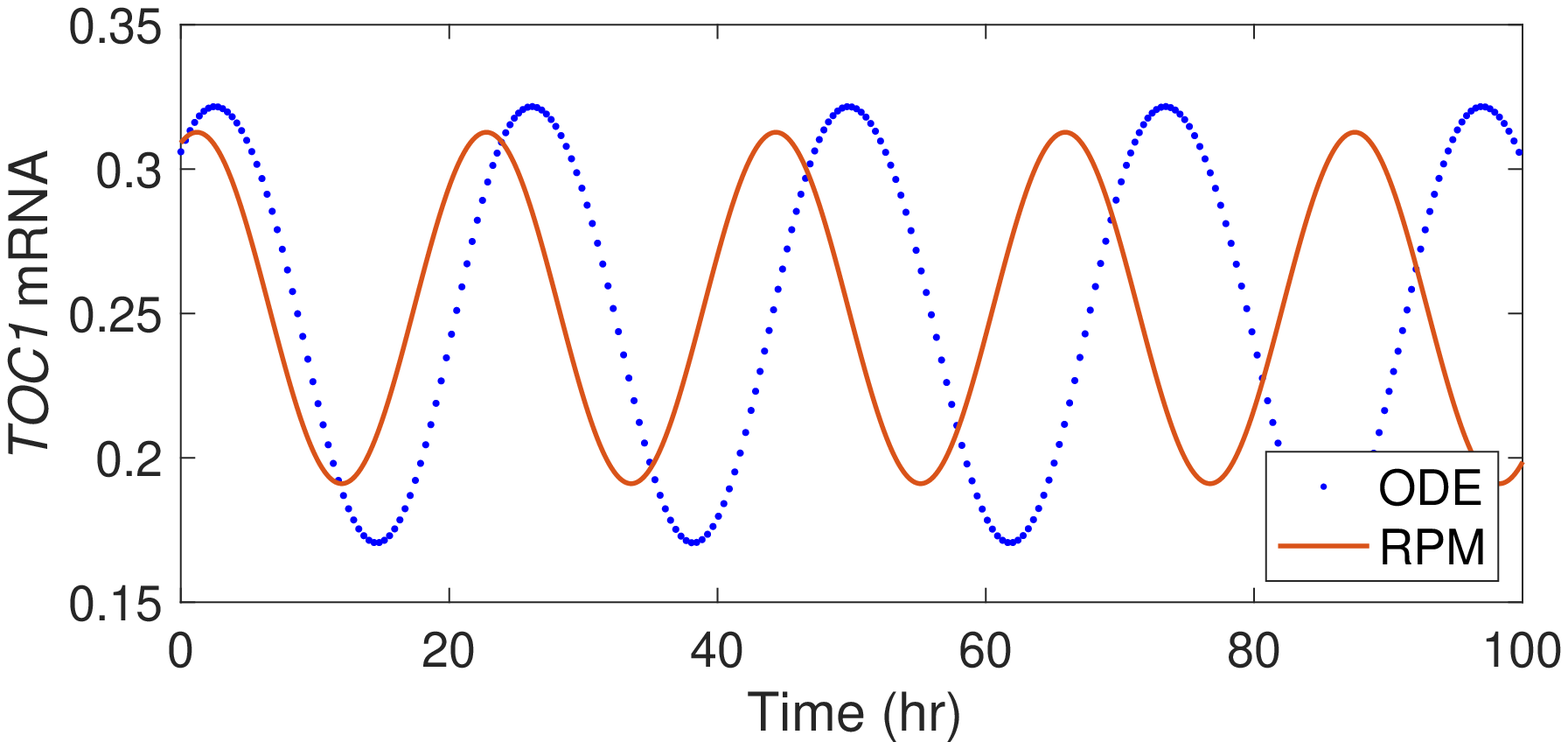}
      \caption{}
      \label{Fogelmark2014:series-TOC1}
    \end{subfigure}
    \begin{subfigure}{0.333\textwidth}
      \includegraphics[width=\textwidth]{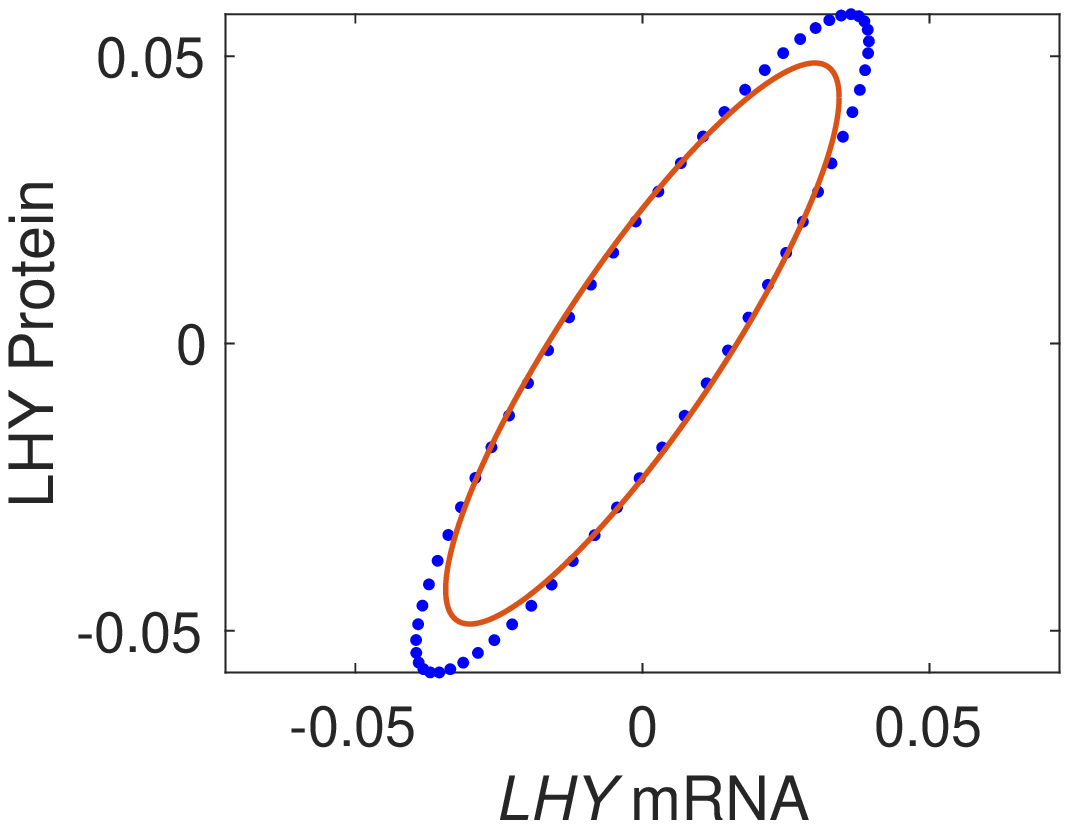}
      \caption{}
      \label{Fogelmark2014:phase-LHY}
    \end{subfigure}%
    \begin{subfigure}{0.333\textwidth}
      \includegraphics[width=\textwidth]{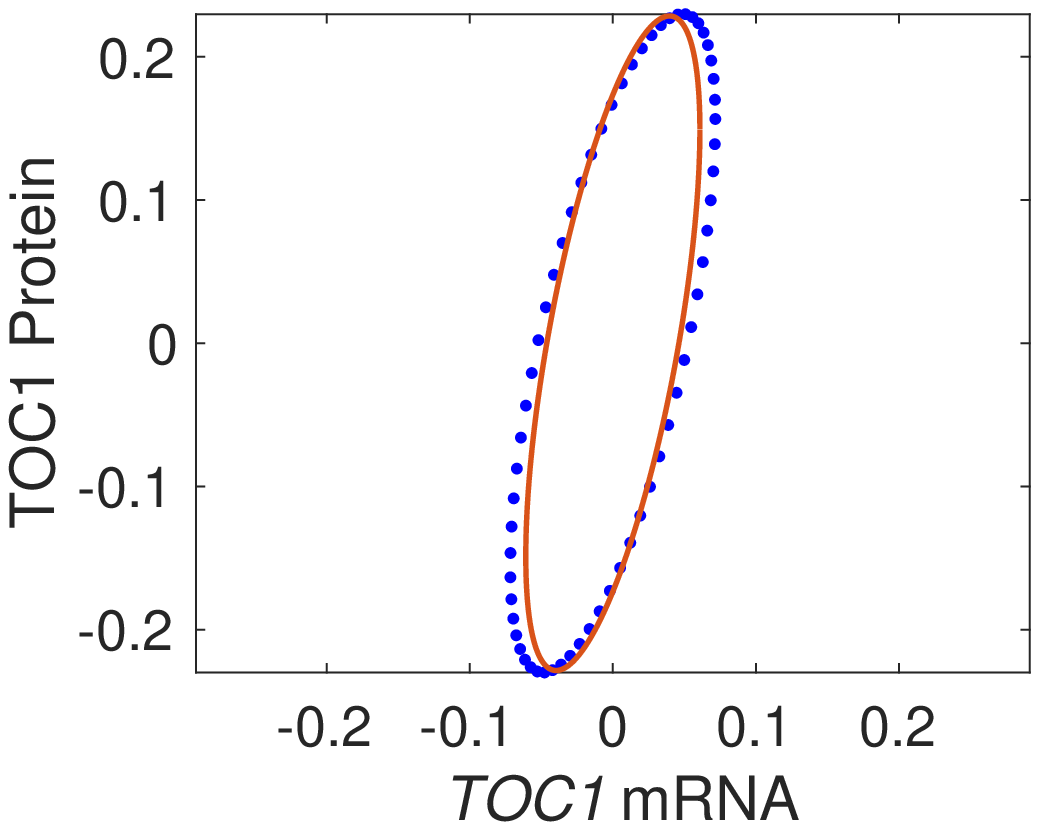}
      \caption{}
      \label{Fogelmark2014:phase-TOC1}
    \end{subfigure}%
    \begin{subfigure}{0.333\textwidth}
      \includegraphics[width=\textwidth]{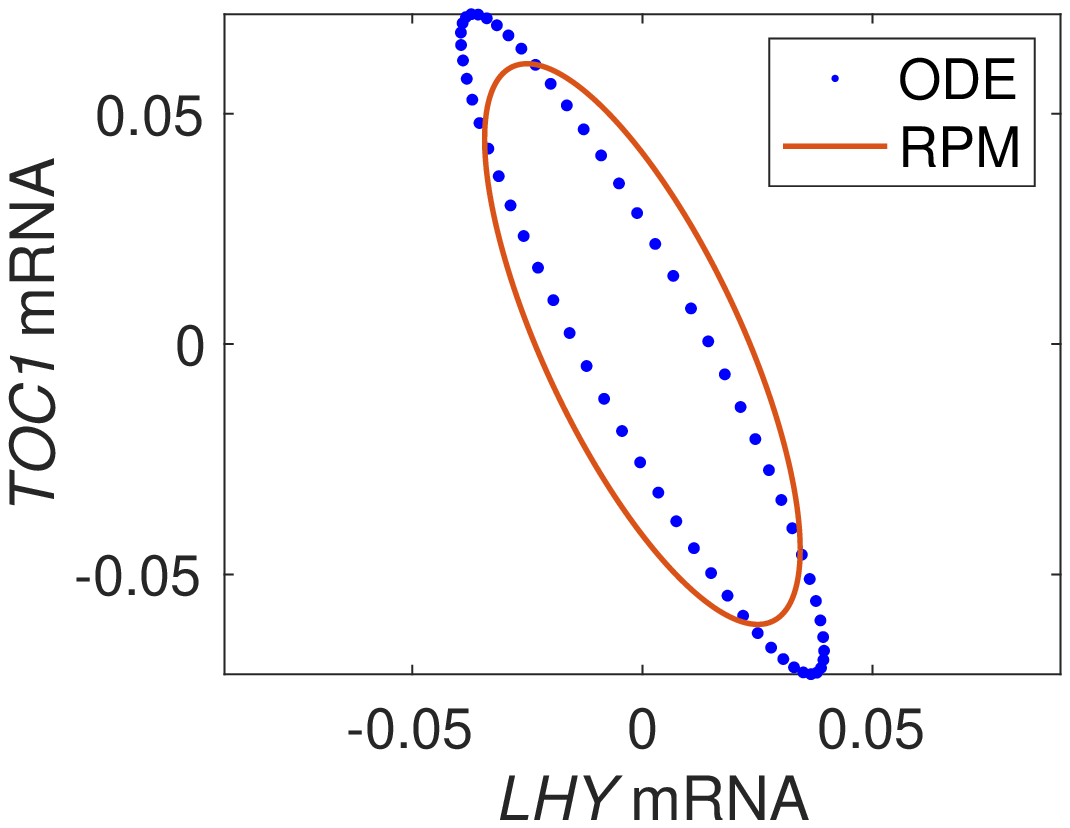}
      \caption{}
      \label{Fogelmark2014:phase-mRNA}
    \end{subfigure}
    \caption{
    A supercritical Hopf bifurcation occurs in F2014 model under perpetual illumination.  Bifurcation diagrams for (a) concentration of \textit{LHY} mRNA and (b) frequency of oscillation, time series generated from both ODE and RPM for concentrations of (c) \textit{LHY} mRNA and (d) \textit{TOC1} mRNA, and (e) - (g) phase diagrams of pairs of LHY and TOC1 protein in the cytoplasm and \textit{LHY} and \textit{TOC1} mRNA oscillations are shown.  The degradation rate in (a) and (b) are normalized so that the biological value given in the original paper is unity.  The amplitude of limit cycle oscillation calculated with RPM matches the numerical solution of the system of ODEs with 1.58 percent difference; and frequency with 9.20 percent difference.  There is a second Hopf bifurcation near a normalized degradation rate of unity, which is excluded due to our criteria as the pre-bifurcation region for this second bifurcation corresponds to lower degradation rate.  
    As fractions of $2\pi$, the absolute values of differences in phase difference are 0.015 for the pair (\textit{LHY} mRNA, LHY protein), 0.006 for the pair (\textit{TOC1} mRNA, TOC1 protein), and 0.059 for the pair (\textit{LHY} mRNA, \textit{TOC1} mRNA).}
    \label{7-panel:Fogelmark2014}
\end{figure}

\begin{figure}[H]
    \centering
    \begin{subfigure}{0.5\textwidth}
      \includegraphics[width=\textwidth]{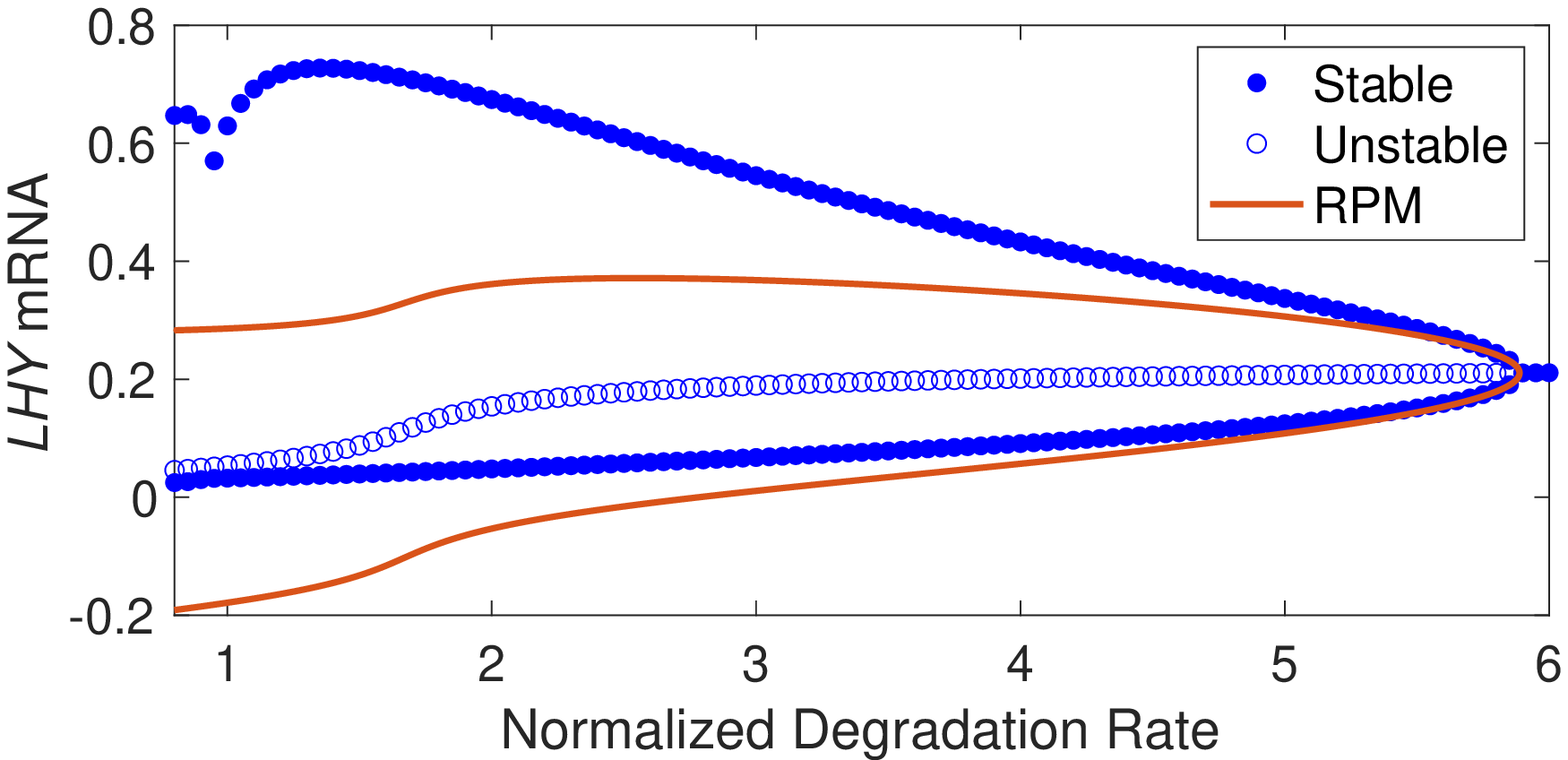}
      \caption{}
      \label{Foo2016:bif-amp}
    \end{subfigure}%
    \begin{subfigure}{0.5\textwidth}
      \includegraphics[width=\textwidth]{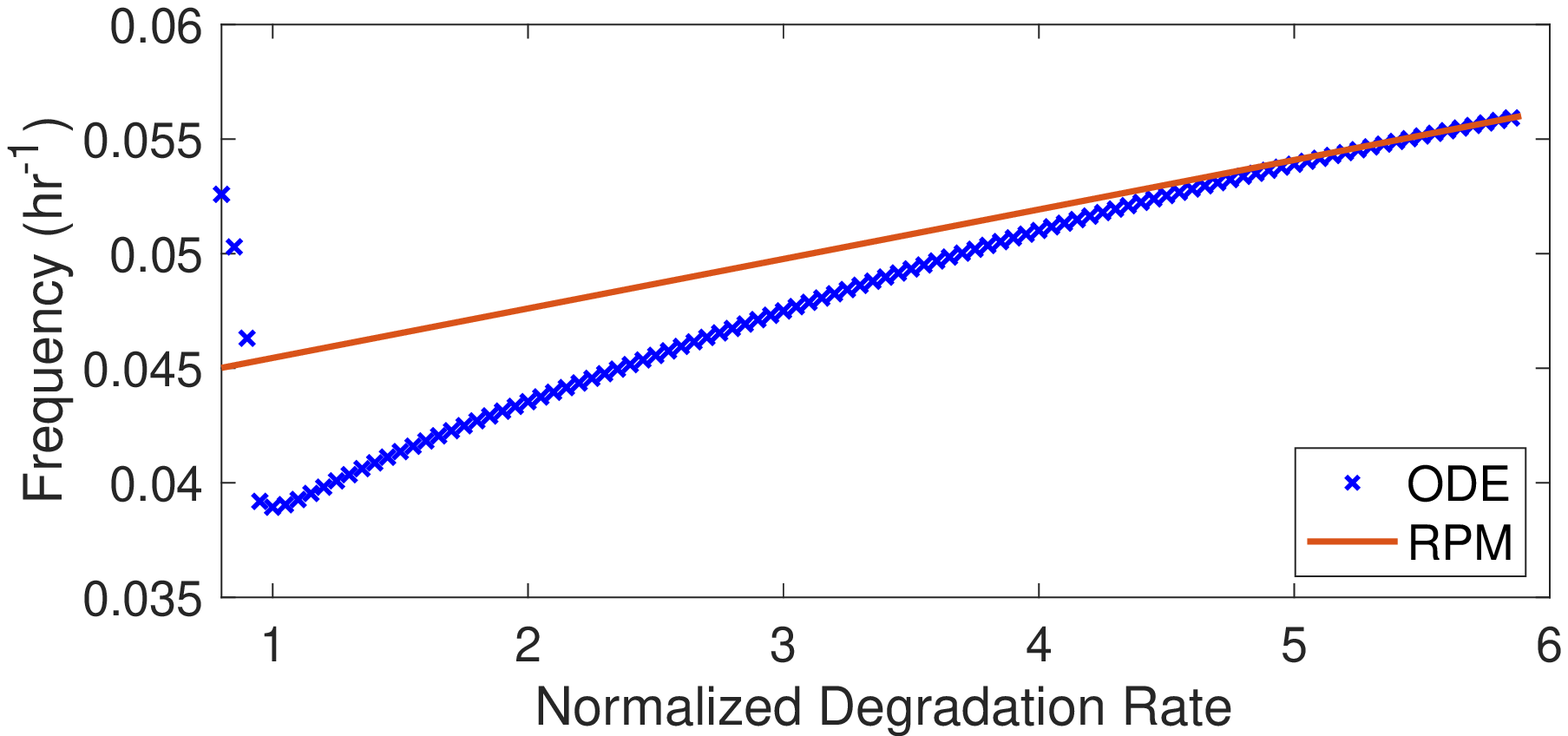}
      \caption{}
      \label{Foo2016:bif-freq}
    \end{subfigure}
    \begin{subfigure}{0.5\textwidth}
      \includegraphics[width=\textwidth]{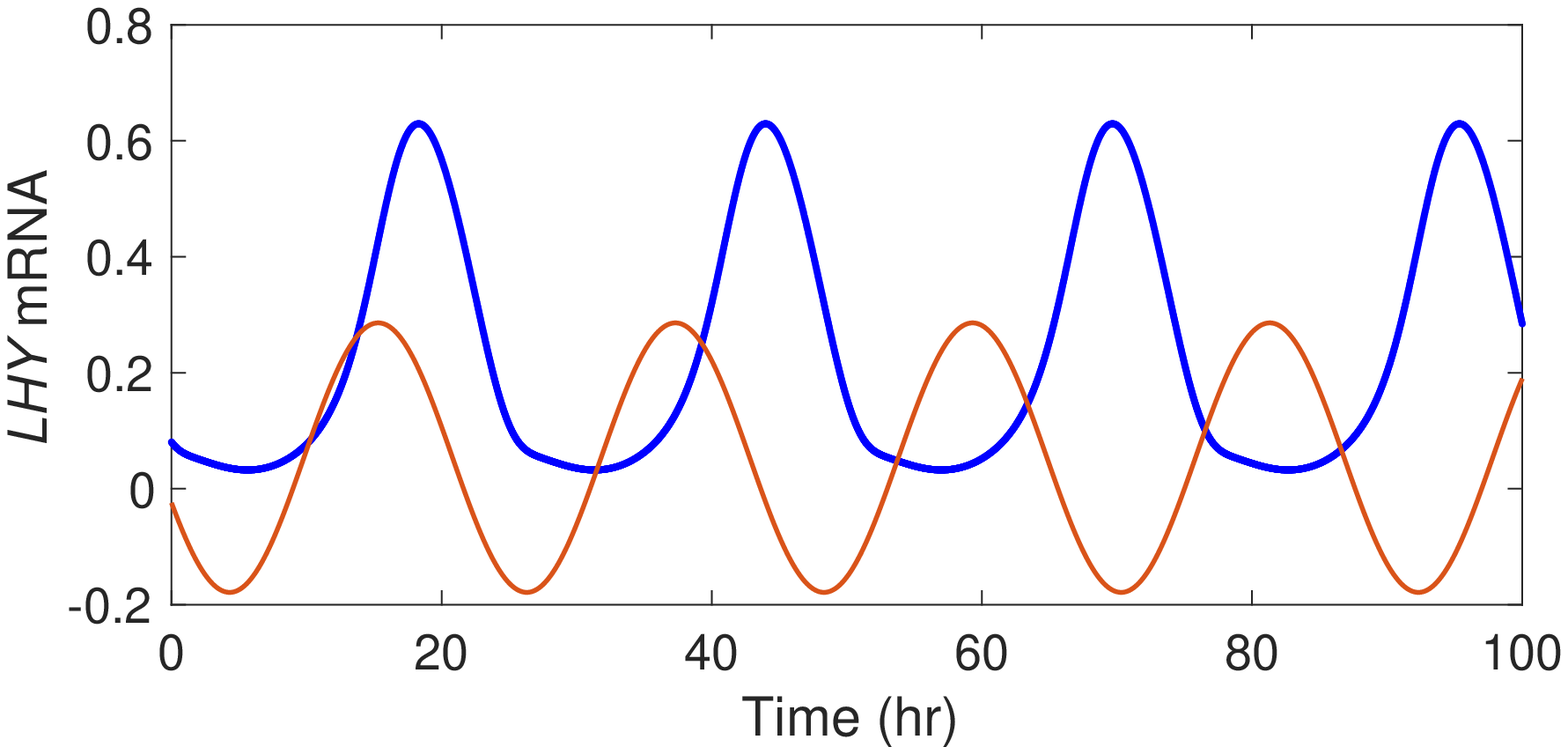}
      \caption{}
      \label{Foo2016:series-LHY}
    \end{subfigure}%
    \begin{subfigure}{0.5\textwidth}
      \includegraphics[width=\textwidth]{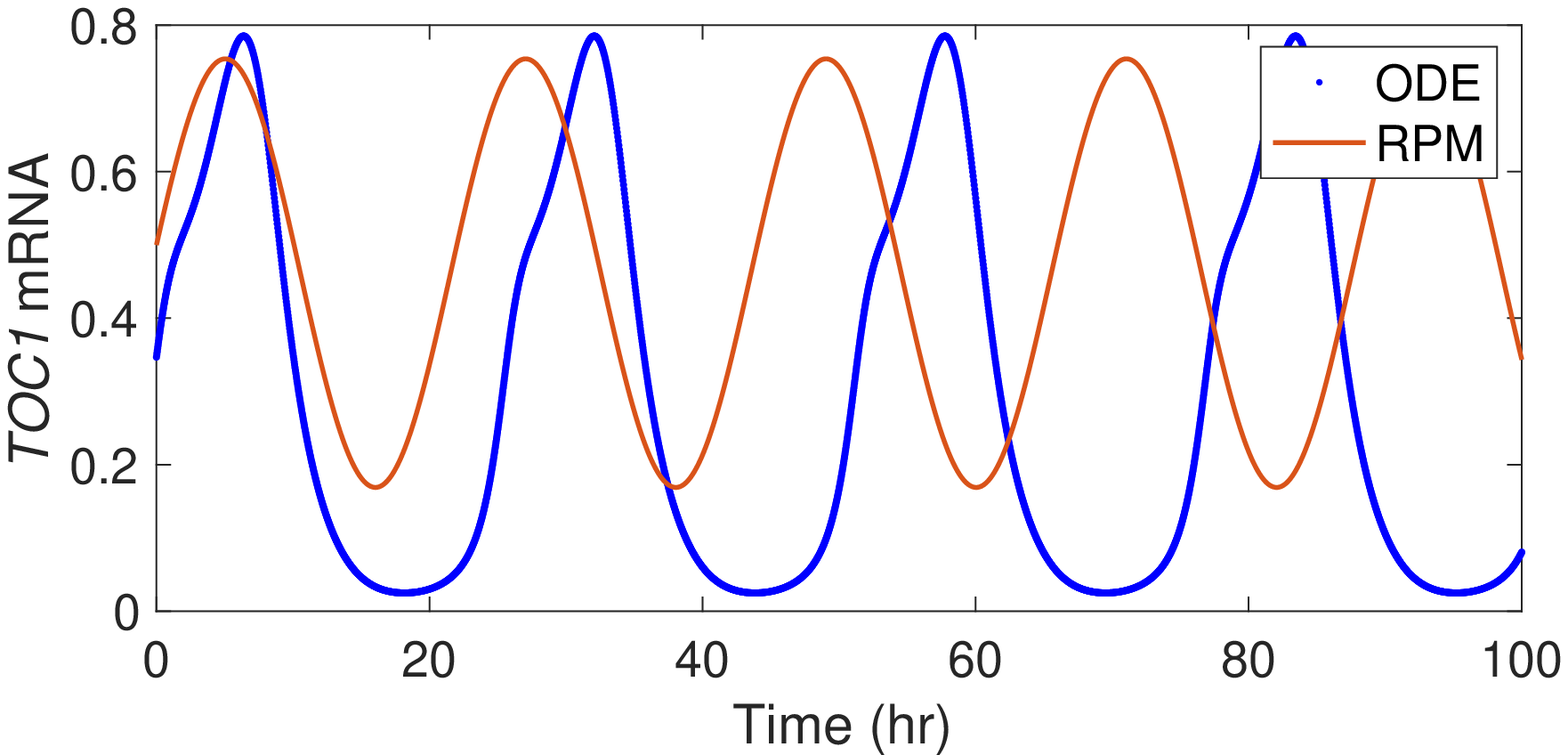}
      \caption{}
      \label{Foo2016:series-TOC1}
    \end{subfigure}
    \begin{subfigure}{0.333\textwidth}
      \includegraphics[width=\textwidth]{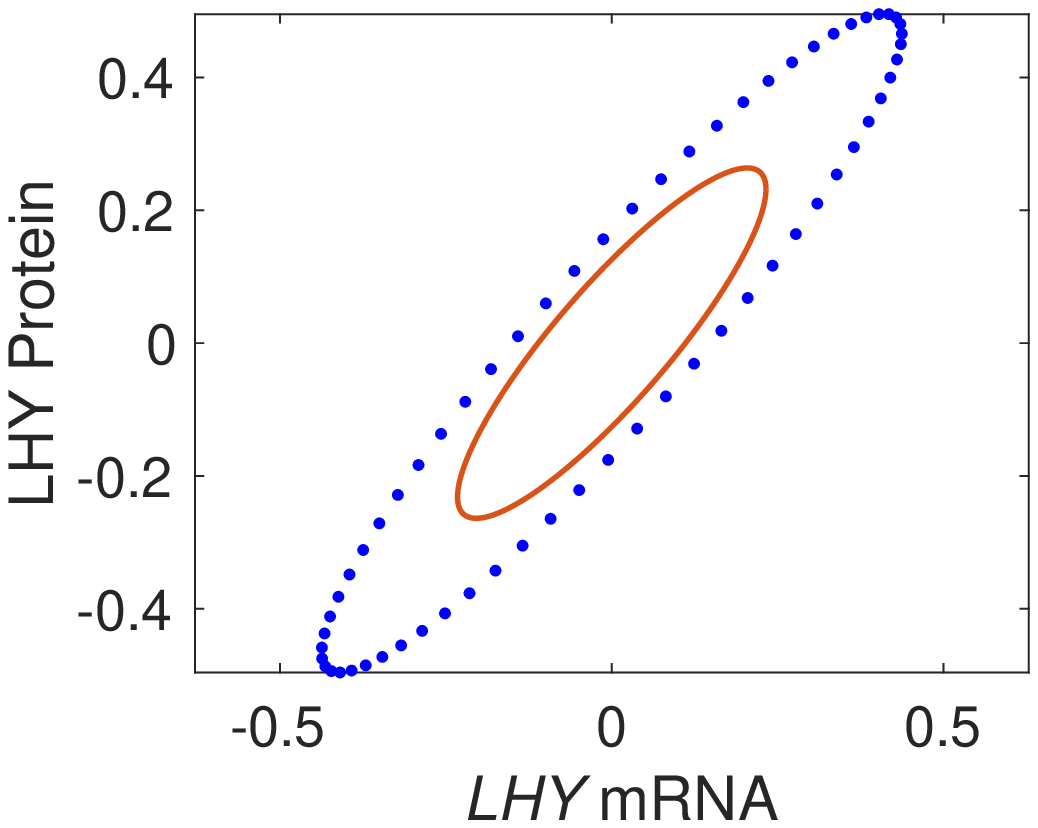}
      \caption{}
      \label{Foo2016:phase-LHY}
    \end{subfigure}%
    \begin{subfigure}{0.333\textwidth}
      \includegraphics[width=\textwidth]{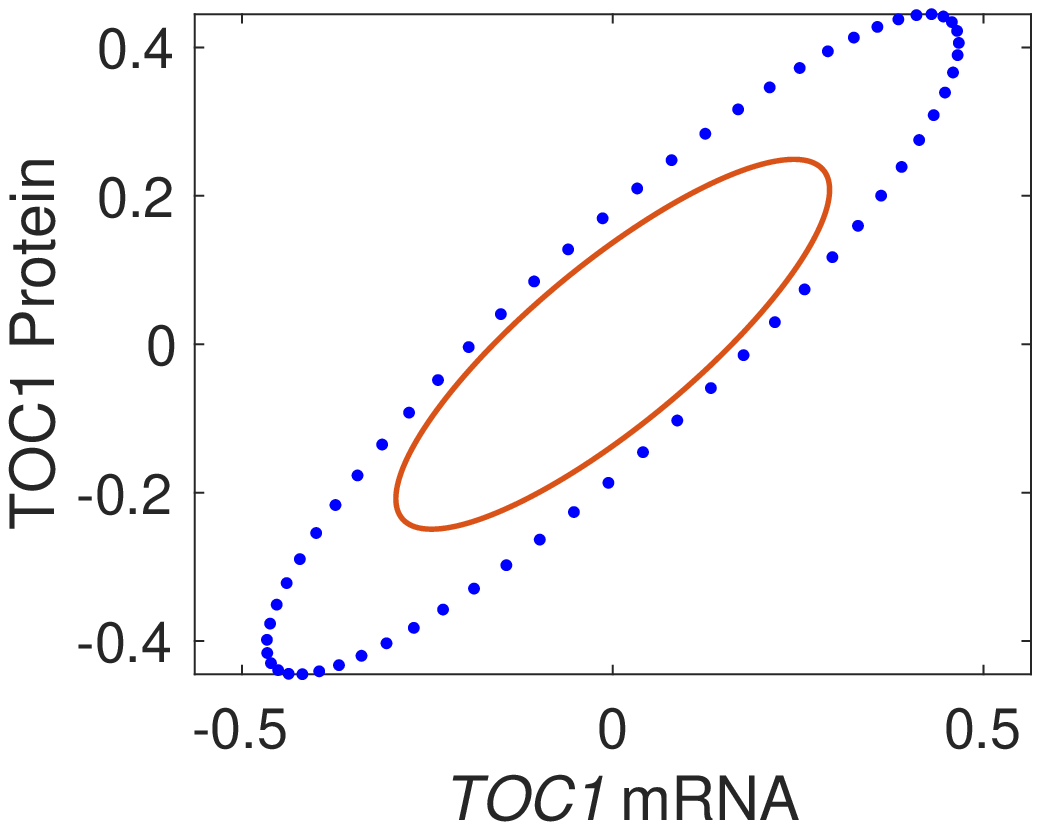}
      \caption{}
      \label{Foo2016:phase-TOC1}
    \end{subfigure}%
    \begin{subfigure}{0.333\textwidth}
      \includegraphics[width=\textwidth]{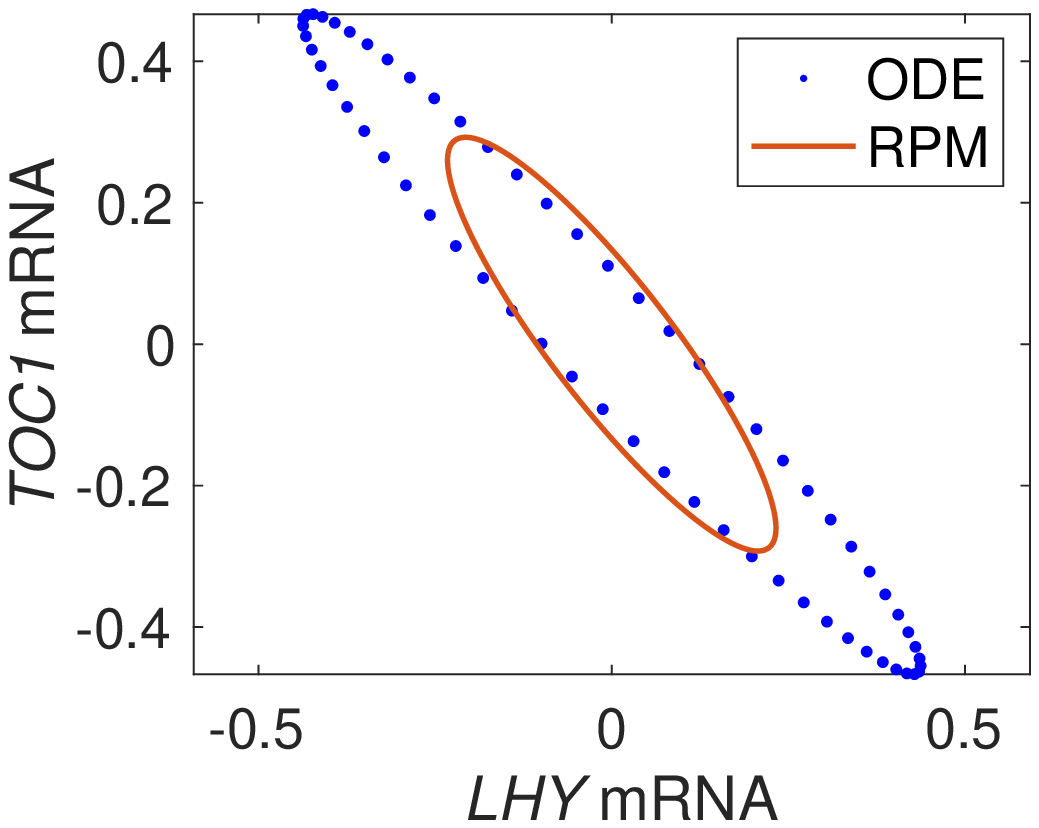}
      \caption{}
      \label{Foo2016:phase-mRNA}
    \end{subfigure}
    \caption{
    A supercritical Hopf bifurcation occurs in F2016 model under perpetual illumination.  Bifurcation diagrams for (a) concentration of \textit{LHY} mRNA and (b) frequency of oscillation, time series generated from both ODE and RPM for concentrations of (c) \textit{LHY} mRNA and (d) \textit{TOC1} mRNA, and (e) - (g) phase diagrams of pairs of LHY and TOC1 protein in the cytoplasm and \textit{LHY} and \textit{TOC1} mRNA oscillations are shown.  The degradation rate in (a) and (b) are normalized so that the biological value given in the original paper is unity.  The amplitude of limit cycle oscillation calculated with RPM matches the numerical solution of the system of ODEs with 22.17 percent difference; and frequency with 16.71 percent difference. 
    As fractions of $2\pi$, the absolute values of differences in phase difference are 0.024 for the pair (\textit{LHY} mRNA, LHY protein), 0.026 for the pair (\textit{TOC1} mRNA, TOC1 protein), and 0.039 for the pair (\textit{LHY} mRNA, \textit{TOC1} mRNA).}
    \label{7-panel:Foo2016}
\end{figure}

\begin{figure}[H]
    \centering
    \begin{subfigure}{0.5\textwidth}
      \includegraphics[width=\textwidth]{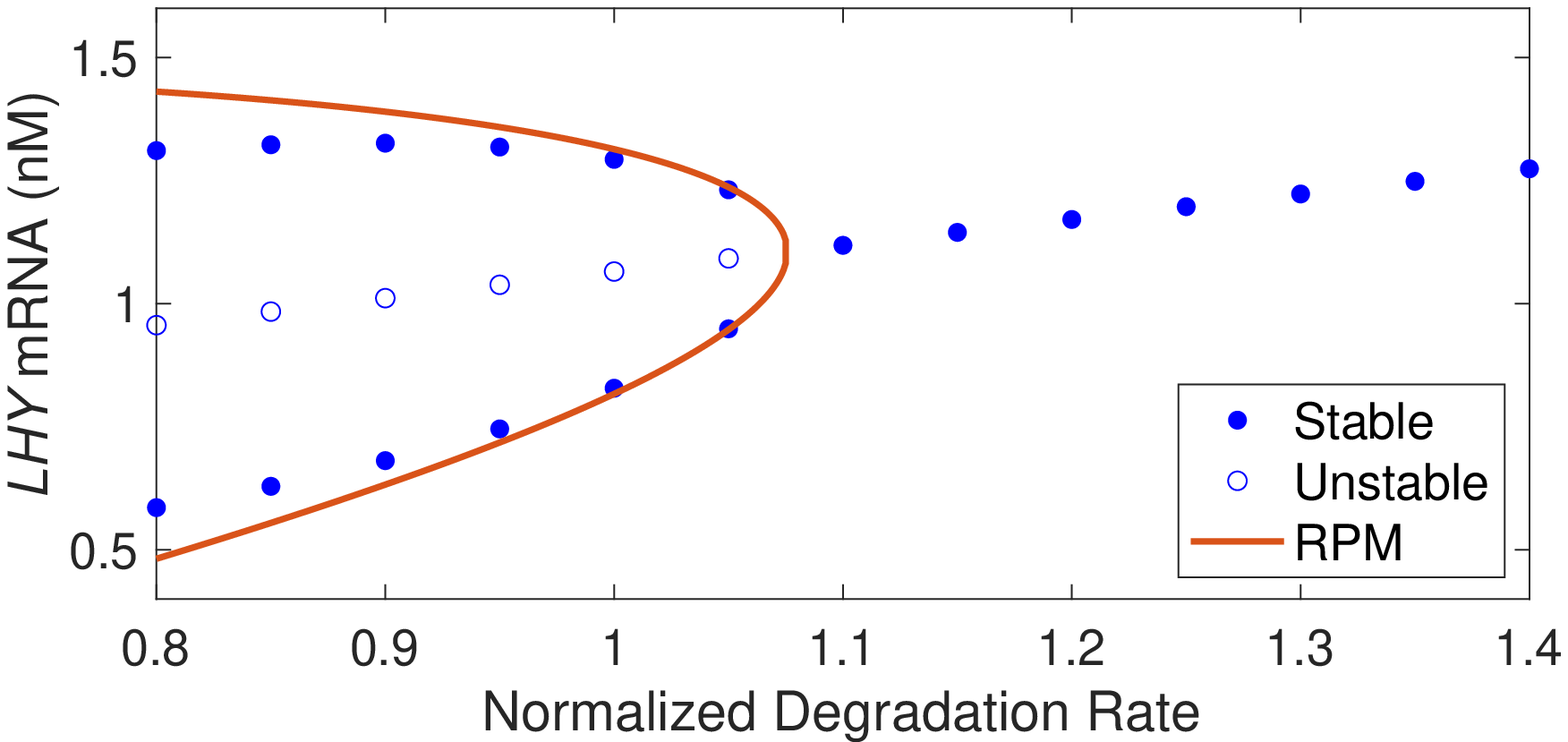}
      \caption{}
      \label{Locke2005a:bif-amp-DD}
    \end{subfigure}%
    \begin{subfigure}{0.5\textwidth}
      \includegraphics[width=\textwidth]{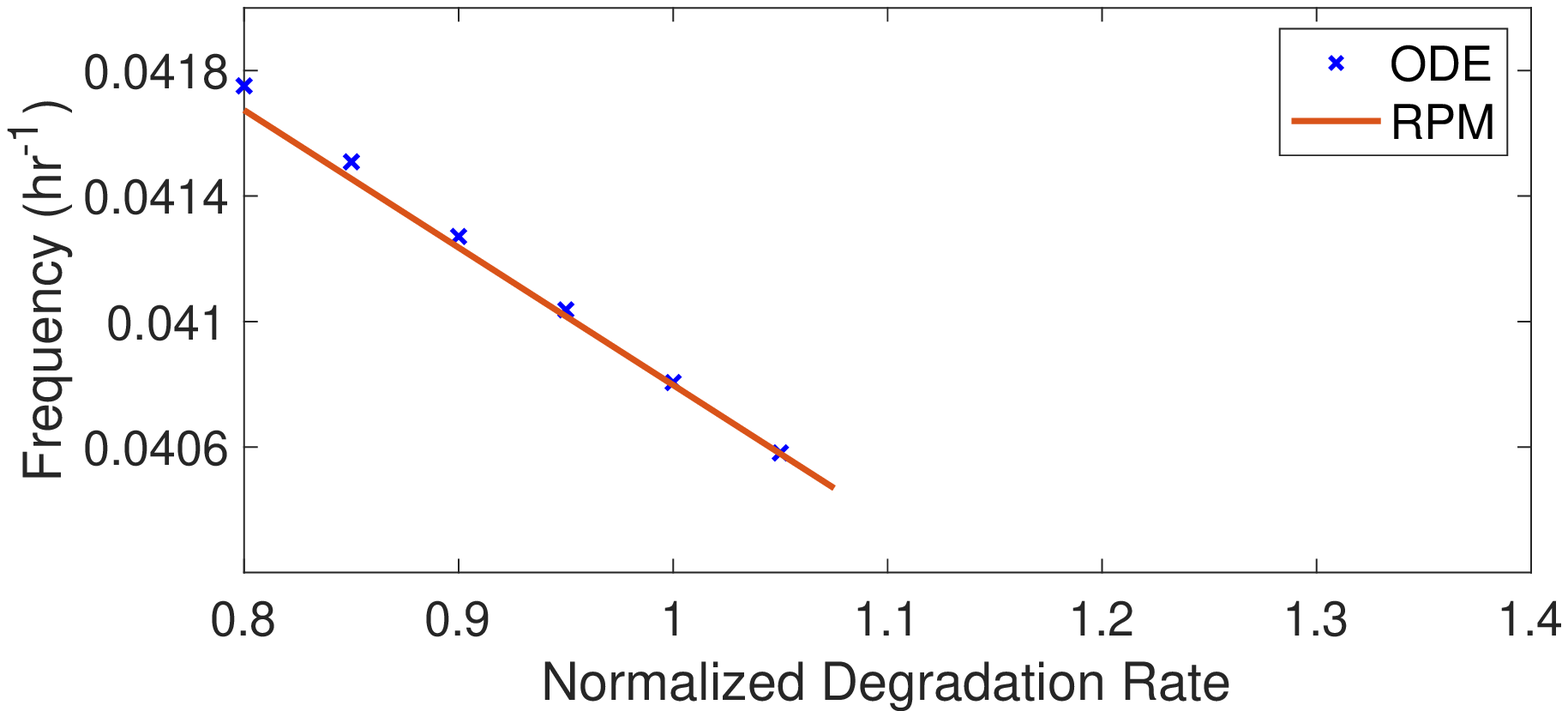}
      \caption{}
      \label{Locke2005a:bif-freq-DD}
    \end{subfigure}
    \begin{subfigure}{0.5\textwidth}
      \includegraphics[width=\textwidth]{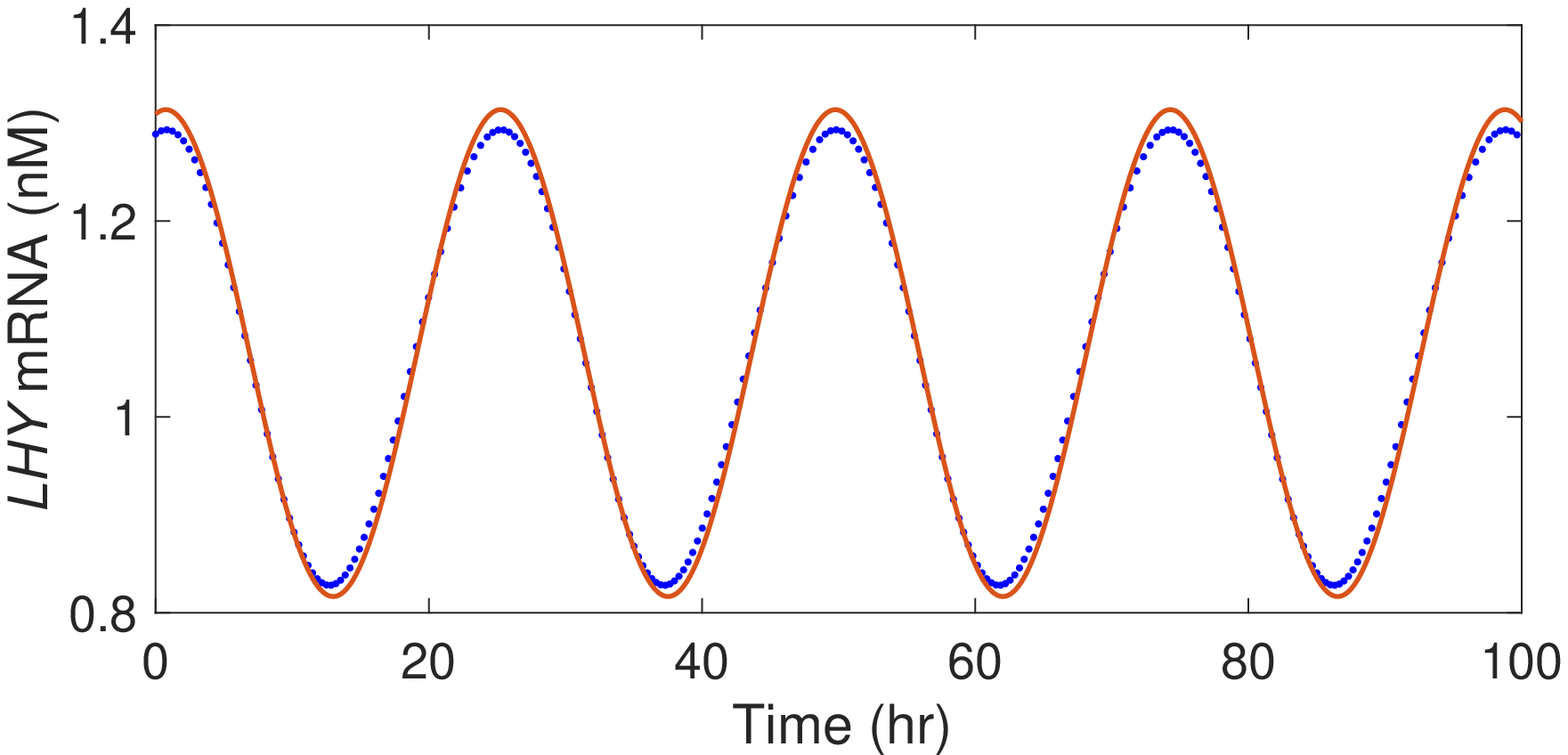}
      \caption{}
      \label{Locke2005a:series-LHY-DD}
    \end{subfigure}%
    \begin{subfigure}{0.5\textwidth}
      \includegraphics[width=\textwidth]{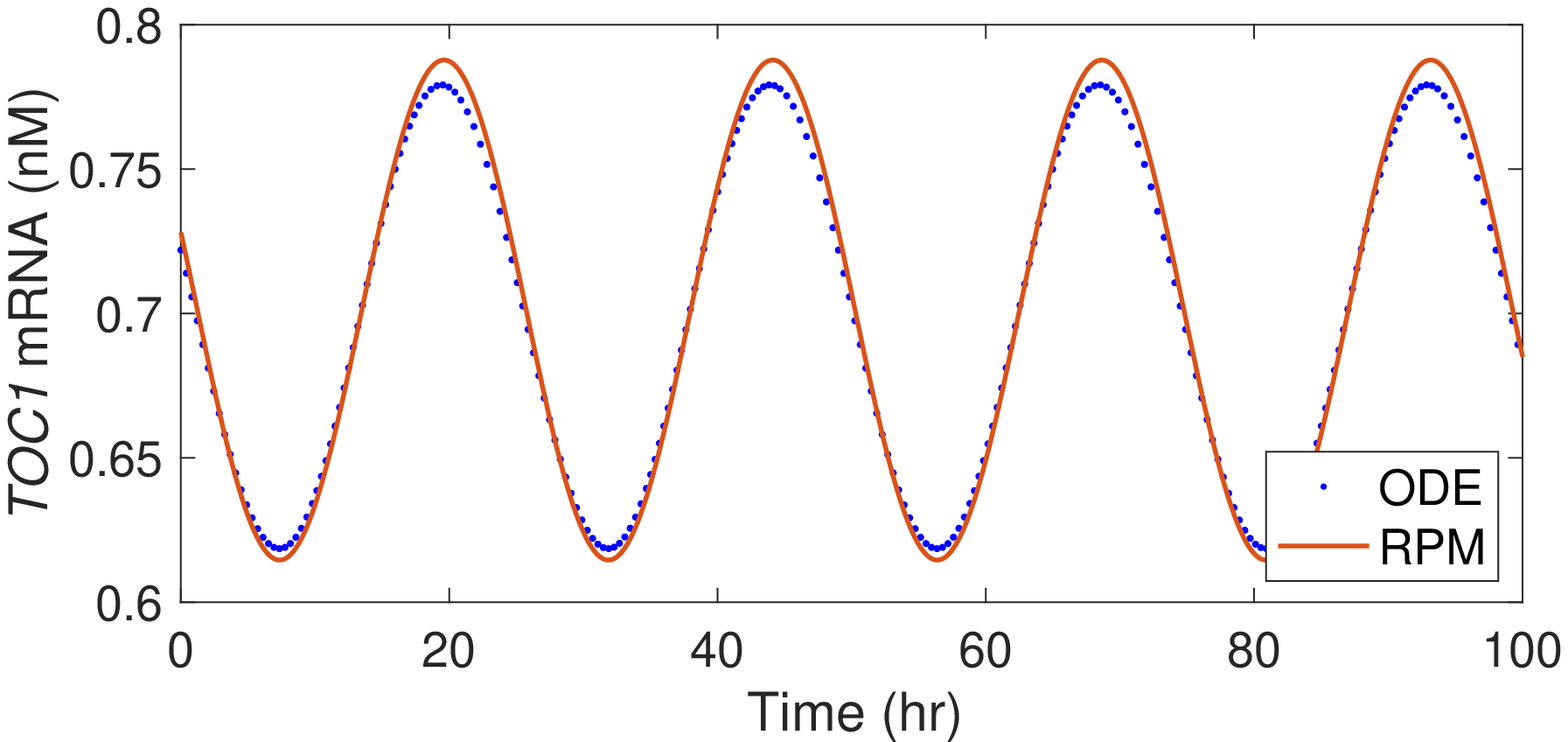}
      \caption{}
      \label{Locke2005a:series-TOC1-DD}
    \end{subfigure}
    \begin{subfigure}{0.333\textwidth}
      \includegraphics[width=\textwidth]{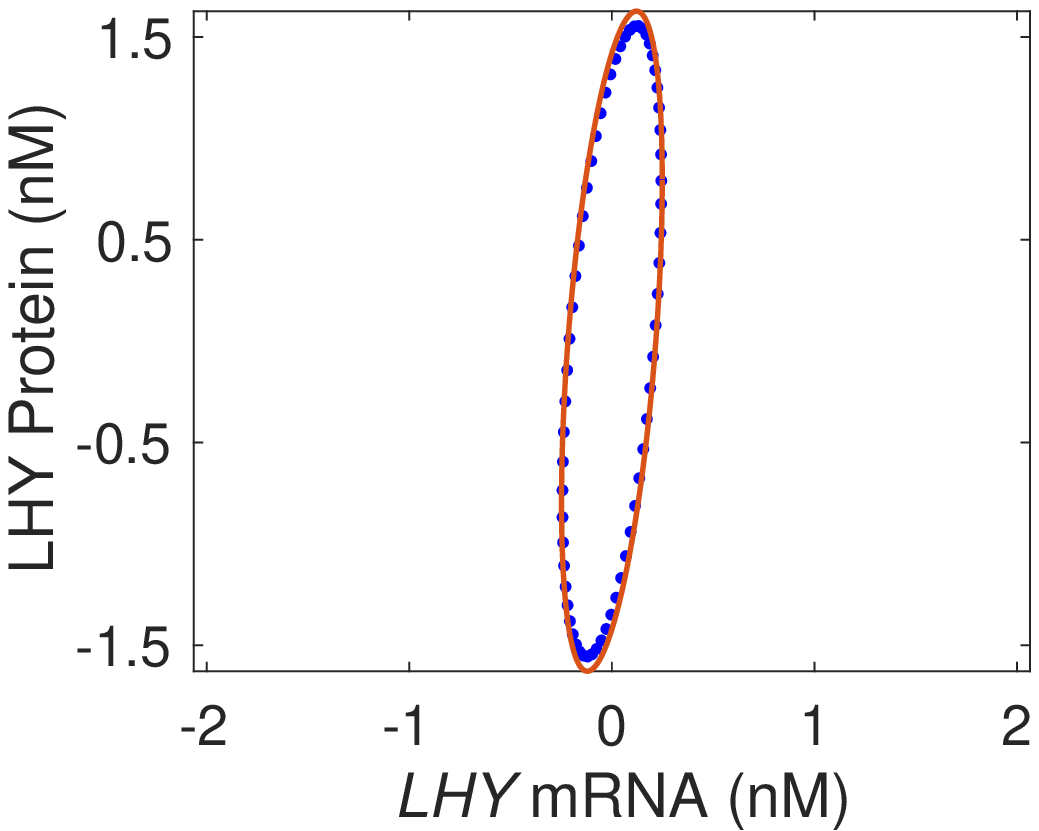}
      \caption{}
      \label{Locke2005a:phase-LHY-DD}
    \end{subfigure}%
    \begin{subfigure}{0.333\textwidth}
      \includegraphics[width=\textwidth]{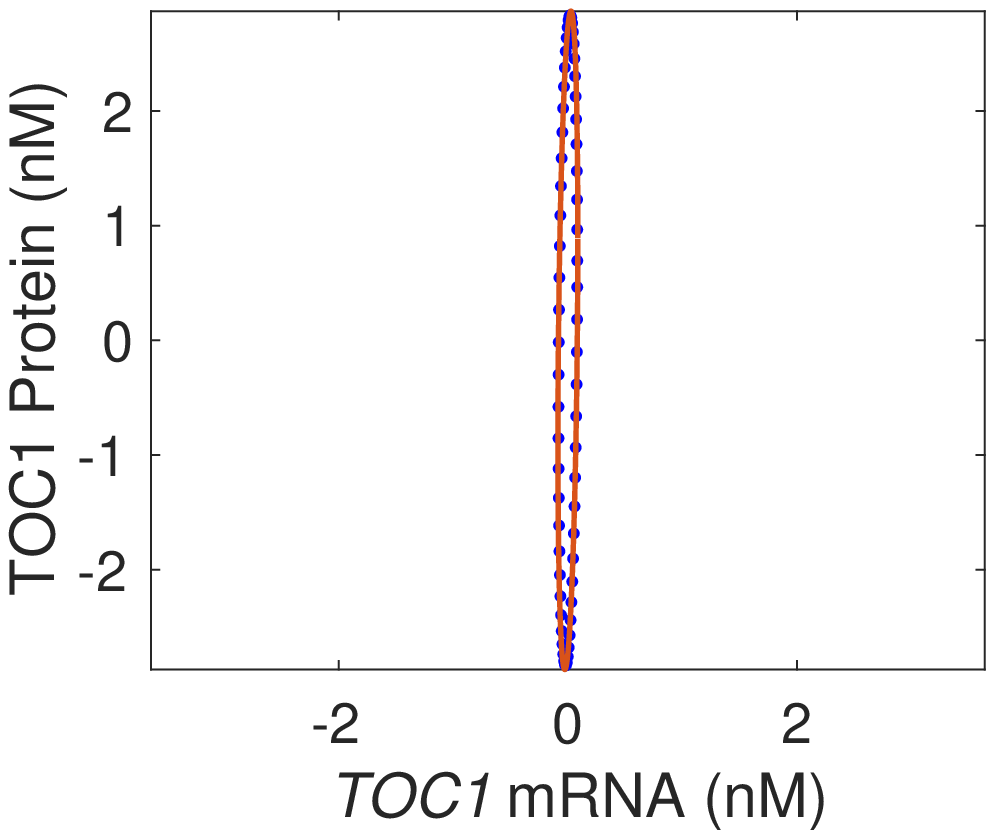}
      \caption{}
      \label{Locke2005a:phase-TOC1-DD}
    \end{subfigure}%
    \begin{subfigure}{0.333\textwidth}
      \includegraphics[width=\textwidth]{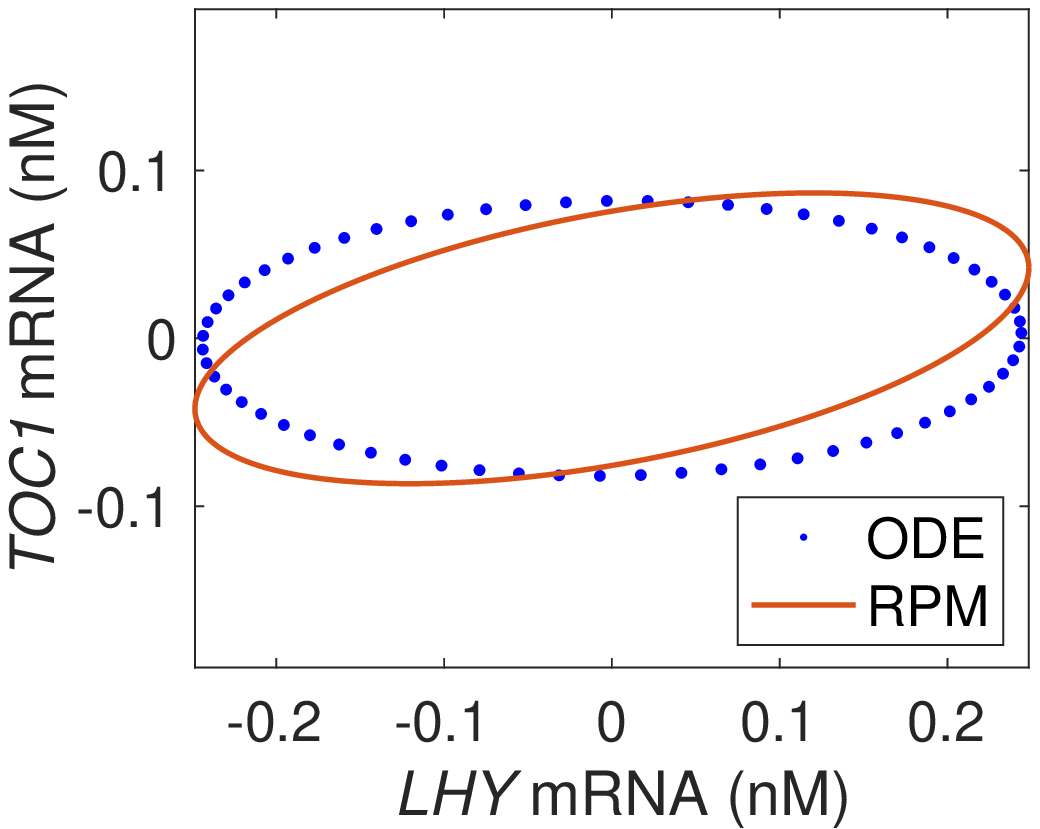}
      \caption{}
      \label{Locke2005a:phase-mRNA-DD}
    \end{subfigure}
    \caption{
    A supercritical Hopf bifurcation occurs in L2005a model under perpetual darkness.  Bifurcation diagrams for (a) concentration of \textit{LHY} mRNA and (b) frequency of oscillation, time series generated from both ODE and RPM for concentrations of (c) \textit{LHY} mRNA and (d) \textit{TOC1} mRNA, and (e) - (g) phase diagrams of pairs of LHY and TOC1 protein in the cytoplasm and \textit{LHY} and \textit{TOC1} mRNA oscillations are shown.  The degradation rate in (a) and (b) are normalized so that the biological value given in the original paper is unity.  The amplitude of limit cycle oscillation calculated with RPM matches the numerical solution of the system of ODEs with 14.78 percent difference; and frequency with 0.02 percent difference. 
    As fractions of $2\pi$, the absolute values of differences in phase difference are 0.004 for the pair (\textit{LHY} mRNA, LHY protein), 0.011 for the pair (\textit{TOC1} mRNA, TOC1 protein), and 0.074 for the pair (\textit{LHY} mRNA, \textit{TOC1} mRNA).}
    \label{7-panel:Locke2005a-DD}
\end{figure}

\begin{figure}[H]
    \centering
    \begin{subfigure}{0.5\textwidth}
      \includegraphics[width=\textwidth]{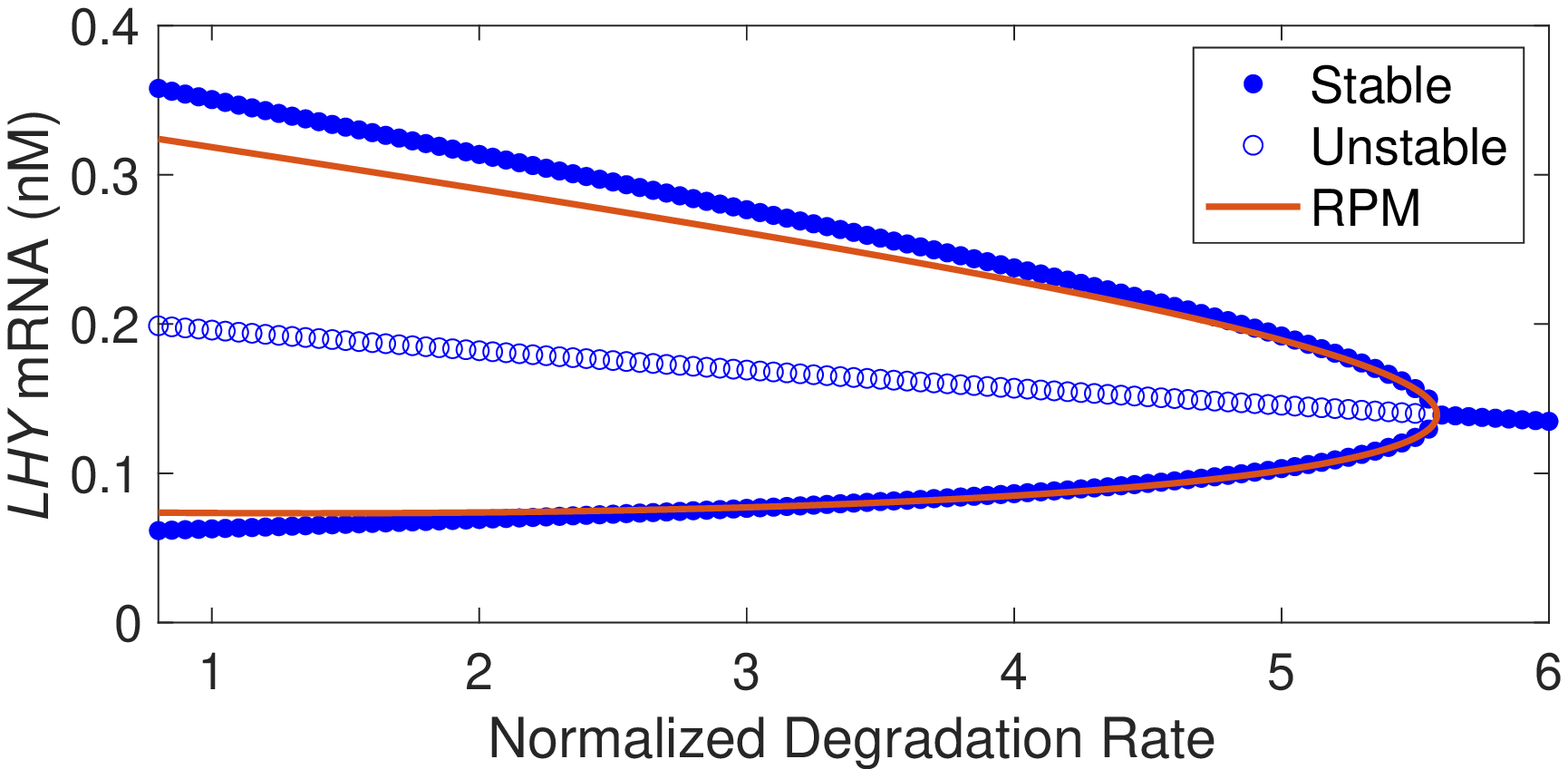}
      \caption{}
      \label{Locke2005b:bif-amp-DD}
    \end{subfigure}%
    \begin{subfigure}{0.5\textwidth}
      \includegraphics[width=\textwidth]{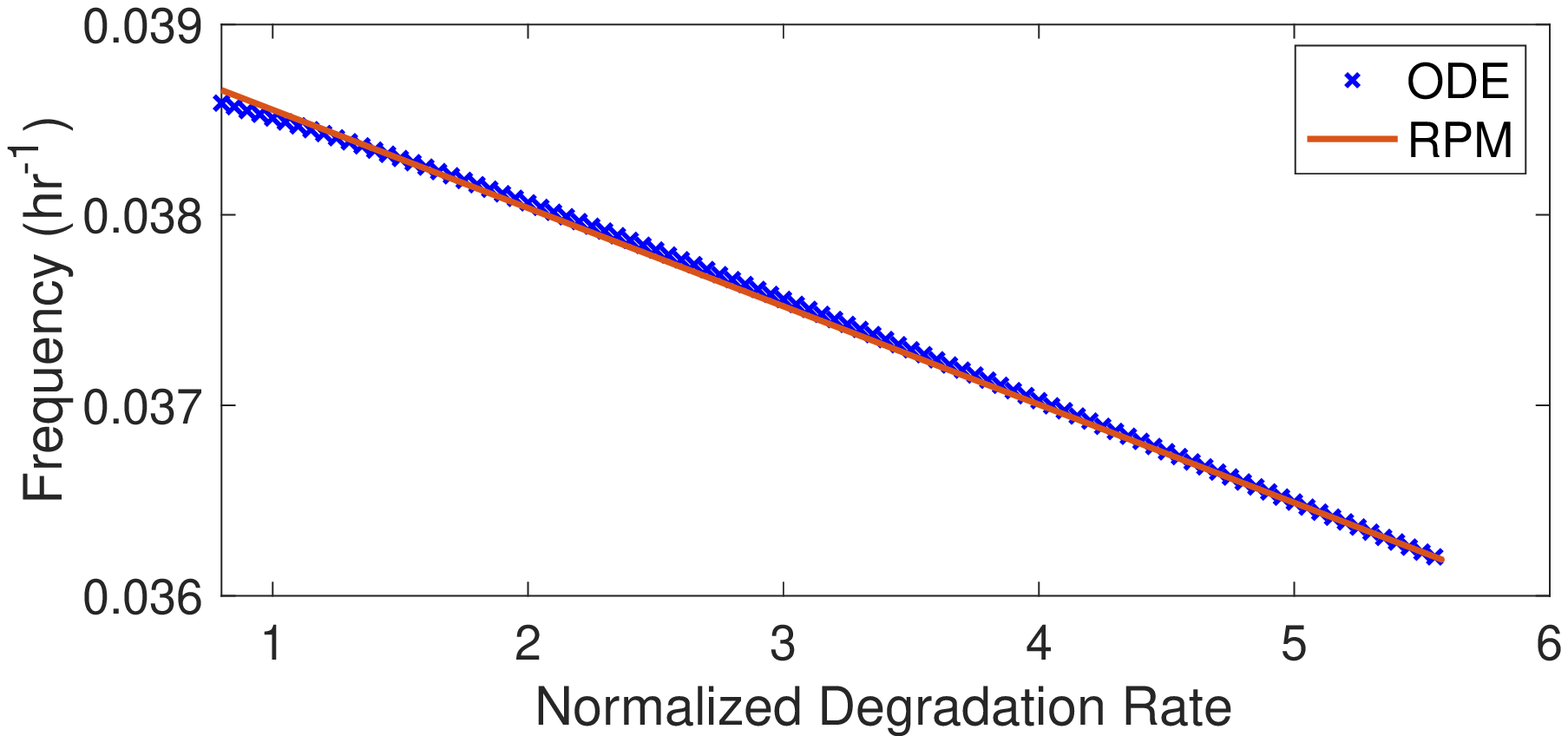}
      \caption{}
      \label{Locke2005b:bif-freq-DD}
    \end{subfigure}
    \begin{subfigure}{0.5\textwidth}
      \includegraphics[width=\textwidth]{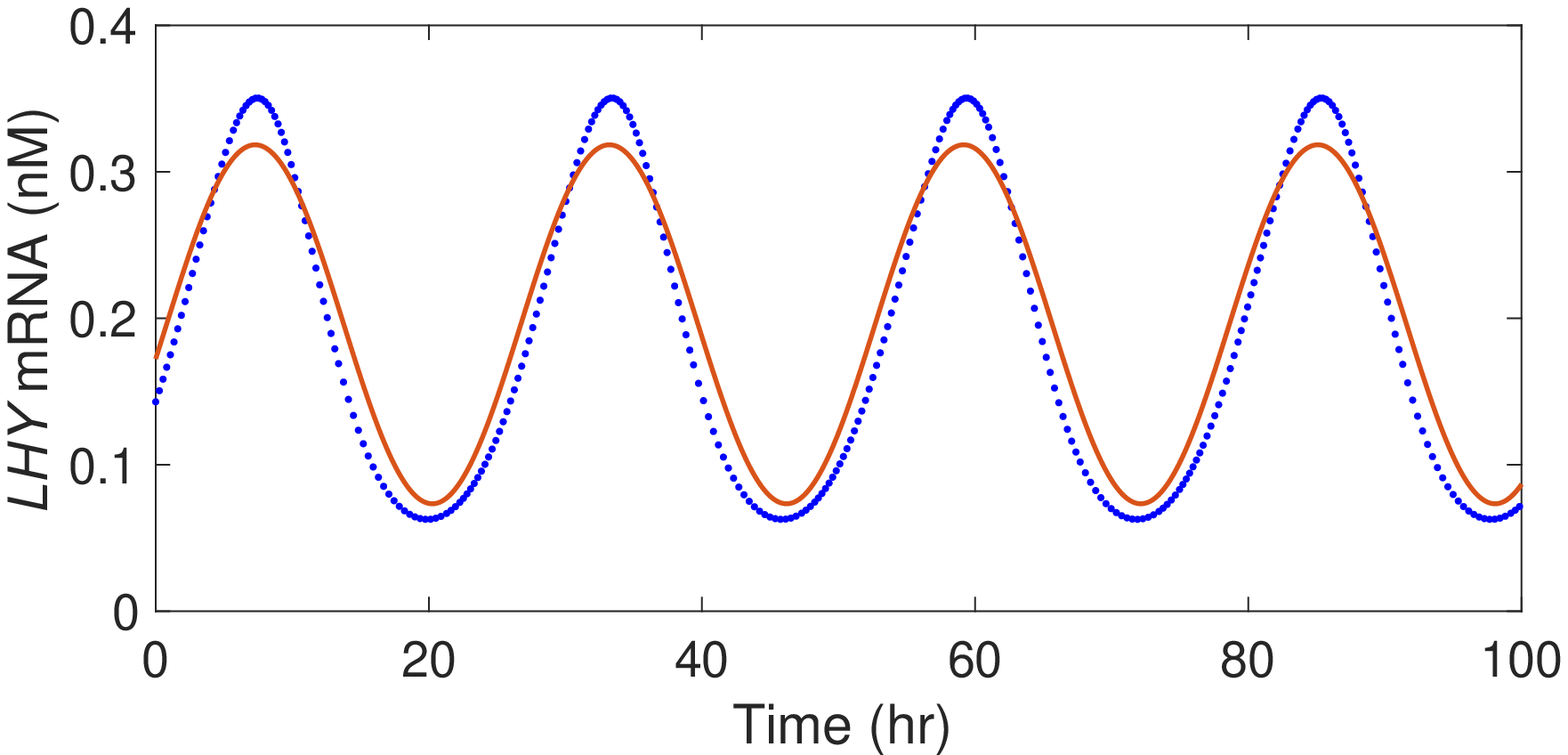}
      \caption{}
      \label{Locke2005b:series-LHY-DD}
    \end{subfigure}%
    \begin{subfigure}{0.5\textwidth}
      \includegraphics[width=\textwidth]{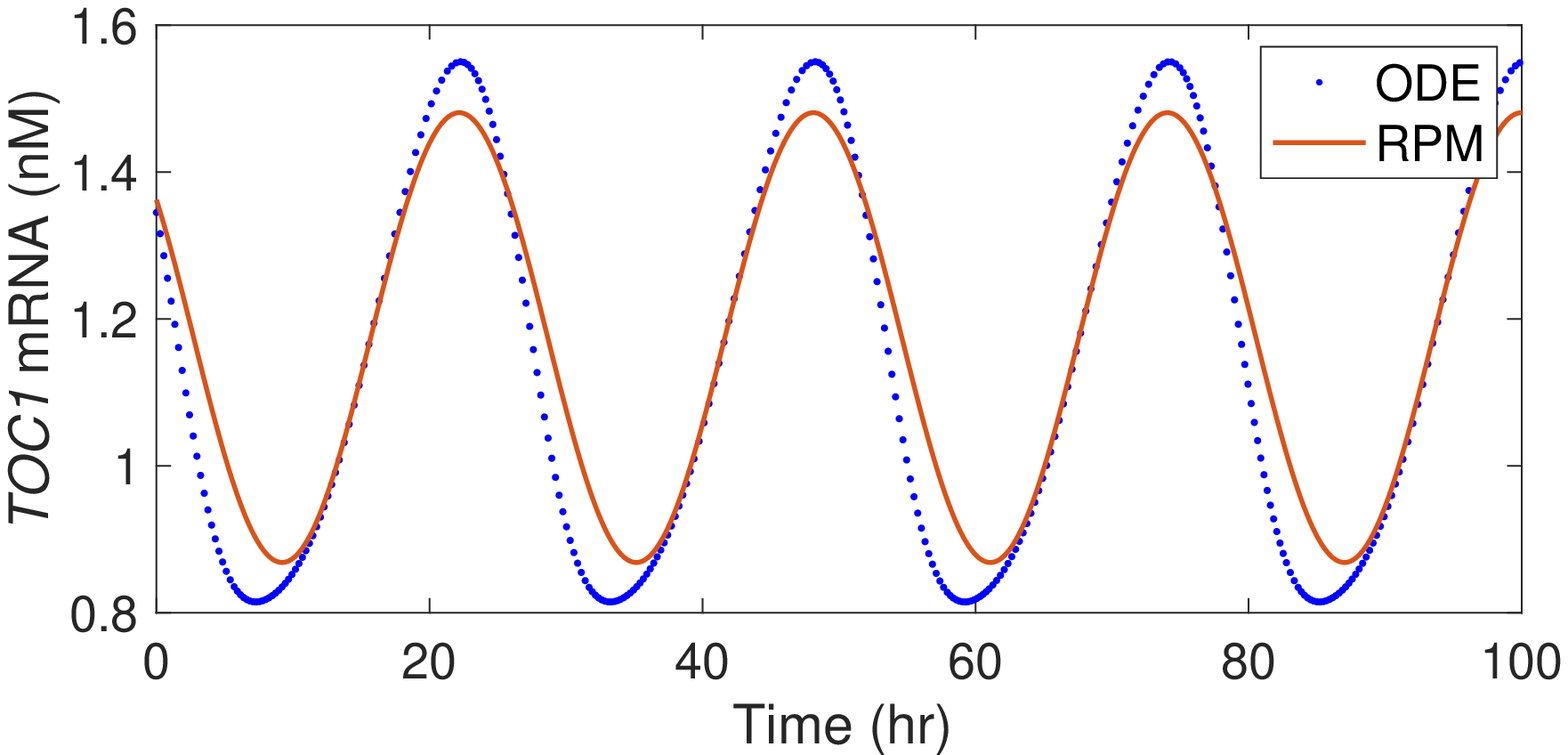}
      \caption{}
      \label{Locke2005b:series-TOC1-DD}
    \end{subfigure}
    \begin{subfigure}{0.333\textwidth}
      \includegraphics[width=\textwidth]{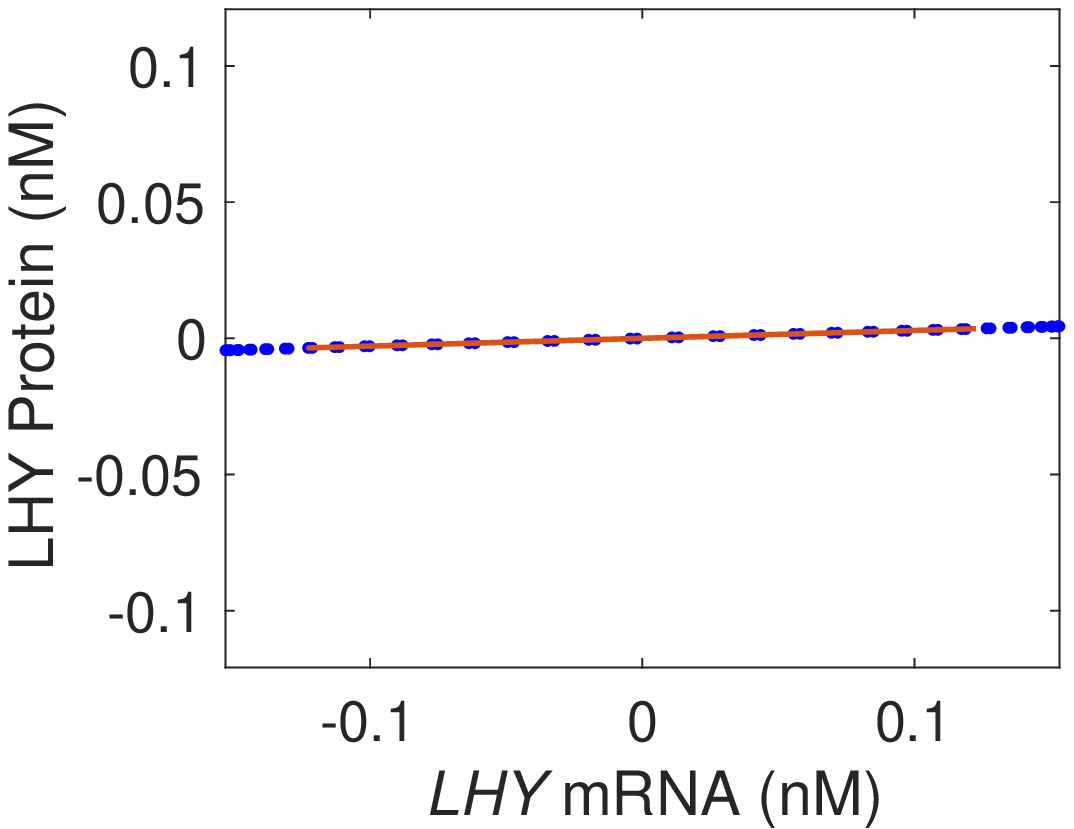}
      \caption{}
      \label{Locke2005b:phase-LHY-DD}
    \end{subfigure}%
    \begin{subfigure}{0.333\textwidth}
      \includegraphics[width=\textwidth]{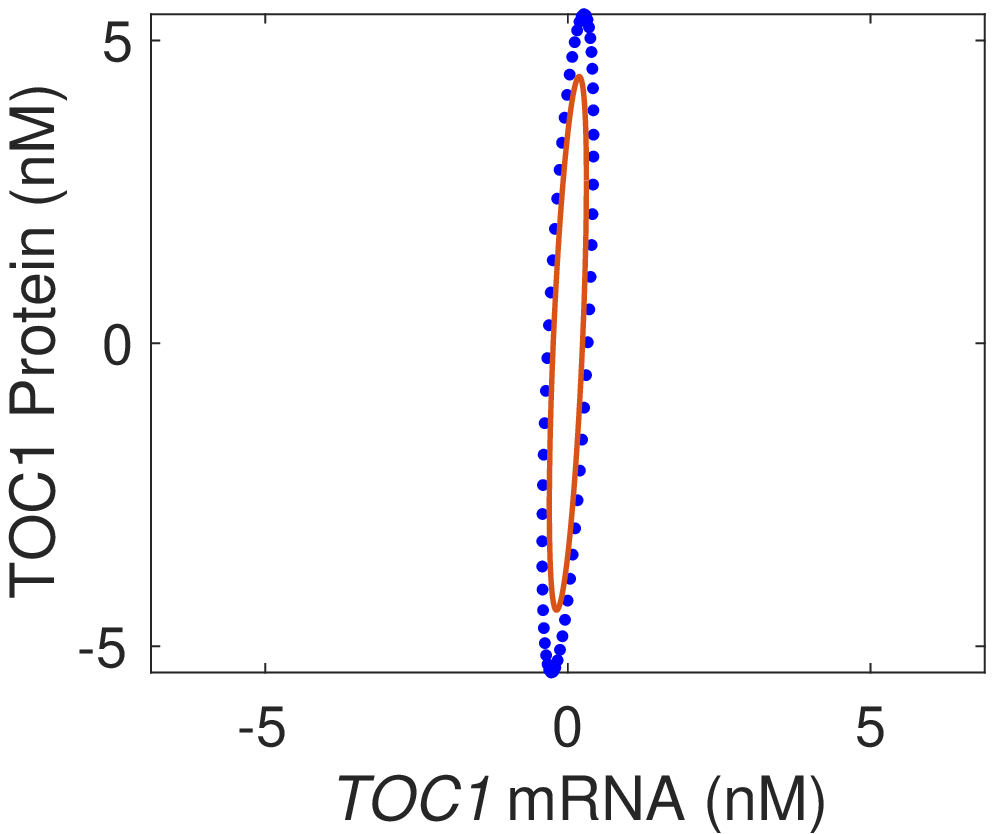}
      \caption{}
      \label{Locke2005b:phase-TOC1-DD}
    \end{subfigure}%
    \begin{subfigure}{0.333\textwidth}
      \includegraphics[width=\textwidth]{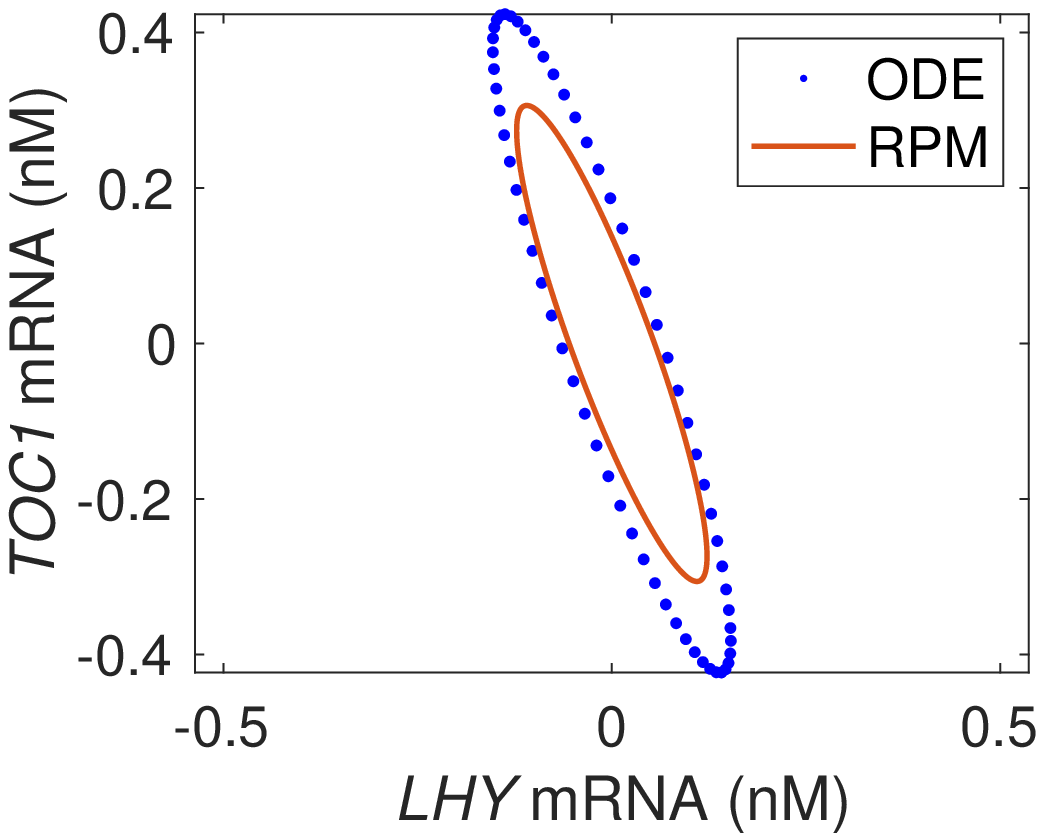}
      \caption{}
      \label{Locke2005b:phase-mRNA-DD}
    \end{subfigure}
    \caption{
    A supercritical Hopf bifurcation occurs in L2005b model under perpetual darkness.  Bifurcation diagrams for (a) concentration of \textit{LHY} mRNA and (b) frequency of oscillation, time series generated from both ODE and RPM for concentrations of (c) \textit{LHY} mRNA and (d) \textit{TOC1} mRNA, and (e) - (g) phase diagrams of pairs of LHY and TOC1 protein in the cytoplasm and \textit{LHY} and \textit{TOC1} mRNA oscillations are shown.  The degradation rate in (a) and (b) are normalized so that the biological value given in the original paper is unity.  The amplitude of limit cycle oscillation calculated with RPM matches the numerical solution of the system of ODEs with 14.78 percent difference; and frequency with 0.26 percent difference. 
    As fractions of $2\pi$, the absolute values of differences in phase difference are 0.0003 for the pair (\textit{LHY} mRNA, LHY protein), 0.004 for the pair (\textit{TOC1} mRNA, TOC1 protein), and 0.003 for the pair (\textit{LHY} mRNA, \textit{TOC1} mRNA).}
    \label{7-panel:Locke2005b-DD}
\end{figure}

\begin{figure}[H]
    \centering
    \begin{subfigure}{0.5\textwidth}
      \includegraphics[width=\textwidth]{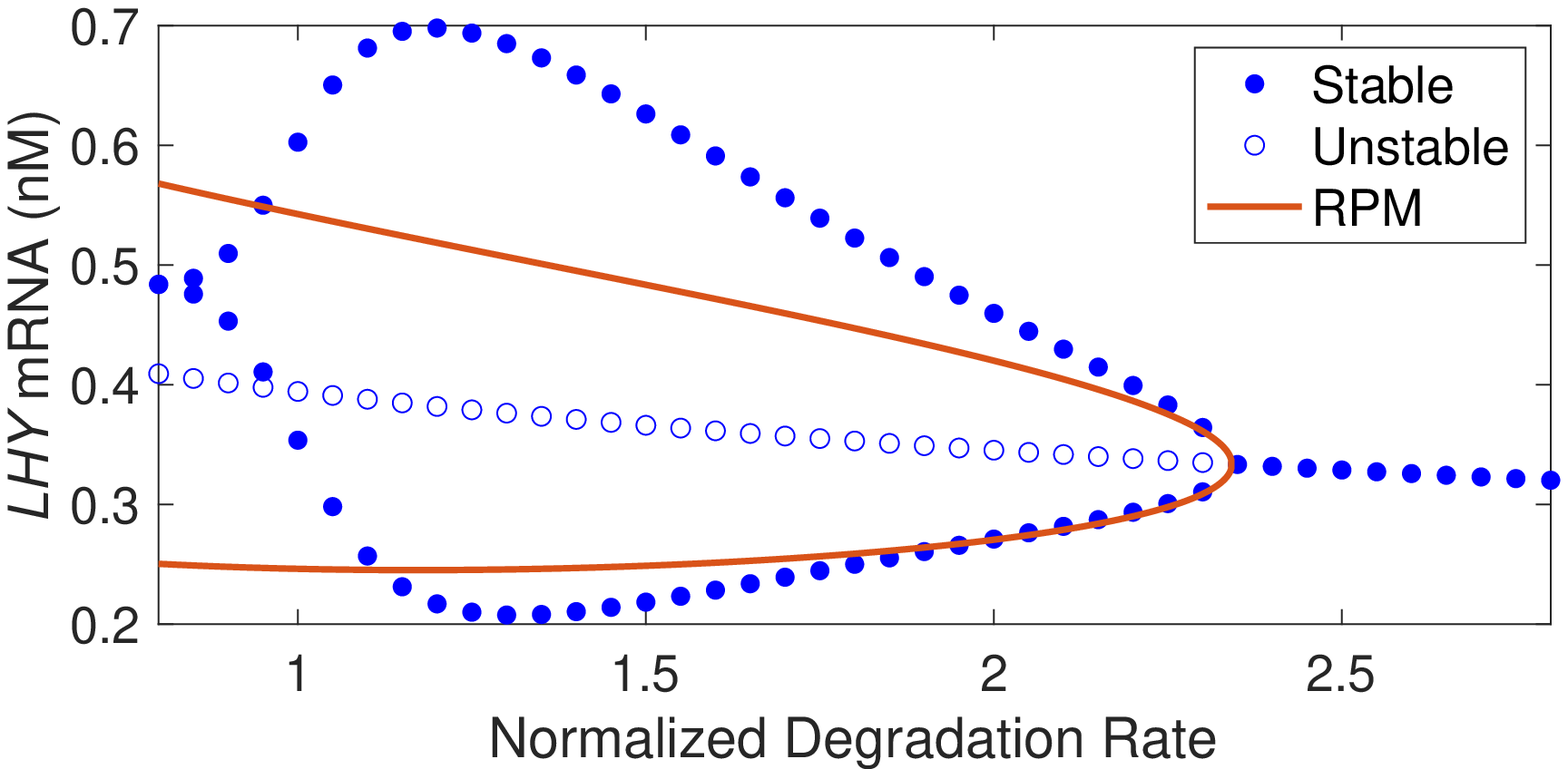}
      \caption{}
      \label{Zeilinger2006:bif-amp-DD}
    \end{subfigure}%
    \begin{subfigure}{0.5\textwidth}
      \includegraphics[width=\textwidth]{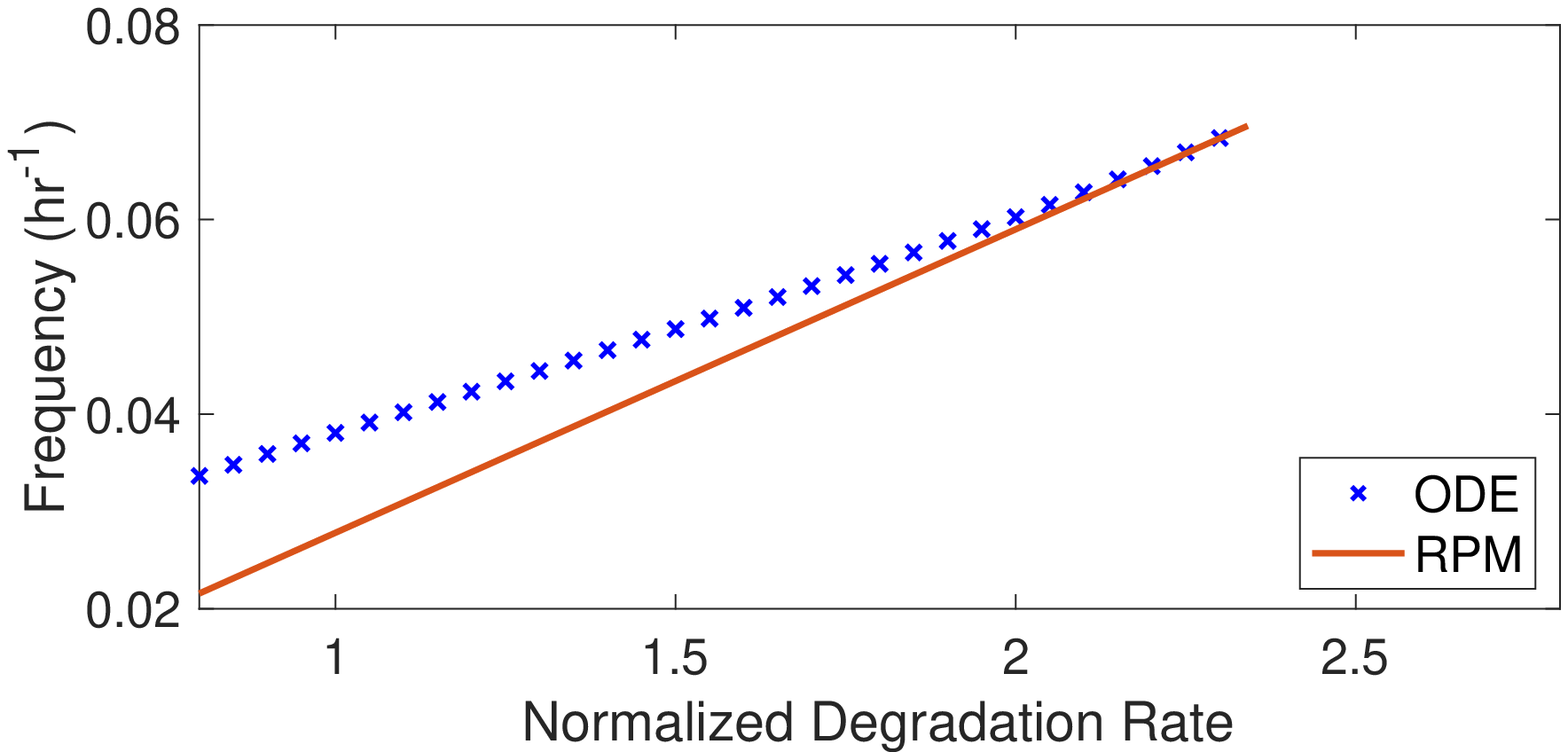}
      \caption{}
      \label{Zeilinger2006:bif-freq-DD}
    \end{subfigure}
    \begin{subfigure}{0.5\textwidth}
      \includegraphics[width=\textwidth]{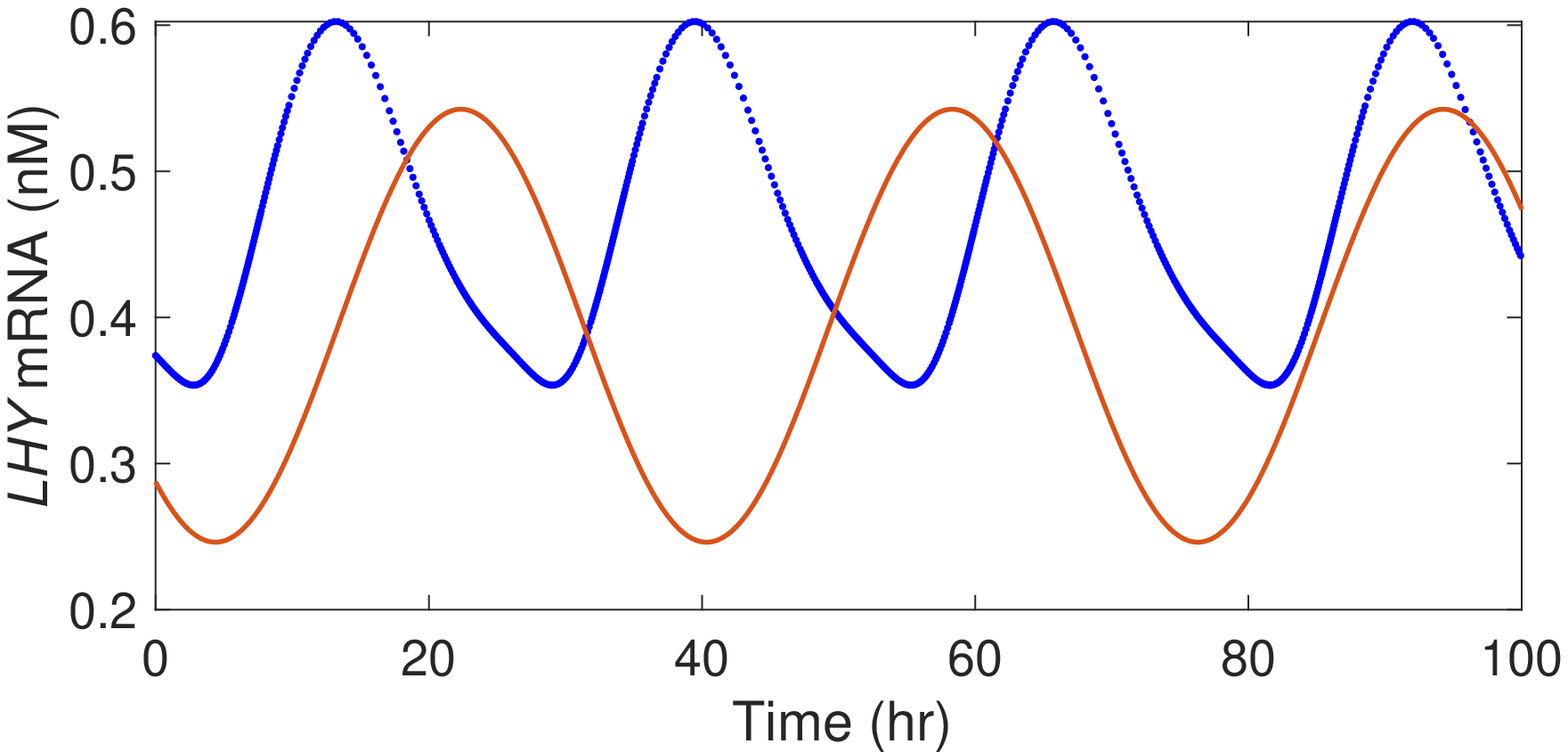}
      \caption{}
      \label{Zeilinger2006:series-LHY-DD}
    \end{subfigure}%
    \begin{subfigure}{0.5\textwidth}
      \includegraphics[width=\textwidth]{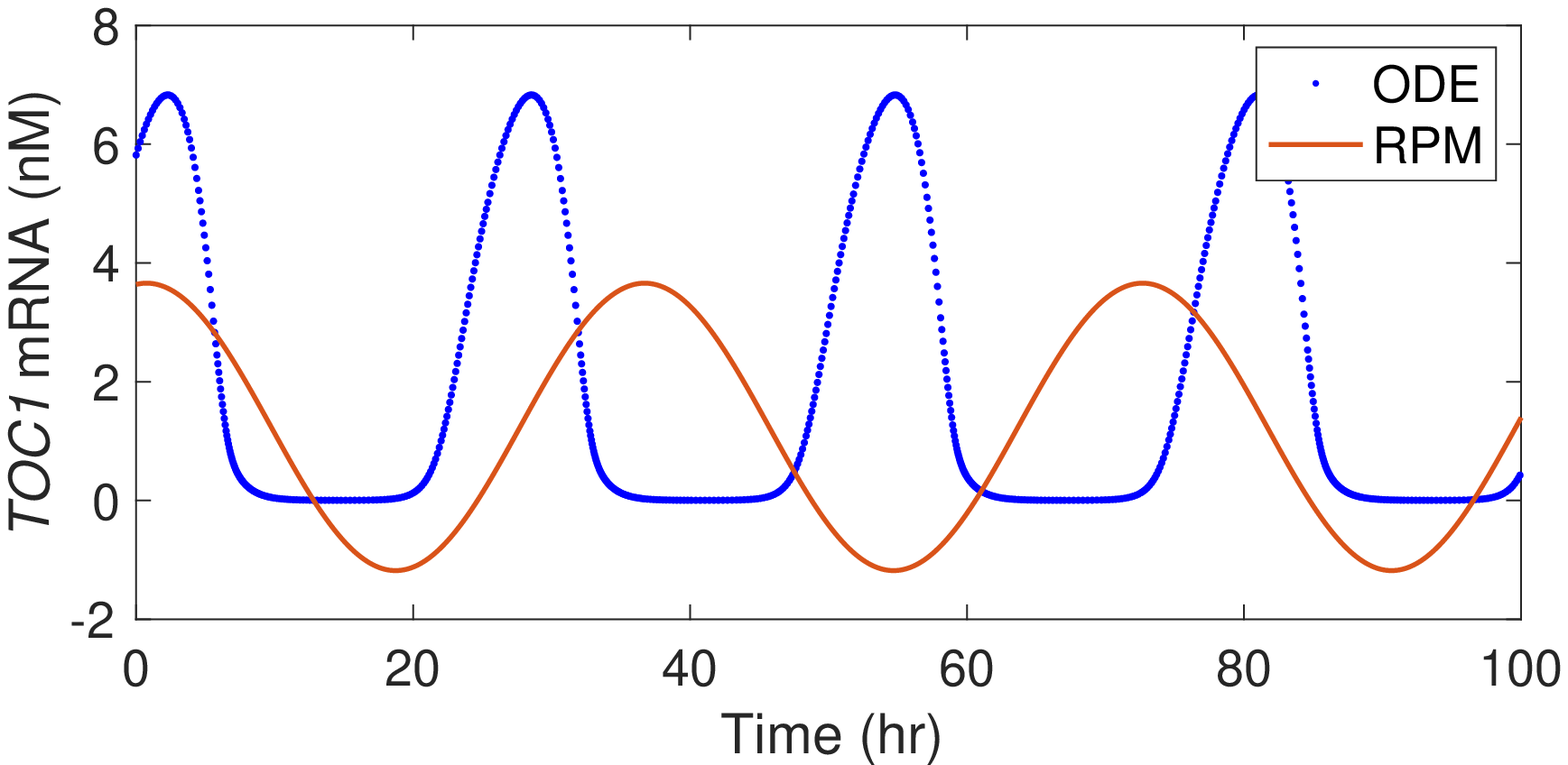}
      \caption{}
      \label{Zeilinger2006:series-TOC1-DD}
    \end{subfigure}
    \begin{subfigure}{0.333\textwidth}
      \includegraphics[width=\textwidth]{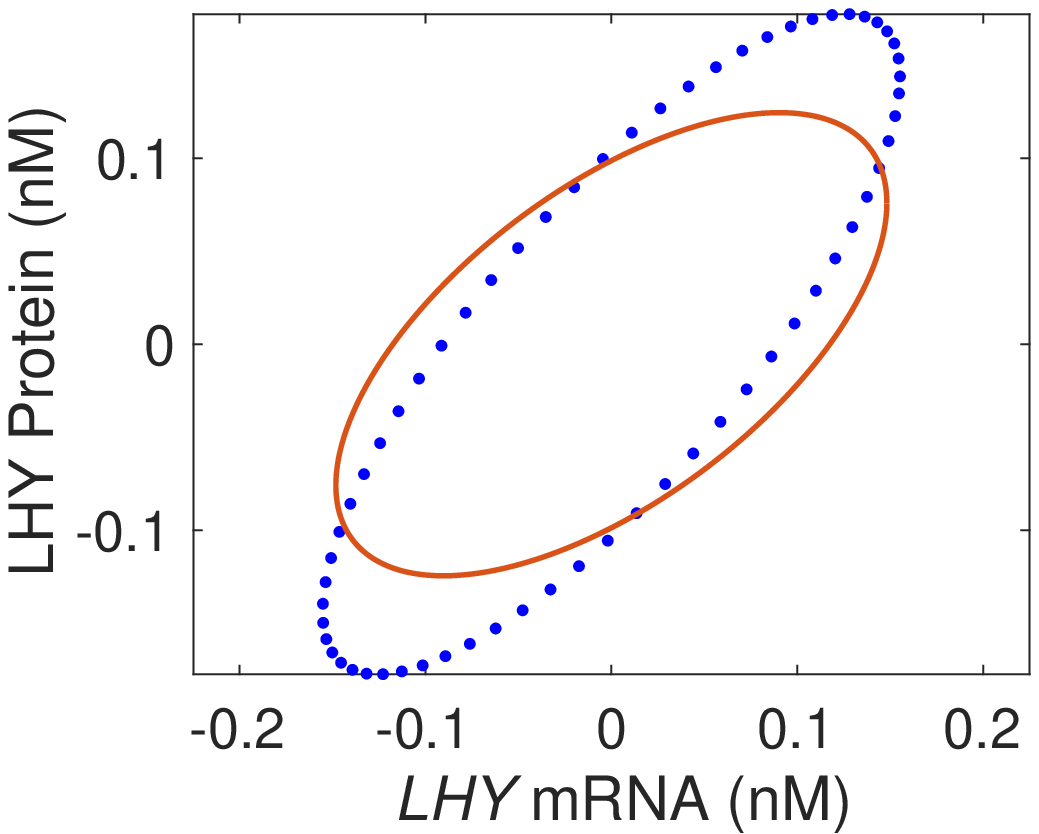}
      \caption{}
      \label{Zeilinger2006:phase-LHY-DD}
    \end{subfigure}%
    \begin{subfigure}{0.333\textwidth}
      \includegraphics[width=\textwidth]{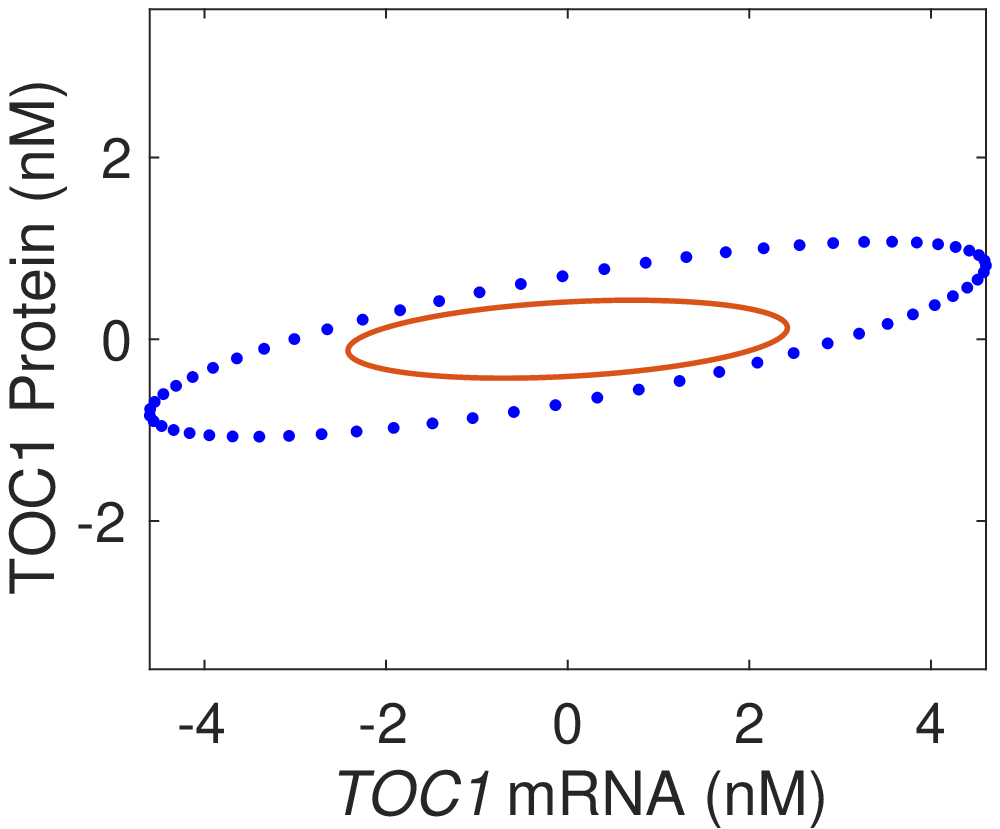}
      \caption{}
      \label{Zeilinger2006:phase-TOC1-DD}
    \end{subfigure}%
    \begin{subfigure}{0.333\textwidth}
      \includegraphics[width=\textwidth]{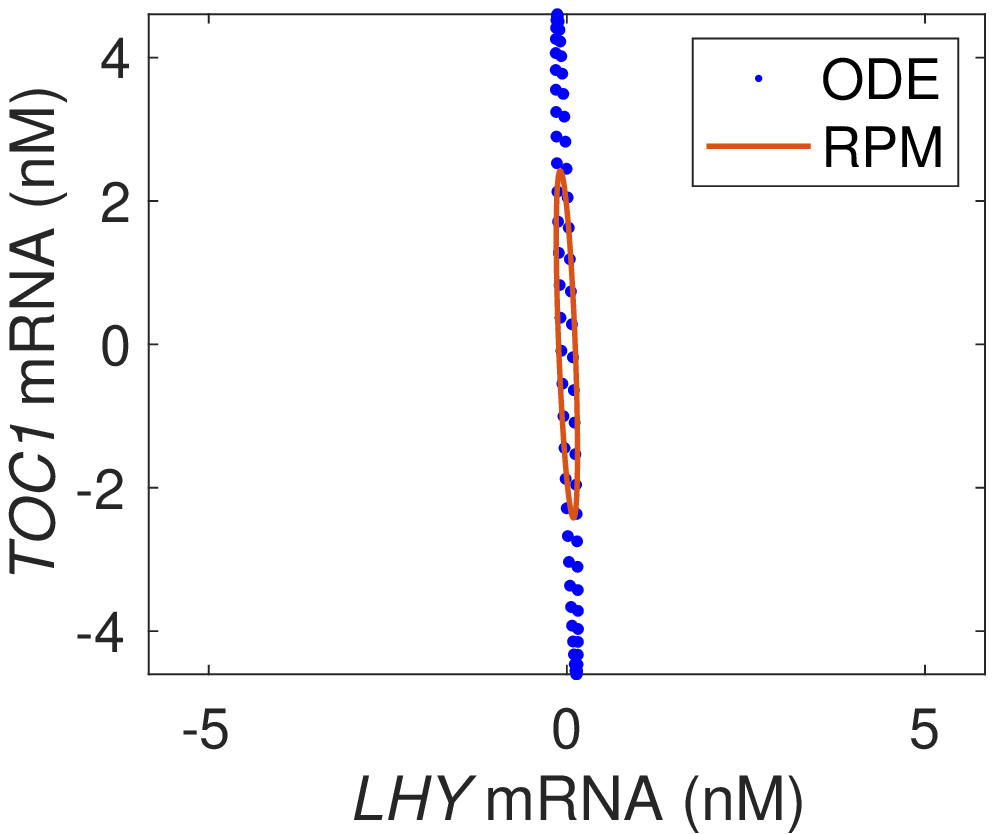}
      \caption{}
      \label{Zeilinger2006:phase-mRNA-DD}
    \end{subfigure}
    \caption{
    A supercritical Hopf bifurcation occurs in Z2006 model under perpetual darkness.  Bifurcation diagrams for (a) concentration of \textit{LHY} mRNA and (b) frequency of oscillation, time series generated from both ODE and RPM for concentrations of (c) \textit{LHY} mRNA and (d) \textit{TOC1} mRNA, and (e) - (g) phase diagrams of pairs of LHY and TOC1 protein in the cytoplasm and \textit{LHY} and \textit{TOC1} mRNA oscillations are shown.  The degradation rate in (a) and (b) are normalized so that the biological value given in the original paper is unity.  The amplitude of limit cycle oscillation calculated with RPM matches the numerical solution of the system of ODEs with 19.04 percent difference; and frequency with 27.03 percent difference. 
    As fractions of $2\pi$, the absolute values of differences in phase difference are 0.046 for the pair (\textit{LHY} mRNA, LHY protein), 0.089 for the pair (\textit{TOC1} mRNA, TOC1 protein), and 0.058 for the pair (\textit{LHY} mRNA, \textit{TOC1} mRNA).}
    \label{7-panel:Zeilinger2006-DD}
\end{figure}

\begin{figure}[H]
    \centering
    \begin{subfigure}{0.5\textwidth}
      \includegraphics[width=\textwidth]{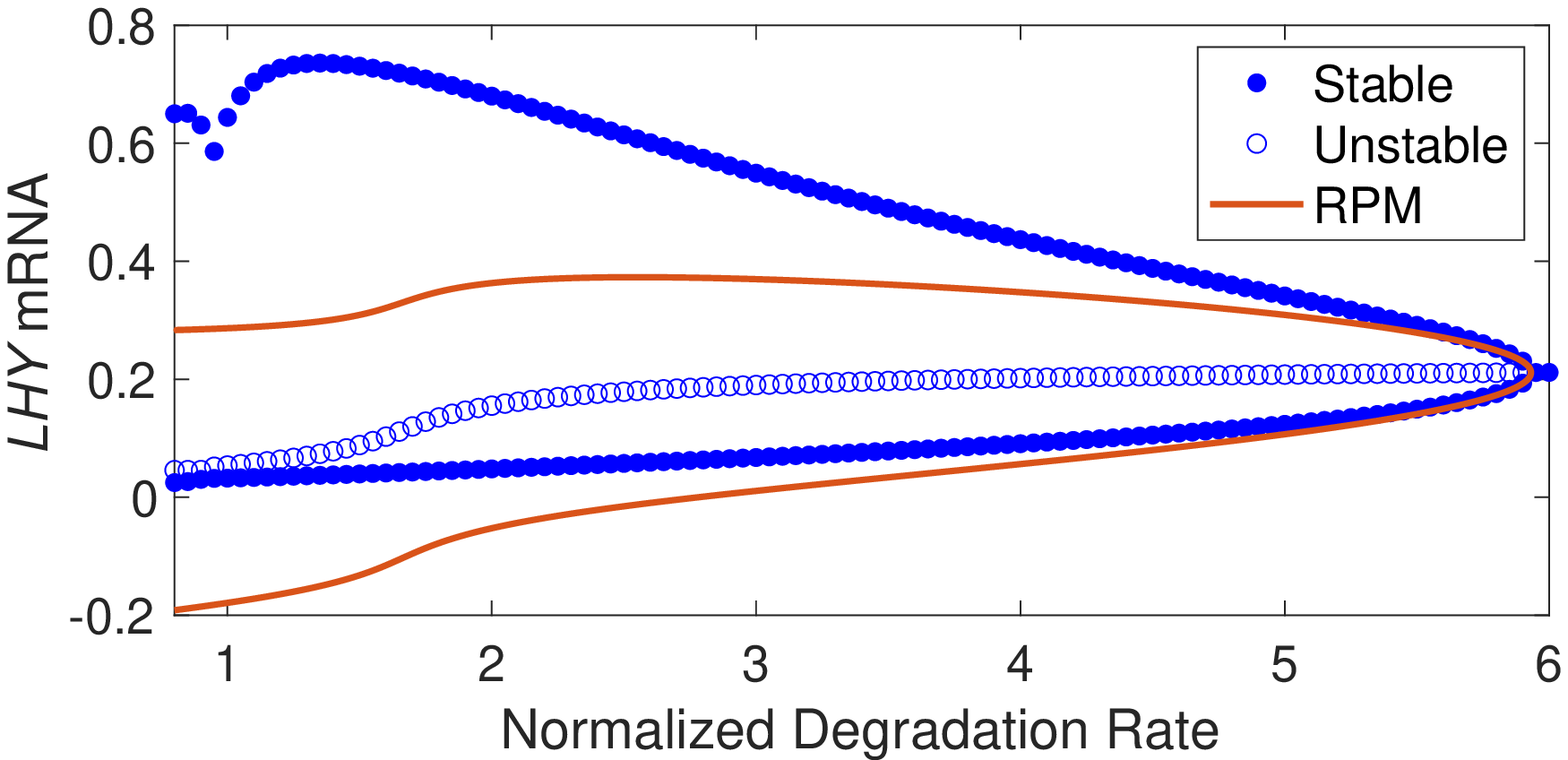}
      \caption{}
      \label{Foo2016:bif-amp-DD}
    \end{subfigure}%
    \begin{subfigure}{0.5\textwidth}
      \includegraphics[width=\textwidth]{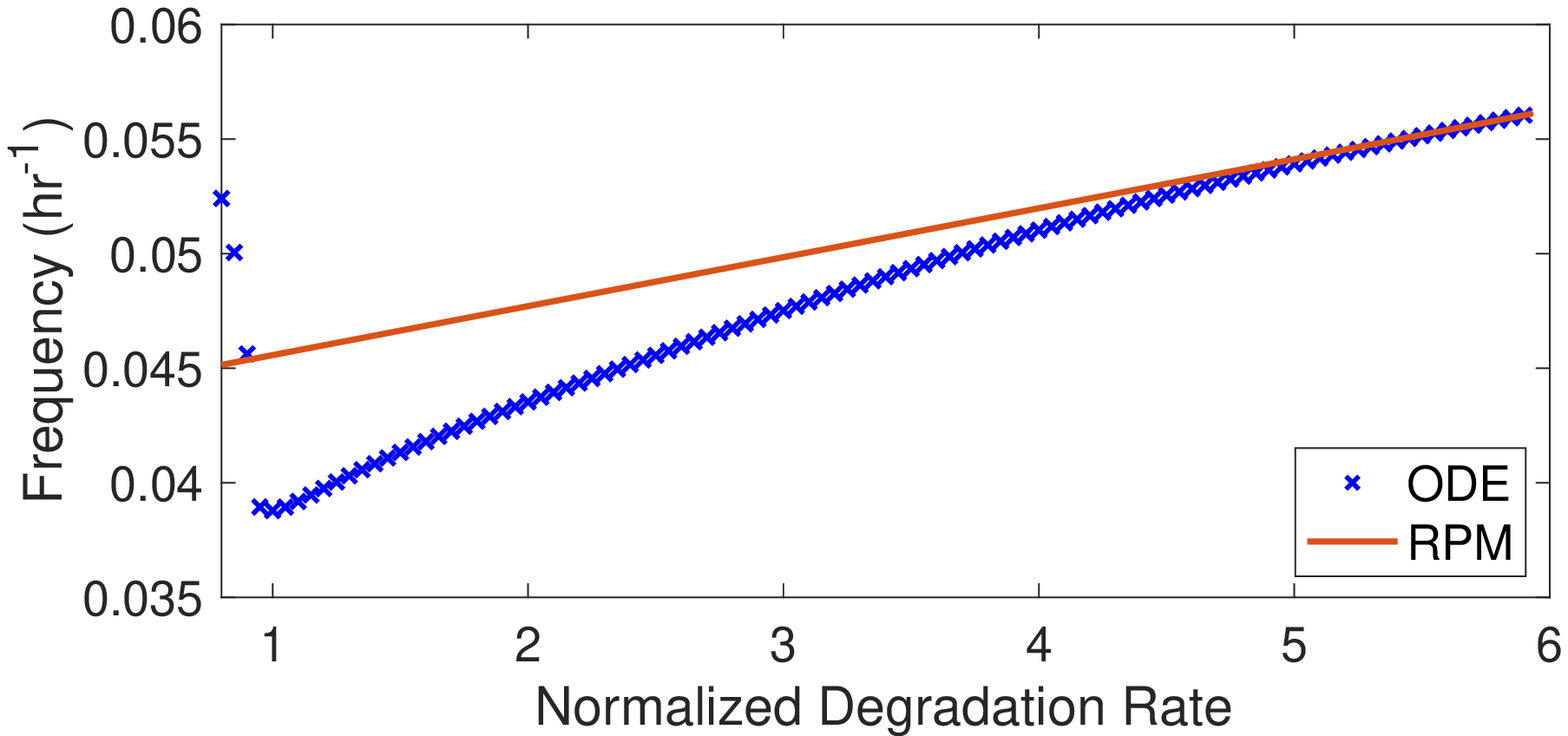}
      \caption{}
      \label{Foo2016:bif-freq-DD}
    \end{subfigure}
    \begin{subfigure}{0.5\textwidth}
      \includegraphics[width=\textwidth]{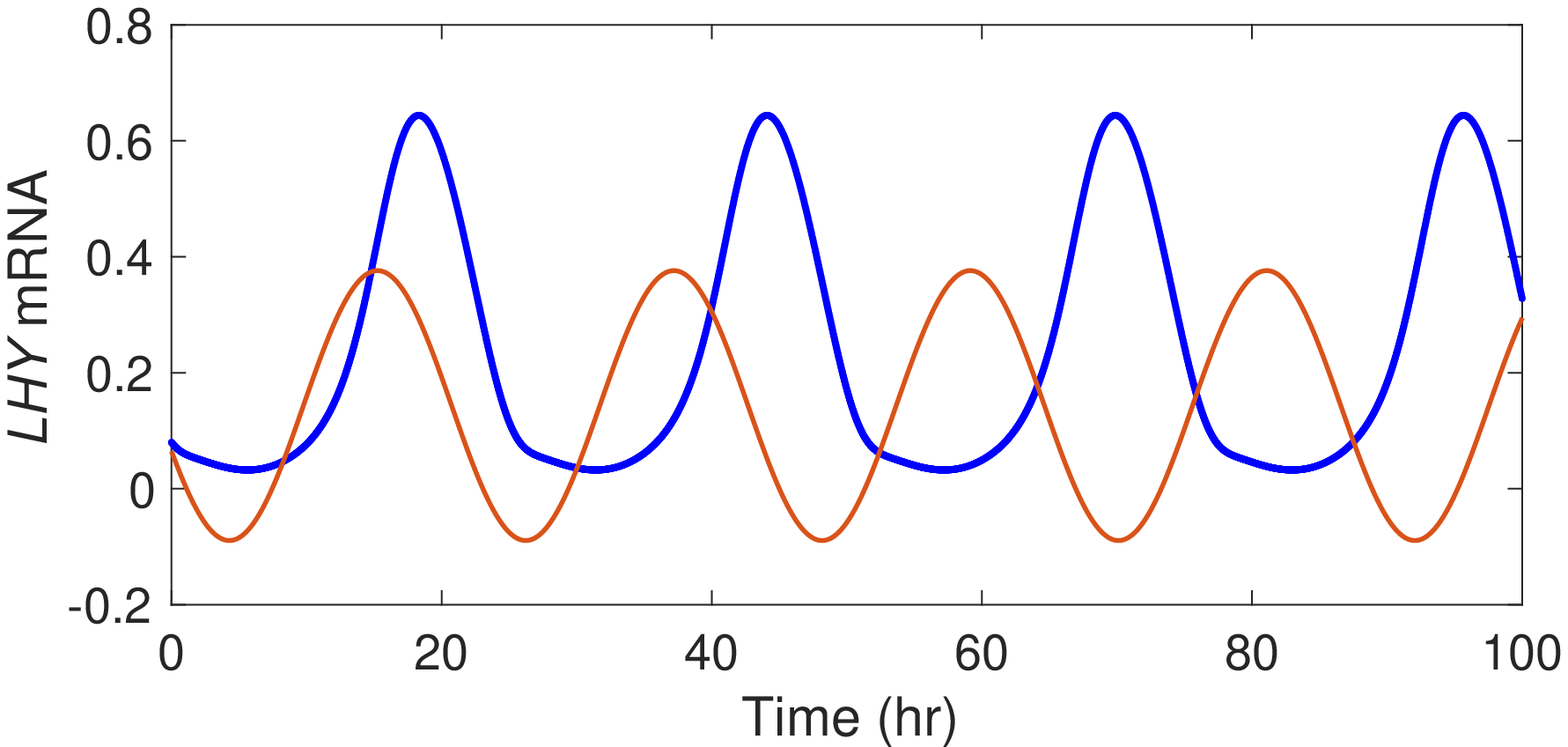}
      \caption{}
      \label{Foo2016:series-LHY-DD}
    \end{subfigure}%
    \begin{subfigure}{0.5\textwidth}
      \includegraphics[width=\textwidth]{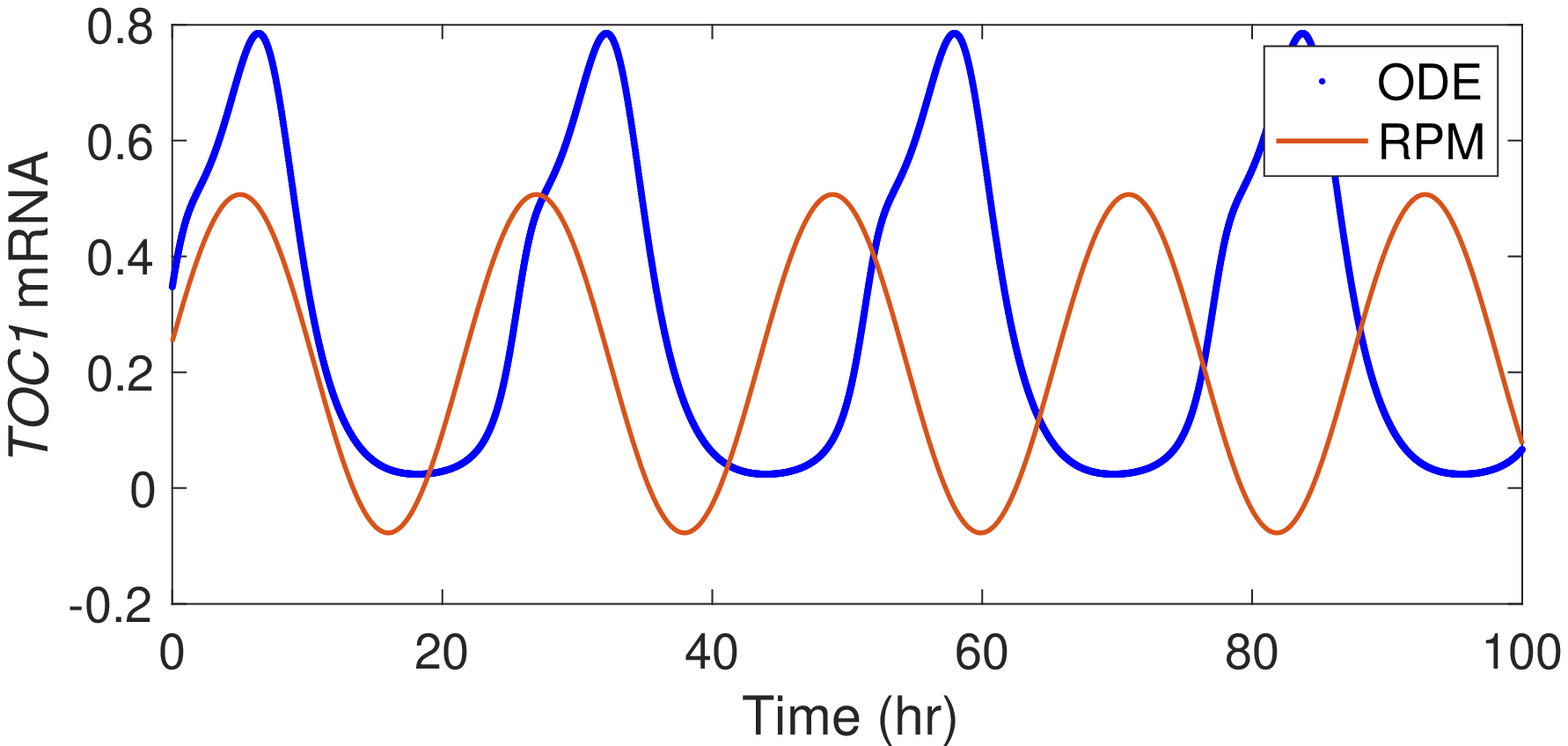}
      \caption{}
      \label{Foo2016:series-TOC1-DD}
    \end{subfigure}
    \begin{subfigure}{0.333\textwidth}
      \includegraphics[width=\textwidth]{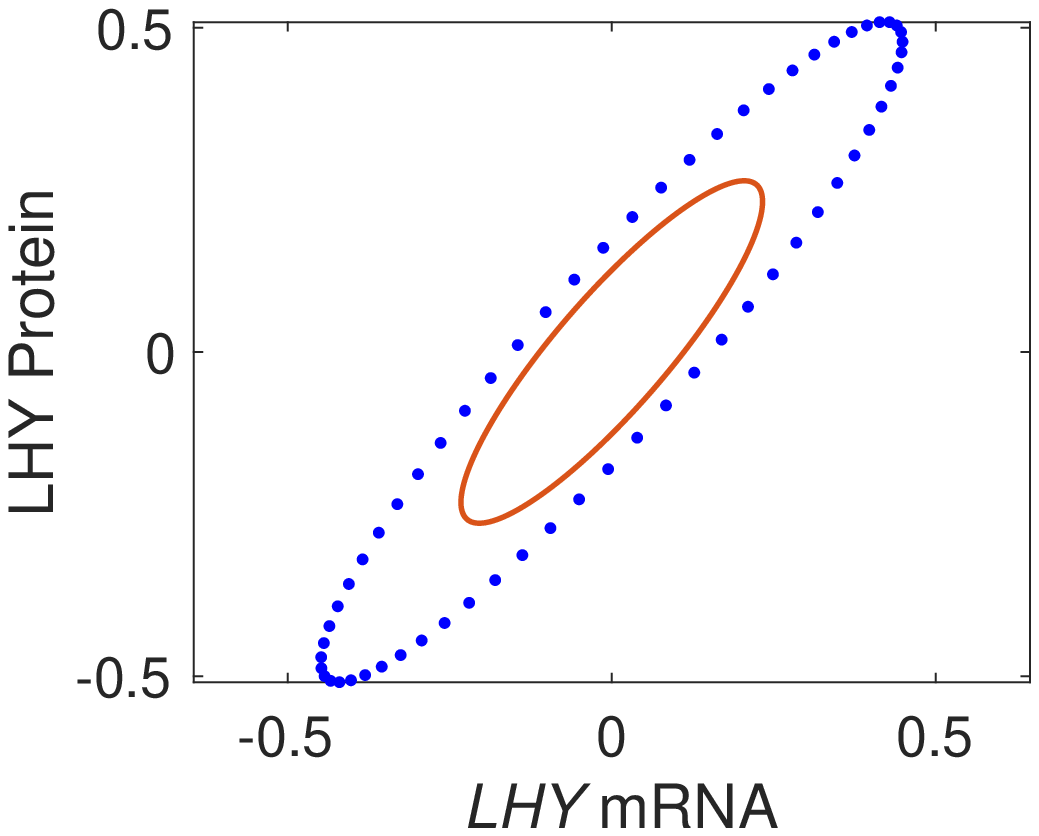}
      \caption{}
      \label{Foo2016:phase-LHY-DD}
    \end{subfigure}%
    \begin{subfigure}{0.333\textwidth}
      \includegraphics[width=\textwidth]{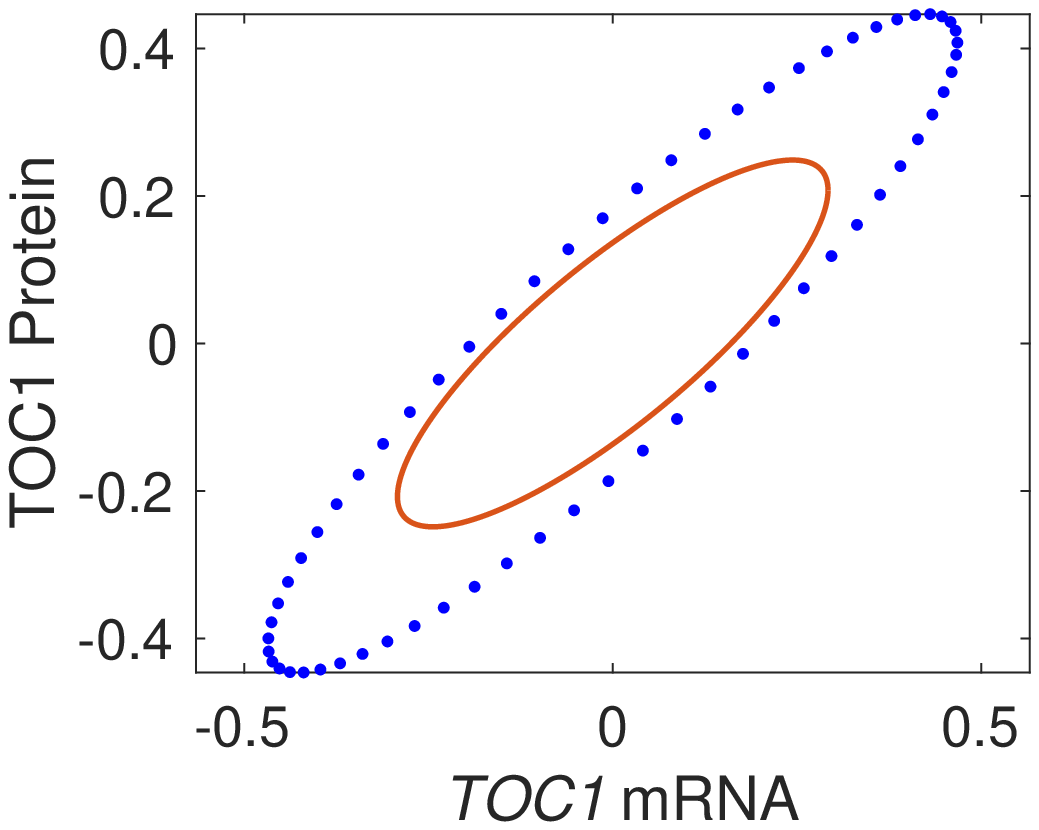}
      \caption{}
      \label{Foo2016:phase-TOC1-DD}
    \end{subfigure}%
    \begin{subfigure}{0.333\textwidth}
      \includegraphics[width=\textwidth]{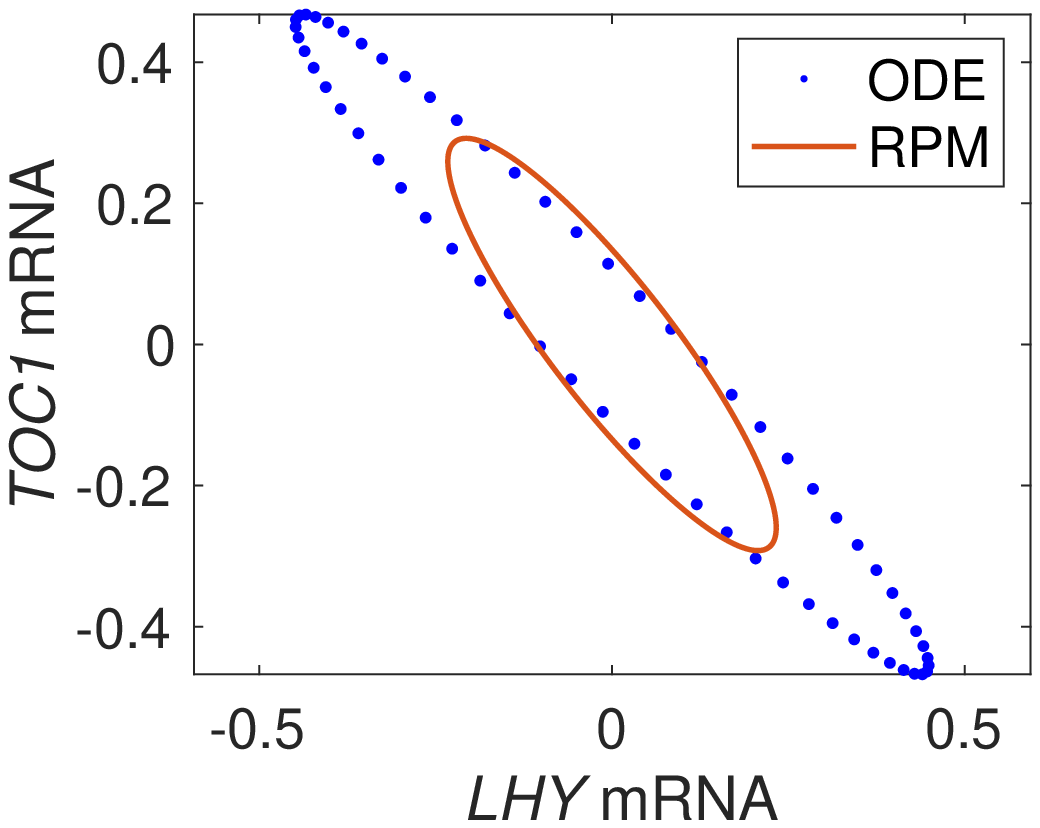}
      \caption{}
      \label{Foo2016:phase-mRNA-DD}
    \end{subfigure}
    \caption{
    A supercritical Hopf bifurcation occurs in F2016 model under perpetual darkness.  Bifurcation diagrams for (a) concentration of \textit{LHY} mRNA and (b) frequency of oscillation, time series generated from both ODE and RPM for concentrations of (c) \textit{LHY} mRNA and (d) \textit{TOC1} mRNA, and (e) - (g) phase diagrams of pairs of LHY and TOC1 protein in the cytoplasm and \textit{LHY} and \textit{TOC1} mRNA oscillations are shown.  The degradation rate in (a) and (b) are normalized so that the biological value given in the original paper is unity.  The amplitude of limit cycle oscillation calculated with RPM matches the numerical solution of the system of ODEs with 23.88 percent difference; and frequency with 17.53 percent difference. 
    As fractions of $2\pi$, the absolute values of differences in phase difference are 0.024 for the pair (\textit{LHY} mRNA, LHY protein), 0.026 for the pair (\textit{TOC1} mRNA, TOC1 protein), and 0.039 for the pair (\textit{LHY} mRNA, \textit{TOC1} mRNA).}
    \label{7-panel:Foo2016-DD}
\end{figure}

\begin{figure}[H]
    \centering
    \begin{subfigure}{0.5\textwidth}
      \includegraphics[width=\textwidth]{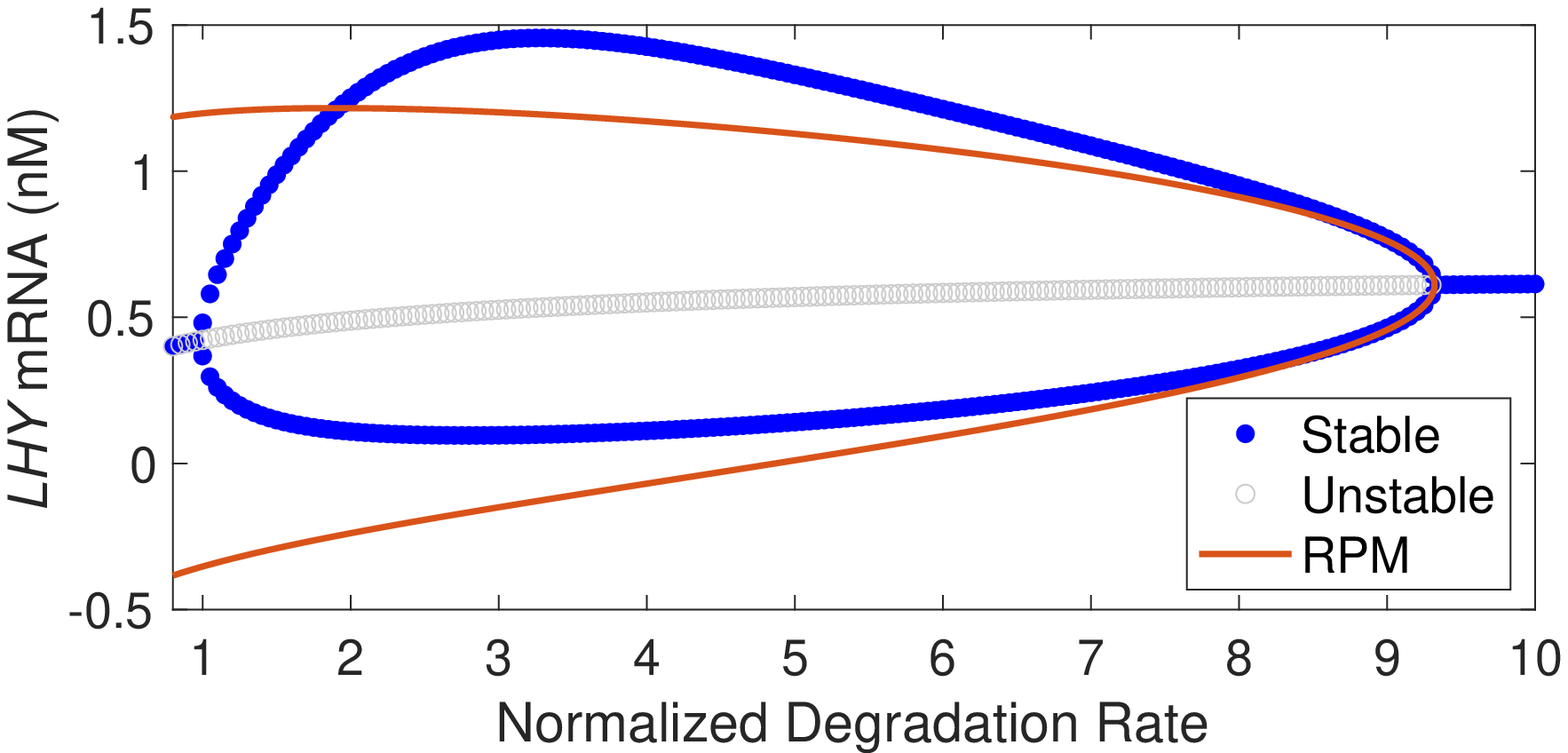}
      \caption{}
      \label{DeCaluwe2016:bif-amp-DD}
    \end{subfigure}%
    \begin{subfigure}{0.5\textwidth}
      \includegraphics[width=\textwidth]{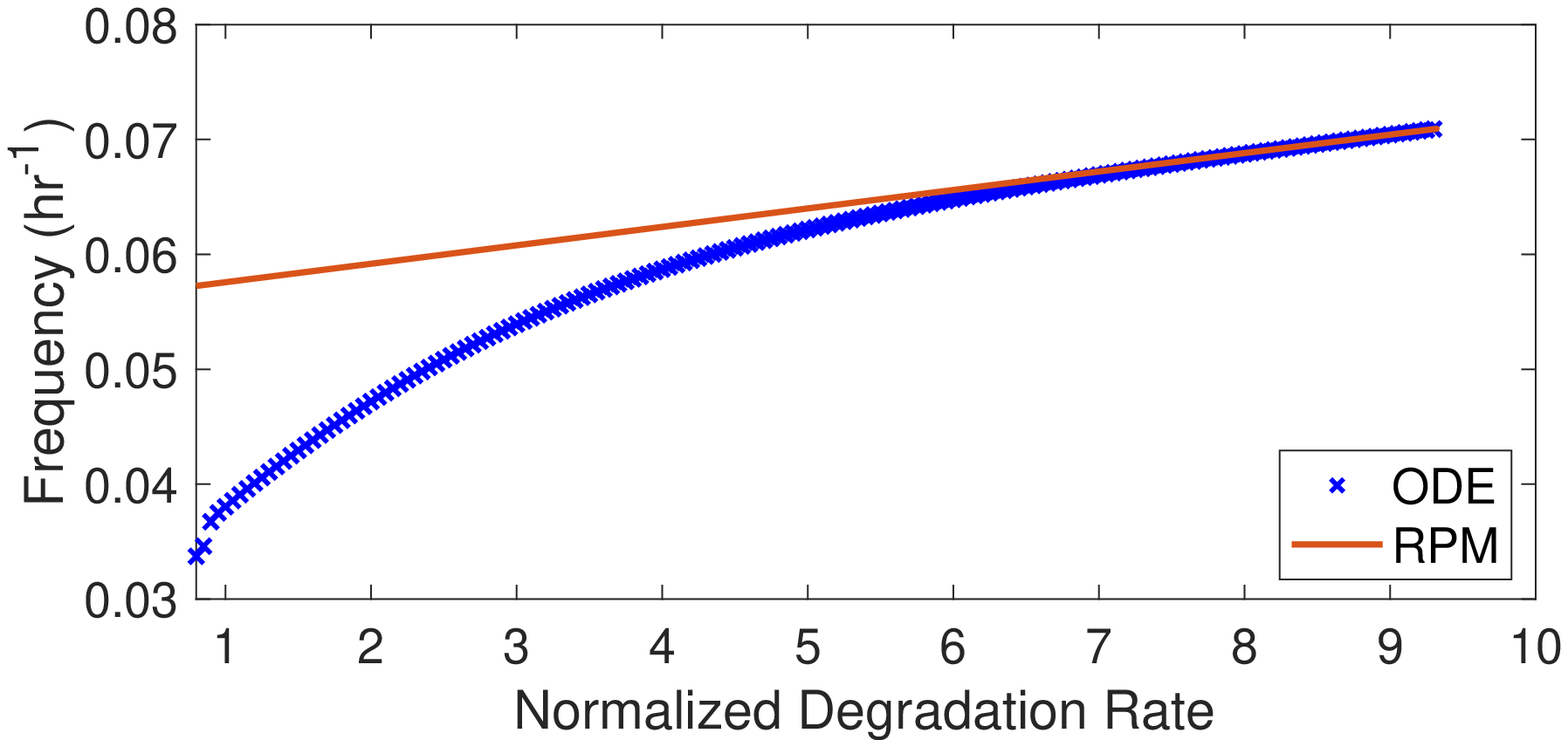}
      \caption{}
      \label{DeCaluwe2016:bif-freq-DD}
    \end{subfigure}
    \begin{subfigure}{0.5\textwidth}
      \includegraphics[width=\textwidth]{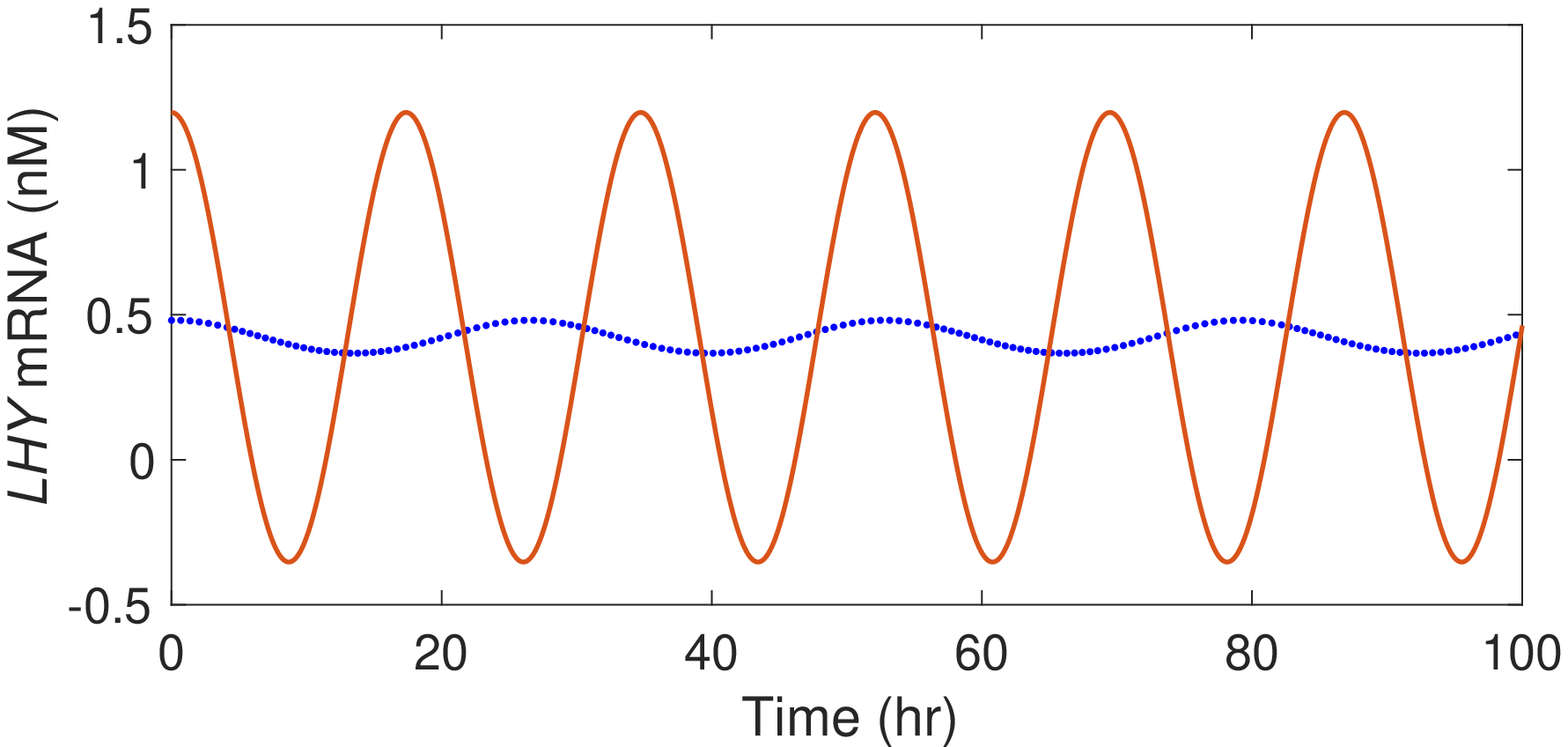}
      \caption{}
      \label{DeCaluwe2016:series-LHY-DD}
    \end{subfigure}%
    \begin{subfigure}{0.5\textwidth}
      \includegraphics[width=\textwidth]{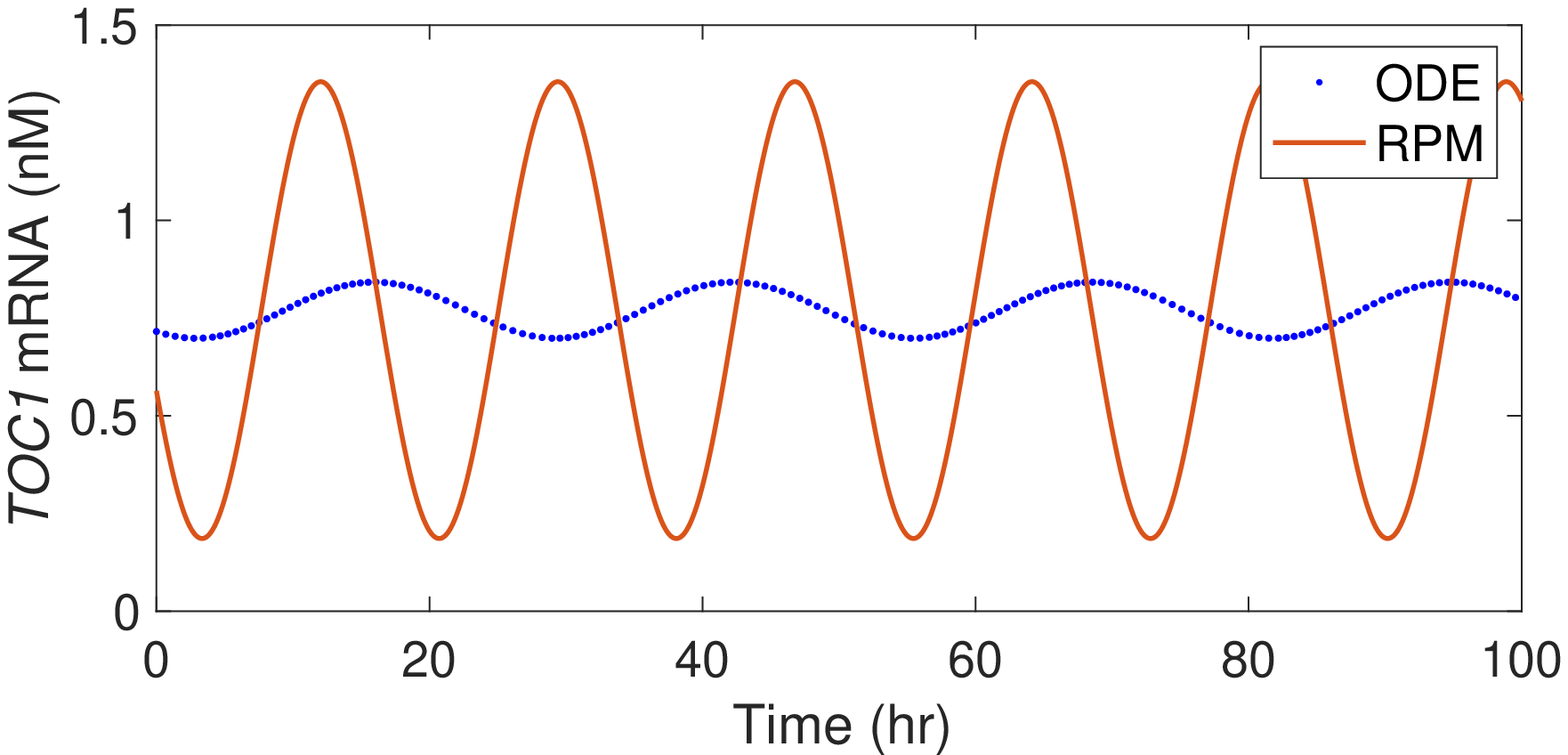}
      \caption{}
      \label{DeCaluwe2016:series-TOC1-DD}
    \end{subfigure}
    \begin{subfigure}{0.333\textwidth}
      \includegraphics[width=\textwidth]{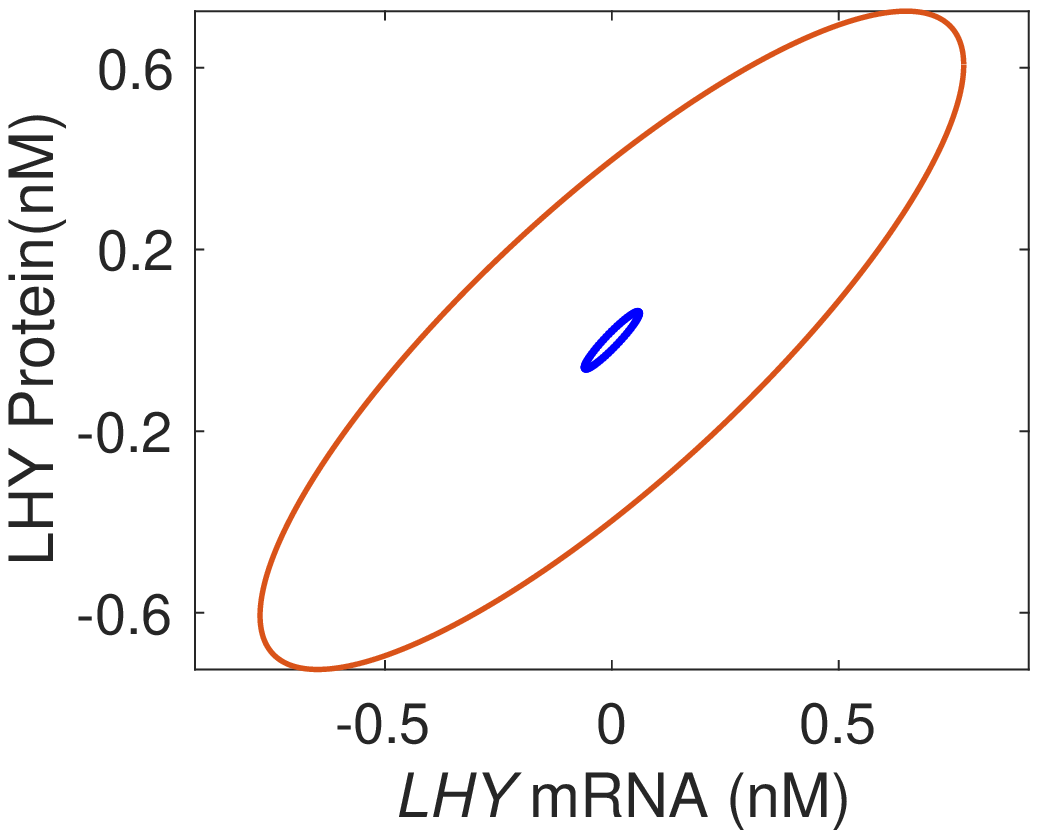}
      \caption{}
      \label{DeCaluwe2016:phase-LHY-DD}
    \end{subfigure}%
    \begin{subfigure}{0.333\textwidth}
      \includegraphics[width=\textwidth]{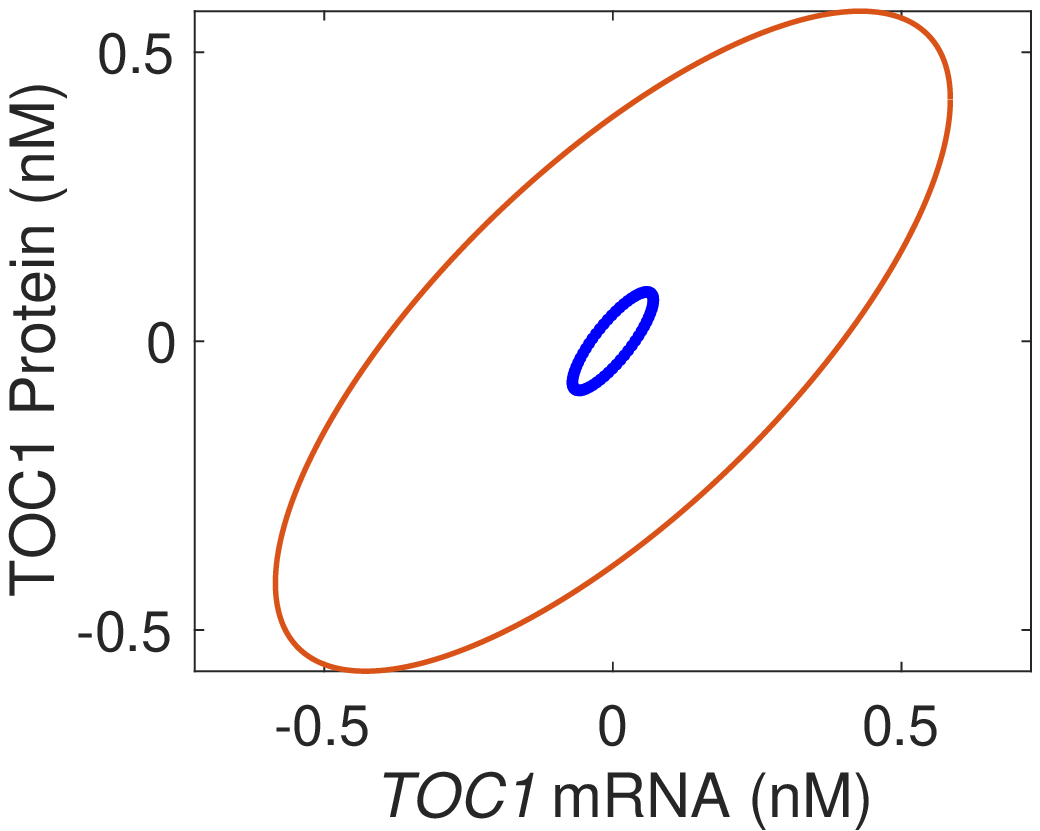}
      \caption{}
      \label{DeCaluwe2016:phase-TOC1-DD}
    \end{subfigure}%
    \begin{subfigure}{0.333\textwidth}
      \includegraphics[width=\textwidth]{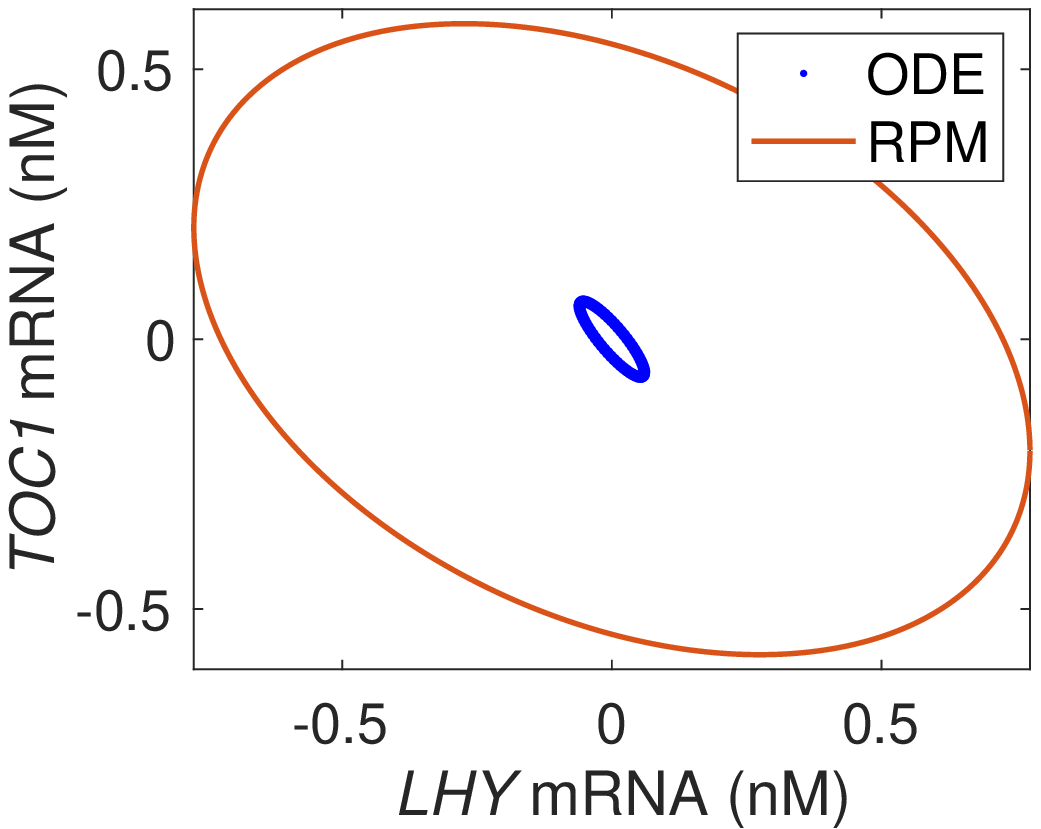}
      \caption{}
      \label{DeCaluwe2016:phase-mRNA-DD}
    \end{subfigure}
    \caption{
    A supercritical Hopf bifurcation occurs in DC2016 model under perpetual darkness.  Bifurcation diagrams for (a) concentration of \textit{LHY} mRNA and (b) frequency of oscillation, time series generated from both ODE and RPM for concentrations of (c) \textit{LHY} mRNA and (d) \textit{TOC1} mRNA, and (e) - (g) phase diagrams of pairs of LHY and TOC1 protein in the cytoplasm and \textit{LHY} and \textit{TOC1} mRNA oscillations are shown.  The degradation rate in (a) and (b) are normalized so that the biological value given in the original paper is unity.  The amplitude of limit cycle oscillation calculated with RPM matches the numerical solution of the system of ODEs with 1267.28 percent difference; and frequency with 51.58 percent difference. There is a second Hopf bifurcation near a normalized degradation rate of unity, which is excluded due to our criteria as the pre-bifurcation region for this second bifurcation corresponds to lower degradation rate.
    As fractions of $2\pi$, the absolute values of differences in phase difference are 0.039 for the pair (\textit{LHY} mRNA, LHY protein), 0.027 for the pair (\textit{TOC1} mRNA, TOC1 protein), and 0.113 for the pair (\textit{LHY} mRNA, \textit{TOC1} mRNA).}
    \label{7-panel:DeCaluwe2016-DD}
\end{figure}

\vspace{0.3cm}
In Figs. \ref{7-panel:Locke2005a}-\ref{7-panel:DeCaluwe2016-DD}, for all the models we studied whose reported optimal parameters give a post-bifurcation system, we show: 1) bifurcation diagrams for limit cycle amplitude and frequency of \textit{LHY} mRNA, 2) time series for \textit{LHY} mRNA and \textit{TOC1} mRNA limit cycle oscillations, and 3) phase space plots for LHY and TOC1 protein oscillations and \textit{LHY} and \textit{TOC1} mRNA oscillations. Each figure compares the numerical results of the full system of differential equations to the result of the \textit{Reductive Perturbation Method} \citep{kuramoto2003chemical}. \vspace{0.3cm}

\begin{figure}[H]
    \centering
    \includegraphics[width=\textwidth]{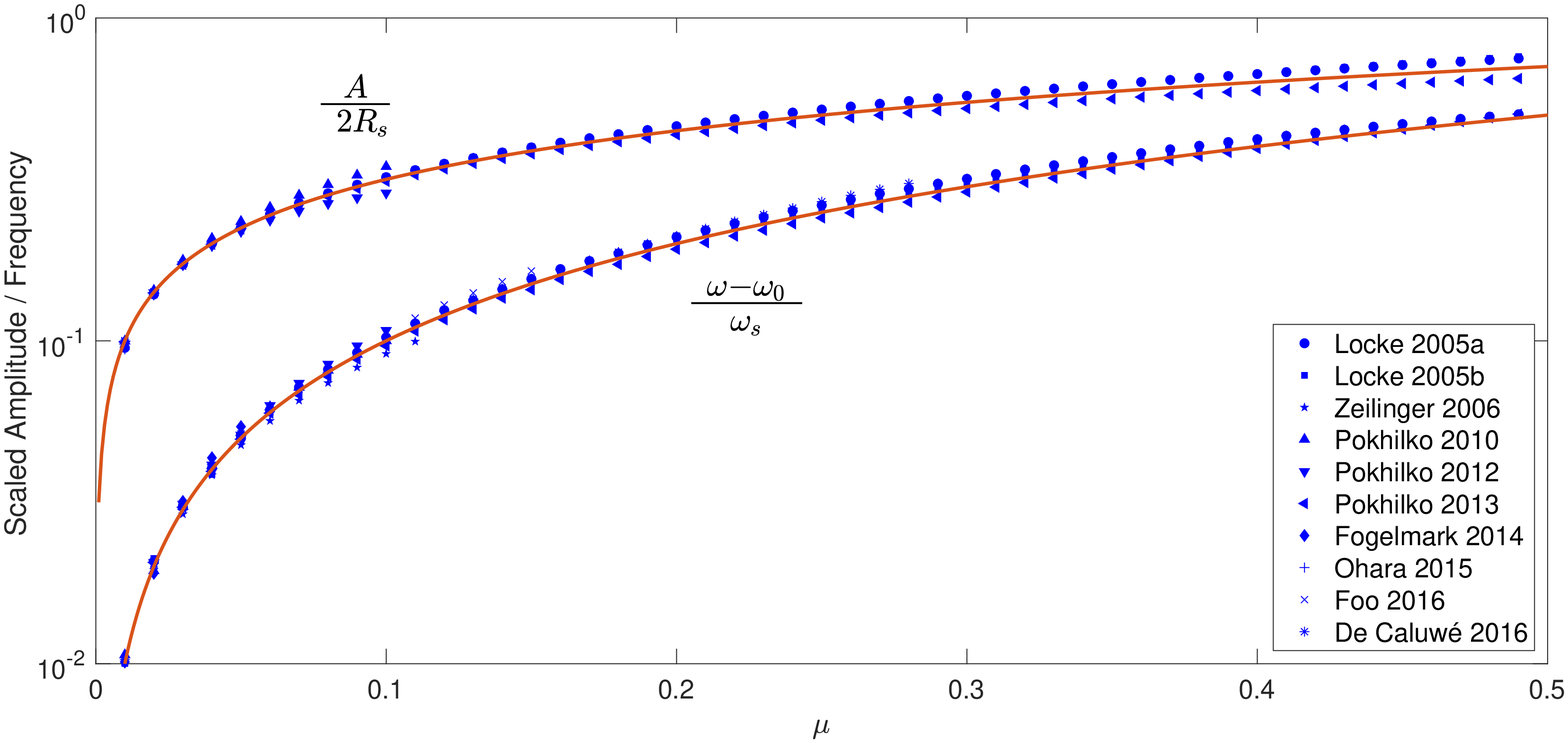}
    \caption{Amplitude and frequency of limit cycle oscillations for ten models of \textit{Arabidopsis} circadian rhythms in perpetual darkness collapsed onto universal functions of the bifurcation parameter $\mu$. The limit cycle amplitude and frequency calculated numerically with ODE solvers are scaled with the asymptotic solutions to the Stuart-Landau equation. Data for each model is shown up to the value of $\mu$ that they diverge from one of the universal curves by 10\%.}
    \label{Results:DD_plots}
\end{figure}

\begin{figure}[H]
    \centering
    \begin{subfigure}{0.5\textwidth}
      \includegraphics[width=\textwidth]{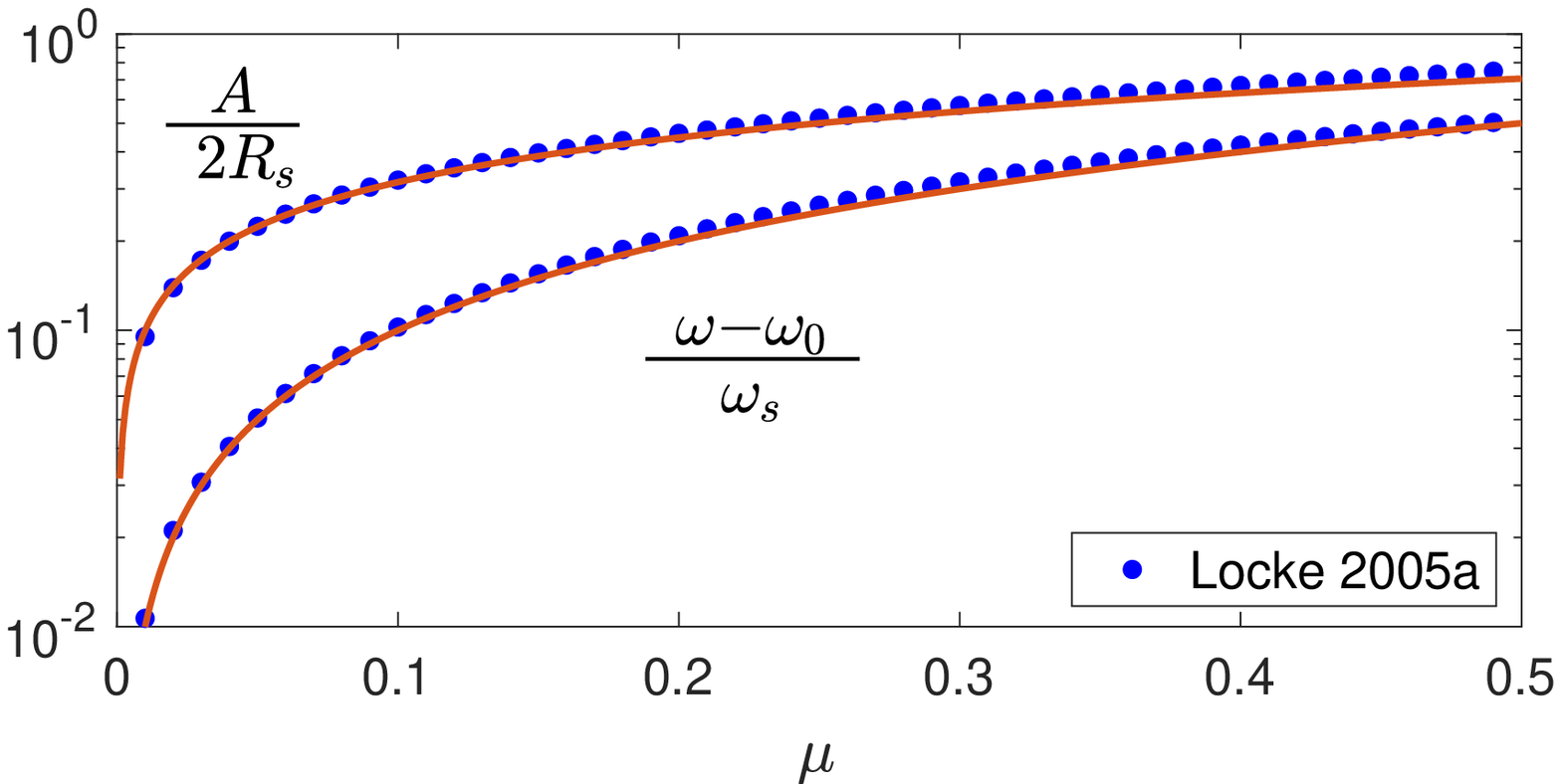}
      \caption{}
      \label{LL-cycle:Locke2005a}
    \end{subfigure}%
    \begin{subfigure}{0.5\textwidth}
      \includegraphics[width=\textwidth]{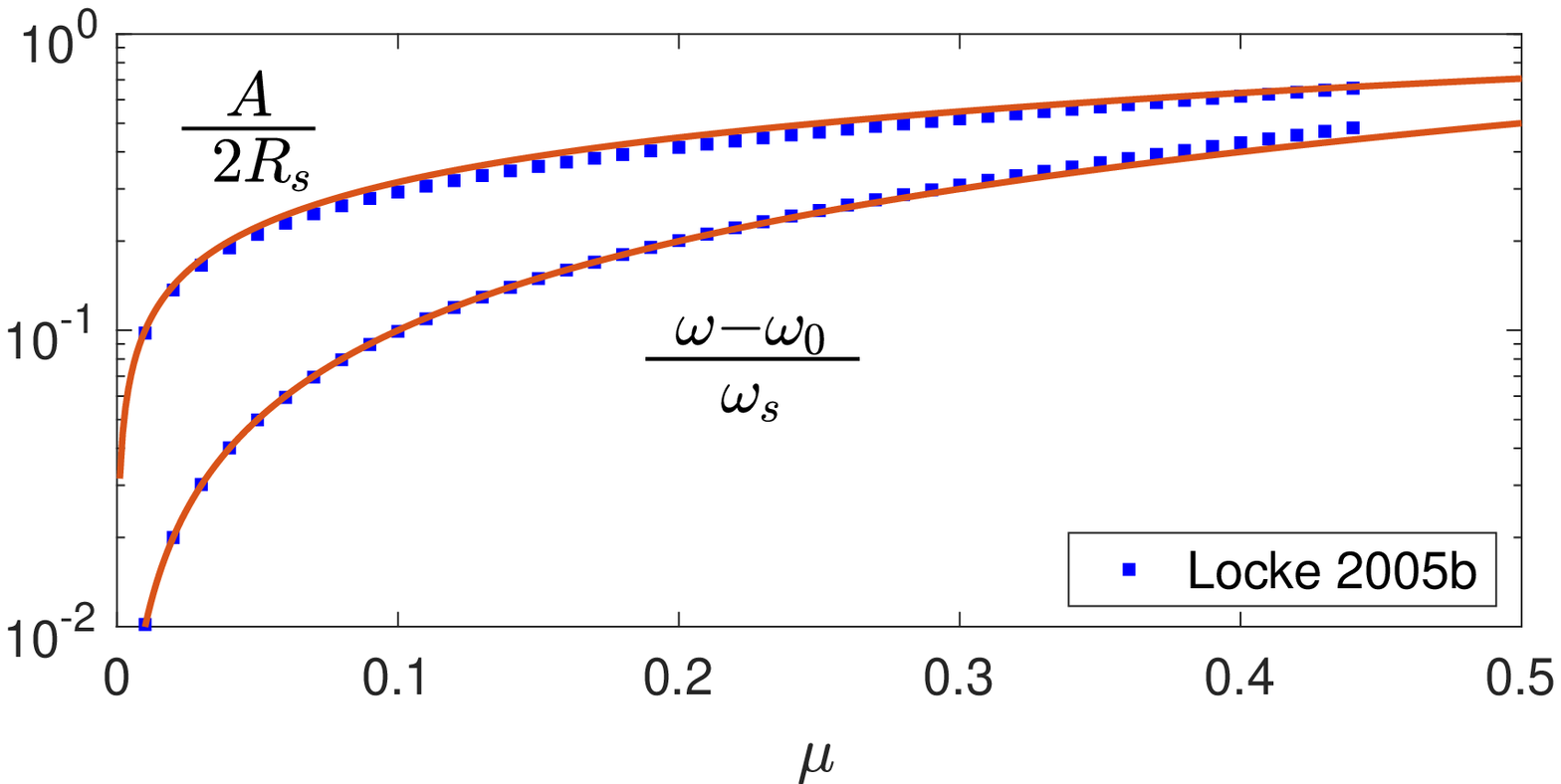}
      \caption{}
      \label{LL-cycle:Locke2005b}
    \end{subfigure}
    \begin{subfigure}{0.5\textwidth}
      \includegraphics[width=\textwidth]{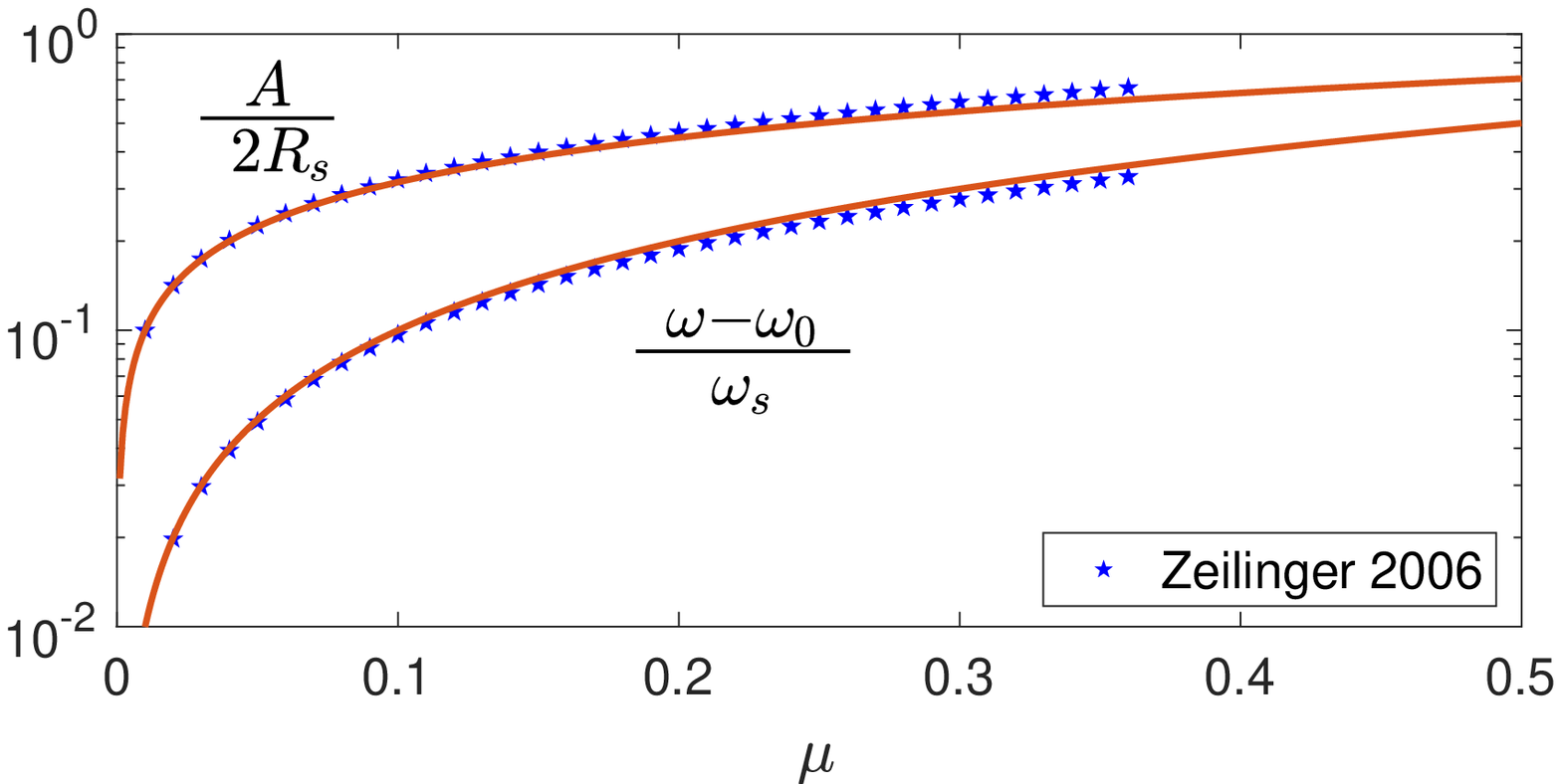}
      \caption{}
      \label{LL-cycle:Zeilinger2006}
    \end{subfigure}%
    \begin{subfigure}{0.5\textwidth}
      \includegraphics[width=\textwidth]{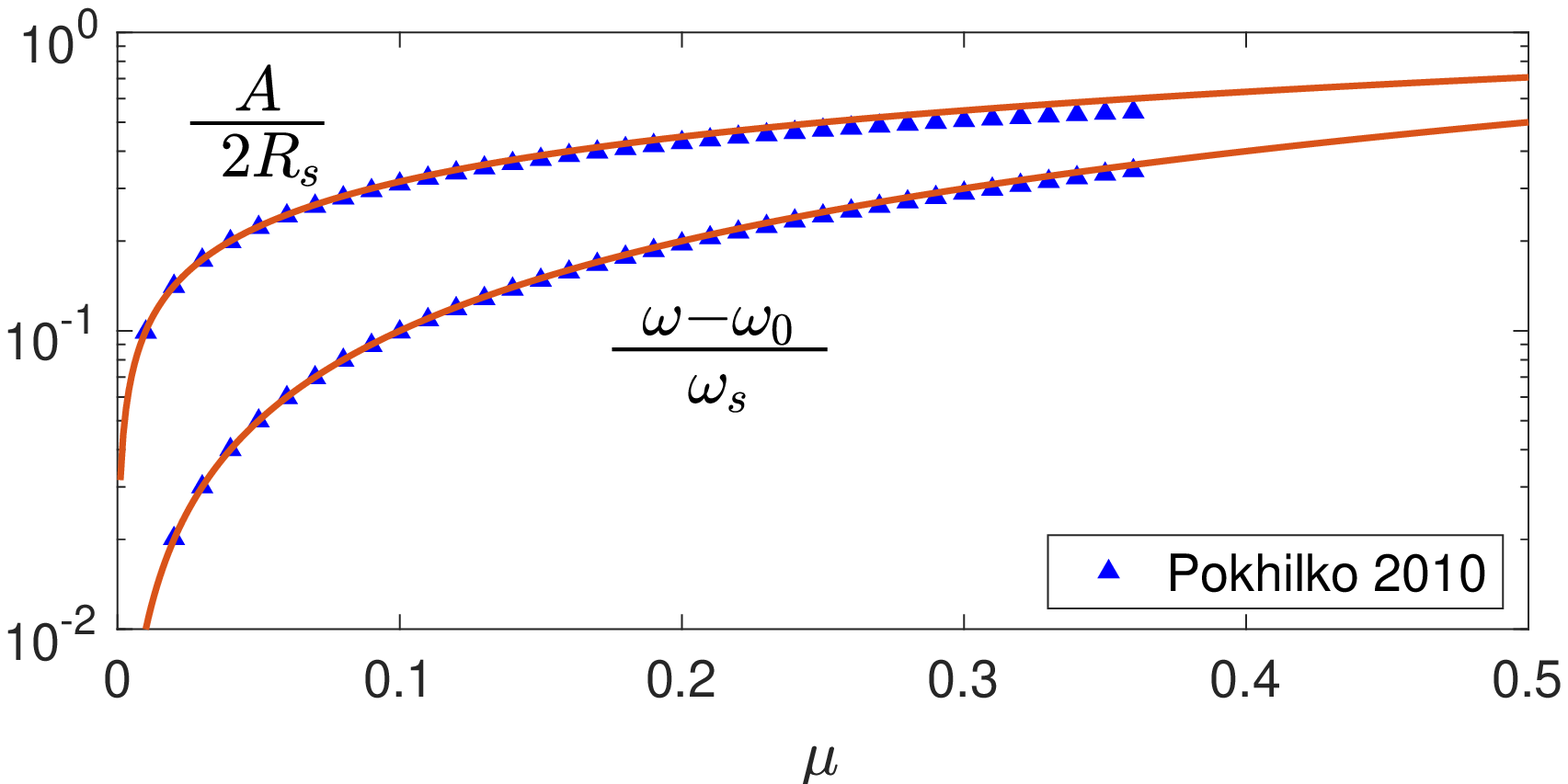}
      \caption{}
      \label{LL-cycle:Pokhilko2010}
    \end{subfigure}
    \begin{subfigure}{0.5\textwidth}
      \includegraphics[width=\textwidth]{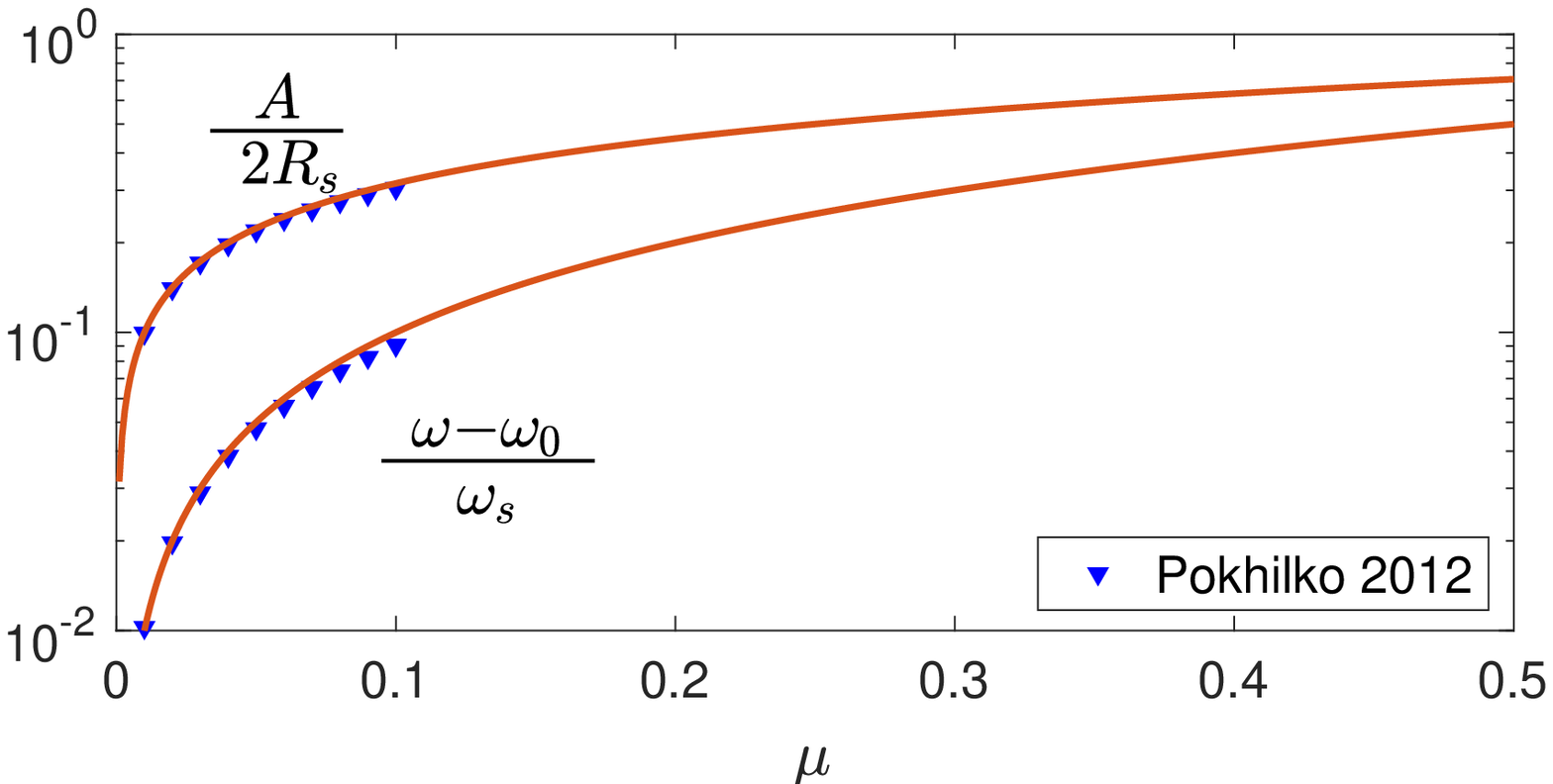}
      \caption{}
      \label{LL-cycle:Pokhilko2012}
    \end{subfigure}%
    \begin{subfigure}{0.5\textwidth}
      \includegraphics[width=\textwidth]{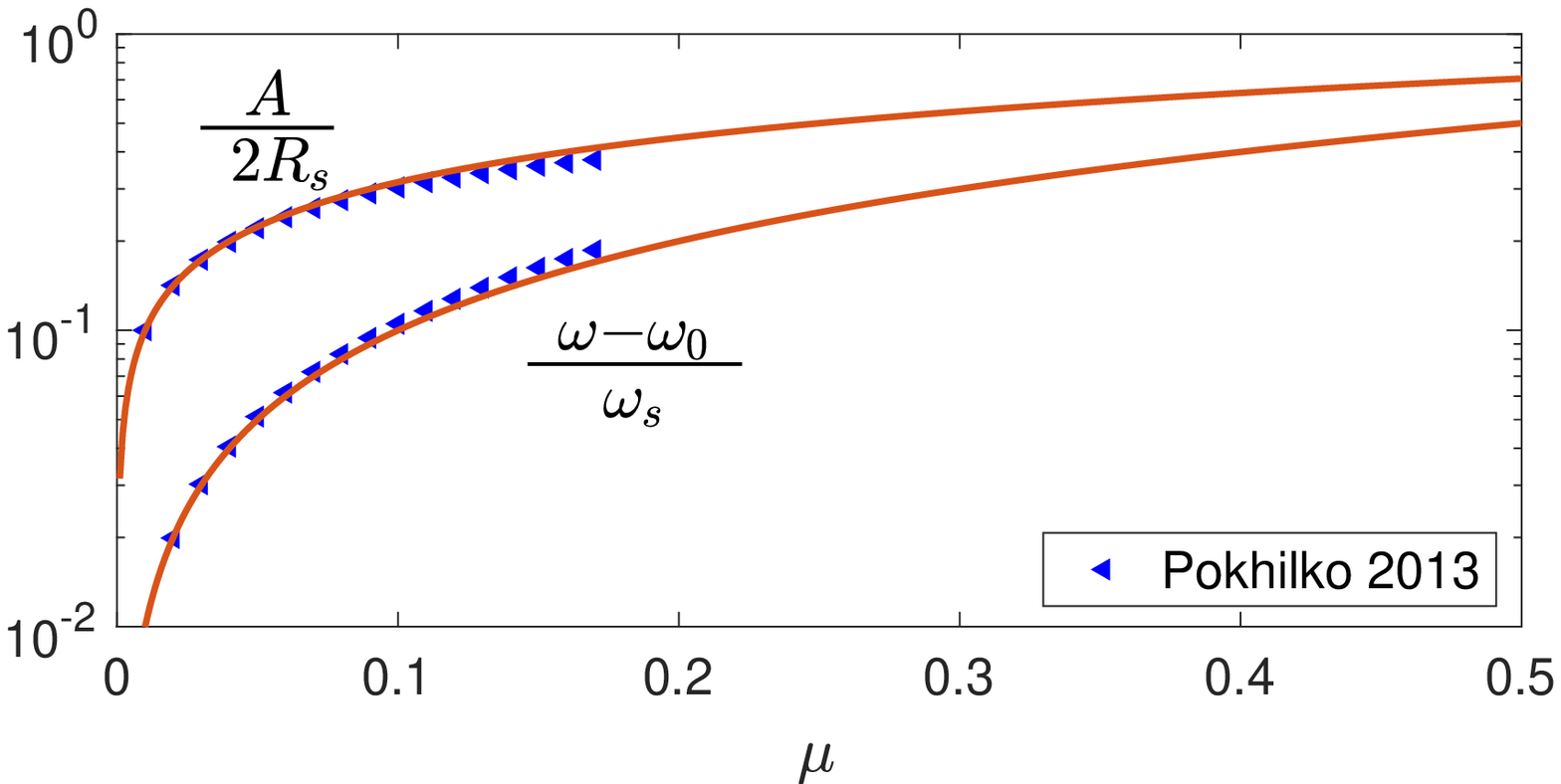}
      \caption{}
      \label{LL-cycle:Pokhilko2013}
    \end{subfigure}
    \begin{subfigure}{0.5\textwidth}
      \includegraphics[width=\textwidth]{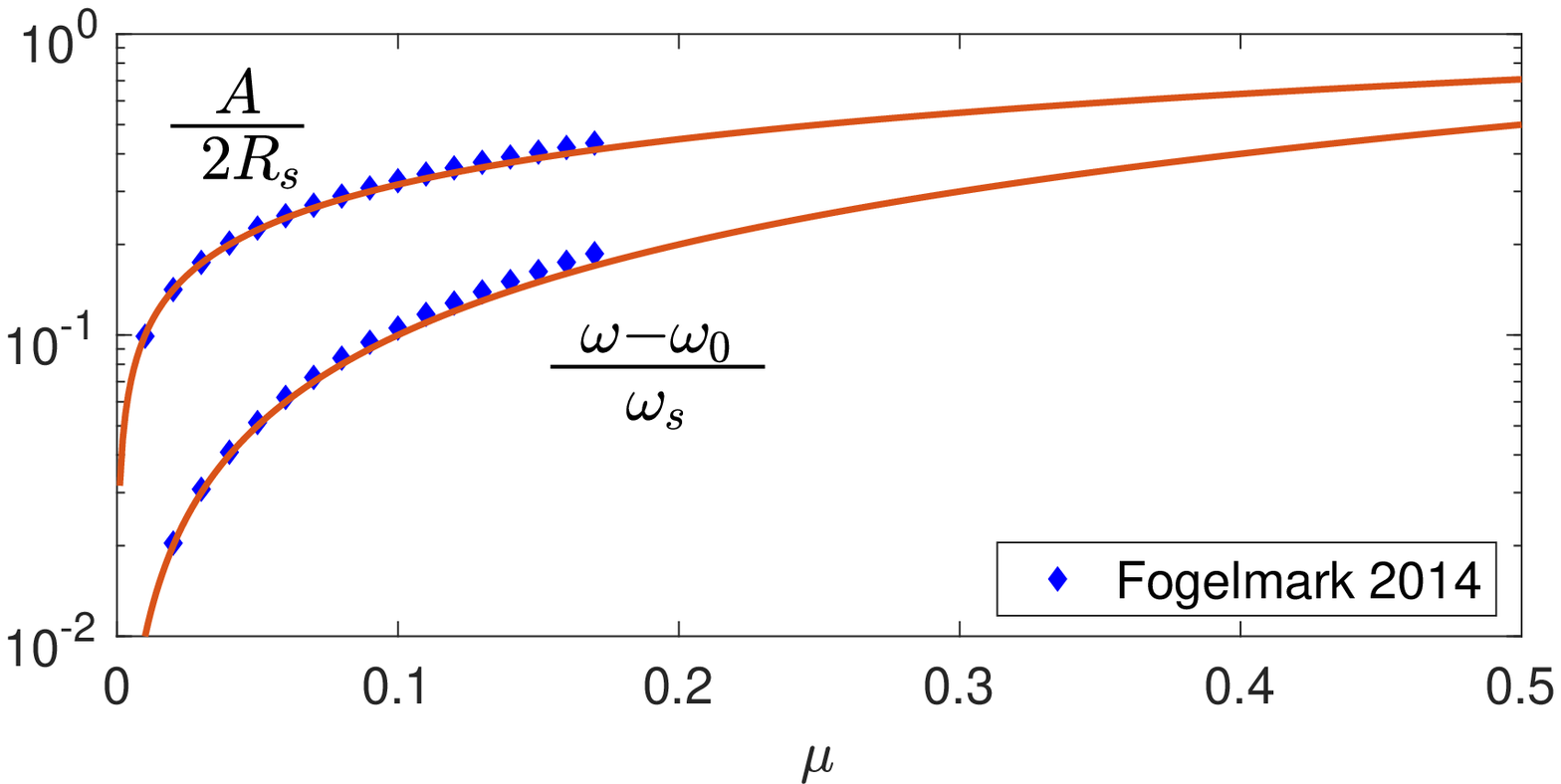}
      \caption{}
      \label{LL-cycle:Fogelmark2014}
    \end{subfigure}%
    \begin{subfigure}{0.5\textwidth}
      \includegraphics[width=\textwidth]{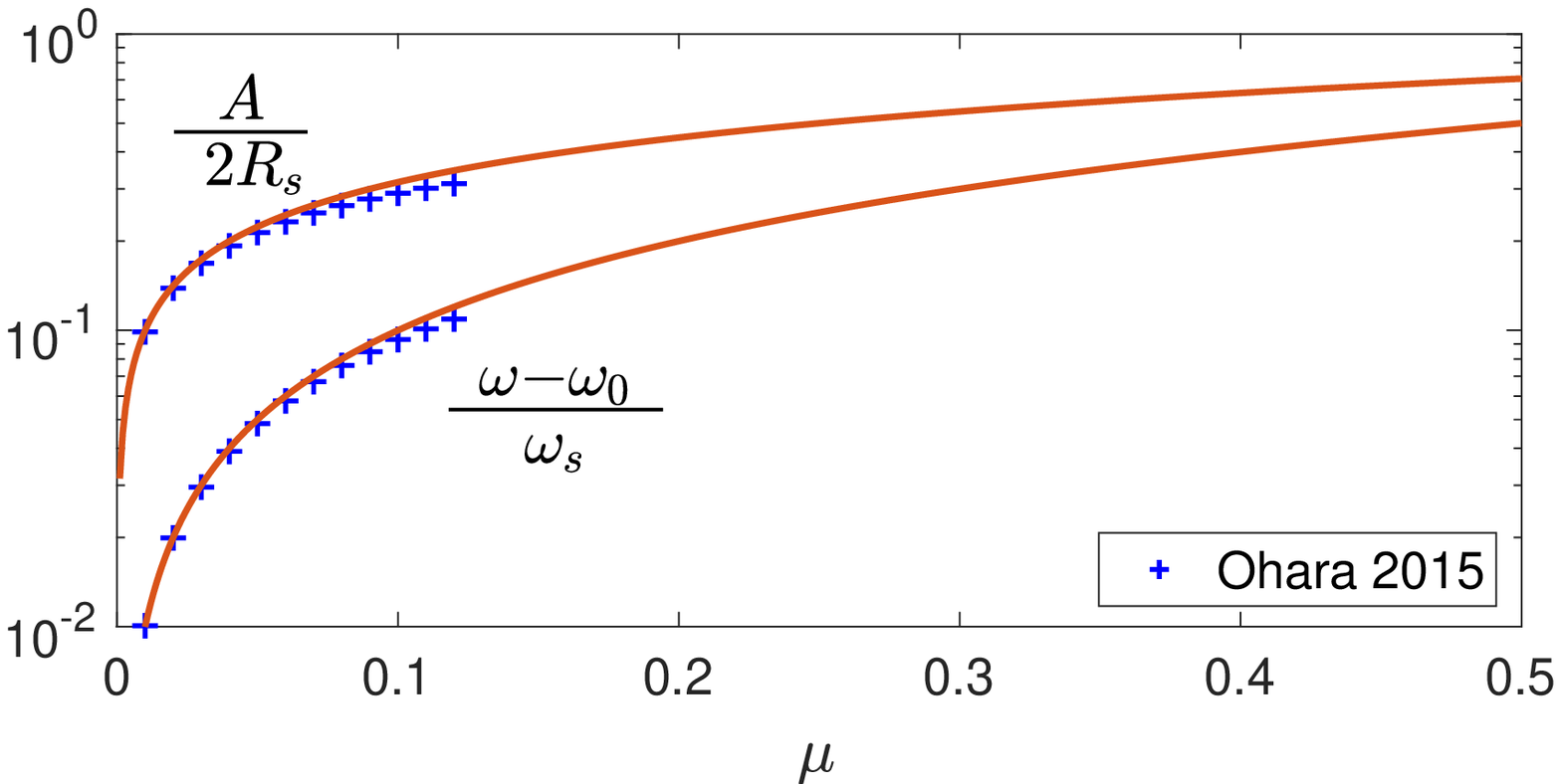}
      \caption{}
      \label{LL-cycle:Ohara2015}
    \end{subfigure}
    \begin{subfigure}{0.5\textwidth}
      \includegraphics[width=\textwidth]{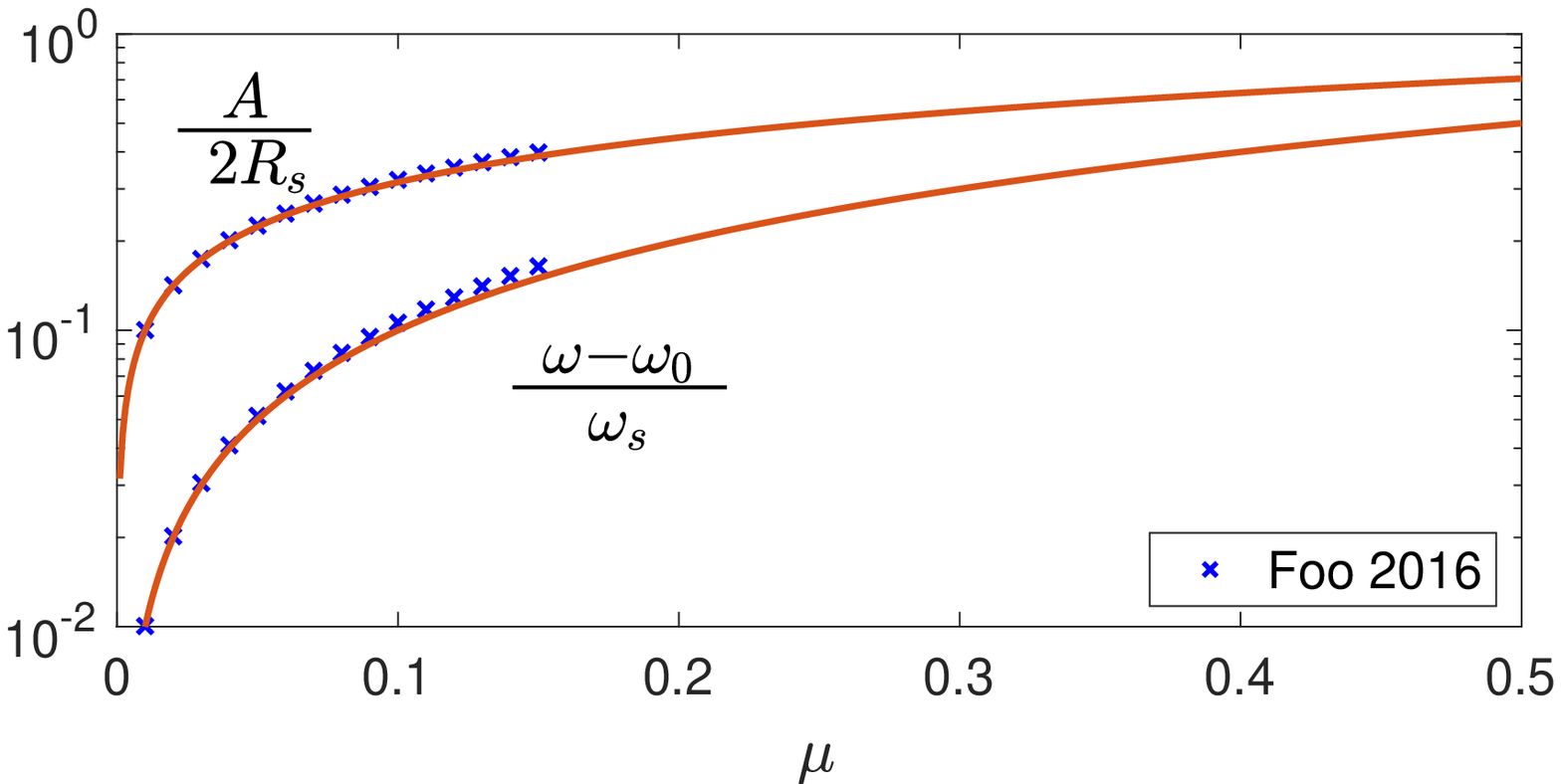}
      \caption{}
      \label{LL-cycle:Foo2016}
    \end{subfigure}%
    \begin{subfigure}{0.5\textwidth}
      \includegraphics[width=\textwidth]{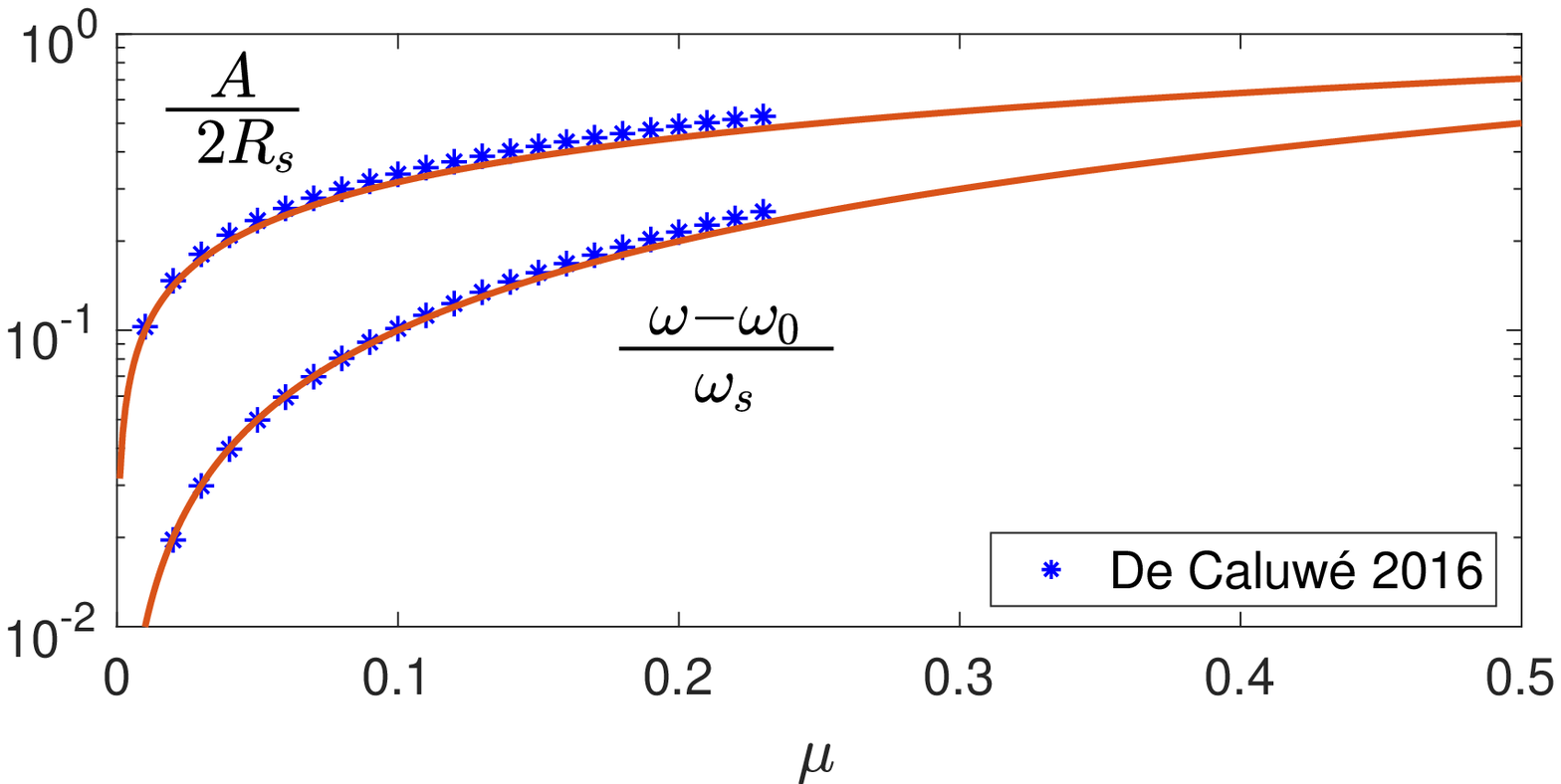}
      \caption{}
      \label{LL-cycle:DeCaluwe2016}
    \end{subfigure}
    \caption{Amplitude and frequency collapse for each model under perpetual illumination. 
    }
    \label{LL-cycle-models}
\end{figure}

\begin{figure}[H]
    \centering
    \begin{subfigure}{0.5\textwidth}
      \includegraphics[width=\textwidth]{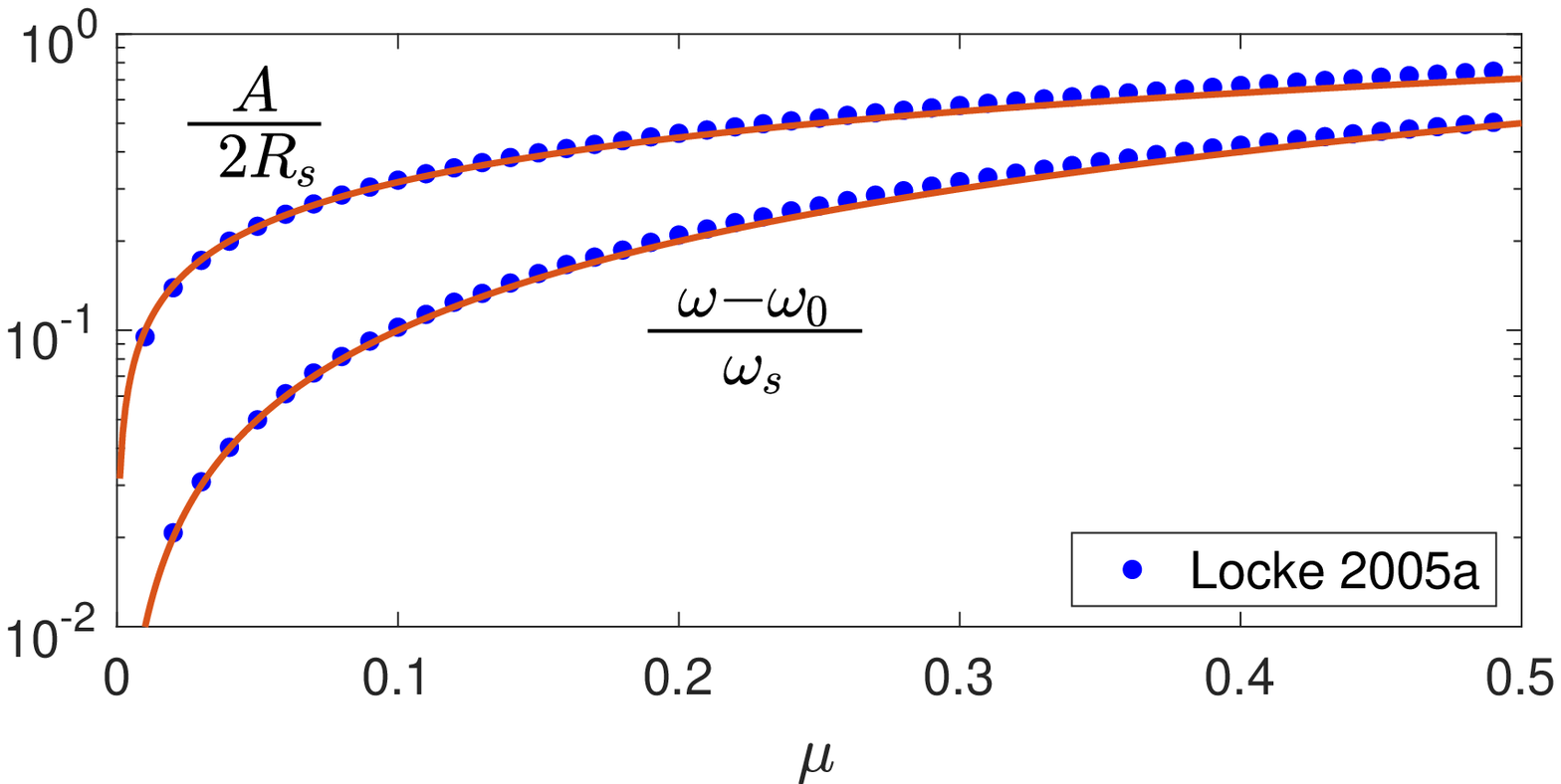}
      \caption{}
      \label{DD-cycle:Locke2005a}
    \end{subfigure}%
    \begin{subfigure}{0.5\textwidth}
      \includegraphics[width=\textwidth]{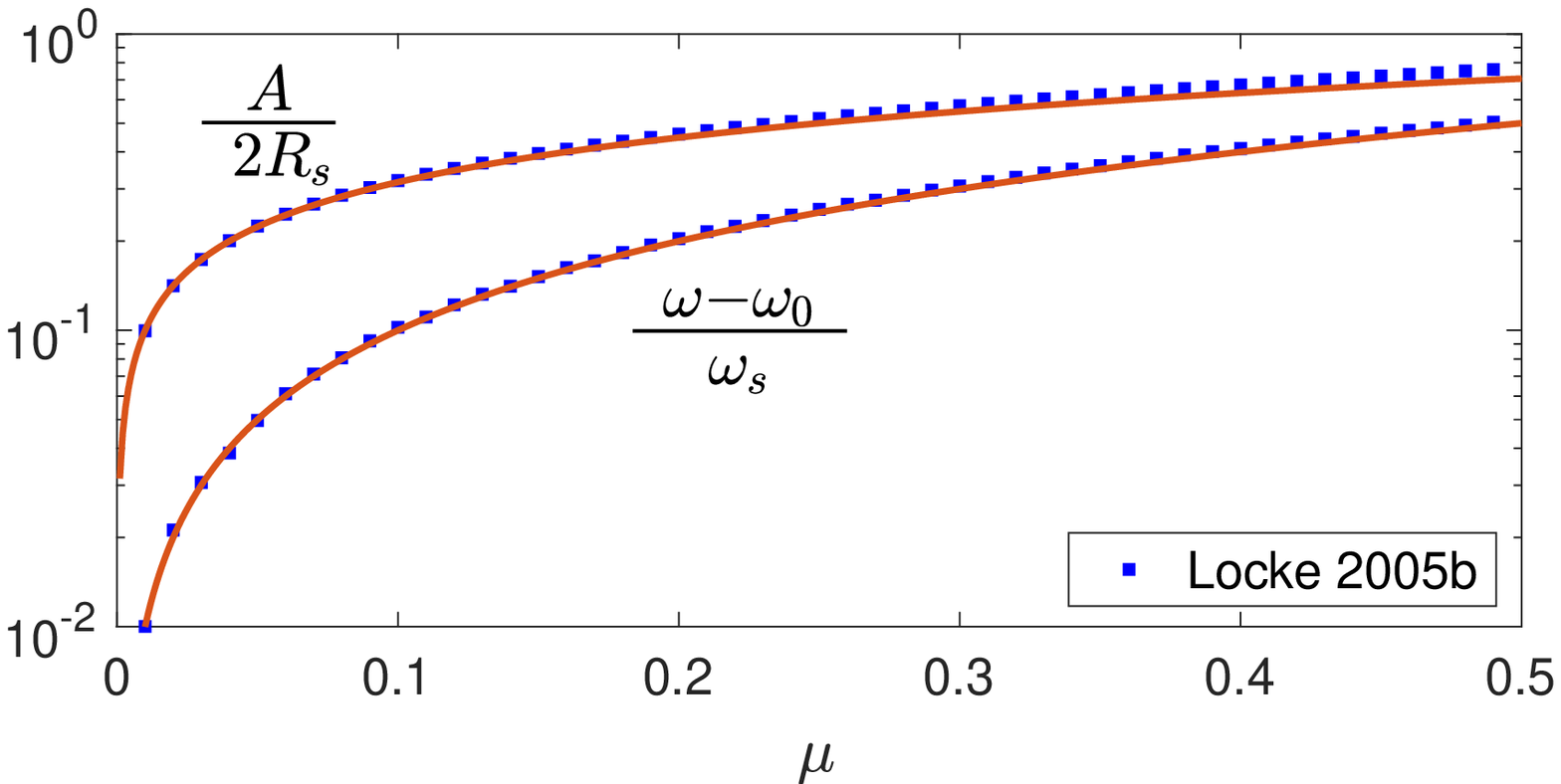}
      \caption{}
      \label{DD-cycle:Locke2005b}
    \end{subfigure}
    \begin{subfigure}{0.5\textwidth}
      \includegraphics[width=\textwidth]{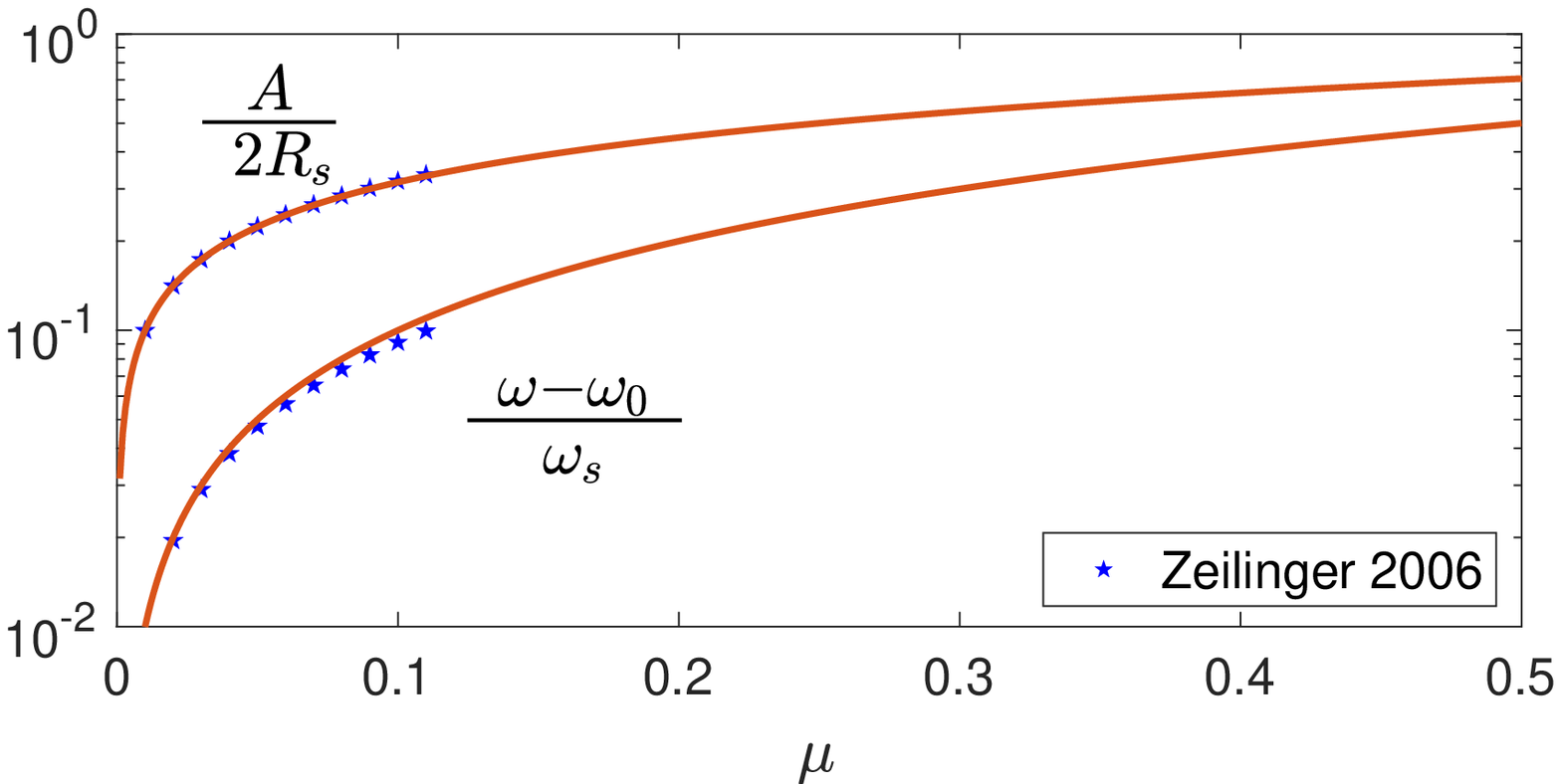}
      \caption{}
      \label{DD-cycle:Zeilinger2006}
    \end{subfigure}%
    \begin{subfigure}{0.5\textwidth}
      \includegraphics[width=\textwidth]{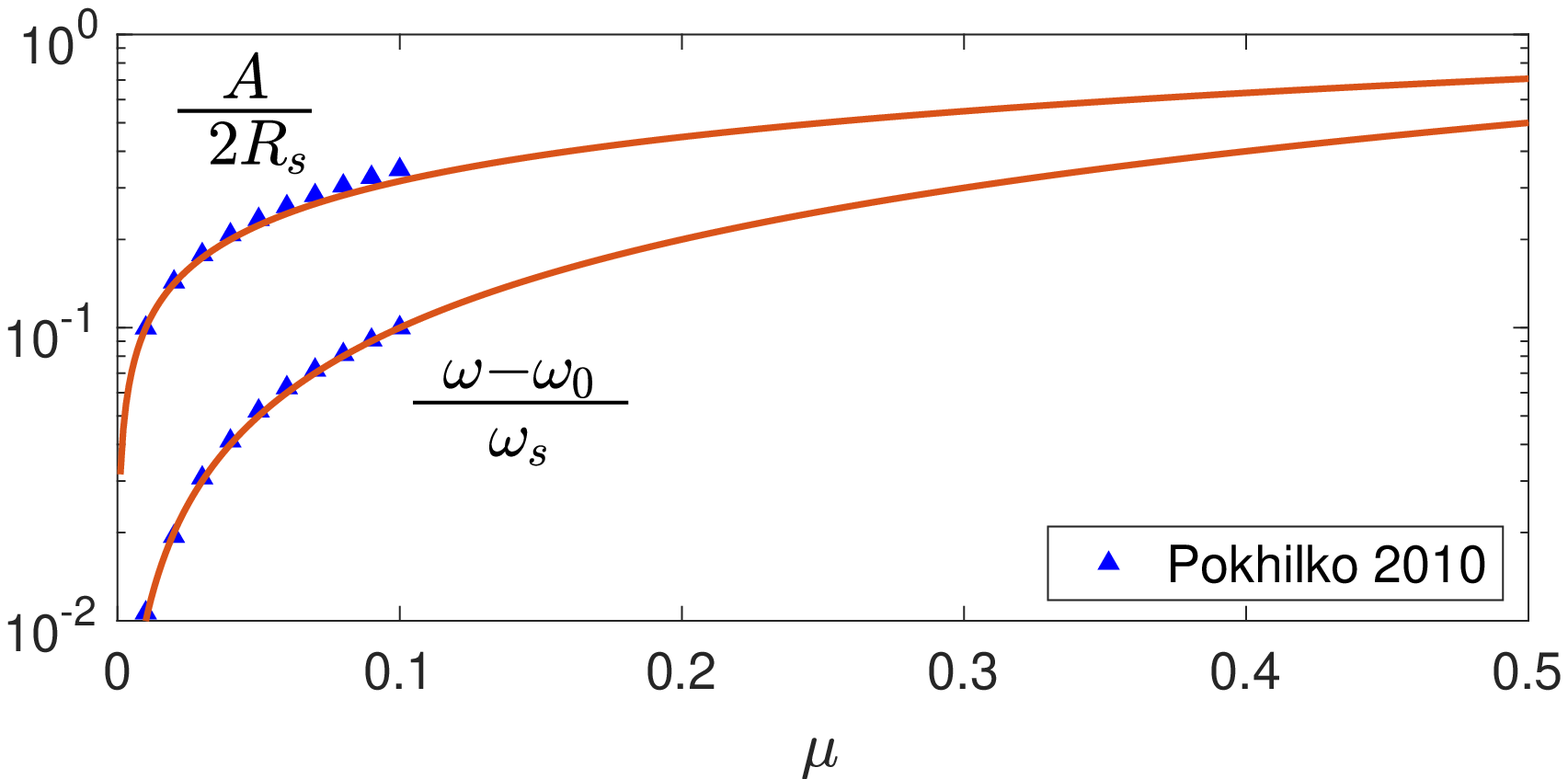}
      \caption{}
      \label{DD-cycle:Pokhilko2010}
    \end{subfigure}
    \begin{subfigure}{0.5\textwidth}
      \includegraphics[width=\textwidth]{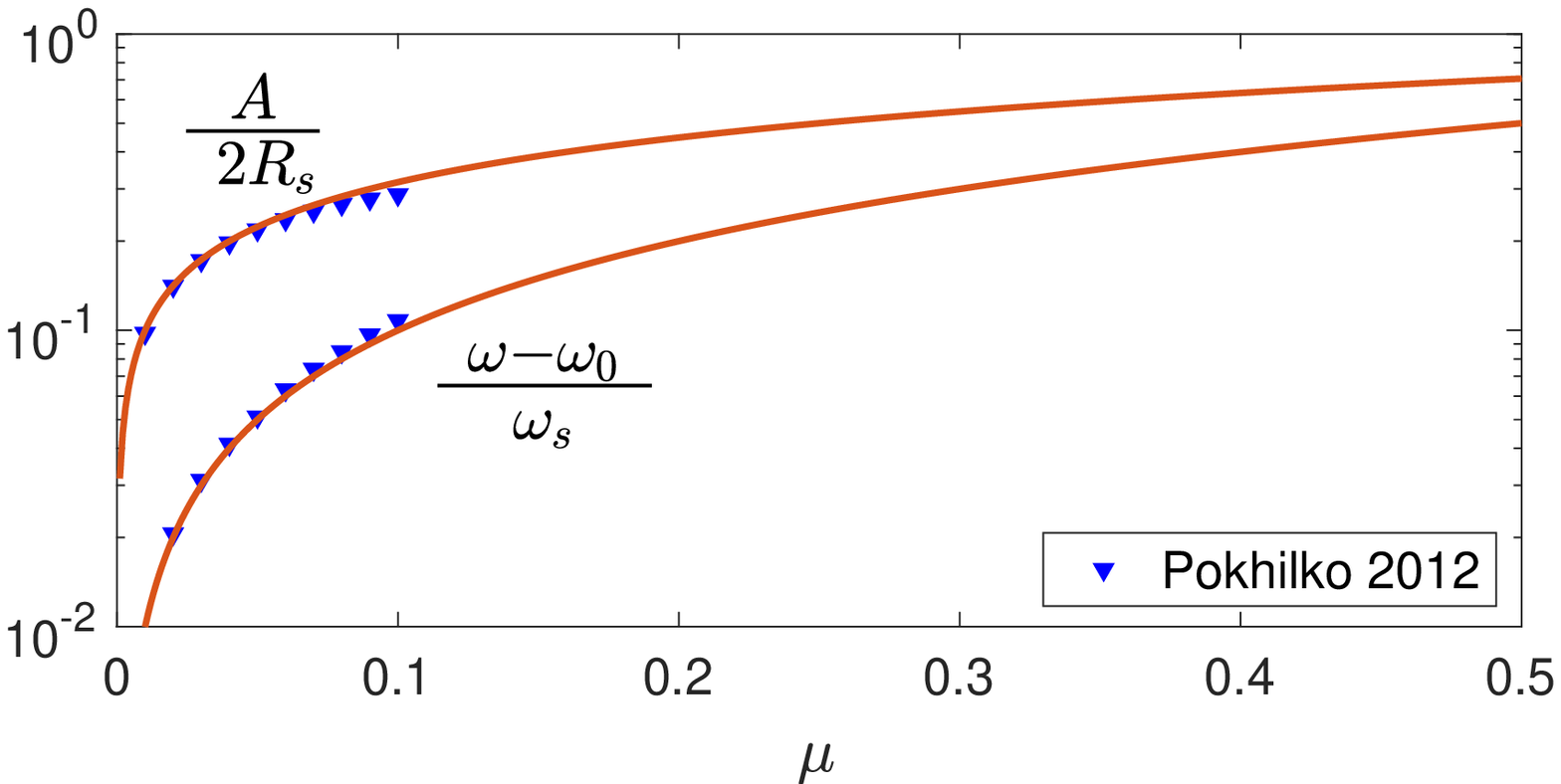}
      \caption{}
      \label{DD-cycle:Pokhilko2012}
    \end{subfigure}%
    \begin{subfigure}{0.5\textwidth}
      \includegraphics[width=\textwidth]{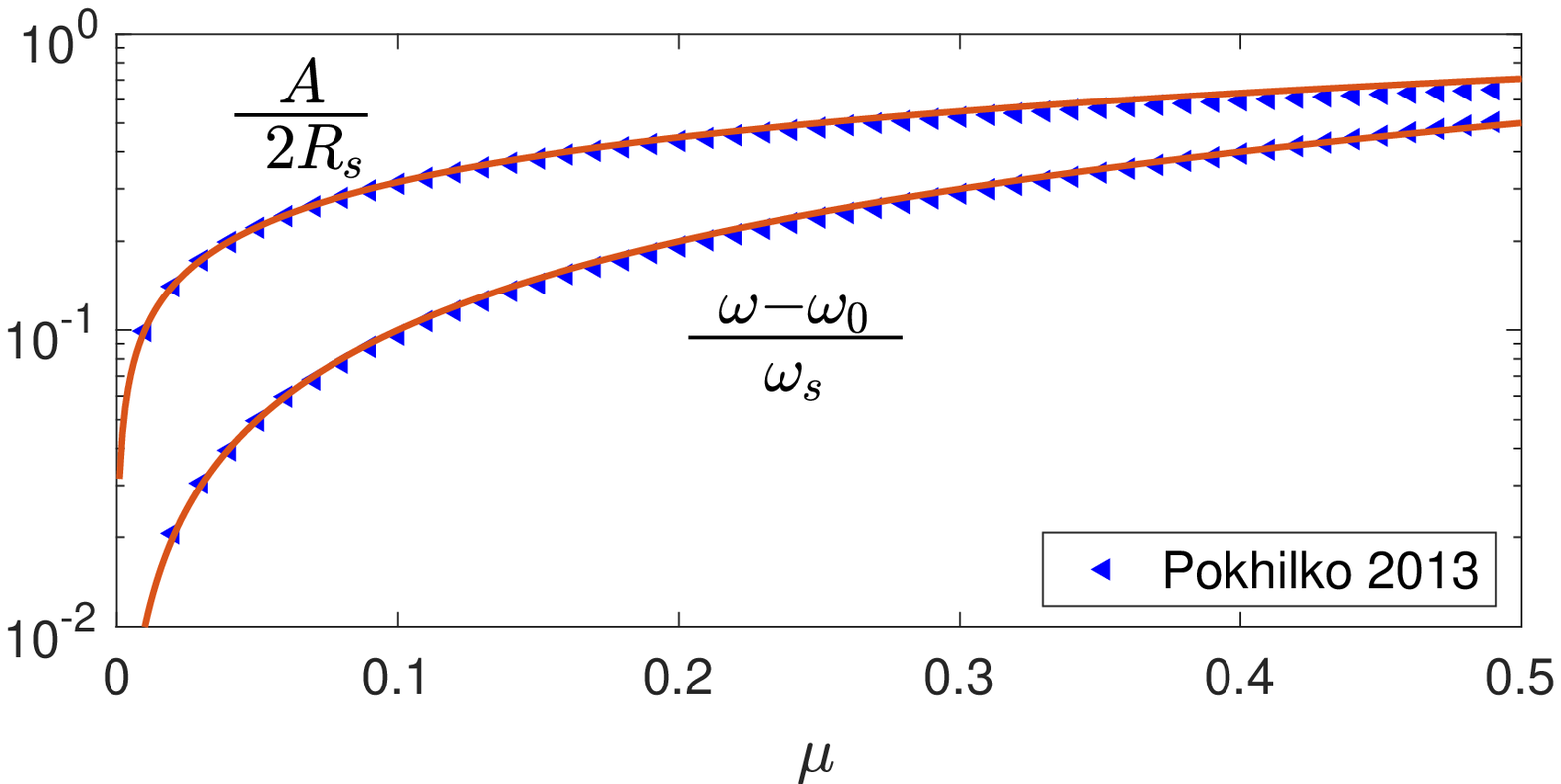}
      \caption{}
      \label{DD-cycle:Pokhilko2013}
    \end{subfigure}
    \begin{subfigure}{0.5\textwidth}
      \includegraphics[width=\textwidth]{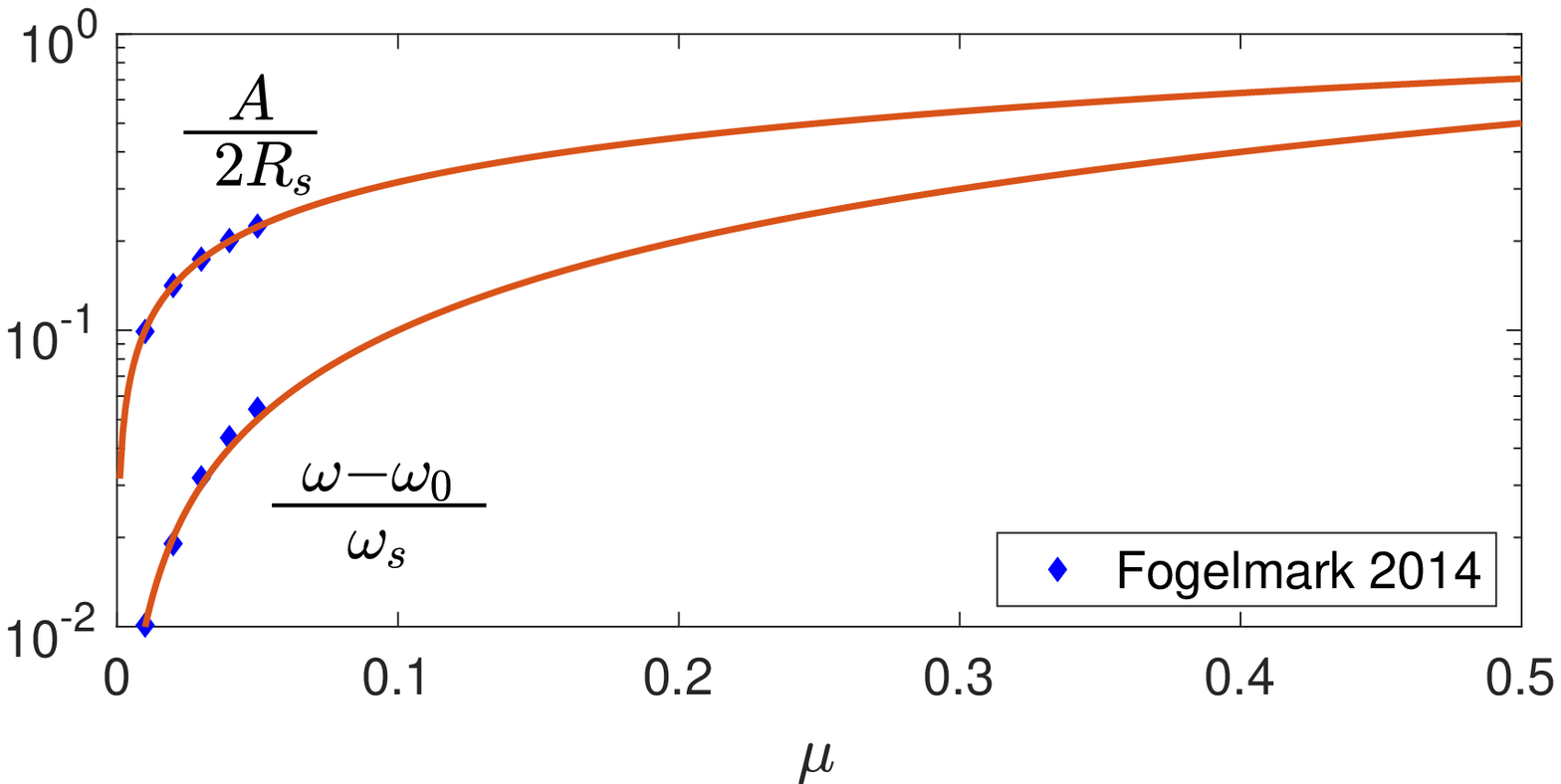}
      \caption{}
      \label{DD-cycle:Fogelmark2014}
    \end{subfigure}%
    \begin{subfigure}{0.5\textwidth}
      \includegraphics[width=\textwidth]{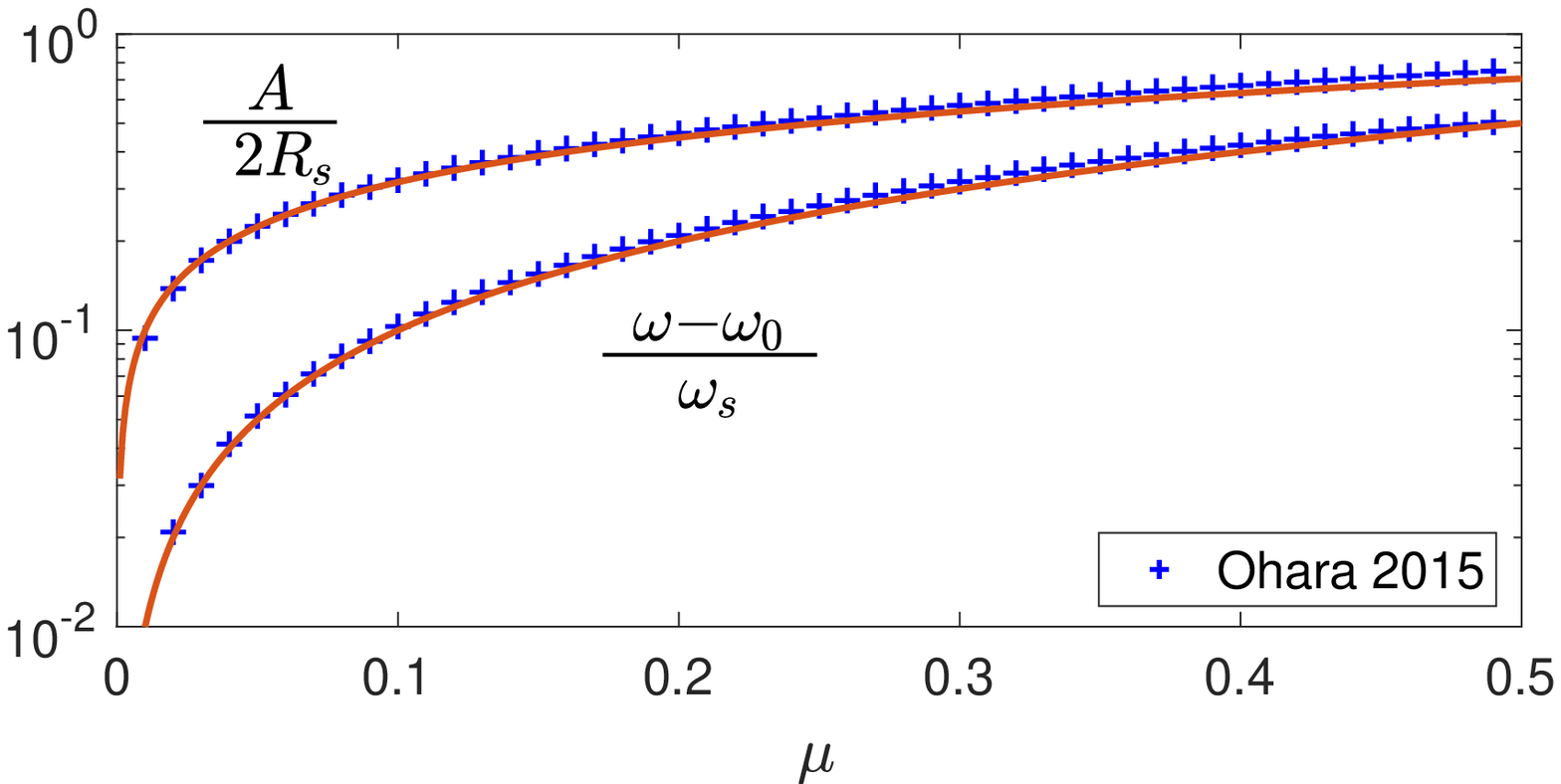}
      \caption{}
      \label{DD-cycle:Ohara2015}
    \end{subfigure}
    \begin{subfigure}{0.5\textwidth}
      \includegraphics[width=\textwidth]{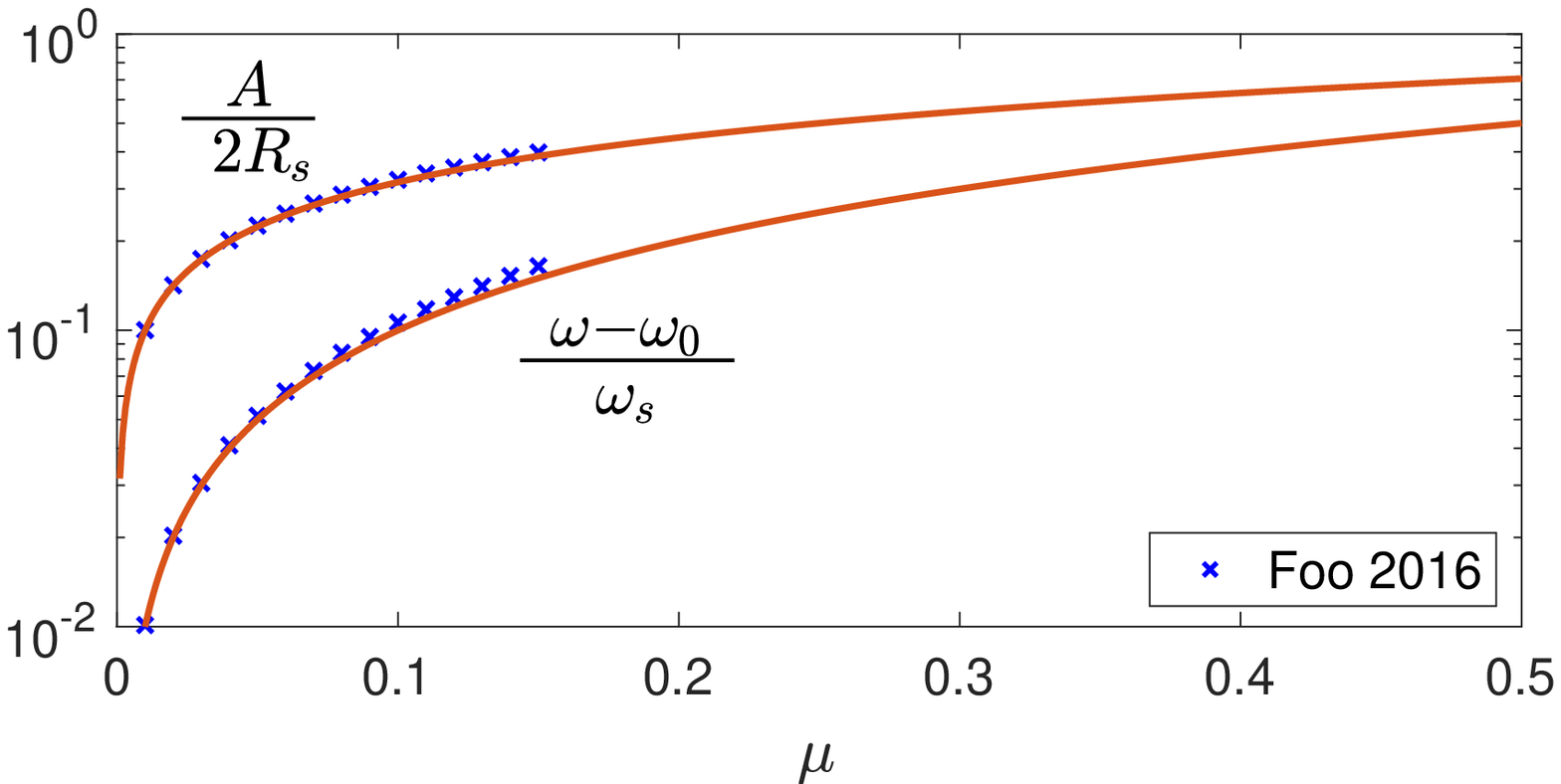}
      \caption{}
      \label{DD-cycle:Foo2016}
    \end{subfigure}%
    \begin{subfigure}{0.5\textwidth}
      \includegraphics[width=\textwidth]{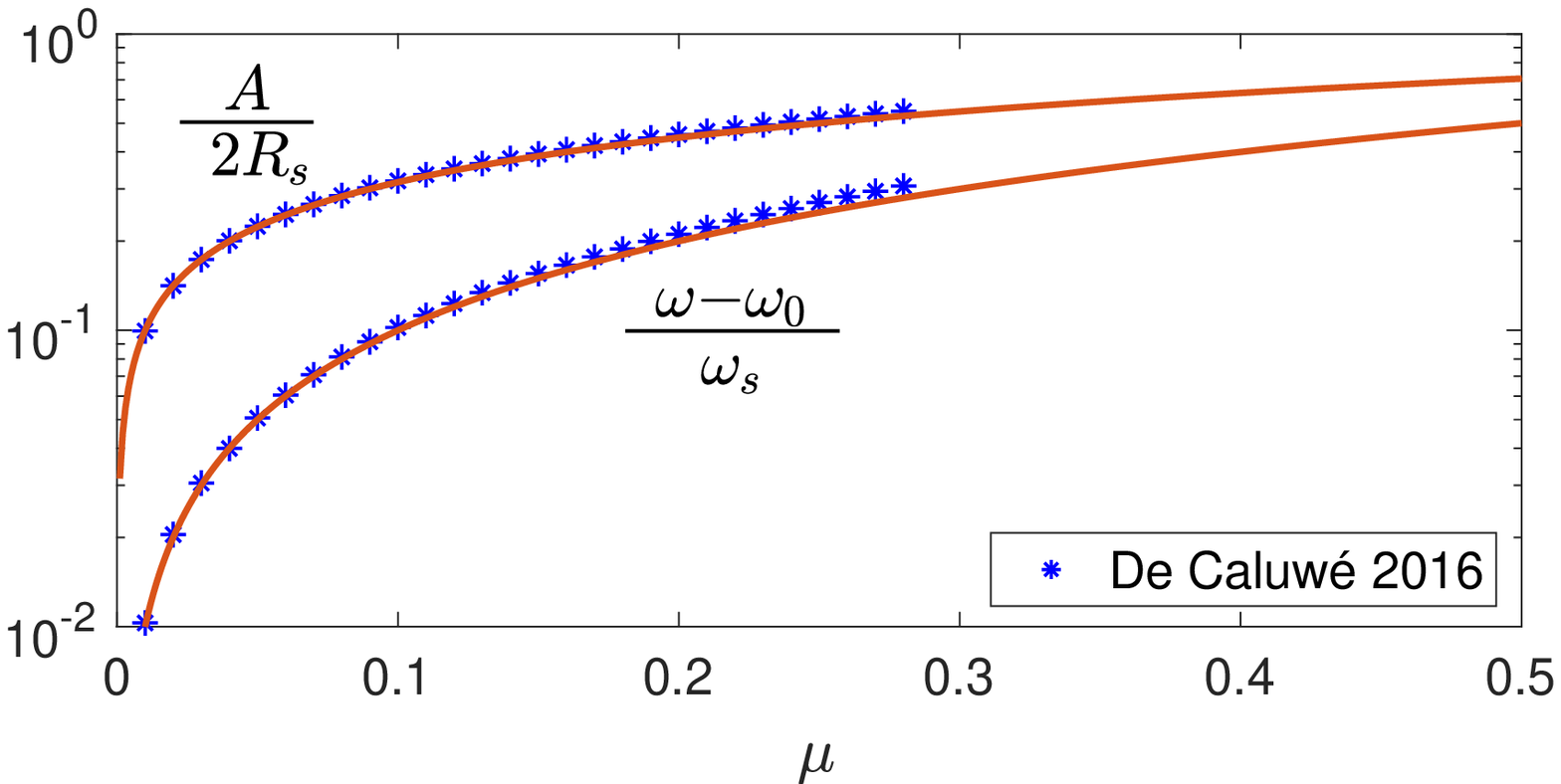}
      \caption{}
      \label{DD-cycle:DeCaluwe2016}
    \end{subfigure}
    \caption{Amplitude and frequency collapse for each model under perpetual darkness. 
    }
    \label{DD-cycle-models}
\end{figure}

\noindent In Figs. \ref{LL-cycle-models} and \ref{DD-cycle-models}, we collapse the limit cycle amplitude and frequency of all the models using their respective natural scales. The data is identical to that plotted in Fig. 3 of the main text and Fig. \ref{Results:DD_plots}, but plotted separately for each model.\vspace{0.3cm}

\noindent \textbf{S2} Bifurcation diagrams, time series of \textit{LHY} and \textit{TOC1} mRNAs, and phase diagrams for Locke et al. 2005a model under perpetual illumination.\vspace{0.3cm}

\noindent \textbf{S3} Bifurcation diagrams, time series of \textit{LHY} and \textit{TOC1} mRNAs, and phase diagrams for Locke et al. 2005b model under perpetual illumination.\vspace{0.3cm}

\noindent \textbf{S4} Bifurcation diagrams, time series of \textit{LHY} and \textit{TOC1} mRNAs, and phase diagrams for Zeilinger et al. 2006 model under perpetual illumination.\vspace{0.3cm}

\noindent \textbf{S5} Bifurcation diagrams, time series of \textit{LHY} and \textit{TOC1} mRNAs, and phase diagrams for Pokhilko et al. 2010 model under perpetual illumination.\vspace{0.3cm}

\noindent \textbf{S6} Bifurcation diagrams, time series of \textit{LHY} and \textit{TOC1} mRNAs, and phase diagrams for Pokhilko et al. 2012 model under perpetual illumination.\vspace{0.3cm}

\noindent \textbf{S7} Bifurcation diagrams, time series of \textit{LHY} and \textit{TOC1} mRNAs, and phase diagrams for Fogelmark et al. 2014 model under perpetual illumination.\vspace{0.3cm}

\noindent \textbf{S8} Bifurcation diagrams, time series of \textit{LHY} and \textit{TOC1} mRNAs, and phase diagrams for Foo et al. 2016 model under perpetual illumination.\vspace{0.3cm}

\noindent \textbf{S9} Bifurcation diagrams, time series of \textit{LHY} and \textit{TOC1} mRNAs, and phase diagrams for Locke et al. 2005a model under perpetual darkness.\vspace{0.3cm}

\noindent \textbf{S10} Bifurcation diagrams, time series of \textit{LHY} and \textit{TOC1} mRNAs, and phase diagrams for Locke et al. 2005b model under perpetual darkness.\vspace{0.3cm}

\noindent \textbf{S11} Bifurcation diagrams, time series of \textit{LHY} and \textit{TOC1} mRNAs, and phase diagrams for Zeilinger et al. 2006 model under perpetual darkness.\vspace{0.3cm}

\noindent \textbf{S12} Bifurcation diagrams, time series of \textit{LHY} and \textit{TOC1} mRNAs, and phase diagrams for Foo et al. 2016 model under perpetual darkness.\vspace{0.3cm}

\noindent \textbf{S13} Bifurcation diagrams, time series of \textit{LHY} and \textit{TOC1} mRNAs, and phase diagrams for De Caluw{\'e} et al. 2016 model under perpetual darkness.\vspace{0.3cm}

\noindent \textbf{S14} Asymptotic amplitude and frequency of oscillation of all models under perpetual darkness.\vspace{0.3cm}

\noindent \textbf{S15} Asymptotic amplitude and frequency of oscillation of each model under perpetual illumination.\vspace{0.3cm}

\noindent \textbf{S16} Asymptotic amplitude and frequency of oscillation of each model under perpetual darkness.\vspace{0.3cm}

\subsection*{Response Curves}
We include the derived Stuart-Landau equation with a forcing function and the response curves for both square-wave forcing and sinusoidal forcing.  These curves may guide determining whether the system is pre-bifurcation or post-bifurcation from experimental results.

\noindent \textbf{S26} Amplitude response curves for systems that are either pre-bifurcation or post-bifurcation.\vspace{0.3cm}

\newpage

\subsection*{List of Pre-bifurcation and Post-Bifurcation Models}
\begin{table}[h!]
    \centering
    \begin{tabular}{|c||c|c|}\hline
        Model  & Perpetual Illumination & Perpetual Darkness    \\\hline\hline
        L2005a & Pre-bifurcation$^\ast$   & Pre-bifurcation$^\ast$  \\\hline
        L2005b & Post-bifurcation       & Post-bifurcation      \\\hline
        Z2006  & Post-bifurcation       & Post-bifurcation      \\\hline
        P2010  & Post-bifurcation       & Pre-bifurcation       \\\hline
        P2012  & Post-bifurcation       & Pre-bifurcation       \\\hline
        P2013  & Pre-bifurcation        & Pre-bifurcation       \\\hline
        F2014  & Post-bifurcation       & Pre-bifurcation       \\\hline
        O2015  & Pre-bifurcation        & Pre-bifurcation       \\\hline
        F2016  & Post-bifurcation       & Post-bifurcation      \\\hline
        DC2016 & Pre-bifurcation        & Post-bifurcation      \\\hline
    \end{tabular}
    \caption{List indicating which models are pre-bifurcation or post-bifurcation with the sets of optimal parameter values reported in the original papers.  Bifurcation diagrams, time series of \textit{LHY} and \textit{TOC1} mRNAs, and the phase diagrams for the post-bifurcation models are shown in Figs. \ref{7-panel:Locke2005a}-\ref{7-panel:DeCaluwe2016-DD}. \\ $^\ast$ The L2005a model is pre-bifurcation with the optimal parameter values but is post-bifurcation with the typical annealed parameter values reported in Fig. 4 of \citet{locke2005modelling}. In the Reductive Perturbation Method and the Stuart-Landau Amplitude Equation subsection of the main text we show results using the typical annealed solution, therefore we show the results for L2005a using the typical annealed parameter values in Figs. S2 and S9 and in Tables S17, S18, S19, and S22.}
    \label{tab:pre-post-models}
\end{table}
\newpage

\subsection*{Post-bifurcation Models Results under Perpetual Illumination}

\newpage

\newpage

\subsection*{Post-bifurcation Models Results under Perpetual Darkness}

\newpage

\newpage

\subsection*{Asymptotic amplitude and frequency of oscillation for all models under Perpetual Darkness}

\newpage

\subsection*{Asymptotic amplitude and frequency of oscillation for each model under Perpetual Illumination}

\subsection*{Asymptotic amplitude and frequency of oscillation for each model under Perpetual Darkness}

\setcounter{table}{16}

\subsection*{Stuart-Landau Parameter Values under Perpetual Illumination}
\begin{table}[H]
    \centering
    \begin{align*}
    &\begin{tabular}{|c|c|c|c|c|c|c|}
        \hline
        Model       & BP            & Scaled By                 & Optimal Value & Critical Value \\ \hline
        L2005a      & $m_2$         & TOC1$_c$                  & 16.9058       & 6.2609         \\
        L2005b      & $m_1$         & TOC1$_c$                  & 1.5283        & 3.6327         \\
        Z2006       & $m_{12}$      & Y$_n$                     & 5.9504        & 9.4460         \\
        P2010       & $m_3$         & Y Protein                 & 0.2           & 0.4195         \\
        P2012       & $m_{35}$      & LUX                       & 0.3           & 1.0493         \\
        P2013       & $m_{15}$      & LUX                       & 0.7           & 0.4099         \\
        F2014       & $m_1$         & $PPR7$ mRNA               & 0.6127        & 2.5971         \\
        O2015       & $m_1$         & TOC1$_c$                  & 9.3383        & 9.0593         \\
        F2016       & $\phi_{36}$   & $PPR7$  mRNA              & 0.37854       & 2.2289         \\
        DC2016      & $d_7$         & ELF4/LUX  Protein         & 0.38          & 0.2232         \\\hline
    \end{tabular}\\
    \\
    &\begin{tabular}{|c|c|c|c|}
        \hline
        Model       & $\omega_0$    & $g = g' + ig''$     & $\lambda_1 = \sigma_1 + i \omega_1$ \\ \hline
        L2005a      & 0.3013        & $0.1686 + 0.1471i$  & $0.0022 + 0.0088i$  \\
        L2005b      & 0.3031        & $0.0003 + 0.0000i$  & $0.0041 - 0.0197i$  \\
        Z2006       & 0.3500        & $0.0015 + 0.0070i$  & $0.0054 - 0.0059i$  \\
        P2010       & 0.3101        & $0.1844 + 0.4475i$  & $0.0929 - 0.0807i$  \\
        P2012       & 0.2193        & $0.0573 + 0.0056i$  & $0.0656 + 0.0857i$  \\
        P2013       & 0.2647        & $0.0472 + 0.0595i$  & $0.0382 - 0.0269i$  \\
        F2014       & 0.3375        & $0.1259 + 0.0817i$  & $0.0131 - 0.0149i$  \\
        O2015       & 0.2614        & $0.1076 + 0.0817i$  & $0.0099 + 0.0206i$  \\
        F2016       & 0.3519        & $0.7441 + 0.3900i$  & $0.0215 - 0.0246i$  \\
        DC2016      & 0.2477        & $0.1318 + 0.1077i$  & $0.0979 - 0.2052i$  \\\hline
    \end{tabular}
    \end{align*}
    \caption{Bifurcation parameters (BP), the chemical species that corresponds to the component in the eigenvectors with the largest modulus (Scaled By), optimal and critical values of the bifurcation parameter for each model, along with important Stuart-Landau parameters.  Calculations and simulations are conducted under perpetual illumination.}
    \label{tab.SLParametersLL}
\end{table}
\newpage

\subsection*{Stuart-Landau Parameter Values under Perpetual Darkness}
\begin{table}[H]
    \centering
    \begin{align*}
    &\begin{tabular}{|c|c|c|c|c|c|}
        \hline
        Model       & BP            & Scaled By                 & Optimal Value & Critical Value \\ \hline
        L2005a      & $m_2$         & TOC1$_c$                  & 16.9058       & 6.2609         \\
        L2005b      & $m_{13}$      & TOC1$_c$                  & 0.1347        & 0.7518         \\
        Z2006       & $m_{12}$      & Y$_n$                     & 5.9504        & 13.9347        \\
        P2010       & $m_{10}$      & Y  Protein                & 0.3           & 0.2751         \\  
        P2012       & $m_{19}$      & LUX                       & 0.2           & 0.0738         \\
        P2013       & $m_{15}$      & LUX                       & 0.7           & 0.6589         \\
        F2014       & $m_{31}$      & LUX                       & 0.3           & 0.1017         \\
        O2015       & $m_2$         & TOC1$_c$                  & 16.9058       & 6.2609         \\
        F2016       & $\phi_{36}$   & $PPR7$  mRNA              & 0.37854       & 2.2453         \\
        DC2016      & $k_2$         & $CCA1/LHY$  mRNA          & 0.21          & 1.9574         \\\hline
    \end{tabular}\\
    \\
    &\begin{tabular}{|c|c|c|c|}
        \hline
        Model       & $\omega_0$    & $g = g' + ig''$     & $\lambda_1 = \sigma_1 + i \omega_1$ \\ \hline
        L2005a      & 0.3013        & $0.1686 + 0.1471i$  & $0.0022 + 0.0088i$  \\
        L2005b      & 0.2274        & $0.0030 - 0.0010i$  & $0.0232 + 0.0160i$  \\
        Z2006       & 0.4377        & $0.0015 + 0.0086i$  & $0.0050 - 0.0035i$  \\
        P2010       & 0.1952        & $1.0472 + 0.2216i$  & $0.1800 + 0.1757i$  \\  
        P2012       & 0.2225        & $0.0691 + 0.0539i$  & $0.1013 - 0.5555i$  \\
        P2013       & 0.2541        & $0.0655 + 0.0725i$  & $0.0245 + 0.0171i$  \\
        F2014       & 0.2842        & $0.0343 + 0.0411i$  & $0.1516 + 0.2601i$  \\
        O2015       & 0.3013        & $0.1686 + 0.1471i$  & $0.0022 + 0.0088i$  \\
        F2016       & 0.3523        & $0.7458 + 0.3953i$  & $0.0213 - 0.0242i$  \\
        DC2016      & 0.4457        & $0.5415 + 0.6186i$  & $0.0466 + 0.0051i$  \\\hline
    \end{tabular}
    \end{align*}
    \caption{Bifurcation parameters (BP), the chemical species that corresponds to the component in the eigenvectors with the largest modulus (Scaled By), optimal and critical values of the bifurcation parameter for each model, along with important Stuart-Landau parameters.  Calculations and simulations are conducted under perpetual darkness.}
    \label{tab.SLParametersDD}
\end{table}
\newpage

\subsection*{Eigenvector Entries under Perpetual Illumination}
\begin{table}[H]
    \centering
    \begin{align*}
    &\addvbuffer[0cm 0.2cm]{\begin{tabular}{|c||c|c|}
        \hline
        Chemical Species        & L2005a            & L2005b            \\ \hline\hline
        \textit{LHY} mRNA       & -0.0065 - 0.1770i & -0.0060 - 0.0171i \\\hline
        LHY protein cytoplasm   & -0.0406 - 0.2550i & -0.0002 - 0.0005i \\\hline
        LHY protein nucleus     & -0.0658 - 0.1958i & -0.0013 - 0.0027i \\\hline
        \textit{TOC1} mRNA      & 0.1283 + 0.2329i  & 0.0302 + 0.0749i  \\\hline
        TOC1 protein cytoplasm  & 1.0000 + 0.0000i  & 1.0000 + 0.0000i  \\\hline
        TOC1 protein nucleus    & 0.0088 - 0.1492i  & 0.0998 - 0.0096i  \\\hline
    \end{tabular}}\\
    &\begin{tabular}{|c||c|c|}
        \hline
        Chemical Species        & Z2006             & P2010             \\ \hline\hline
        \textit{LHY} mRNA       & -0.0150 + 0.0116i & -0.1086 - 0.3031i \\\hline
        LHY protein cytoplasm   & -0.0019 + 0.0180i & -0.1496 - 0.1325i \\\hline
        LHY protein nucleus     & -0.0009 + 0.0090i & -0.0759 + 0.0213i \\\hline
        \textit{TOC1} mRNA      & 0.1978 - 0.0181i  & 0.1504 - 0.1471i  \\\hline
        TOC1 protein cytoplasm  & 0.0060 - 0.0443i  & 0.0048 - 0.0832i  \\\hline
        TOC1 protein nucleus    & 0.0241 - 0.2542i  & -0.0196 - 0.0198i \\\hline
    \end{tabular}
    \end{align*}
    \caption{Eigenvector entries for the mRNA and proteins of $LHY/CCA1$ and $TOC1$ genes under perpetual illumination.  For P2010 model, the second and fifth entries are protein (LHY and TOC1 in Pokhilko et al. 2010), and the third and sixth entries are modified protein (LHY$_{\text{mod}}$ and TOC1$_{\text{mod}}$ in Pokhilko et al. 2010).}
    \label{tab:Eigenvectors1_LL}
\end{table}

\begin{table}[H]
    \centering
    \begin{align*}
    &\addvbuffer[0cm 0.2cm]{\begin{tabular}{|c||c|c|}
        \hline
        Chemical Species        & P2012             & P2013             \\ \hline\hline
        \textit{LHY} mRNA       & -0.0629 - 0.0056i & -0.0999 - 0.1054i \\\hline
        LHY protein cytoplasm   & -0.0601 + 0.0402i & -0.1464 - 0.0116i \\\hline
        LHY protein nucleus     & -0.0026 + 0.0173i & -0.0275 + 0.0310i \\\hline
        \textit{TOC1} mRNA      & 0.2803 - 0.0578i  & 0.1630 + 0.0013i  \\\hline
        TOC1 protein cytoplasm  & 0.1770 - 0.0606i  & 0.0938 - 0.0310i  \\\hline
        TOC1 protein nucleus    &                   &                   \\\hline
    \end{tabular}}\\
    &\begin{tabular}{|c||c|c|}
        \hline
        Chemical Species        & F20014             & O2015            \\ \hline\hline
        \textit{LHY} mRNA       & -0.0347 + 0.0147i & -0.0479 - 0.3371i \\\hline
        LHY protein cytoplasm   & -0.0333 + 0.0422i & -0.0392 - 0.2089i \\\hline
        LHY protein nucleus     &                   & -0.0530 - 0.1568i \\\hline
        \textit{TOC1} mRNA      & 0.0630 - 0.0229i  & 0.1083 + 0.2022i  \\\hline
        TOC1 protein cytoplasm  & 0.0887 - 0.2355i  & 1.0000 + 0.0000i  \\\hline
        TOC1 protein nucleus    & 0.0127 - 0.0313i  & 0.0061 - 0.1723i  \\\hline
    \end{tabular}
    \end{align*}
    \caption{Eigenvector entries for the mRNA and proteins of $LHY/CCA1$ and $TOC1$ genes under perpetual illumination continued.  Empty entries are due to different definitions of variables.}
    \label{tab:Eigenvectors2_LL}
\end{table}

\begin{table}[H]
    \centering
    \begin{tabular}{|c||c|c|}
        \hline
        Chemical Species    & F2016             & DC2016 \\ \hline\hline
        \textit{LHY} mRNA   & -0.1688 + 0.4729i & -0.0813 - 0.3151i \\\hline
        LHY protein         & 0.0881 + 0.5630i  & -0.3003 - 0.4373i \\\hline
        \textit{TOC1} mRNA  & 0.0831 - 0.6266i  & 0.6043 + 0.4132i \\\hline
        TOC1 protein        & -0.3526 - 0.4070i & 0.5482 + 0.1650i \\\hline
    \end{tabular}
    \caption{Eigenvector entries for the mRNA and proteins of $LHY/CCA1$ and $TOC1$ genes under perpetual illumination continued.  F2016 and DC2016 models use only one variable for proteins of $LHY$ and $TOC1$.  The $TOC1$ entries represent $PRR5/TOC1$ gene group.}
    \label{tab:Eigenvectors3_LL}
\end{table}

\subsection*{Eigenvector Entries under Perpetual Darkness}
\begin{table}[H]
    \centering
    \begin{align*}
    &\addvbuffer[0cm 0.2cm]{\begin{tabular}{|c||c|c|}
        \hline
        Chemical Species        & L2005a            & L2005b            \\ \hline\hline
        \textit{LHY} mRNA       & -0.0065 - 0.1770i & -0.0054 - 0.0273i \\\hline
        LHY protein cytoplasm   & -0.0406 - 0.2550i & -0.0002 - 0.0008i \\\hline
        LHY protein nucleus     & -0.0658 - 0.1958i & -0.0013 - 0.0044i \\\hline
        \textit{TOC1} mRNA      & 0.1283 + 0.2329i  & 0.0426 + 0.0550i  \\\hline
        TOC1 protein cytoplasm  & 1.0000 + 0.0000i  & 1.0000 + 0.0000i  \\\hline
        TOC1 protein nucleus    & 0.0088 - 0.1492i  & 0.0815 - 0.0048i  \\\hline
    \end{tabular}}\\
    &\begin{tabular}{|c||c|c|}
        \hline
        Chemical Species        & Z2006             & P2010             \\ \hline\hline
        \textit{LHY} mRNA       & -0.0102 + 0.0098i & -0.5879 - 0.2651i \\\hline
        LHY protein cytoplasm   & 0.0013 + 0.0118i  & -0.4563 + 0.0538i \\\hline
        LHY protein nucleus     & 0.0007 + 0.0059i  & -0.1291 + 0.1596i \\\hline
        \textit{TOC1} mRNA      & 0.2296 - 0.0305i  & 0.2702 - 0.1906i  \\\hline
        TOC1 protein cytoplasm  & 0.0067 - 0.0405i  & 0.0825 - 0.2191i  \\\hline
        TOC1 protein nucleus    & 0.0270 - 0.2324i  & -0.0712 - 0.1066i \\\hline
    \end{tabular}
    \end{align*}
    \caption{Eigenvector entries for the mRNA and proteins of $LHY/CCA1$ and $TOC1$ genes under perpetual darkness.  For P2010 model, the second and fifth entries are protein (LHY and TOC1 in Pokhilko et al. 2010), and the third and sixth entries are modified protein (LHY$_{\text{mod}}$ and TOC1$_{\text{mod}}$ in Pokhilko et al. 2010).}
    \label{tab:Eigenvectors1_DD}
\end{table}

\begin{table}[H]
    \centering
    \begin{align*}
    &\addvbuffer[0cm 0.2cm]{\begin{tabular}{|c||c|c|}
        \hline
        Chemical Species        & P2012             & P2013             \\ \hline\hline
        \textit{LHY} mRNA       & -0.2115 - 0.1598i & -0.0851 - 0.1892i \\\hline
        LHY protein cytoplasm   & -0.1884 - 0.0041i & -0.1261 - 0.0620i \\\hline
        LHY protein nucleus     & -0.0466 + 0.0496i & -0.0420 + 0.0201i \\\hline
        \textit{TOC1} mRNA      & 0.1737 + 0.0253i  & 0.0827 + 0.0256i  \\\hline
        TOC1 protein cytoplasm  & 0.0727 - 0.0015i  & 0.0689 - 0.0083i  \\\hline
        TOC1 protein nucleus    &                   &                   \\\hline
    \end{tabular}}\\
    &\begin{tabular}{|c||c|c|}
        \hline
        Chemical Species        & F20014            & O2015             \\ \hline\hline
        \textit{LHY} mRNA       & -0.0966 - 0.0447i & -0.0472 - 0.3212i \\\hline
        LHY protein cytoplasm   & -0.0678 - 0.0001i & -0.0379 - 0.1987i \\\hline
        LHY protein nucleus     &                   & -0.0500 - 0.1490i \\\hline
        \textit{TOC1} mRNA      & 0.1067 + 0.0126i  & 0.1031 + 0.1951i  \\\hline
        TOC1 protein cytoplasm  & 0.2552 - 0.1780i  & 1.0000 + 0.0000i  \\\hline
        TOC1 protein nucleus    & 0.0338 - 0.0228i  & 0.0035 - 0.1787i  \\\hline
    \end{tabular}
    \end{align*}
    \caption{Eigenvector entries for the mRNA and proteins of $LHY/CCA1$ and $TOC1$ genes under perpetual darkness continued.  Empty entries are due to different definitions of variables.}
    \label{tab:Eigenvectors2_DD}
\end{table}

\begin{table}[H]
    \centering
    \begin{tabular}{|c||c|c|}
        \hline
        Chemical Species    & F2016             & DC2016 \\ \hline\hline
        \textit{LHY} mRNA   & -0.1706 + 0.4737i & 1.0000 + 0.0000i  \\\hline
        LHY protein         & 0.0871 + 0.5645i  & 0.7818 - 0.5124i  \\\hline
        \textit{TOC1} mRNA  & 0.0835 - 0.6265i  & -0.2667 + 0.7053i \\\hline
        TOC1 protein        & -0.3524 - 0.4064i & 0.2779 + 0.6823i  \\\hline
    \end{tabular}
    \caption{Eigenvector entries for the mRNA and proteins of $LHY/CCA1$ and $TOC1$ genes under perpetual darkness continued.  F2016 and DC2016 models use only one variable for proteins of $LHY$ and $TOC1$.  The $TOC1$ entries represent $PRR5/TOC1$ gene group.}
    \label{tab:Eigenvectors3_DD}
\end{table}
\newpage

\subsection*{Model Idiosyncrasies}
\begin{table}[H]
    \begin{tabular}{|p{2cm}|p{1.5cm}|p{13cm}|}\hline
        Model   & \shortstack{ODE \\ Solver} & Comments  \\\hline
        L2005a \citep{locke2005modelling}  & ode15s     & The set of parameter values for optimal solution from Fig. 5 in Locke et al. 2005a is used for obtaining data in Fig. 3 in the main text and supplementary Figs. 
        \ref{Results:DD_plots}, \ref{LL-cycle:Locke2005a} and \ref{DD-cycle:Locke2005a}.  The set of parameter values for a typical annealed solution from Fig. 4 in Locke et al. 2005a is used for obtaining data in supplementary Figs. \ref{7-panel:Locke2005a} and \ref{7-panel:Locke2005a-DD}.  Parameters $q_1$ and $q_2$ are used with unit h$^{-1}$, and $p_3$ with unit nM/h. \\\hline
        L2005b \citep{locke2005extension}  & ode15s     & Model Two (the interlocked feedback look model) is used in our analysis, since it adds an extra loop to Model One as an improvement. \\\hline
        Z2006 \citep{zeilinger2006novel}   & ode15s     & $PRR7-PRR9light-Y'$ model is used in our analysis. 
        In $PRR7-PRR9light-Y'$ model, in equation of $\frac{dc_Y^{(m)}}{dt}$, term ${c_L^{(n)}}^{fi}$ is interpreted as ${c_L^{(n)}}^{i}$ as in $PRR7-PRR9-Y$ model. \\\hline
        P2010 \citep{pokhilko2010data}   & ode15s     & Model uses dimensionless chemical levels.  \textit{L} is set to 1 for perpetual illumination, and 0 for perpetual darkness in our analysis. \\\hline
        P2012 \citep{pokhilko2012clock}   & ode15s     & Model uses dimensionless chemical levels.  \textit{L} is set to 1 for perpetual illumination, and 0 for perpetual darkness in our analysis.  We added $c_{Ltot} = c_L + c_{L\text{ mod}}$, and interpret $c_{G}$ in Eqs.(25)(26) as $c_{Gc}$. \\\hline
        P2013 \citep{pokhilko2013modelling}  & ode15s     & Use dimensionless chemical levels.  \textit{L} is set to 1 for perpetual illumination, and 0 for perpetual darkness in our analysis.  Equations for HY5 and HFR1 proteins are not included since they are only used for optimization of $COP1$ parameters and are decoupled from other equations.  We added $c_{Ltot} = c_L + c_{L\text{ mod}}$.  We redefined $c_{Gn}$ to be $c_{Gn} = p_{28}c_{Gc}/(p_{29}+m_{19}+p_{17}c_{E3n})$; and $c_{AR}$ to be $c_{AR} = 0.5 \cdot (A_0 + c^m_{ABAR}+g_{29} - \sqrt{(A_0 + c^m_{ABAR}+g_{29}) + 4A_0c^m_{ABAR}})$.\\\hline
        F2014 \citep{fogelmark2014rethinking}   & ode15s     & Use dimensionless chemical levels.  The model is situated very close to another bifurcation at a lower degradation rate value under perpetual illumination. 
        We interpret $c_Tn$ in Eq. (18) as $c_{Tn}$.\\\hline
        O2015 \citep{ohara2015extended}   & ode15s     & All the $\Theta$ terms are set to 1 for perpetual illumination, and 0 for perpetual darkness in our analysis. \\\hline
        F2016 \citep{foo2016kernel}   & ode23      & Use dimensionless chemical levels.  The kernel model is used in our analysis.  The indexing and notations of parameters used in code given and the ones in main body of the original paper differs, and we followed the convention in the code provided ($\phi_7 \rightarrow \phi_{75}$, $\phi_8 \rightarrow \phi_{76}$, $\phi_9 \rightarrow \phi_{77}$, and $\phi_n \rightarrow \phi_{n-3}$ for $n\geq 10$.  $\theta_{144} \rightarrow \phi_{71}$, $\theta_{145} \rightarrow \phi_{72}$). \\\hline
        DC2016 \citep{de2016compact}  & ode15s     & We did not include the $PIF$ gene, which controls the hypocotyl growth, since this gene is decoupled from the rest of the network and is not of special interests for our purposes.  The model is situated very close to another bifurcation at a lower degradation rate value under perpetual darkness.  We interpret $[P]_p$ in Eq. (3) as $[P]$; and $P$ in Eq. (9) as the concentration $[P]$. \\\hline
    \end{tabular}
    \caption{Details and modifications to each model in our simulations and analysis.}
    \label{tab:my_label}
\end{table}

\subsection*{S26 Response Curves}
In this subsection we show amplitude response curves for systems that are either pre-bifurcation or post-bifurcation. 
To enable comparisons to experiments, results for both sinusoidal and square-wave forcing functions are shown.

To put the Stuart-Landau equation in a form useful to compute periodic forcing \citep{le2001hysteresis} and to generalize Eqs. (7) and (8) of the main text to the pre- and post-bifurcation regions, we make the a variable changes
\begin{equation}
    \label{e.scale1}
    \begin{split}
        \mu &\rightarrow \chi\mu \\
        \lambda_1 &\rightarrow \chi\lambda_1\\
        W &\rightarrow \left(\chi\mu\right)^{-\frac{1}{2}}W'e^{-i\omega_0t}
    \end{split}
\end{equation}
where $\chi \equiv \text{sign}(\mu)$. Upon substitution of the scaled variables in Eqs. \eqref{e.scale1} and the addition of a forcing function $F(\omega t)$ with frequency $\omega$, we find 
\begin{equation}
    \label{e.SLfull}
    \frac{dW'}{dt} = \left[\mu\sigma_1 + i\left(\omega_0 + \mu\omega_1\right)\right]W' - \left(g' + ig''\right)\left|W'\right|^2W' + F(\omega t)
\end{equation}
In the absence of forcing, the asymptotic free-running frequency is given by
\begin{equation}
    \omega_s = 
    \begin{cases}
      \omega_0 + \mu\left(\omega_1 - \frac{g''}{g'}\sigma_1\right) & \text{for $\mu>0$}, \\
      \omega_0+\mu\omega_1 & \text{for $\mu<0$}
    \end{cases}
\end{equation}
Making another set of variable changes
\begin{equation}
    \label{e.scale2}
    \begin{split}
        t &\rightarrow \sigma_1^{-1}t' \\
        \omega & \rightarrow 
        \sigma_1\omega' \\
        \omega_s & \rightarrow \sigma_1\omega'_s\\
        W' & \rightarrow \left({\sigma_1}/{g'}\right)^{\frac{1}{2}}W''e^{i\omega_st} \\
        F & \rightarrow \left({\sigma_1^3}/{g'}\right)^{\frac{1}{2}}F'  
    \end{split}
\end{equation}
non-dimensionalizes Eq. \eqref{e.SLfull}:
\begin{equation}
    \label{e.SLsplit}
    \frac{dW''}{dt} = 
    \begin{cases}
      \left(1 + i \frac{g''}{g'}\right)\left(\mu W'' - \left|W''\right|^2W''\right) + F'(\omega' t')e^{-i\omega'_st'} & \text{for $\mu>0$} \\
      \mu W'' - \left(1 + i\frac{g'}{g''}\right)\left|W''\right|^2W'' + F'(\omega' t')e^{-i\omega_s't'}& \text{for $\mu<0$}
    \end{cases}
\end{equation}

To predict the response of the \textit{Arabidopsis} circadian rhythm, we numerically compute the solutions to Eq. \eqref{e.SLsplit}. 
Then, as a measure of the size of the response, we calculate the quantity $R\equiv\lim_{t\rightarrow\infty}\sqrt{\left<|W|^2\right>}$, where the angle brackets denote time average. 
Results are shown for sinusoidal forcing: $F'(\omega't')=|F'|e^{i\omega't'}$ and for square wave forcing: $F'(\omega't') = 2|F'|\left[\lfloor \omega't'/\pi\rfloor\pmod2\right]$, for a range of $|F'|$ from $10^{-3}$ to $10^{3}$. 
We set the constants in Eq. \eqref{e.SLsplit} to correspond to typical values from the models: $\mu=0.5$, $g'/g''=1$, $\omega_s'=30$. We also set $\omega'=30$ to force the system at the asymptotic free-running frequency.
The resulting response curves are shown in Fig. \ref{fig:response_curve}. 
Fig. \ref{fig:response_curve-step} shows the results for square-wave forcing, and Fig. \ref{fig:response_curve-exp} for sinusoidal forcing. 
The results indicate for weak forcing, i.e., where the dimensionless amplitude of the forcing function is small relative to the dimensionless limit-cycle amplitude, a pre-bifurcation system exhibits a linear response while a post-bifurcation exhibits no response. 
The maximum forcing amplitude that constitutes the upper bound of the ``weak'' regime is significantly larger for a square-wave forcing function. 
An experiment may be able to measure the response of the system to weak values of forcing to determine whether the system is pre-bifurcation or post-bifurcation. 

\setcounter{figure}{25}
\begin{figure}[h!]
    \centering
        \begin{subfigure}{0.48\textwidth}
        \centering
        \includegraphics[width=\textwidth]{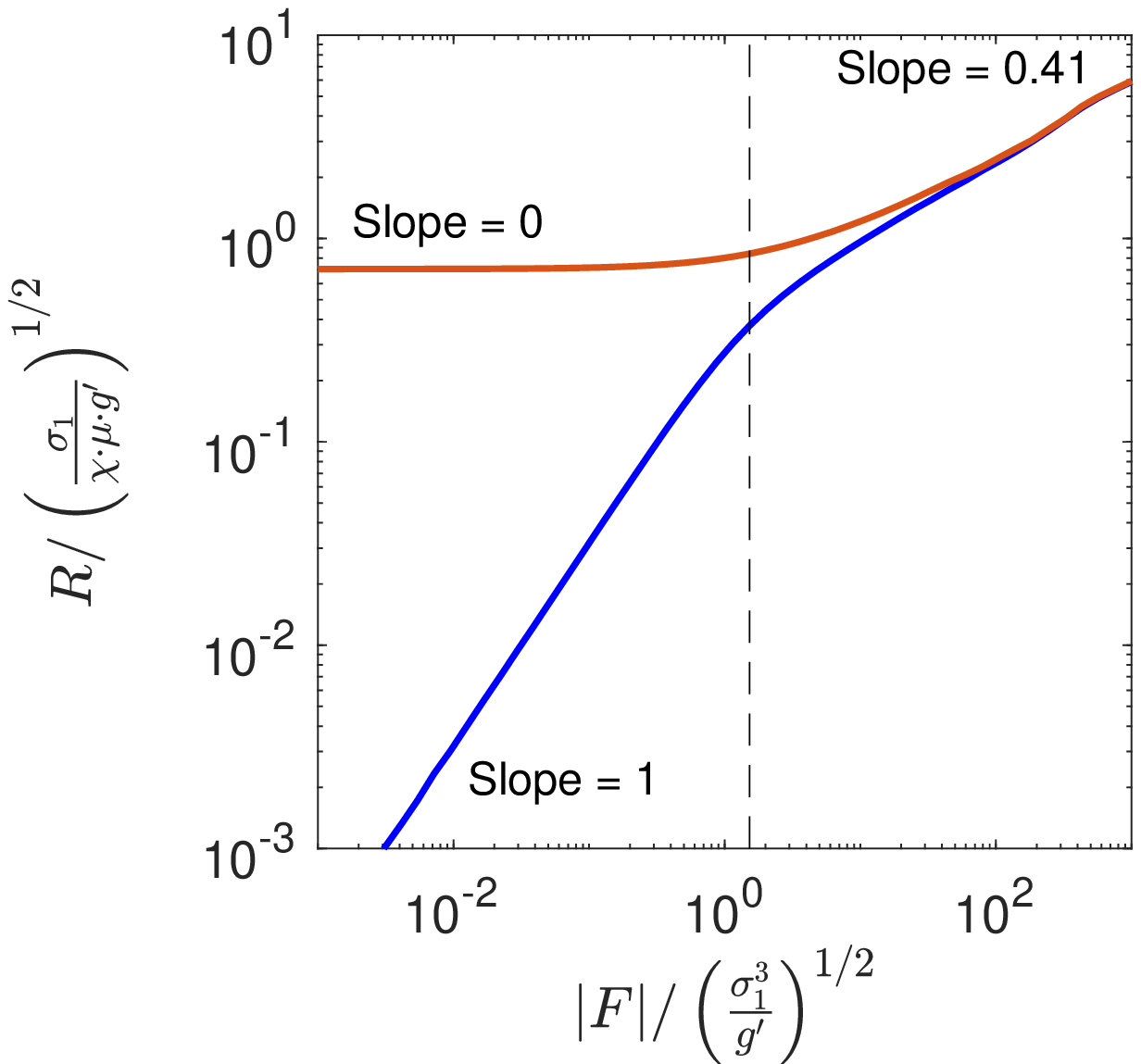}
        \caption{}
        \label{fig:response_curve-step}
    \end{subfigure}%
    \begin{subfigure}{0.48\textwidth}
        \centering
        \includegraphics[width=\textwidth]{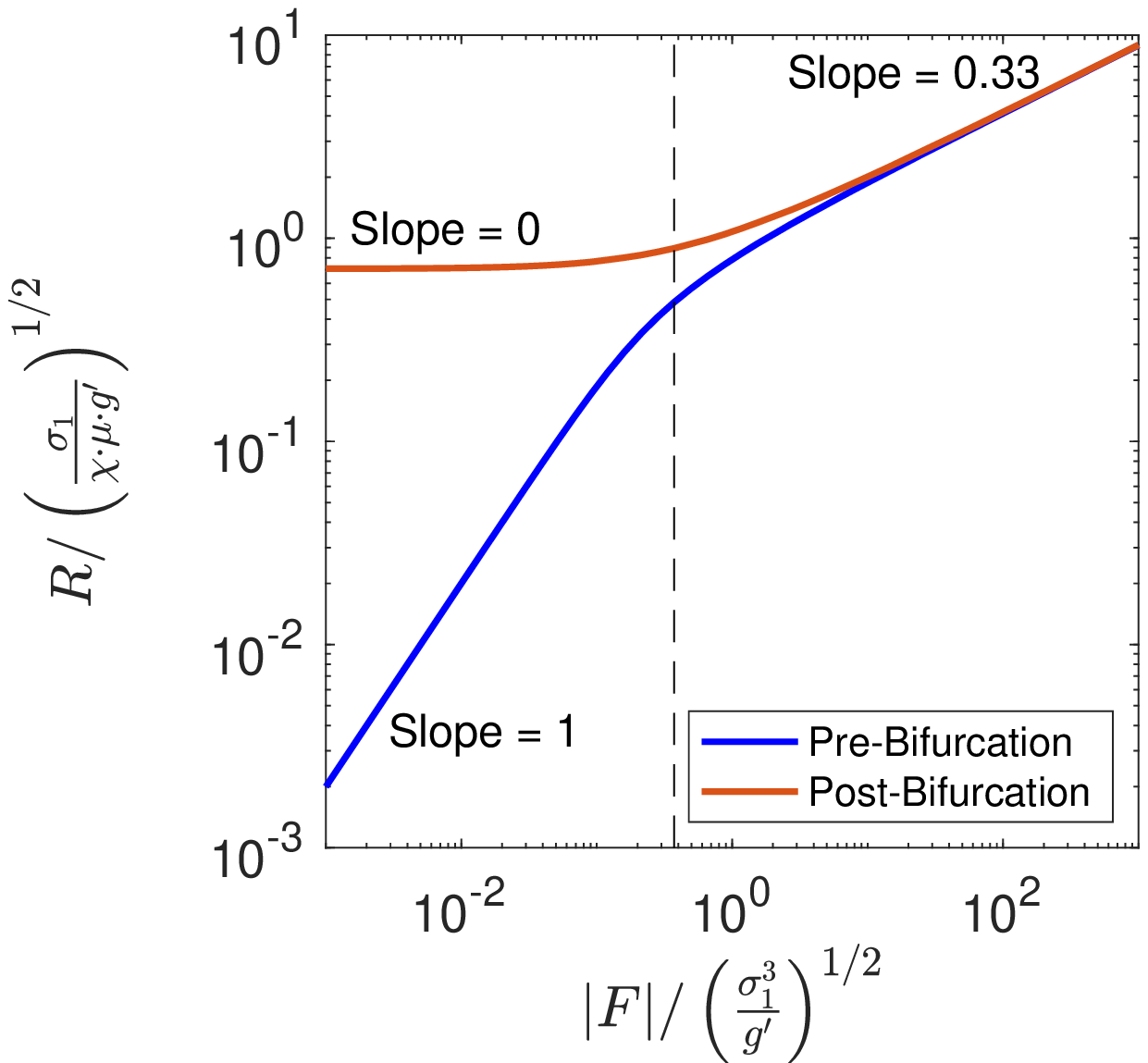}
        \caption{}
        \label{fig:response_curve-exp}
    \end{subfigure}
    \caption{Amplitude response curves for (a) square-wave forcing and (b) sinusoidal forcing. The vertical black dotted line indicates the visually estimated crossover value where the ``weak-forcing'' regime where $R \propto F^1$ (pre-bifurcation) or $R \propto F^0$ (post-bifurcation) changes to a ``strong-forcing'' regime were $R \propto 
    F^n$ where $n = 0.41$ (square-wave) or $n = 0.33$ (sinusoidal).}
    \label{fig:response_curve}
\end{figure}

\input{supplement.bbl}

%% file: ms.bbl
%

%% file: supplement.bbl
%